\documentclass[10pt,epsfig]{article}

\usepackage{enumerate}
\usepackage{float}
\usepackage{subfig}
\usepackage{color}
\usepackage[table]{xcolor}
\usepackage{slashed}
\usepackage{multirow}
\usepackage{cite}

\let\counterwithin\relax
\usepackage{chngcntr}
\usepackage{amssymb, amsmath,mathrsfs}
\allowdisplaybreaks
\usepackage{breqn}

\usepackage{graphics}
\usepackage{graphicx}
\usepackage{epsf}
\usepackage{epsfig}
\usepackage{float}
\usepackage{makecell}
\usepackage{multirow}
\usepackage{color}
\usepackage{xcolor}
\usepackage{simplewick}

\usepackage[utf8]{inputenc}
\usepackage{amsmath}
\usepackage{amssymb}
\usepackage{subfig}
\usepackage[normalem]{ulem}

\usepackage{mathtools}
\usepackage{amsmath}
\usepackage{diagbox}

\newcommand\undermat[2]{
	\makebox[0.5pt][l]{$\smash{\underbrace{\phantom{%
					\begin{matrix}#2\end{matrix}}}_{ \let\scriptstyle\textstyle\text{\large $#1$}}}$}#2}
\newcommand\overmat[2]{
	\makebox[-1pt][l]{$\smash{\overbrace{\phantom{%
					\begin{matrix}#2\end{matrix}}}^{ \let\scriptstyle\textstyle\text{\large $#1$}}}$}#2}    
\usepackage{tikz}
\usepackage[vcentermath]{youngtab}
\usepackage{slashed} 
\usepackage[font=small]{caption} 
\usepackage{times} 

\usepackage{bm}       
\usepackage{bbm} 

\usepackage{relsize}  

\usepackage{autobreak}  

\usepackage[colorlinks=true,
linkcolor=blue,
urlcolor=red,
citecolor=red]{hyperref}

\usepackage{makeidx} 
\usepackage{bbding} 
\usepackage{listings} 
\usepackage{ytableau}
\usepackage[utf8]{inputenc}
\usepackage[T1]{fontenc}
\usepackage{lmodern}

\ytableausetup
{boxsize=1.0em}

\usepackage{mmacells}

\mmaSet{
  morefv={gobble=2},
  linklocaluri=mma/symbol/definition:#1,
  morecellgraphics={yoffset=1.9ex}
}

\long\def\rpl#1!!#2!!{\textcolor{red}{#1} \textcolor{blue}{#2}}

\usepackage[top=1in, left=0.95in, bottom=1.1in, right=0.95in]{geometry}
\def\baselinestretch{1.27}
\usepackage[toc]{appendix}

\newcommand{\beq}{\begin {equation}}  
\newcommand{\eeq}{\end   {equation}} 
\newcommand{\bea}{\begin {eqnarray}} 
\newcommand{\eea}{\end   {eqnarray}}  
\newcommand{\baa}{\begin {array}   } 
\newcommand{\eaa}{\end   {array}   }     
\newcommand{\bit}{\begin {itemize} }
\newcommand{\eit}{\end   {itemize} }
\newcommand{\be }{\begin {equation}} 
\newcommand{\ee }{\end   {equation}}
\newcommand{\nn }{\nonumber        }

\newcommand{\mc}[1]{\mathcal{#1}}

\newcommand{\vev}[1]{ \left\langle {#1}  \right\rangle }

\newcommand{\eq}[1]{\begin{equation}\begin{split} #1 \end{split}\end{equation}}
\newcommand{\eqs}[1]{\begin{align} #1 \end{align}}

\newcommand{\comment}[1]{}

\newcommand{\mlx}[2]{\sout{#1}~{\textcolor{cyan}{#2}}}

\newcolumntype{M}[1]{>{\centering\arraybackslash}m{#1}}
\newcolumntype{N}{@{}m{0pt}@{}}

\allowdisplaybreaks

\begin{document}

\begin{center}

{\Large \textbf{
On-Shell Renormalization of Dim-8 SMEFT from Complete Amplitude Basis: I. Four-Fermion Operators}}\\[10mm]

Chao Wu$^{b}$\footnote{wuch7@itp.ac.cn}, Ming-Lei Xiao$^{a}$\footnote{xiaomlei@mail.sysu.edu.cn}, Jiang-Hao Yu$^{b, c, d, e}$\footnote{jhyu@itp.ac.cn}, Yu-Hui Zheng$^{f}$\footnote{zhengyuhui@kias.re.kr}\\[10mm]

\noindent 
$^a${\em \small School of Science, Sun Yat-sen University, Shenzhen 518100, China} \\
$^b${\em \small Institute of Theoretical Physics, Chinese Academy of Sciences, Beijing 100190, P. R. China}  \\
$^c${\em \small School of Physical Sciences, University of Chinese Academy of Sciences, Beijing 100049, P.R. China}   \\
$^d${\em \small School of Fundamental Physics and Mathematical Sciences, Hangzhou Institute for Advanced Study, UCAS, Hangzhou 310024, China} \\
$^e${\em \small International Centre for Theoretical Physics Asia-Pacific, Beijing/Hangzhou, China}\\
$^f${\em \small School of Physics, Korea Institute for Advanced Study (KIAS), 85 Hoegi-ro, Seoul 02455, Korea}\\[10mm]

\date{\today}   
          
\end{center}

\begin{abstract}

We compute the complete one-loop renormalization group equations (RGEs) for all the four-fermion operators at dimension-8 Standard Model Effective Field Theory (SMEFT).  
We adopt the on-shell framework, where the RGEs are obtained from the unitarity cuts of the bubble integrals. To construct a consistent set of RGEs without redundancy, we utilize the Young Tensor amplitude/operator basis~\cite{Li:2020gnx} as the building blocks of the tree-level amplitudes that constitute the unitarity cuts, which incorporates the full flavor structures of the effective operators.
Due to the large number of effective operators for a dimension-8 type, it is crucial to reduce the integrated cuts to the same RGE amplitude basis, which is also made possible by the algorithm in the Young Tensor method.
With the Mathematica package \texttt{ABC4EFT} that implements the method, we obtain the full result of the dimension-8 four-fermion RGEs with the output file attached as the supplementary material.

\end{abstract}

\newpage

\setcounter{tocdepth}{4}
\setcounter{secnumdepth}{4}

\tableofcontents

\setcounter{footnote}{0}

\def\baselinestretch{1.5}
\counterwithin{equation}{section}

\newpage

\section{Introduction}

The Standard Model (SM) of particle physics stands as one of the most successful scientific frameworks ever developed, having withstood experimental scrutiny up to energy scales of several TeV. 
Yet, despite its empirical success, compelling theoretical and observational motivations point unequivocally toward new physics beyond the Standard Model (BSM). 
Given the absence of direct evidence for new particles at the Large Hadron Collider (LHC), a significant mass gap likely separates the electroweak scale ($\sim$ 246 GeV) from the scale of new physics ($\Lambda \gg $ TeV). This hierarchy makes the framework of Effective Field Theory (EFT) an indispensable tool for the systematic and model-independent parametrization of BSM effects defined at ultraviolate (UV) scales. 
The Standard Model Effective Field Theory (SMEFT) implements this philosophy by extending the SM Lagrangian with a tower of higher-dimensional effective operators, organized as a power series $\Lambda$:
\begin{eqnarray}
{\mathcal L}_{\rm SMEFT} = {\mathcal L}_{\rm SM} + \sum_i \frac{C_i^{(5)}}{\Lambda} O_i^{(5)} + \sum_i \frac{C_i^{(6)}}{\Lambda^2} O_i^{(6)} + \cdots
\end{eqnarray}
Each operator $O_i^{(d)}$ of mass dimension $d > 4$ is accompanied by a Wilson coefficient $C_i^{(d)}$, which encapsulates the unknown couplings from the UV theory.

There are two major calculations in the framework of EFT: matching and running. The former aims to establish the mapping between the Wilson coefficients in the EFT to the parameters in its UV origin; the latter involves the renormalization group (RG) evolution of the effective operators. The RG not only renders the logarithmic running of the Wilson coefficients from the scale $\Lambda$ towards the experimental energies, but also mix various operators, obscuring the connection of the experiment data to the UV physics. The renormalization group equations (RGE) are typically characterized by the anomalous dimension matrix (ADM) $\gamma$ defined as
\begin{eqnarray}
    \mu\frac{d}{d\mu} C_i(\mu) \sim \gamma_{ij} C_j(\mu) \ .
\end{eqnarray}
There are also cases at higher dimensions that the running depends non-linearly on the Wilson coefficients. For example, the renormalization of dimension-8 operators may have quadratic contributions from lower dimensional operators
\eq{
    \mu\frac{d}{d\mu} C^{(d8)}_i(\mu) \sim \gamma_{ijk} C^{(d6)}_j(\mu)C^{(d6)}_k(\mu) \ .
}
Consequently, precise phenomenological predictions—whether for Higgs couplings, electroweak precision observables, or flavor-changing processes—require knowledge of the full one-loop (and often two-loop) anomalous dimension matrix. Translating the vast catalogues of SMEFT operators into their corresponding renormalization-group equations (RGEs) has thus become a frontier area of intense research.

A systematic study of the RGEs naturally requires an complete and independent set of effective operators, also known as an operator basis. The construction of such a basis is highly non-trivial, due to the complex interplay of various redundancy relations among the operators, including but not limited to the Equation of Motion (EOM), Integration By Part (IBP) and the Fierz Identity. Decades after Weinberg first proposed the dimension-5 effective operator for neutrino masses in the SMEFT \cite{Weinberg:1979sa}, the dimension-6 operator basis, now known as the Warsaw Basis \cite{Grzadkowski:2010es}, was constructed with great endeavor. Beyond dimension-6, the task becomes enormous and forces the development of new methods. Among all the approaches, the on-shell method \cite{Shadmi:2018xan,Ma:2019gtx,Durieux:2019eor} reinterpret the problem as the construction of an independent amplitude basis via the spinor-helicity formalism. In this work, we study the RGEs based on the dimension-8 operator basis constructed in \cite{Li:2020gnx}, which uses the Young Tensor method initiated by Henning and Melia \cite{Henning:2019enq,Henning:2019mcv} and completed in the series of papers \cite{Li:2020gnx,Li:2020xlh,Li:2020tsi,Li:2020lba,Li:2021tsq,Li:2022tec}. There are two crucial features of the Young Tensor method: one is the systematic treatment of the flavor structures, where the independent operators are written as irreducible flavor tensors under the symmetric group of the repeated fields; the other is a reduction algorithm of on-shell amplitudes, which allows for a decomposition of any local amplitude into a linear combination of the amplitude basis.

Following the pioneering calculations of the full dimension-6 one-loop RGEs using traditional Feynman diagram techniques \cite{Jenkins:2013zja,Jenkins:2013wua,Alonso:2013hga}, the focus has recently shifted to the next order: the dimension-8 one-loop RGEs. Chala et al. \cite{Chala:2021pll,DasBakshi:2022mwk} computed the one-loop running for bosonic dimension-8 operators generated from two insertions of dimension-6 sources. Bakshi et al. \cite{Bakshi:2024wzz} extended this to include operators containing two fermions. Boughezal et al. \cite{Boughezal:2024zqa} provided the one-loop running of dimension-8 operators within the four-fermion sector under the assumption of Minimal Flavor Violation. These calculations represent major advances, but also highlight the growing computational burden of the traditional method.

In recent years, the on-shell philosophy has been extended to the calculation of RGEs, offering a compelling and elegant alternative \cite{Caron-Huot:2016cwu,Baratella:2020lzz,EliasMiro:2020tdv,Bern:2020ikv,Bern:2019wie,Jiang:2020mhe,Bresciani:2023jsu,Aebischer:2025zxg,Aebischer:2025ddl}. The calculation entirely bypasses the Feynman loop integral, whose UV-divergence, which is responsible for the RGEs, solely comes from the scalar bubble integrals $I_2(p^2)$ and can be extracted as the two-particle unitarity cuts 
\begin{align}
    \text{cut}\,\mc{A}_{\rm 1-loop} \equiv \int {\rm d}\Phi_2\,\mathcal{A}_L\times \mathcal{A}_R\ .
\end{align}
Thus, the RGE calculation involves two essential ingredients:
\begin{itemize}

\item The tree-level amplitudes $\mathcal{A}_L$ and $\mathcal{A}_R$ can be directly obtained via the operator-amplitude correspondence~\cite{Li:2020gnx,Li:2022tec}. This allows for a systematic algorithm to automatically calculate the RGEs based on the amplitude basis. In practice, once all the tree-level amplitudes are listed, it is quite convenient to automatically loop over all the amplitudes in the algorithm.  

\item The phase space integral of the product of two tree-level amplitudes is performed over on-shell variables, and exhibits selection rules for angular momentum conservation \cite{Jiang:2020rwz,Baratella:2020dvw,Shu:2021qlr}, dramatically reducing the complexity of the problem. It is also essential to project the cut result onto the operator basis in the case of multiple independent operators of the same type, especially for dimension-8 operators. This could be made simply by the reduction algorithm built in the Young Tensor method, but is extremely cumbersome otherwise. 

\end{itemize} 
Therefore, we employ these modern on-shell techniques to close a significant remaining gap in the SMEFT RGE program based on the Young Tensor operator basis. 
We present the calculation of the complete one-loop RGEs for all the four-fermion operators up to dimension 8 in a fully flavor-general SMEFT, including contributions from both fermion loops and bosonic loops. 
As a powerful consistency check, our results automatically reproduce all previously known anomalous dimensions in the dimension-6 sector. To maximize the utility of our results for the high-energy physics community, we provide the complete set of effective operators and encode the full RGEs in a comprehensive Mathematica file. This allows theorists and phenomenologists to readily incorporate the RG evolution of these crucial dimension-8 effects into precision studies, thereby enhancing the sensitivity of the current and future experiments to the subtle fingerprints of BSM physics.

The paper is organized as follows. In section 2, the spinor helicity formalism and the amplitude basis for the EFT operators are introduced. Section 3 provides the calculation on the one-loop renormalization using the helicity spinors and gives the collinear anomalous dimensions. In section 4, the RGEs for the 4-fermion effective operators are listed and a mathematica code is provided. Finally, we summarize in section 5.

\section{Spinor Helicity Formalism and On-shell Amplitude Basis}
To compute the one-loop renormalization of higher-dimensional SMEFT operators entirely within the on-shell framework, it is essential to express both Standard Model amplitudes and EFT operator insertions in a language where gauge invariance and Lorentz structure are manifest.
The spinor-helicity formalism provides precisely such a representation.

In four dimensions, any massless momentum $p^{\mu}$ can be expressed as an outer product of two 2-component Weyl spinors $\lambda_{\alpha}$ and $\tilde{\lambda}_{\dot\alpha}$, where $\alpha=1,2$ and $\dot\alpha=1,2$. For a massless particle $i$, we have $p^{\mu}_i\sigma_{\mu\alpha\dot\alpha}=\lambda_{i\alpha}\tilde{\lambda}_{i\dot\alpha}=|i\rangle_{\alpha} [i|_{\dot\alpha}$. This form is obtained via the Pauli matrix $\sigma_{\mu\alpha\dot\alpha}$, which is a submatrix of the 4D Gamma matrix
\begin{align}
     \gamma^{\mu}=\left(\begin{array}{cc}0 & \sigma^{\mu}_{\alpha\dot\alpha}\\\bar{\sigma}^{\mu\dot\alpha\alpha}&0\end{array}\right),
\end{align}
Thus, the 4D momentum vector $p_i$ is reconstructed by mapping spinors $\lambda_{i\alpha},\tilde{\lambda}_{i\dot\alpha}$ to a $2\times 2$ matrix that in turn corresponds to a 4-vector via the Pauli matrices. Lorentz-invariant contractions are then expressed in terms of angle and square brackets,
\begin{align}
    \langle ij\rangle=\langle i|^{\alpha}|j\rangle_{\alpha}=\epsilon^{\alpha\beta}\lambda_{i\beta}\lambda_{j\alpha},\quad [ij]=[i|_{\dot\alpha}|j]^{\dot\alpha}=\epsilon^{\dot\alpha\dot\beta}\tilde{\lambda}_{i\dot\alpha}\tilde{\lambda}_{j\dot\beta}.
\end{align}
Thus, the Levi-Civita tensor $\epsilon^{\alpha\beta}$ ($\epsilon^{\dot\alpha\dot\beta}$) can be used to raise the spinor index $\alpha$ ($\dot{\alpha}$). By contracting these Weyl spinors with the Pauli matrix, we can construct the Lorentz vector
\begin{align}
    \langle i|\sigma^{\mu}|j]=\langle i|^{\alpha}\sigma^{\mu}_{\alpha\dot\alpha}|j]^{\dot\alpha},\quad [i|\sigma^{\mu}|j\rangle=[i|_{\dot\alpha}\bar{\sigma}^{\mu\dot\alpha\alpha}|j\rangle_{\alpha}.
\end{align}

In this notation, all external wavefunctions and polarization tensors can be written compactly in terms of the spinor helicity variables \cite{Elvang:2015rqa},
\begin{align}
    u^-_i=\left(\begin{array}{c}|i\rangle_{\alpha}\\0\end{array}\right),\quad u^+_i=\left(\begin{array}{c}0\\|i]^{\dot{\alpha}}\end{array}\right),\quad \bar{v}_i^-=\left(\langle i|^{\alpha},\;0\right),\quad \bar{v}_i^+=\left(0,\;[i|_{\dot{\alpha}}\right).
\end{align}
\begin{align}
    \epsilon_i^{+,\mu}=\frac{\langle \eta_i|\sigma^{\mu}|i]}{\sqrt{2}\langle i\eta_i\rangle},\quad \epsilon_i^{-,\mu}=\frac{\langle i|\sigma^{\mu}|\eta_i]}{\sqrt{2}[ i\eta_i]},
\end{align}
where $\langle\eta_i|$ and $[\eta_i|$ are arbitrary reference spinors and the superscripts $\pm$ indicate the helicities of the particles. Here, we treat all the momenta as incoming, so that we use only $u$ and $\bar{v}$ associated with incoming fermions and anti-fermions. 

For effective operators, instead of the polarization vector $\epsilon_\mu$, gauge field strengths are most conveniently expressed in terms of chiral spinor tensors,
\begin{align}\label{eq:FLFR}
    F_{{\rm L}\mu\nu}=\frac{F_{\mu\nu}-i\tilde{F}_{\mu\nu}}{2},\quad F_{{\rm R}\mu\nu}=\frac{F_{\mu\nu}+i\tilde{F}_{\mu\nu}}{2},
\end{align}
where $\tilde{F}_{\mu\nu}=\epsilon_{\mu\nu\rho\sigma}F^{\rho\sigma}$. In the spinor helicity formalism, they are defined as
\begin{align}
    F_{{\rm L}\alpha\beta}=\frac{i}{2}F_{\mu\nu}\sigma^{\mu\nu}_{\alpha\beta}=|i\rangle_{\alpha}|i\rangle_{\beta},\quad F_{{\rm R}\dot\alpha\dot\beta}=-\frac{i}{2}F_{\mu\nu}\bar{\sigma}^{\mu\nu}_{\dot\alpha\dot\beta}=[i|_{\dot\alpha}[i|_{\dot\beta}.
\end{align}
where $\sigma^{\mu\nu}_{\alpha\beta}=\epsilon_{\beta\gamma}\sigma^{\mu}_{\alpha\dot\alpha}\bar{\sigma}^{\nu\dot\alpha\gamma}$ and $\bar{\sigma}^{\mu\nu}_{\dot\alpha\dot\beta}=\epsilon_{\dot\alpha\dot\gamma}\bar{\sigma}^{\nu\dot\gamma\alpha}\sigma^{\mu}_{\alpha\dot\beta}$.
Then the Lorentz factor of the amplitude becomes the product of the spinor brackets, while the little group representations of each particle constrain the Lorentz structures of the amplitude as follows:
\begin{align}
    \text{helicity }h\text{ massless particle with spinor variables }(\lambda_i,\tilde{\lambda}_i):\quad \mathcal{A}\sim\left\{\begin{array}{cc}\lambda^{r-2h}_{i\{\alpha\}}\tilde{\lambda}^r_{i\{\dot\alpha\}}, & h\leq 0,\\
    \lambda^{r}_{i\{\alpha\}}\tilde{\lambda}^{r+2h}_{i\{\dot\alpha\}}, & h\geq 0\end{array}, \right.
\end{align}
where $\{.\}$ denotes totally symmetric indices. This corresponds to the fact that all antisymmetric spinor indices are related to the equation of motion(EOM), and such contributions vanish in the on-shell amplitude, $\lambda_{i[\alpha}\lambda_{i\beta]}=\langle ii\rangle\epsilon_{\alpha\beta}=0$, where $[.]$ denotes antisymmetric spinor indices. At the operator level, it follows from the relations
\begin{align}
\begin{split}
    &D_{[\alpha\dot\alpha}D_{\beta]\dot\beta}=D_{\mu}D_{\nu}\sigma^{\mu}_{[\alpha\dot\alpha}\sigma^{\nu}_{\beta]\dot\beta}=-D^2\epsilon_{\alpha\beta}\epsilon_{\dot\alpha\dot\beta}+\frac{i}{2}[D_{\mu},D_{\nu}]\epsilon_{\alpha\beta}\bar{\sigma}^{\mu\nu}_{\dot\alpha\dot\beta},\\
    &D_{[\alpha\dot\alpha}\psi_{\beta]}=D_{\mu}\sigma^{\mu}_{[\alpha\dot\alpha}\psi_{\beta]}=-\epsilon_{\alpha\beta}(D\!\!\!\!/\psi)_{\dot\alpha},\\
    &D_{[\alpha\dot\alpha}F_{L\beta]\gamma}=D_{\mu}F_{\nu\rho}\sigma^{\mu}_{[\alpha\dot\alpha}\sigma^{\nu\rho}_{\beta]\gamma}=2D^{\mu}F_{\mu\nu}\epsilon_{\alpha\beta}\sigma^{\nu}_{\gamma\dot\alpha}.
\end{split}
\end{align}
So operators differing by EOM or total derivatives do not contribute to physical amplitudes.
This observation underlies the operator–amplitude correspondence: every local operator corresponds to a unique on-shell contact amplitude \cite{Shadmi:2018xan,Ma:2019gtx}, and conversely, every allowed on-shell amplitude corresponds to an operator modulo EOM and IBP.
This motivates constructing EFT operator bases directly from on-shell amplitudes, rather than from off-shell Lagrangian monomials.

Furthermore, on-shell contact amplitudes with $N$ external legs transform as totally symmetric Lorentz tensors, and their independent structures correspond to semi-standard Young tableaux (SSYT).
This “y-basis” provides a complete and minimal parametrization of possible local EFT operators, as shown in \cite{Henning:2019enq,Henning:2019mcv,Li:2020gnx,Li:2020xlh,Li:2020tsi,Li:2020zfq}.

The construction proceeds by reducing arbitrary spinor structures to the SSYT basis using:
\begin{enumerate}
    \item \textbf{Momentum conservation}, which removes lower-index momenta and corresponds to IBP at the operator level,
    \item \textbf{Schouten identities}, which reduce multi-bracket products without introducing new momenta.
\end{enumerate}
These relations ensure that any contact amplitude can be systematically rewritten in terms of a finite set of y-basis elements.

The Y-basis plays a central role in our renormalization procedure: every unitarity cut produces spinor structures which are reduced to the Y-basis, and the anomalous dimensions are extracted by matching onto these basis amplitudes.
Section 2.1 presents the reduction rules.

\begin{figure}[htbp]
\centering
\includegraphics[width=0.6\linewidth]{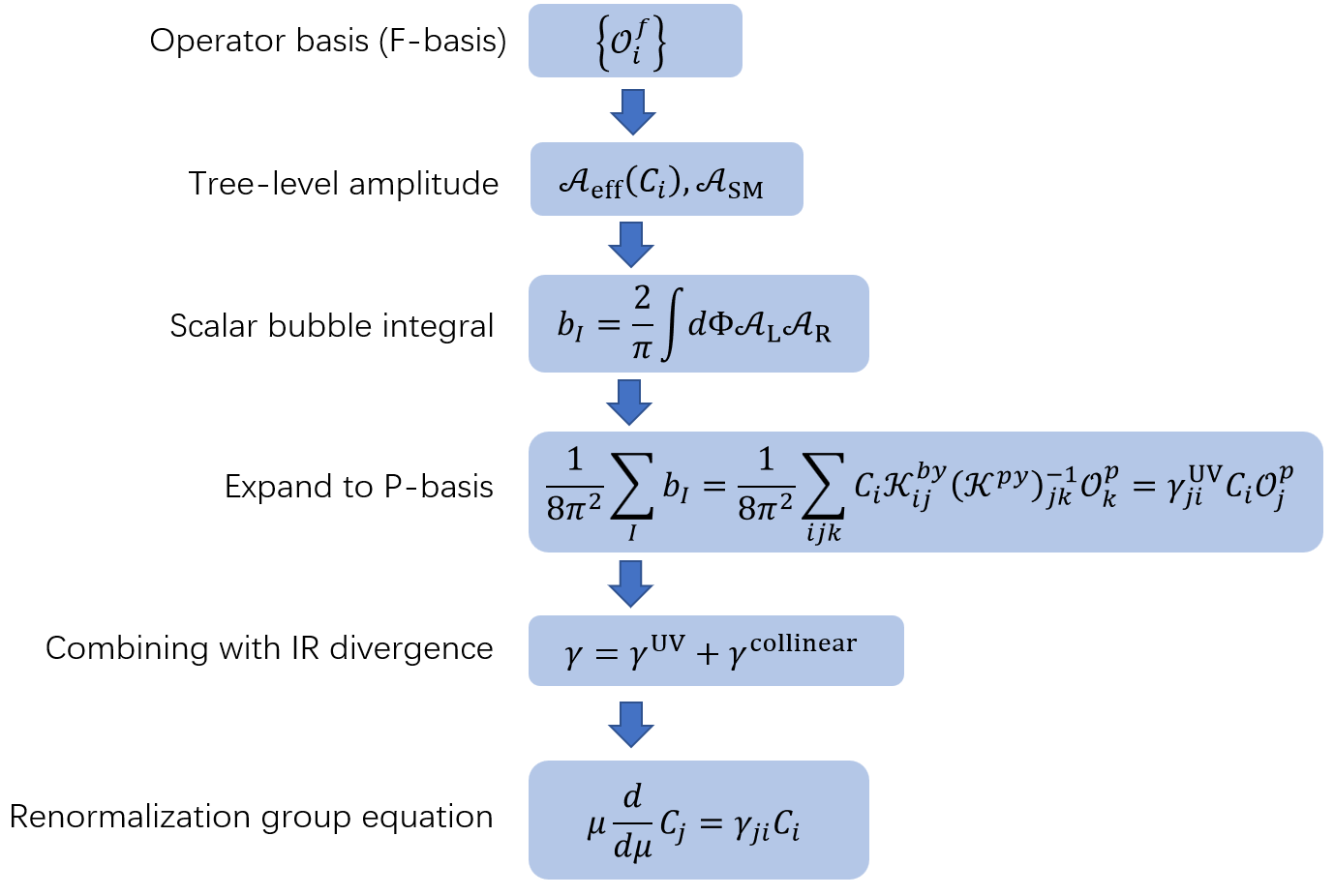}
\caption{Procedure for calculating the renormalization group equation using on-shell amplitudes.}
\label{fig:procedure}
\end{figure}

\subsection{Operator Y-basis and Basis Reduction Rules}

In our previous work \cite{Ma:2019gtx,Li:2020gnx,Li:2020xlh,Li:2022tec}, a one-to-one correspondence between local EFT operators and on-shell contact amplitudes was established, leading to the definition of several closely related amplitude bases: the y-basis (Young Tableau basis), the p-basis (permutation basis), and the f-basis (flavor basis). 
Among these, the y-basis is the most fundamental. It is defined directly from semi-standard Young tableaux (SSYT) and provides a complete and minimal parametrization of all independent Lorentz structures of local amplitudes. All other bases can be expanded in terms of the y-basis with well-defined coefficients.

Any on-shell amplitude that does not initially take the SSYT form can be systematically reduced to the y-basis by using momentum conservation and Schouten identities.
The reduction depends on the ordering of external particles; different orderings lead to different y-basis representations. In practice, we adopt the helicity-non-decreasing ordering proposed in \cite{Henning:2019enq}, although the choice is not unique.

Consider the amplitude with $N$ particles, the reduction can be implemented in two steps: 
    \begin{itemize}
        \item \textbf{Step 1: Momentum conservation reduction.} 
        
        momentum conservation is used to eliminate momenta with lower labels in favor of higher-label momenta
            \begin{align}
                &\qquad \qquad \langle i1\rangle[1j]=-\sum^N_{k=2}\langle ik\rangle[kj].\label{eq:rulep1}\\
                & [1|p_2|i\rangle=-\sum_{k=3}^N[1|p_k|i\rangle,\quad \langle 1|p_2|i]=-\sum_{k=3}^N\langle 1|p_k|i],   \\
                & [1|p_3|2\rangle=-\sum_{k=4}^N[1|p_k|2\rangle,\quad \langle 1|p_3|2]=-\sum_{k=4}^N\langle 1|p_k|2], \\
                &\qquad\qquad p_2\cdot p_3=\sum_{\substack{i,j\neq 1\\ \{i,j\}\neq\{2,3\} }} -p_i\cdot p_j .
            \end{align}
            At the operator level, this corresponds to integration-by-parts (IBP) relations and the use of equations of motion (EOM). While intermediate expressions may involve EOM terms, the final reduction remains valid for physical operators
            \begin{align}
                \left(D^n\Psi_1\right)\cdots \simeq (-)^n\Psi_1 D^n\left(\cdots\right),
            \end{align}
            where $\Psi$ denotes a general field.
    
        \item \textbf{Setp 2: Schouten reduction.}

        When two same-type brackets contain 4 distinct particles with orderd labels $i<j<k<l$, we use the Schouten Identity to apply the following replacements
        \begin{align}
            \langle il\rangle\langle jk\rangle &= \langle ik\rangle\langle jl\rangle - \langle ij\rangle\langle kl\rangle, \label{eq:ASch}\\
            [il][jk] &= [ik][jl] - [ij][kl].\label{eq:SSch}
        \end{align}
        These relations do not generate new momentum labels, so no further momentum-conservation identities are required.
    \end{itemize}
After these two steps, any local on-shell amplitude can be uniquely reduced to a linear combination of those Lorentz y-basis elements.

To construct the independent basis of gauge factors, we define an inner product for group tensors of the same type:
\begin{align}
 (T,T')\equiv \sum_{\{a_i\}}(T^{a_1a_2\cdots})^*T'{}^{a_1a_2\cdots}.
\end{align}
And introduce the metric $\bar{g}_{ij}$ of an overcomplete gauge tensor basis $\{T_i\}$ 
\begin{align}
    \bar{g}_{ij}\equiv(T_i,T_j).
\end{align}
In general case, $\bar{g}_{ij}$ is a singular matrix, $\det \bar{g}=0$. We can find the redundancy relation by solving the null space $\{\bm{n}^j\}$ of $\bar{g}$, which satisfies $\bar{g}_{ij}\bm{n}^j=0$, and obtain a maximal independent subset $\mc{I}$ that $g_{ij}=\bar{g}_{ij}$, $\det g\neq 0$, for $i,j\in\mc{I}$. The independent gauge tensor basis is thus defined in this subset $\{T^y_i\}=\{T_i\}$, for $i\in\mc{I}$.
An arbitrary gauge factor $T$ can then be projected onto this basis using
\eq{\label{eq:ReduceGF}
    (T^y_i,T) = \sum_j \mc{K}_{T,j} (T^y_i,T^y_j) = \sum_j g_{ij} \mc{K}_{T,j}, \quad \Rightarrow \mc{K}_{T,i} = \sum_j g^{ij} (T^y_j,T),
}
where $g^{ij}$ is the inverse metric $g^{ij}g_{jk} = \delta^i_{\ k}$, the invertibility guaranteed by the independence of this basis $\{T^y_i\}$. $\mc{K}_{T}$ is the expansion coefficient under the gauge group basis, and we can also define $\mc{K}_{\mc{Y}}$ as the expansion coefficient under the kinetics y-basis $\mc{B}=\sum_i\mc{K}_{\mc{Y},i}\mc{B}^y_{i}$. 
Combining the gauge and Lorentz structures, any local operator (up to EOM) admits a unique decomposition in the y-basis,
\begin{align}
    {\cal O} = T \times {\cal B}
    =\sum_{i,j}\mc{K}_{T,j} \mc{K}_{\mc{Y},i} (T^y_j\times \mc{B}^y_{i}).
\end{align}
The y-basis thus provides the universal intermediate basis used throughout this work: all unitarity cuts are reduced to the y-basis before imposing permutation symmetries and flavor structures, which will be discussed in the next subsection.

\subsection{Flavor Basis}\label{sec:f-basis}
Then we need to substitute the explicit field representations. We adopt the chiral basis to express the Standard Model fields, as shown in Tab.~\ref{tab:SMEFT-field-content}, where we omit the conjugate fields for simplicity. 
\begin{table}[t]
	\begin{center}
		\begin{tabular}{|c|cc|ccc|ccc|}
			\hline
			\text{Fields} & $SU(2)_{l}\times SU(2)_{r}$	& $h$ & $SU(3)_{C}$ & $SU(2)_{W}$ & $U(1)_{Y}$ &  Flavor & $B$ & $L$ \tabularnewline
			\hline
			$G_{{\rm L}\alpha\beta}^A$   & $\left(1,0\right)$  & $-1$    & $\boldsymbol{8}$ & $\boldsymbol{1}$ & 0  & $1$ & 0 & 0 \tabularnewline
			$W_{{\rm L}\alpha\beta}^I$   & $\left(1,0\right)$  & $-1$           & $\boldsymbol{1}$ & $\boldsymbol{3}$ & 0  & $1$ & 0 & 0 \tabularnewline
			$B_{{\rm L}\alpha\beta}$   & $\left(1,0\right)$    & $-1$        & $\boldsymbol{1}$ & $\boldsymbol{1}$ & 0  & $1$ & 0 & 0 \tabularnewline
			\hline
			$L_{\alpha i}$     & $\left(\frac{1}{2},0\right)$  & $-\frac12$  & $\boldsymbol{1}$ & $\boldsymbol{2}$ & $-\frac12$  & $n_f$ & 0 & $1$ \tabularnewline
			$e_{\mathbb{C}\alpha}$ & $\left(\frac{1}{2},0\right)$ & $-\frac12$   & $\boldsymbol{1}$ & $\boldsymbol{1}$ & $1$  & $n_f$ & 0 & $-1$ \tabularnewline
			$Q_{\alpha ai}$     & $\left(\frac{1}{2},0\right)$ & $-\frac12$   & $\boldsymbol{3}$ & $\boldsymbol{2}$ & $\frac16$  & $n_f$ & $\frac13$ & 0 \tabularnewline
			$u_{\mathbb{C}\alpha}^a$ & $\left(\frac{1}{2},0\right)$ & $-\frac12$   & $\overline{\boldsymbol{3}}$ & $\boldsymbol{1}$ & $-\frac23$  & $n_f$ & $-\frac13$ & 0 \tabularnewline
			$d_{\mathbb{C}\alpha}^a$ & $\left(\frac{1}{2},0\right)$ & $-\frac12$   & $\overline{\boldsymbol{3}}$ & $\boldsymbol{1}$ & $\frac13$  & $n_f$ & $-\frac13$ & $0$ \tabularnewline
			\hline
			$H_i$     & $\left(0,0\right)$&  0     & $\boldsymbol{1}$ & $\boldsymbol{2}$ & $\frac12$  & $1$ & 0 & 0 \tabularnewline
			\hline
		\end{tabular}
		\caption{\label{tab:SMEFT-field-content}
			The field content of the standard model, along with their representations under the Lorentz and gauge symmetries. The representation under Lorentz group is denoted by $(j_l,j_r)$, while the helicity of the field is given by $h = j_r-j_l$ .
			The number of fermion flavors is denoted as $n_f$, which is 3 in the standard model. We also list their global charges, the baryon number $B$ and the lepton number $L$. All of the fields are accompanied with their Hermitian conjugates that are omitted, $(F_{{\rm L} \alpha\beta})^\dagger = F_{{\rm R} \dot\alpha\dot\beta}$ for gauge bosons, $(\psi_\alpha)^\dagger = (\psi^\dagger)_{\dot\alpha}$ for fermions, and $H^\dagger$ for the Higgs, which are under the conjugate representations of all the groups. }
	\end{center}
\end{table}

In our previous study, we constructed the flavor-blind operators where all fields are distinguished by a label $i=1,\dots,N$. Now we impose the permutation symmetry of flavor indices for the repeated fields in the operator to obtain the p-basis\cite{Li:2020gnx,Li:2020xlh,Li:2022tec}. 
\begin{align}
    \underbrace{\pi\circ {\cal O}^{\{f_{k},...\}}}_{\rm permute\ flavor} =& \underbrace{\left(\pi\circ T_{{\rm G_1}}^{\{g_k,...\}}\right)\left(\pi\circ T_{{\rm G2}}^{\{h_k,...\}}\right)\cdots}_{\rm permute\ gauge}\underbrace{\left(\pi\circ{\cal B}^{\{f_k,...\}}_{\{g_{k},...\},\{h_{k},...\}}\right)}_{\rm permute\ Lorentz},
\label{eq:tperm}
\end{align}
Note that if we exchange two repeated fermionic fields, we obtain an additional minus sign. The operators obtained in this way serving as the basis vectors of a irreducible representation of the symmetric group, are called p-basis, with ``p'' for permutation. Then we treat each operator in the p-basis as flavor tensors of the corresponding SU($n_f$) group, i.e., the flavor-specified operators, the spaces spanned by these flavor tensors are identical if they were the bases of the same irreducible representation of the symmetric group. Thus, redundancies appear in the p-basis when irreducible representations of the symmetric group of dimension larger than one exist, which only happens for operators with three or more repeated fields with flavor number larger than one. After removing these redundancies, the remaining operators form our f-basis, with ``f'' for flavor. These redundancies are related to the so-called \textit{flavor relation} in other literature. 

To illustrate this symmetry, we use the type $LQ^3 D^2$ as an example. Labeling the fields as $L_1 Q_2 Q_3 Q_4$, the y-basis of $LQ^3 D^2$ are as follows:
\begin{align}\label{eq:LQQQybasis}
    \begin{array}{c|c}
        \text{y-basis} &  \\
        \hline
        \mc{O}^{(y)}_{LQ^3 D^2,1} & \epsilon^{abc} \epsilon^{ik} \epsilon^{jl} \left(L_{1i} Q_{2aj}\right) \left(D_{\mu}Q_{3bk} D^{\mu}Q_{4cl}\right) \\
        \mc{O}^{(y)}_{LQ^3 D^2,2} & \epsilon^{abc} \epsilon^{ik} \epsilon^{jl} \left(L_{1i} Q_{3bk}\right) \left(D_{\mu}Q_{2aj} D^{\mu}Q_{4cl}\right) \\
        \mc{O}^{(y)}_{LQ^3 D^2,3} & i\epsilon^{abc} \epsilon^{ik} \epsilon^{jl} \left(L_{1i}\sigma_{\mu\nu} Q_{2bj}\right) \left(D^{\mu}Q_{3bk} D^{\nu}Q_{4cl}\right) \\
        \mc{O}^{(y)}_{LQ^3 D^2,4} & \epsilon^{abc} \epsilon^{ij} \epsilon^{kl} \left(L_{1i} Q_{2aj}\right) \left(D_{\mu}Q_{3bk} D^{\mu}Q_{4cl}\right) \\
        \mc{O}^{(y)}_{LQ^3 D^2,5} & \epsilon^{abc} \epsilon^{ij} \epsilon^{kl} \left(L_{1i} Q_{3bk}\right) \left(D_{\mu}Q_{2aj} D^{\mu}Q_{4cl}\right) \\
        \mc{O}^{(y)}_{LQ^3 D^2,6} & i\epsilon^{abc} \epsilon^{ij} \epsilon^{kl} \left(L_{1i}\sigma_{\mu\nu} Q_{2bj}\right) \left(D^{\mu}Q_{3bk} D^{\nu}Q_{4cl}\right)
    \end{array}
\end{align}
The permutations of the three repeated fields $\{Q_1,Q_2,Q_3\}$ form a representation of the symmetric group $S_3$. We can use the $S_3$ group generators $(1\;2)$ and $(1\;2\;3)$ and the matrix representations of the two generators to generate matrix representations of all Young symmetrizers of the $S_3$.

The y-basis of this type is in dimension-6. The matrix representations of group algebra projectors $b^\lambda_i$ ~\cite{Li:2020xlh} served as basis vectors for the irreducible representations $\lambda=[3]$, $[2,1]$, $[1,1,1]$ of the $S_3$ symmetric group,
which can be defined by applying the corresponding permutations to the basis in eq.~\eqref{eq:LQQQybasis} and re-expanding the result in the y-basis using the reduction rules of the previous subsection.
\begin{eqnarray}\label{eq:youngsymofS3}
	&&b_1^{[3]}=\mc{Y}^{[3]}_1=\left(
	\begin{array}{cccccc}
		-\frac{1}{6} & 0 & -\frac{1}{3} & \frac{1}{3} & 0 & \frac{1}{6} \\
		\frac{1}{6} & 0 & \frac{1}{3} & -\frac{1}{3} & 0 & -\frac{1}{6} \\
		\frac{1}{3} & 0 & \frac{2}{3} & -\frac{2}{3} & 0 & -\frac{1}{3} \\
		-\frac{1}{3} & 0 & -\frac{2}{3} & \frac{2}{3} & 0 & \frac{1}{3} \\
		0 & 0 & 0 & 0 & 0 & 0 \\
		\frac{1}{6} & 0 & \frac{1}{3} & -\frac{1}{3} & 0 & -\frac{1}{6} \\
	\end{array}
	\right), \quad
	b_1^{[2,1]}=\mc{Y}^{[2,1]}=\left(
	\begin{array}{cccccc}
		\frac{1}{3} & 0 & \frac{2}{3} & \frac{1}{3} & 0 & -\frac{1}{3} \\
		-\frac{1}{3} & \frac{4}{3} & -\frac{2}{3} & \frac{1}{3} & -\frac{2}{3} & \frac{1}{3} \\
		0 & 0 & 0 & 0 & 0 & 0 \\
		\frac{1}{3} & 0 & \frac{2}{3} & \frac{1}{3} & 0 & -\frac{1}{3} \\
		-\frac{1}{3} & \frac{2}{3} & -\frac{2}{3} & 0 & -\frac{1}{3} & \frac{1}{3} \\
		-\frac{1}{3} & 0 & -\frac{2}{3} & -\frac{1}{3} & 0 & \frac{1}{3} \\
	\end{array}
	\right), \nn \\
	&&b_2^{[2,1]}=(2\ 3)\mc{Y}^{[2,1]}=\left(
	\begin{array}{cccccc}
		\frac{1}{3} & 0 & -\frac{1}{3} & -\frac{2}{3} & 0 & -\frac{1}{3} \\
		\frac{1}{3} & -\frac{2}{3} & \frac{1}{3} & -\frac{2}{3} & \frac{4}{3} & -1 \\
		0 & 0 & 0 & 0 & 0 & 0 \\
		\frac{1}{3} & 0 & -\frac{1}{3} & -\frac{2}{3} & 0 & -\frac{1}{3} \\
		0 & -\frac{1}{3} & \frac{1}{3} & 0 & \frac{2}{3} & -\frac{1}{3} \\
		-\frac{1}{3} & 0 & \frac{1}{3} & \frac{2}{3} & 0 & \frac{1}{3} \\
	\end{array}
	\right), \quad
	b_1^{[1,1,1]}=\mc{Y}^{[1,1,1]}=\left(
	\begin{array}{cccccc}
		\frac{1}{2} & 0 & 0 & 0 & 0 & \frac{1}{2} \\
		\frac{1}{6} & 0 & 0 & 0 & 0 & \frac{1}{6} \\
		0 & 0 & 0 & 0 & 0 & 0 \\
		0 & 0 & 0 & 0 & 0 & 0 \\
		\frac{1}{3} & 0 & 0 & 0 & 0 & \frac{1}{3} \\
		\frac{1}{2} & 0 & 0 & 0 & 0 & \frac{1}{2} \\
	\end{array}
	\right),\nonumber\\
\end{eqnarray}
where one can see that by convention the first basis vector $b_1^{\lambda}$ in a irreducible representation always corresponds to the Young symmetrizer $\mc{Y}^{\lambda}$, and for the multi-dimensional representations such as $[2,1]$, the rest of the basis vectors are related to the Young symmetrizer by certain permutations, such as $(2 \ 3)$ denoting the permutation of the 2nd and 3rd repeated fields. In the following, we express the Young symmetrizer explicitly as $\mc{Y}[\text{tableau}]$ where the Young tableau is filled by the flavor indices, such as
\eq{
    \mc{Y}^{[2,1]} \mc{O}_{rst} \equiv \mc{Y}[{\tiny\young(rs,t)}] \mc{O}_{rst}.
}

From the above matrix representations of projectors $b^\lambda$, we find that there are 6 independent basis vectors in the p-basis,
\begin{align} \label{eq:pbasis_LQ3D2}
    \begin{array}{c|c}
        \text{p-basis} &  \\
        \hline
        \mc{O}^{(p)}_{LQ^3 D^2,1} & \dfrac{1}{6}\mc{Y}\left[\tiny{\young(rst)}\right] \epsilon^{abc} \epsilon^{ik} \epsilon^{jl} \left(L_{pi} Q_{raj}\right) \left(D_{\mu}Q_{sbk} D^{\mu}Q_{tcl}\right) \\
        \mc{O}^{(p)}_{LQ^3 D^2,2} & \dfrac{1}{3}\mc{Y}\left[\tiny{\young(rs,t)}\right] \epsilon^{abc} \epsilon^{ik} \epsilon^{jl} \left(L_{pi} Q_{raj}\right) \left(D_{\mu}Q_{sbk} D^{\mu}Q_{tcl}\right) \\
        \mc{O}^{(p)}_{LQ^3 D^2,3} & \dfrac{1}{3} (s \ t)\mc{Y}\left[\tiny{\young(rs,t)}\right] \epsilon^{abc} \epsilon^{ik} \epsilon^{jl} \left(L_{pi} Q_{raj}\right) \left(D_{\mu}Q_{sbk} D^{\mu}Q_{tcl}\right) \\
        \mc{O}^{(p)}_{LQ^3 D^2,4} & \dfrac{1}{3}\mc{Y}\left[\tiny{\young(rs,t)}\right] \epsilon^{abc} \epsilon^{il} \epsilon^{jk} \left(L_{pi} Q_{raj}\right) \left(D_{\mu}Q_{sbk} D^{\mu}Q_{tcl}\right) \\
        \mc{O}^{(p)}_{LQ^3 D^2,5} & \dfrac{1}{3}(s \ t)\mc{Y}\left[\tiny{\young(rs,t)}\right] \epsilon^{abc} \epsilon^{il} \epsilon^{jk} \left(L_{pi} Q_{raj}\right) \left(D_{\mu}Q_{sbk} D^{\mu}Q_{tcl}\right) \\
        \mc{O}^{(p)}_{LQ^3 D^2,6} & \dfrac{1}{6}\mc{Y}\left[\tiny{\young(r,s,t)}\right] \epsilon^{abc} \epsilon^{ik} \epsilon^{jl} \left(L_{pi} Q_{raj}\right) \left(D_{\mu}Q_{sbk} D^{\mu}Q_{tcl}\right)
    \end{array}
\end{align}
with the expansion coefficient from the p-basis to the y-basis
\beq
\mc{K}^{(py)}=\left(
\begin{array}{cccccc}
	-\frac{1}{6} & 0 & -\frac{1}{3} & \frac{1}{3} & 0 & \frac{1}{6} \\
	\frac{1}{3} & 0 & \frac{2}{3} & \frac{1}{3} & 0 & -\frac{1}{3} \\
	\frac{1}{3} & 0 & -\frac{1}{3} & -\frac{2}{3} & 0 & -\frac{1}{3} \\
	-\frac{1}{3} & \frac{4}{3} & -\frac{2}{3} & \frac{1}{3} & -\frac{2}{3} & \frac{1}{3} \\
	\frac{1}{3} & -\frac{2}{3} & \frac{1}{3} & -\frac{2}{3} & \frac{4}{3} & -1 \\
	\frac{1}{2} & 0 & 0 & 0 & 0 & \frac{1}{2} \\
\end{array}
\right).
\eeq
Among the 6 p-basis elements constructed above, the third and fifth ones are obtained simply by exchanging the $s$ and $t$ indices of the second and fourth p-basis elements, respectively. Accordingly, their associated Wilson-coefficient tensors $C^{prst}_i$ are related by a $s \leftrightarrow t$ permutation, thus carrying the same set of parameters despite that they are linearly independent tensors in the p-basis. Therefore, to find an operator basis with independent flavor components, we should choose a subset from them called the f-basis. In the above example, we have only four independent f-basis operators, 
\begin{align}
    \begin{array}{c|c}
        \text{f-basis} &  \\
        \hline
        \mc{O}^{(f)}_{LQ^3 D^2,1} & \dfrac{1}{6}\mc{Y}\left[\tiny{\young(rst)}\right] \epsilon^{abc} \epsilon^{ik} \epsilon^{jl} \left(L_{pi} Q_{raj}\right) \left(D_{\mu}Q_{sbk} D^{\mu}Q_{tcl}\right) \\
        \mc{O}^{(f)}_{LQ^3 D^2,2} & \dfrac{1}{3}\mc{Y}\left[\tiny{\young(rs,t)}\right] \epsilon^{abc} \epsilon^{ik} \epsilon^{jl} \left(L_{pi} Q_{raj}\right) \left(D_{\mu}Q_{sbk} D^{\mu}Q_{tcl}\right) \\
        \mc{O}^{(f)}_{LQ^3 D^2,3} & \dfrac{1}{3}\mc{Y}\left[\tiny{\young(rs,t)}\right] \epsilon^{abc} \epsilon^{il} \epsilon^{jk} \left(L_{pi} Q_{raj}\right) \left(D_{\mu}Q_{sbk} D^{\mu}Q_{tcl}\right) \\
        \mc{O}^{(f)}_{LQ^3 D^2,4} & \dfrac{1}{6}\mc{Y}\left[\tiny{\young(r,s,t)}\right] \epsilon^{abc} \epsilon^{ik} \epsilon^{jl} \left(L_{pi} Q_{raj}\right) \left(D_{\mu}Q_{sbk} D^{\mu}Q_{tcl}\right)
    \end{array}
\end{align}
Later, we will need to write a generic operator, or its corresponding amplitude, as a linear combination of the f-basis in the sense of flavor components: we will first expand the flavor tensor onto the p-basis, while the combination of p-basis tensors involves permutations of the f-basis operators, such as
\eq{
    c_2 \mc{O}^{(p)}_{LQ^3 D^2,2} + c_3 \mc{O}^{(p)}_{LQ^3 D^2,3} = \big(c_2 + c_3 (s \ t)\big) \mc{O}^{(f)}_{LQ^3 D^2,2} \ .
}
The relation between the p-basis and the f-basis, as well as the distinction between operator-level redundancy and coefficient-level linear independence, has been discussed in detail in our previous work \cite{Li:2020xlh}, to which we refer the reader for a complete derivation.

Furthermore, we impose the same symmetry to Wilson coefficients, which are also flavor tensors. The permutation symmetry is labeled in the subscript of repeated fields:
\begin{align}
    C^{prst}_{LQ^3_{[3]}D^2i}&\equiv \frac16\left(C^{prst}_{LQ^3D^2i}+C^{prts}_{LQ^3D^2i}+C^{psrt}_{LQ^3D^2i}+C^{pstr}_{LQ^3D^2i}+C^{ptsr}_{LQ^3D^2i}+C^{ptrs}_{LQ^3D^2i}\right),\label{eq:CLQ31}\\
    C^{prst}_{LQ^3_{[1,1,1]}D^2i}&\equiv \frac16\left(C^{prst}_{LQ^3D^2i}-C^{prts}_{LQ^3D^2i}-C^{psrt}_{LQ^3D^2i}+C^{pstr}_{LQ^3D^2i}-C^{ptsr}_{LQ^3D^2i}+C^{ptrs}_{LQ^3D^2i}\right),\\
    C^{prst}_{LQ^3_{[2,1]}D^2i}&\equiv \frac13\left(C^{prst}_{LQ^3D^2i}+C^{psrt}_{LQ^3D^2i}-C^{ptsr}_{LQ^3D^2i}-C^{pstr}_{LQ^3D^2i}\right),\label{eq:CLQ33}
    
\end{align}
where the flavor indices in the superscript correspond one by one according to the fermion order of the subscript. $i$ shows that this is the $i$th operator with this symmetry.

\section{One-loop Renormalization}
In dimensional regularization $D=4-2\epsilon$, 1-loop amplitudes may contain ultraviolet poles of the form
\begin{align}
    \mc{A}^{\text{1-loop}}_i=\frac{1}{\epsilon}Z^{(1)}_{ij}\mc{A}^{\text{tree}}_{j}+O(\epsilon)\ ,
\end{align}
with $Z^{(1)}_{ij}$ denoting the one-loop ultraviolet(UV) pole of the renormalization constant. This notation, which matches the anomalous dimension $\gamma_{ij}$, is defined as
\eq{
\gamma_{ij}=-2Z^{(1)}_{ij}\ ,\quad \text{while }\mu\frac{{\rm d}}{{\rm d}\mu}C_{i}=\sum_j\gamma_{ij}C_j\ .
}
The coefficient of $1/\epsilon$ in a given amplitude encodes how the associated operator coefficients $C_i$ must renormalize to cancel the divergence. 
In particular, the divergences contain both UV parts, which encode the operator mixing, and collinear infrared (IR) singularities, which arise from massless external legs.
\eq{
\mu\frac{{\rm d}}{{\rm d}\mu}C_{i}=\sum_j(\gamma_{ij}^{\text{UV}}+\delta_{ij}\gamma_j^{\text{coll}})C_j\ . \label{eq:gammadefine}
}
We will discuss the $\gamma^{\text{coll}}$ later and look at the more complex UV divergence terms for now. To extract the UV part, we perform Passarino-Veltman(PV) reduction on the 1-loop amplitude. The reduced amplitude takes the form
\eq{
\mathcal{A}^{\text{1-loop}}&=\sum_{\mathcal{I}}b_{\mathcal{I}}I_2(s_{\mathcal{I}})+\text{UV finite terms}\ ,
}
where $I_2(s_{\mc{I}})$ is the scalar bubble integral in channel $\mc{I}$, depending on $s_{\mc{I}}$. In dimensional regularization,
\begin{align}
    I_2(s_{\mc{I}})&=\frac{1}{16\pi^2}\left(\frac{1}{\epsilon}-\ln\frac{-s_{\mc{I}}}{\mu^2}+2\right)+O(\epsilon)\ .
\end{align}
Since boxes and triangles do not contain UV poles, the entire UV divergence is controlled by the bubble coefficients $b_{\mc{I}}$.
Matching the $1/\epsilon$ terms then gives
\begin{align}
    \gamma^{\text{UV}}_{ji}C_i\mathcal{O}_j=\frac{1}{8\pi^2}\sum_{\mathcal{I}}b_{\mathcal{I}}\ ,
\end{align}
in which we expand the tree-level amplitudes $\mc{A}^{\text{tree}}$ by our amplitude basis $\mc{O}_i$. 
Instead of computing one-loop corrections using Feynman diagrams, we directly extract the UV-divergent part using the on-shell unitarity cut method
\begin{align}
    \text{cut}_{\mathcal{I}}\mathcal{A}^{\text{1-loop}} = \sum_{1'2'}\int {\rm d}\Phi_{1'2'}\mathcal{A}_{\rm L}\mathcal{A}_{\rm R} = \frac{\pi}{2}b_{\mathcal{I}} + \text{contributions from $I_3$ and $I_4$} \ ,\label{eq:DoubleCut}
\end{align}
where we have summed over all possible intermediate two-particle states $(1',2')$ and integrate over the Lorentz invariant phase space measure ${\rm d}\Phi_{1'2'}$. Suppose the momenta of the two particles are $k_1',k_2'$, with $P=k_1'+k_2'$. Define $(k_1')_{\alpha\dot\alpha}=t|l\rangle_{\alpha}[l|_{\dot\alpha}$, so that the integral can be expressed as
\begin{align}
    \int {\rm d}\Phi_{1'2'}=\int {\rm d}^4k_1'\delta\left((k_1')^2\right)\delta\left((P-k_1')^2 \right)=\frac{i}{4}\int t{\rm d}t \langle l\,dl\rangle [dl\,l]\delta\left(2t\langle lx\rangle [lx]+P^2\right).
\end{align}
It can be converted to a complex integral by expanding the loop spinors in a fixed basis $|l\rangle=|1\rangle+z|2\rangle$, $|l]=|1]+\bar{z}|2]$ with $P=|1\rangle [1|+|2\rangle [2|$, as
\begin{align}
    \int {\rm d}\Phi_{1'2'}=\frac{i}{4}\oint{\rm d}z\int {\rm d}\bar{z}{\rm d}t\, t^2 \delta\left(t-\frac{1}{1+z\bar{z}}\right).
\end{align}
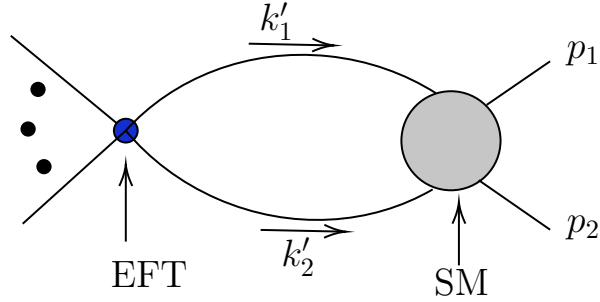
\begin{figure}[htbp]
    \centering
\tikzset{every picture/.style={line width=0.75pt}}     

\begin{tikzpicture}[x=0.75pt,y=0.75pt,yscale=-1,xscale=1]

\draw    (59,29) -- (117,76.9) ;
\draw  [fill={rgb, 255:red, 18; green, 47; blue, 210 }  ,fill opacity=1 ] (111,76.9) .. controls (111,73.59) and (113.69,70.9) .. (117,70.9) .. controls (120.31,70.9) and (123,73.59) .. (123,76.9) .. controls (123,80.21) and (120.31,82.9) .. (117,82.9) .. controls (113.69,82.9) and (111,80.21) .. (111,76.9) -- cycle ;
\draw    (117,76.9) -- (65,123.9) ;
\draw  [draw opacity=0] (117,76.9) .. controls (140.91,50.34) and (178.58,35.12) .. (219.1,39.32) .. controls (239.27,41.41) and (257.76,48.07) .. (273.46,58.03) -- (208.01,146.28) -- cycle ; \draw   (117,76.9) .. controls (140.91,50.34) and (178.58,35.12) .. (219.1,39.32) .. controls (239.27,41.41) and (257.76,48.07) .. (273.46,58.03) ;  
\draw  [draw opacity=0] (271.78,106.55) .. controls (240.92,124.57) and (200.41,127.61) .. (163.08,111.27) .. controls (144.51,103.14) and (128.92,91.17) .. (117,76.9) -- (206.2,12.76) -- cycle ; \draw   (271.78,106.55) .. controls (240.92,124.57) and (200.41,127.61) .. (163.08,111.27) .. controls (144.51,103.14) and (128.92,91.17) .. (117,76.9) ;  
\draw  [fill={rgb, 255:red, 198; green, 198; blue, 198 }  ,fill opacity=1 ] (256,82) .. controls (256,68.19) and (267.19,57) .. (281,57) .. controls (294.81,57) and (306,68.19) .. (306,82) .. controls (306,95.81) and (294.81,107) .. (281,107) .. controls (267.19,107) and (256,95.81) .. (256,82) -- cycle ;
\draw    (331,41.9) -- (299,63.9) ;
\draw    (295,101.9) -- (331,126.9) ;
\draw  [fill={rgb, 255:red, 0; green, 0; blue, 0 }  ,fill opacity=1 ] (69.4,56.33) .. controls (69.26,54.5) and (70.63,52.89) .. (72.47,52.75) .. controls (74.31,52.61) and (75.91,53.98) .. (76.06,55.82) .. controls (76.2,57.66) and (74.82,59.26) .. (72.99,59.4) .. controls (71.15,59.55) and (69.54,58.17) .. (69.4,56.33) -- cycle ;
\draw  [fill={rgb, 255:red, 0; green, 0; blue, 0 }  ,fill opacity=1 ] (64.4,76.33) .. controls (64.26,74.5) and (65.63,72.89) .. (67.47,72.75) .. controls (69.31,72.61) and (70.91,73.98) .. (71.06,75.82) .. controls (71.2,77.66) and (69.82,79.26) .. (67.99,79.4) .. controls (66.15,79.55) and (64.54,78.17) .. (64.4,76.33) -- cycle ;
\draw  [fill={rgb, 255:red, 0; green, 0; blue, 0 }  ,fill opacity=1 ] (72.4,95.33) .. controls (72.26,93.5) and (73.63,91.89) .. (75.47,91.75) .. controls (77.31,91.61) and (78.91,92.98) .. (79.06,94.82) .. controls (79.2,96.66) and (77.82,98.26) .. (75.99,98.4) .. controls (74.15,98.55) and (72.54,97.17) .. (72.4,95.33) -- cycle ;
\draw    (179,33) -- (220,33.86) ;
\draw [shift={(222,33.9)}, rotate = 181.2] [color={rgb, 255:red, 0; green, 0; blue, 0 }  ][line width=0.75]    (10.93,-3.29) .. controls (6.95,-1.4) and (3.31,-0.3) .. (0,0) .. controls (3.31,0.3) and (6.95,1.4) .. (10.93,3.29)   ;
\draw    (185,127) -- (226,127.86) ;
\draw [shift={(228,127.9)}, rotate = 181.2] [color={rgb, 255:red, 0; green, 0; blue, 0 }  ][line width=0.75]    (10.93,-3.29) .. controls (6.95,-1.4) and (3.31,-0.3) .. (0,0) .. controls (3.31,0.3) and (6.95,1.4) .. (10.93,3.29)   ;
\draw    (117,132.9) -- (117,95.9) ;
\draw [shift={(117,93.9)}, rotate = 90] [color={rgb, 255:red, 0; green, 0; blue, 0 }  ][line width=0.75]    (10.93,-3.29) .. controls (6.95,-1.4) and (3.31,-0.3) .. (0,0) .. controls (3.31,0.3) and (6.95,1.4) .. (10.93,3.29)   ;
\draw    (285,144.9) -- (285,114.9) ;
\draw [shift={(285,112.9)}, rotate = 90] [color={rgb, 255:red, 0; green, 0; blue, 0 }  ][line width=0.75]    (10.93,-3.29) .. controls (6.95,-1.4) and (3.31,-0.3) .. (0,0) .. controls (3.31,0.3) and (6.95,1.4) .. (10.93,3.29)   ;

\draw (183,10.4) node [anchor=north west][inner sep=0.75pt]  [font=\Large]  {$k_{1} '$};
\draw (194,129.4) node [anchor=north west][inner sep=0.75pt]  [font=\Large]  {$k_{2} '$};
\draw (338,30.4) node [anchor=north west][inner sep=0.75pt]  [font=\Large]  {$p_{1}$};
\draw (338,118.4) node [anchor=north west][inner sep=0.75pt]  [font=\Large]  {$p_{2}$};
\draw (108,141) node [anchor=north west][inner sep=0.75pt]  [font=\Large] [align=left] {EFT};
\draw (271,145) node [anchor=north west][inner sep=0.75pt]  [font=\Large] [align=left] {SM};

\end{tikzpicture}
\caption{One-loop unitarity cut for effective operators renormalization. While the renormalization contribution is from the same dimension.}
\label{fig:UnitarityCut}
\end{figure}
It should be noted that the unitary cut integrals here may contain infrared (IR) divergences, which would be canceled with $\text{cut}I_3$ and $\text{cut}I_4$. A straightforward way to handle this, as done in \cite{Mastrolia:2009dr}, is to simply discard the logarithmic terms in the result and keep only the rational terms, since IR divergences manifest only through the logarithmic terms in the integrals. A more rigorous approach, as used in \cite{Arkani-Hamed:2008owk}, is to first separate the $I_2$ integrals from the $I_3$ and $I_4$ integrals using a BCFW shift, $\hat{k}'_1= k'_1+wq$, $\hat{k}'_2=k'_2-wq$, so that the $I_2$ coefficient can be extracted as
\begin{align}
    b_{\mathcal{I}}=\frac{2}{\pi}\sum_{1'2'}\int {\rm d}\Phi_{1'2'}\oint_{\mathcal{C}_\infty} \frac{{\rm d}w}{w}\mathcal{A}_{\rm L}(\hat{k}'_1(w),\hat{k}'_2(w))\mathcal{A}_{\rm R}(\hat{k}'_1(w),\hat{k}'_2(w)), \label{eq:b_master}
\end{align}
where the contour $\mathcal{C}_\infty$ encircles the pole at infinity. In four dimensions, ultraviolet (UV) divergences arise solely from the $I_2$ integrals, which do not contain any IR-divergent contributions.

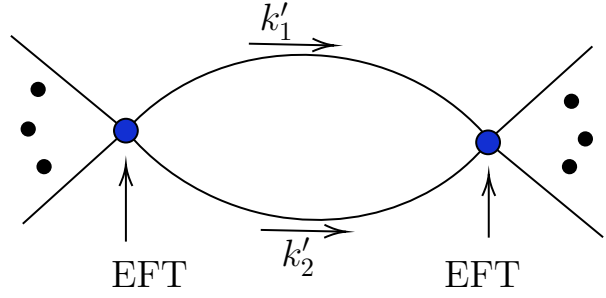
\begin{figure}[tbp]
\centering

\tikzset{every picture/.style={line width=0.75pt}} 

\begin{tikzpicture}[x=0.75pt,y=0.75pt,yscale=-1,xscale=1]

\draw  [draw opacity=0] (117,76.9) .. controls (140.91,50.34) and (178.58,35.12) .. (219.1,39.32) .. controls (252.41,42.78) and (281.13,58.67) .. (300.27,81.37) -- (208.01,146.28) -- cycle ; \draw   (117,76.9) .. controls (140.91,50.34) and (178.58,35.12) .. (219.1,39.32) .. controls (252.41,42.78) and (281.13,58.67) .. (300.27,81.37) ;  
\draw  [draw opacity=0] (299.67,82.81) .. controls (269.17,120.19) and (213.25,133.23) .. (163.08,111.27) .. controls (144.51,103.14) and (128.92,91.17) .. (117,76.9) -- (206.2,12.76) -- cycle ; \draw   (299.67,82.81) .. controls (269.17,120.19) and (213.25,133.23) .. (163.08,111.27) .. controls (144.51,103.14) and (128.92,91.17) .. (117,76.9) ;  
\draw    (117,76.9) -- (65,123.9) ;
\draw    (352.27,34.37) -- (300.27,81.37) ;
\draw    (299.67,82.81) -- (357.67,130.71) ;
\draw    (59,29) -- (117,76.9) ;
\draw  [fill={rgb, 255:red, 18; green, 47; blue, 210 }  ,fill opacity=1 ] (111,76.9) .. controls (111,73.59) and (113.69,70.9) .. (117,70.9) .. controls (120.31,70.9) and (123,73.59) .. (123,76.9) .. controls (123,80.21) and (120.31,82.9) .. (117,82.9) .. controls (113.69,82.9) and (111,80.21) .. (111,76.9) -- cycle ;
\draw  [fill={rgb, 255:red, 0; green, 0; blue, 0 }  ,fill opacity=1 ] (69.4,56.33) .. controls (69.26,54.5) and (70.63,52.89) .. (72.47,52.75) .. controls (74.31,52.61) and (75.91,53.98) .. (76.06,55.82) .. controls (76.2,57.66) and (74.82,59.26) .. (72.99,59.4) .. controls (71.15,59.55) and (69.54,58.17) .. (69.4,56.33) -- cycle ;
\draw  [fill={rgb, 255:red, 0; green, 0; blue, 0 }  ,fill opacity=1 ] (64.4,76.33) .. controls (64.26,74.5) and (65.63,72.89) .. (67.47,72.75) .. controls (69.31,72.61) and (70.91,73.98) .. (71.06,75.82) .. controls (71.2,77.66) and (69.82,79.26) .. (67.99,79.4) .. controls (66.15,79.55) and (64.54,78.17) .. (64.4,76.33) -- cycle ;
\draw  [fill={rgb, 255:red, 0; green, 0; blue, 0 }  ,fill opacity=1 ] (72.4,95.33) .. controls (72.26,93.5) and (73.63,91.89) .. (75.47,91.75) .. controls (77.31,91.61) and (78.91,92.98) .. (79.06,94.82) .. controls (79.2,96.66) and (77.82,98.26) .. (75.99,98.4) .. controls (74.15,98.55) and (72.54,97.17) .. (72.4,95.33) -- cycle ;
\draw    (179,33) -- (220,33.86) ;
\draw [shift={(222,33.9)}, rotate = 181.2] [color={rgb, 255:red, 0; green, 0; blue, 0 }  ][line width=0.75]    (10.93,-3.29) .. controls (6.95,-1.4) and (3.31,-0.3) .. (0,0) .. controls (3.31,0.3) and (6.95,1.4) .. (10.93,3.29)   ;
\draw    (185,127) -- (226,127.86) ;
\draw [shift={(228,127.9)}, rotate = 181.2] [color={rgb, 255:red, 0; green, 0; blue, 0 }  ][line width=0.75]    (10.93,-3.29) .. controls (6.95,-1.4) and (3.31,-0.3) .. (0,0) .. controls (3.31,0.3) and (6.95,1.4) .. (10.93,3.29)   ;
\draw    (117,132.9) -- (117,95.9) ;
\draw [shift={(117,93.9)}, rotate = 90] [color={rgb, 255:red, 0; green, 0; blue, 0 }  ][line width=0.75]    (10.93,-3.29) .. controls (6.95,-1.4) and (3.31,-0.3) .. (0,0) .. controls (3.31,0.3) and (6.95,1.4) .. (10.93,3.29)   ;
\draw    (300,130.9) -- (300,100.9) ;
\draw [shift={(300,98.9)}, rotate = 90] [color={rgb, 255:red, 0; green, 0; blue, 0 }  ][line width=0.75]    (10.93,-3.29) .. controls (6.95,-1.4) and (3.31,-0.3) .. (0,0) .. controls (3.31,0.3) and (6.95,1.4) .. (10.93,3.29)   ;
\draw  [fill={rgb, 255:red, 18; green, 47; blue, 210 }  ,fill opacity=1 ] (293.67,82.81) .. controls (293.67,79.5) and (296.36,76.81) .. (299.67,76.81) .. controls (302.99,76.81) and (305.67,79.5) .. (305.67,82.81) .. controls (305.67,86.12) and (302.99,88.81) .. (299.67,88.81) .. controls (296.36,88.81) and (293.67,86.12) .. (293.67,82.81) -- cycle ;
\draw  [fill={rgb, 255:red, 0; green, 0; blue, 0 }  ,fill opacity=1 ] (338.4,62.33) .. controls (338.26,60.5) and (339.63,58.89) .. (341.47,58.75) .. controls (343.31,58.61) and (344.91,59.98) .. (345.06,61.82) .. controls (345.2,63.66) and (343.82,65.26) .. (341.99,65.4) .. controls (340.15,65.55) and (338.54,64.17) .. (338.4,62.33) -- cycle ;
\draw  [fill={rgb, 255:red, 0; green, 0; blue, 0 }  ,fill opacity=1 ] (345.4,81.33) .. controls (345.26,79.5) and (346.63,77.89) .. (348.47,77.75) .. controls (350.31,77.61) and (351.91,78.98) .. (352.06,80.82) .. controls (352.2,82.66) and (350.82,84.26) .. (348.99,84.4) .. controls (347.15,84.55) and (345.54,83.17) .. (345.4,81.33) -- cycle ;
\draw  [fill={rgb, 255:red, 0; green, 0; blue, 0 }  ,fill opacity=1 ] (337.4,95.33) .. controls (337.26,93.5) and (338.63,91.89) .. (340.47,91.75) .. controls (342.31,91.61) and (343.91,92.98) .. (344.06,94.82) .. controls (344.2,96.66) and (342.82,98.26) .. (340.99,98.4) .. controls (339.15,98.55) and (337.54,97.17) .. (337.4,95.33) -- cycle ;

\draw (183,10.4) node [anchor=north west][inner sep=0.75pt]  [font=\Large]  {$k_{1} '$};
\draw (194,129.4) node [anchor=north west][inner sep=0.75pt]  [font=\Large]  {$k_{2} '$};
\draw (108,141) node [anchor=north west][inner sep=0.75pt]  [font=\Large] [align=left] {EFT};
\draw (276,141) node [anchor=north west][inner sep=0.75pt]  [font=\Large] [align=left] {EFT};

\end{tikzpicture}

\caption{One-loop unitarity cut for effective operators renormalization, while the renormalization contribution are from the loeer-dimension.}
\label{fig:FFS_all}
\end{figure}

As shown in Fig.~\ref{fig:UnitarityCut}, the vertex on the left corresponds to the effective operator expressed in the f-basis, while the amplitude on the right represents the Standard Model contribution. Similarly, in Fig.~\ref{fig:FFS_all},  when computing the renormalization of higher-dimensional operators induced by the squared contributions of lower-dimensional ones, we also perform the calculation by expressing the effective operators in terms of the f-basis.

One might worry that some operators in effective field theory can be reduced using equations of motion or field redefinitions, and their on-shell amplitudes vanish. A natural question is whether such operators might still contribute to the renormalization of other operators. This issue has been addressed in \cite{Jenkins:2013zja}, and it has been shown that these operators do not affect the renormalization of physical operators. For example, if we divide the operator basis into two parts, $O_i^{(d)}$ are the usual independent operator basis which we obtain from on-shell amplitude basis in $d$-dimension, and $E_r$ are generated by field redefinitions in the SM Lagrangian, $E_i^{(d)}=0$ under the on-shell condition or the classical SM equations of motion.
\begin{align}
    \mathcal{L}^{(d)}=\sum_i C_i O_i^{(d)}+\sum_r D_r E_r^{(d)}.
\end{align}
Calculate the RGEs for both $O_i$ and $E_r$, 
\begin{align}
    \mu\frac{{\rm d}}{{\rm d}\mu}\left[\begin{array}{c}O_i\\E_r\end{array}\right]=\left[\begin{array}{cc}-\gamma_{ji}&-a_{si}\\0&-b_{sr}\end{array}\right]\left[\begin{array}{c}O_j\\E_s\end{array}\right].
\end{align}
The EOM operators do not contribute to the S-matrix elements, so they do not contribute to the renormalization of $O_i$ either. Result in the anomalous dimension matrix for the coefficients has the form
\begin{align}
    \mu\frac{{\rm d}}{{\rm d}\mu}\left[\begin{array}{c}C_i\\D_r\end{array}\right]=\left[\begin{array}{cc}\gamma_{ij}&0\\a_{rj}&b_{rs}\end{array}\right]\left[\begin{array}{c}C_j\\D_s\end{array}\right].
\end{align}
Therefore, the EOM parts $E_r$ can be neglected in on-shell calculation as in eq.~\eqref{eq:gammadefine}.

We then substitute the previously introduced f-basis into the renormalization
\begin{align}\label{eq:mudmu}
    \gamma_{ji}^{\text{UV}}C^f_i\mathcal{O}^f_j=\frac{1}{8\pi^2}\sum_{\mathcal{I}}b_{\mathcal{I}}.
\end{align}
We compute the coefficients of the scalar bubble integrals $b_{\mathcal{I}}$ by the unitarity cut eq.~\eqref{eq:DoubleCut} and express the result as a combination of the f-basis amplitudes
\begin{align}
    \sum_{\mathcal{I}}b_{\mathcal{I}}=\sum_{ij} C_i\mathcal{K}^{bf}_{ij}\mathcal{O}^{f}_j \ .
\end{align}
Therefore, we have the anomalous dimension
\begin{align}
    \gamma_{ji}^{\rm UV} = \frac{1}{8\pi^2}\mc{K}^{bf}_{ij} \ .
\end{align}
In practice, the direct reduction to the f-basis is sometimes hard to do. We need to make use of the standard reduction rules to the y-basis
\begin{align}
    \sum_{\mathcal{I}}b_{\mathcal{I}}=\sum_{i,j} C_i\mathcal{K}^{by}_{ij}\mathcal{O}^y_j,\quad \mathcal{O}^p_j=\sum_k\mathcal{K}^{py}_{jk}\mathcal{O}^y_k,
    \quad\Rightarrow\quad \sum_{\mathcal{I}}b_{\mathcal{I}}=\sum_{i,j,k} C_i \mathcal{K}^{by}_{ij}\left(\mathcal{K}^{py}\right)^{-1}_{jk}\mathcal{O}^p_k\ .
\end{align}
We use p-basis rather than f-basis because $\mathcal{K}^{py}$ is a invertible square matrix while $\mathcal{K}^{fy}$ is not. But after the Schur-Weyl duality, they are basically the same. Hence, in this scenario we have
\begin{align}
    \gamma_{ki}^{\text{UV}}=\frac{1}{8\pi^2}\mathcal{K}^{by}_{ij}\left(\mathcal{K}^{py}\right)^{-1}_{jk}.
\end{align}

For example, when calculating $\mathcal{O}_{d_{\mathbb{C}}d_{\mathbb{C}}^{\dagger}LL^{\dagger}}\to\mathcal{O}_{L^2L^{\dagger 2}}$, $\mathcal{K}^{py}_{ij}=\left(\begin{array}{cc} 2 & 2 \\2 & -2
    \end{array}\right)$ is the y-basis expansion for $\mathcal{O}_{L^2L^{\dagger 2}}$, and $C_i\mathcal{K}^{by}_{ij}$ comes from the y-basis expansion for the unitarity cut $\sum_{\mathcal{I}}b_{I}$,
\begin{align}
    16\pi^2\mu\frac{{\rm d}}{{\rm d}\mu}\left(\begin{array}{c} C^{f_1f_2f_3f_4}_{L^2_{[2]}L^{\dagger}{}^2_{[2]}}\\C^{f_1f_2f_3f_4}_{L^2_{[1,1]}L^{\dagger}{}^2_{[1,1]}}
    \end{array}\right)^T=
    \left(\begin{array}{c} -4y_dy_lg_1^2\delta_{f_2f_4}C^{f'f_1f'f_3}_{d_{\mathbb{C}}Ld_{\mathbb{C}}^{\dagger}L^{\dagger}}-4y_dy_lg_1^2\delta_{f_1f_3}C^{f'f_2f'f_4}_{d_{\mathbb{C}}Ld_{\mathbb{C}}^{\dagger}L^{\dagger}} \\
    -4y_dy_lg_1^2\delta_{f_2f_3}C^{f'f_1f'f_4}_{d_{\mathbb{C}}Ld_{\mathbb{C}}^{\dagger}L^{\dagger}}-4y_dy_lg_1^2\delta_{f_1f_4}C^{f'f_2f'f_3}_{d_{\mathbb{C}}Ld_{\mathbb{C}}^{\dagger}L^{\dagger}}
    \end{array}\right)^T
    \left(\begin{array}{cc} 2 & 2 \\2 & -2
    \end{array}\right)^{-1}\ .
\end{align}
Similarly, we can consider the case that both $\mathcal{A}_L$ and $\mathcal{A}_R$ involve effective operators, which feature non-linear dependence on the Wilson coefficients. In our case, these are typically quadratic contributions
\begin{align}
    \sum_{\mathcal{I}}b_{\mathcal{I}}=\sum_{ijk} C_iC_j\mathcal{K}^{by}_{ijk}\mathcal{O}^y_k.
\end{align}
Hence, they contribute to the RGEs as
\begin{align}
    \mu\frac{{\rm d}}{{\rm d}\mu}C_i \supset \sum_{j,k,l}C_j C_k \mathcal{K}^{by}_{jkl} \left(\mathcal{K}^{py}\right)^{-1}_{li}.
\end{align}
We also illustrate this formula with an example  $\mathcal{O}_{d_{\mathbb{C}}d_{\mathbb{C}}^{\dagger}LL^{\dagger}}\times\mathcal{O}_{d_{\mathbb{C}}d_{\mathbb{C}}^{\dagger}LL^{\dagger}}\to\mathcal{O}_{L^2L^{\dagger 2}D^2}$ 
\begin{align}
    16\pi^2\mu\frac{{\rm d}}{{\rm d}\mu}\left(\begin{array}{c} C^{f_1f_2f_3f_4}_{L^2_{[2]}L^{\dagger}{}^2_{[2]}D^2,1}\\C^{f_1f_2f_3f_4}_{L^2_{[2]}L^{\dagger}{}^2_{[2]}D^2,2}\\C^{f_1f_2f_3f_4}_{L^2_{[1,1]}L^{\dagger}{}^2_{[1,1]}D^2,1}\\C^{f_1f_2f_3f_4}_{L^2_{[1,1]}L^{\dagger}{}^2_{[1,1]}D^2,2}
    \end{array}\right)^T \supset
    \underbrace{\left(\begin{array}{c} 0 \\
    4C_{d_{\mathbb{C}}Ld_{\mathbb{C}}^{\dagger}L^{\dagger}}^{f'f_2f''f_4}C_{d_{\mathbb{C}}Ld_{\mathbb{C}}^{\dagger}L^{\dagger}}^{f''f_1f'f_3} \\
    4C_{d_{\mathbb{C}}Ld_{\mathbb{C}}^{\dagger}L^{\dagger}}^{f'f_2f''f_3}C_{d_{\mathbb{C}}Ld_{\mathbb{C}}^{\dagger}L^{\dagger}}^{f''f_1f'f_4} \\
    -4C_{d_{\mathbb{C}}Ld_{\mathbb{C}}^{\dagger}L^{\dagger}}^{f'f_2f''f_3}C_{d_{\mathbb{C}}Ld_{\mathbb{C}}^{\dagger}L^{\dagger}}^{f''f_1f'f_4}
    \end{array}\right)^T}_{\sum_{j,k}C_j C_k \mathcal{K}^{by}_{jkl}}
    \underbrace{\left(\begin{array}{cccc} 1 & 0 & 1 & 0 \\ 1 & -2 & -1 & 2 \\1 & 0 & -1 & 0 \\ 1 & -2 & 1 & -2
    \end{array}\right)^{-1}}_{\left(\mathcal{K}^{py}\right)^{-1}}\ .
\end{align}
where the quadratic coefficients $\sum_{j,k}C_j C_k \mathcal{K}^{by}_{jkl}$ are to be extracted from the unitarity cuts via the amplitude reduction.

\subsection{Calculating the One-Loop Renormalization with Helicity Spinor}
We will illustrate the detailed calculation process using an example where the $L^2L^{\dagger 2}$ renormalization  from itself. Here are the corresponding operators f-basis. 
\begin{align}
\begin{array}{c|c}
\text{f-basis} & \\
\hline
\mc{O}^{(f)}_{L^2L^{\dagger 2},1} & \frac14 \mc{Y}\left[\tiny{\young(pr)},\tiny{\young(st)}\right](L_{pi}L_{rj})(L^{\dagger}{}_{s}^i L^{\dagger}{}_{t}^j)\\
\mc{O}^{(f)}_{L^2L^{\dagger 2},2} & \frac14 \mc{Y}\left[\tiny{\young(p,r)},\tiny{\young(s,t)}\right](L_{pi}L_{rj})(L^{\dagger}{}_{s}^i L^{\dagger}{}_{t}^j)
\end{array}.
\end{align}
First, we translate the operators f-basis to on-shell amplitudes:
\begin{align}\begin{split}\label{eq:Fb:L4}
\mc{A}_{\text{eff}}\left(L_{f_1i_1}(p_1),L_{f_2i_2}(p_2),L^{\dagger i_3}_{f_3}(p_3),L^{\dagger i_4}_{f_4}(p_4)\right)&=2\vev{12}[34]\left(C_{L^2_{[2]}L^{\dagger}{}^2_{[2]}}^{f_1f_2f_3f_4}T_1 + C_{L^2_{[1,1]}L^{\dagger}{}^2_{[1,1]}}^{f_1f_2f_3f_4}T_2\right)\\
T_1=\delta^{i_1}_{i_3}\delta^{i_2}_{i_4}+\delta^{i_1}_{i_4}\delta^{i_2}_{i_3}\ ,\quad &T_2=\delta^{i_1}_{i_3}\delta^{i_2}_{i_4}-\delta^{i_1}_{i_4}\delta^{i_2}_{i_3}\ ,
\end{split}
\end{align}
where $f$ are the flavor indices and $i$ are the $SU(2)_L$ fundamental indices. Note that the subscripts of $L$ and $L^{\dagger}$ in Wilson coefficients indicate the permutation symmetries of the identical particles via the partition representation of Young diagrams: $[2]\equiv{\tiny \yng(2)}$ indicates they are symmetric, and $[1,1]\equiv{\tiny \yng(1,1)}$ indicates they are anti-symmetric. Similar to eqs.~(\ref{eq:CLQ31}-\ref{eq:CLQ33}), these subscripts indicate the following flavor relations
\eq{\label{eq:frelation}
   C_{L^2_{[2]}L^{\dagger}{}^2_{[2]}}^{f_1f_2f_3f_4}=C_{L^2_{[2]}L^{\dagger}{}^2_{[2]}}^{f_2f_1f_3f_4}&=C_{L^2_{[2]}L^{\dagger}{}^2_{[2]}}^{f_1f_2f_4f_3}=C_{L^2_{[2]}L^{\dagger}{}^2_{[2]}}^{f_2f_1f_4f_3},\\
   C_{L^2_{[1,1]}L^{\dagger}{}^2_{[1,1]}}^{f_1f_2f_3f_4}=-C_{L^2_{[1,1]}L^{\dagger}{}^2_{[1,1]}}^{f_2f_1f_3f_4}&=-C_{L^2_{[1,1]}L^{\dagger}{}^2_{[1,1]}}^{f_1f_2f_4f_3}=C_{L^2_{[1,1]}L^{\dagger}{}^2_{[1,1]}}^{f_2f_1f_4f_3},
}
while the order of the flavor indices is consistent with the order of the fermions in the subscripts. In the following calculation, we will adjust the ordering of the fermions to align with the ordering used later in the loop diagram calculations. The specific combination in eq.~\eqref{eq:Fb:L4} reflects the requirement of the spin-statistics theorem. Note that permuting the fermion order introduces an overall negative sign in the amplitude.
\begin{align}
    \mc{A}_{\text{eff}}\left(L_{f_1i_1}(p_1),L^{\dagger i_2}_{f_2}(p_2),L_{f_3i_3}(p_3),L^{\dagger i_4}_{f_4}(p_4)\right)=-\mc{A}_{\text{eff}}\left(L_{f_1i_1}(p_1),L_{f_2i_2}(p_2),L^{\dagger i_3}_{f_3}(p_3),L^{\dagger i_4}_{f_4}(p_4)\right)\Big|_{2\leftrightarrow 3}.
\end{align}
\begin{figure}[htbp]
    \centering

\tikzset{every picture/.style={line width=0.75pt}} 

\begin{tikzpicture}[x=0.75pt,y=0.75pt,yscale=-1,xscale=1]

\draw    (391.75,13.03) -- (352.03,64.49) ;
\draw [shift={(368.84,42.72)}, rotate = 307.66] [fill={rgb, 255:red, 0; green, 0; blue, 0 }  ][line width=0.08]  [draw opacity=0] (8.93,-4.29) -- (0,0) -- (8.93,4.29) -- cycle    ;
\draw  [fill={rgb, 255:red, 18; green, 47; blue, 210 }  ,fill opacity=1 ] (351.88,59.29) .. controls (354.73,59.21) and (357.11,61.46) .. (357.2,64.33) .. controls (357.29,67.2) and (355.05,69.59) .. (352.19,69.68) .. controls (349.34,69.77) and (346.95,67.51) .. (346.87,64.64) .. controls (346.78,61.78) and (349.02,59.38) .. (351.88,59.29) -- cycle ;
\draw    (352.03,64.49) -- (310.18,20.72) ;
\draw [shift={(327.65,38.99)}, rotate = 46.28] [fill={rgb, 255:red, 0; green, 0; blue, 0 }  ][line width=0.08]  [draw opacity=0] (8.93,-4.29) -- (0,0) -- (8.93,4.29) -- cycle    ;
\draw  [draw opacity=0] (352.19,64.63) .. controls (374.15,83.39) and (387.86,112.83) .. (387.38,145.38) -- (294.69,145.08) -- cycle ; \draw    (352.19,64.63) .. controls (373.49,82.83) and (387.03,111.07) .. (387.38,142.46) ; \draw [shift={(387.38,145.38)}, rotate = 267.37] [fill={rgb, 255:red, 0; green, 0; blue, 0 }  ][line width=0.08]  [draw opacity=0] (8.93,-4.29) -- (0,0) -- (8.93,4.29) -- cycle    ; 
\draw  [draw opacity=0] (330.51,199.09) .. controls (318.66,180.01) and (313.5,156.31) .. (316.61,132.23) -- (409.63,139.99) -- cycle ; \draw    (330.51,199.09) .. controls (319.14,180.77) and (313.93,158.2) .. (316.28,135.12) ; \draw [shift={(316.61,132.23)}, rotate = 93.9] [fill={rgb, 255:red, 0; green, 0; blue, 0 }  ][line width=0.08]  [draw opacity=0] (8.93,-4.29) -- (0,0) -- (8.93,4.29) -- cycle    ; 
\draw  [fill={rgb, 255:red, 198; green, 198; blue, 198 }  ,fill opacity=1 ] (351.32,184.91) .. controls (363.21,184.55) and (373.14,193.94) .. (373.51,205.89) .. controls (373.87,217.84) and (364.53,227.82) .. (352.64,228.18) .. controls (340.75,228.55) and (330.81,219.16) .. (330.45,207.21) .. controls (330.08,195.26) and (339.43,185.28) .. (351.32,184.91) -- cycle ;
\draw    (387.83,248.76) -- (368.04,221.65) ;
\draw [shift={(374.99,231.17)}, rotate = 53.87] [fill={rgb, 255:red, 0; green, 0; blue, 0 }  ][line width=0.08]  [draw opacity=0] (8.93,-4.29) -- (0,0) -- (8.93,4.29) -- cycle    ;
\draw    (335.21,219.19) -- (314.63,251) ;
\draw [shift={(322.21,239.29)}, rotate = 302.9] [fill={rgb, 255:red, 0; green, 0; blue, 0 }  ][line width=0.08]  [draw opacity=0] (8.93,-4.29) -- (0,0) -- (8.93,4.29) -- cycle    ;
\draw [color={rgb, 255:red, 74; green, 144; blue, 226 }  ,draw opacity=1 ]   (391.48,116.99) -- (391.82,152.22) ;
\draw [shift={(391.84,154.22)}, rotate = 269.44] [color={rgb, 255:red, 74; green, 144; blue, 226 }  ,draw opacity=1 ][line width=0.75]    (10.93,-3.29) .. controls (6.95,-1.4) and (3.31,-0.3) .. (0,0) .. controls (3.31,0.3) and (6.95,1.4) .. (10.93,3.29)   ;
\draw [color={rgb, 255:red, 74; green, 144; blue, 226 }  ,draw opacity=1 ]   (310.69,124.65) -- (311.03,159.89) ;
\draw [shift={(311.05,161.89)}, rotate = 269.44] [color={rgb, 255:red, 74; green, 144; blue, 226 }  ,draw opacity=1 ][line width=0.75]    (10.93,-3.29) .. controls (6.95,-1.4) and (3.31,-0.3) .. (0,0) .. controls (3.31,0.3) and (6.95,1.4) .. (10.93,3.29)   ;
\draw [color={rgb, 255:red, 74; green, 144; blue, 226 }  ,draw opacity=1 ]   (395.58,20.3) -- (376.9,43.89) ;
\draw [shift={(375.66,45.45)}, rotate = 308.38] [color={rgb, 255:red, 74; green, 144; blue, 226 }  ,draw opacity=1 ][line width=0.75]    (10.93,-3.29) .. controls (6.95,-1.4) and (3.31,-0.3) .. (0,0) .. controls (3.31,0.3) and (6.95,1.4) .. (10.93,3.29)   ;
\draw [color={rgb, 255:red, 74; green, 144; blue, 226 }  ,draw opacity=1 ]   (305.01,20.59) -- (325.07,42.71) ;
\draw [shift={(326.42,44.19)}, rotate = 227.79] [color={rgb, 255:red, 74; green, 144; blue, 226 }  ,draw opacity=1 ][line width=0.75]    (10.93,-3.29) .. controls (6.95,-1.4) and (3.31,-0.3) .. (0,0) .. controls (3.31,0.3) and (6.95,1.4) .. (10.93,3.29)   ;
\draw [color={rgb, 255:red, 74; green, 144; blue, 226 }  ,draw opacity=1 ]   (388.55,244.12) -- (374.48,225.4) ;
\draw [shift={(373.28,223.8)}, rotate = 53.07] [color={rgb, 255:red, 74; green, 144; blue, 226 }  ,draw opacity=1 ][line width=0.75]    (10.93,-3.29) .. controls (6.95,-1.4) and (3.31,-0.3) .. (0,0) .. controls (3.31,0.3) and (6.95,1.4) .. (10.93,3.29)   ;
\draw [color={rgb, 255:red, 74; green, 144; blue, 226 }  ,draw opacity=1 ]   (312.74,245.57) -- (326.47,225.12) ;
\draw [shift={(327.58,223.46)}, rotate = 123.87] [color={rgb, 255:red, 74; green, 144; blue, 226 }  ,draw opacity=1 ][line width=0.75]    (10.93,-3.29) .. controls (6.95,-1.4) and (3.31,-0.3) .. (0,0) .. controls (3.31,0.3) and (6.95,1.4) .. (10.93,3.29)   ;
\draw  [draw opacity=0] (387.39,143.91) .. controls (387.38,146.54) and (387.29,149.19) .. (387.09,151.86) .. controls (385.82,169.39) and (380.57,185.58) .. (372.39,199.43) -- (294.69,145.08) -- cycle ; \draw    (387.39,143.91) .. controls (387.38,146.54) and (387.29,149.19) .. (387.09,151.86) .. controls (385.82,169.39) and (380.57,185.58) .. (372.39,199.43) ;  
\draw  [draw opacity=0] (316.07,137.47) .. controls (316.92,126.75) and (319.4,115.89) .. (323.65,105.27) .. controls (329.75,90) and (338.83,76.92) .. (349.81,66.53) -- (409.63,139.99) -- cycle ; \draw    (316.07,137.47) .. controls (316.92,126.75) and (319.4,115.89) .. (323.65,105.27) .. controls (329.75,90) and (338.83,76.92) .. (349.81,66.53) ;  
\draw    (489.49,254.57) -- (530.7,204.3) ;
\draw [shift={(513.27,225.57)}, rotate = 129.34] [fill={rgb, 255:red, 0; green, 0; blue, 0 }  ][line width=0.08]  [draw opacity=0] (8.93,-4.29) -- (0,0) -- (8.93,4.29) -- cycle    ;
\draw  [fill={rgb, 255:red, 18; green, 47; blue, 210 }  ,fill opacity=1 ] (530.71,209.5) .. controls (527.85,209.5) and (525.53,207.18) .. (525.53,204.31) .. controls (525.53,201.44) and (527.84,199.11) .. (530.69,199.11) .. controls (533.55,199.11) and (535.87,201.43) .. (535.87,204.3) .. controls (535.87,207.17) and (533.56,209.5) .. (530.71,209.5) -- cycle ;
\draw    (530.7,204.3) -- (571.25,249.28) ;
\draw [shift={(554.32,230.51)}, rotate = 227.96] [fill={rgb, 255:red, 0; green, 0; blue, 0 }  ][line width=0.08]  [draw opacity=0] (8.93,-4.29) -- (0,0) -- (8.93,4.29) -- cycle    ;
\draw  [draw opacity=0] (530.54,204.16) .. controls (509.15,184.76) and (496.31,154.93) .. (497.74,122.41) -- (590.38,125.43) -- cycle ; \draw    (530.54,204.16) .. controls (509.79,185.34) and (497.08,156.71) .. (497.65,125.33) ; \draw [shift={(497.74,122.41)}, rotate = 89.05] [fill={rgb, 255:red, 0; green, 0; blue, 0 }  ][line width=0.08]  [draw opacity=0] (8.93,-4.29) -- (0,0) -- (8.93,4.29) -- cycle    ; 
\draw  [draw opacity=0] (556.17,70.39) .. controls (567.44,89.81) and (571.9,113.64) .. (568.09,137.63) -- (475.34,127.14) -- cycle ; \draw    (556.17,70.39) .. controls (566.99,89.03) and (571.54,111.74) .. (568.51,134.75) ; \draw [shift={(568.09,137.63)}, rotate = 275.58] [fill={rgb, 255:red, 0; green, 0; blue, 0 }  ][line width=0.08]  [draw opacity=0] (8.93,-4.29) -- (0,0) -- (8.93,4.29) -- cycle    ; 
\draw  [fill={rgb, 255:red, 198; green, 198; blue, 198 }  ,fill opacity=1 ] (534.95,83.95) .. controls (523.05,83.96) and (513.4,74.29) .. (513.38,62.33) .. controls (513.37,50.38) and (523,40.67) .. (534.89,40.66) .. controls (546.79,40.64) and (556.44,50.32) .. (556.46,62.28) .. controls (556.47,74.23) and (546.84,83.94) .. (534.95,83.95) -- cycle ;
\draw    (500.32,19.06) -- (519.31,46.74) ;
\draw [shift={(512.64,37.02)}, rotate = 235.55] [fill={rgb, 255:red, 0; green, 0; blue, 0 }  ][line width=0.08]  [draw opacity=0] (8.93,-4.29) -- (0,0) -- (8.93,4.29) -- cycle    ;
\draw    (552.05,50.16) -- (573.55,18.97) ;
\draw [shift={(565.64,30.45)}, rotate = 124.58] [fill={rgb, 255:red, 0; green, 0; blue, 0 }  ][line width=0.08]  [draw opacity=0] (8.93,-4.29) -- (0,0) -- (8.93,4.29) -- cycle    ;
\draw [color={rgb, 255:red, 74; green, 144; blue, 226 }  ,draw opacity=1 ]   (492.81,150.67) -- (493.5,115.44) ;
\draw [shift={(493.54,113.44)}, rotate = 91.12] [color={rgb, 255:red, 74; green, 144; blue, 226 }  ,draw opacity=1 ][line width=0.75]    (10.93,-3.29) .. controls (6.95,-1.4) and (3.31,-0.3) .. (0,0) .. controls (3.31,0.3) and (6.95,1.4) .. (10.93,3.29)   ;
\draw [color={rgb, 255:red, 74; green, 144; blue, 226 }  ,draw opacity=1 ]   (573.79,145.38) -- (574.48,110.14) ;
\draw [shift={(574.52,108.14)}, rotate = 91.12] [color={rgb, 255:red, 74; green, 144; blue, 226 }  ,draw opacity=1 ][line width=0.75]    (10.93,-3.29) .. controls (6.95,-1.4) and (3.31,-0.3) .. (0,0) .. controls (3.31,0.3) and (6.95,1.4) .. (10.93,3.29)   ;
\draw [color={rgb, 255:red, 74; green, 144; blue, 226 }  ,draw opacity=1 ]   (483.58,249.45) -- (502.94,226.43) ;
\draw [shift={(504.23,224.9)}, rotate = 130.06] [color={rgb, 255:red, 74; green, 144; blue, 226 }  ,draw opacity=1 ][line width=0.75]    (10.93,-3.29) .. controls (6.95,-1.4) and (3.31,-0.3) .. (0,0) .. controls (3.31,0.3) and (6.95,1.4) .. (10.93,3.29)   ;
\draw [color={rgb, 255:red, 74; green, 144; blue, 226 }  ,draw opacity=1 ]   (576.42,249.56) -- (557.01,226.86) ;
\draw [shift={(555.71,225.34)}, rotate = 49.47] [color={rgb, 255:red, 74; green, 144; blue, 226 }  ,draw opacity=1 ][line width=0.75]    (10.93,-3.29) .. controls (6.95,-1.4) and (3.31,-0.3) .. (0,0) .. controls (3.31,0.3) and (6.95,1.4) .. (10.93,3.29)   ;
\draw [color={rgb, 255:red, 74; green, 144; blue, 226 }  ,draw opacity=1 ]   (499.46,23.67) -- (512.98,42.8) ;
\draw [shift={(514.13,44.44)}, rotate = 234.75] [color={rgb, 255:red, 74; green, 144; blue, 226 }  ,draw opacity=1 ][line width=0.75]    (10.93,-3.29) .. controls (6.95,-1.4) and (3.31,-0.3) .. (0,0) .. controls (3.31,0.3) and (6.95,1.4) .. (10.93,3.29)   ;
\draw [color={rgb, 255:red, 74; green, 144; blue, 226 }  ,draw opacity=1 ]   (575.28,24.45) -- (560.96,44.48) ;
\draw [shift={(559.8,46.11)}, rotate = 305.55] [color={rgb, 255:red, 74; green, 144; blue, 226 }  ,draw opacity=1 ][line width=0.75]    (10.93,-3.29) .. controls (6.95,-1.4) and (3.31,-0.3) .. (0,0) .. controls (3.31,0.3) and (6.95,1.4) .. (10.93,3.29)   ;
\draw  [draw opacity=0] (497.69,123.88) .. controls (497.77,121.25) and (497.95,118.6) .. (498.22,115.94) .. controls (500,98.46) and (505.72,82.43) .. (514.3,68.82) -- (590.38,125.43) -- cycle ; \draw    (497.69,123.88) .. controls (497.77,121.25) and (497.95,118.6) .. (498.22,115.94) .. controls (500,98.46) and (505.72,82.43) .. (514.3,68.82) ;  
\draw  [draw opacity=0] (568.79,132.41) .. controls (567.62,143.1) and (564.82,153.88) .. (560.27,164.37) .. controls (553.72,179.45) and (544.26,192.27) .. (532.98,202.33) -- (475.34,127.14) -- cycle ; \draw    (568.79,132.41) .. controls (567.62,143.1) and (564.82,153.88) .. (560.27,164.37) .. controls (553.72,179.45) and (544.26,192.27) .. (532.98,202.33) ;  
\draw    (21.39,90.29) -- (68.22,129.47) ;
\draw [shift={(48.64,113.09)}, rotate = 219.92] [fill={rgb, 255:red, 0; green, 0; blue, 0 }  ][line width=0.08]  [draw opacity=0] (8.93,-4.29) -- (0,0) -- (8.93,4.29) -- cycle    ;
\draw  [fill={rgb, 255:red, 18; green, 47; blue, 210 }  ,fill opacity=1 ] (63.38,129.47) .. controls (63.38,126.76) and (65.55,124.56) .. (68.22,124.56) .. controls (70.9,124.56) and (73.07,126.76) .. (73.07,129.47) .. controls (73.07,132.18) and (70.9,134.38) .. (68.22,134.38) .. controls (65.55,134.38) and (63.38,132.18) .. (63.38,129.47) -- cycle ;
\draw    (68.22,129.47) -- (26.24,167.92) ;
\draw [shift={(43.54,152.07)}, rotate = 317.52] [fill={rgb, 255:red, 0; green, 0; blue, 0 }  ][line width=0.08]  [draw opacity=0] (8.93,-4.29) -- (0,0) -- (8.93,4.29) -- cycle    ;
\draw  [draw opacity=0] (67.86,129.86) .. controls (85.94,109.26) and (113.96,96.88) .. (144.52,98.27) -- (141.7,186.23) -- cycle ; \draw    (67.86,129.86) .. controls (85.4,109.88) and (112.28,97.63) .. (141.77,98.18) ; \draw [shift={(144.52,98.27)}, rotate = 178.77] [fill={rgb, 255:red, 0; green, 0; blue, 0 }  ][line width=0.08]  [draw opacity=0] (8.93,-4.29) -- (0,0) -- (8.93,4.29) -- cycle    ; 
\draw  [draw opacity=0] (193.51,153.55) .. controls (175.39,164.38) and (153.1,168.69) .. (130.66,165.1) -- (140.24,77) -- cycle ; \draw    (193.51,153.55) .. controls (176.11,163.94) and (154.87,168.34) .. (133.35,165.49) ; \draw [shift={(130.66,165.1)}, rotate = 5.37] [fill={rgb, 255:red, 0; green, 0; blue, 0 }  ][line width=0.08]  [draw opacity=0] (8.93,-4.29) -- (0,0) -- (8.93,4.29) -- cycle    ; 
\draw  [fill={rgb, 255:red, 198; green, 198; blue, 198 }  ,fill opacity=1 ] (180.45,133.64) .. controls (180.45,122.35) and (189.48,113.19) .. (200.63,113.19) .. controls (211.78,113.19) and (220.82,122.35) .. (220.82,133.64) .. controls (220.82,144.94) and (211.78,154.1) .. (200.63,154.1) .. controls (189.48,154.1) and (180.45,144.94) .. (180.45,133.64) -- cycle ;
\draw    (241,100.84) -- (215.17,118.84) ;
\draw [shift={(223.98,112.7)}, rotate = 325.14] [fill={rgb, 255:red, 0; green, 0; blue, 0 }  ][line width=0.08]  [draw opacity=0] (8.93,-4.29) -- (0,0) -- (8.93,4.29) -- cycle    ;
\draw    (211.94,149.92) -- (241,170.37) ;
\draw [shift={(230.56,163.03)}, rotate = 215.13] [fill={rgb, 255:red, 0; green, 0; blue, 0 }  ][line width=0.08]  [draw opacity=0] (8.93,-4.29) -- (0,0) -- (8.93,4.29) -- cycle    ;
\draw [color={rgb, 255:red, 74; green, 144; blue, 226 }  ,draw opacity=1 ]   (118.28,93.56) -- (151,94.25) ;
\draw [shift={(153,94.3)}, rotate = 181.21] [color={rgb, 255:red, 74; green, 144; blue, 226 }  ,draw opacity=1 ][line width=0.75]    (10.93,-3.29) .. controls (6.95,-1.4) and (3.31,-0.3) .. (0,0) .. controls (3.31,0.3) and (6.95,1.4) .. (10.93,3.29)   ;
\draw [color={rgb, 255:red, 74; green, 144; blue, 226 }  ,draw opacity=1 ]   (123.12,170.46) -- (155.84,171.15) ;
\draw [shift={(157.84,171.19)}, rotate = 181.21] [color={rgb, 255:red, 74; green, 144; blue, 226 }  ,draw opacity=1 ][line width=0.75]    (10.93,-3.29) .. controls (6.95,-1.4) and (3.31,-0.3) .. (0,0) .. controls (3.31,0.3) and (6.95,1.4) .. (10.93,3.29)   ;
\draw [color={rgb, 255:red, 74; green, 144; blue, 226 }  ,draw opacity=1 ]   (18.7,95.11) -- (40.06,113.45) ;
\draw [shift={(41.58,114.75)}, rotate = 220.64] [color={rgb, 255:red, 74; green, 144; blue, 226 }  ,draw opacity=1 ][line width=0.75]    (10.93,-3.29) .. controls (6.95,-1.4) and (3.31,-0.3) .. (0,0) .. controls (3.31,0.3) and (6.95,1.4) .. (10.93,3.29)   ;
\draw [color={rgb, 255:red, 74; green, 144; blue, 226 }  ,draw opacity=1 ]   (25.97,172.83) -- (47.06,154.51) ;
\draw [shift={(48.57,153.2)}, rotate = 139.03] [color={rgb, 255:red, 74; green, 144; blue, 226 }  ,draw opacity=1 ][line width=0.75]    (10.93,-3.29) .. controls (6.95,-1.4) and (3.31,-0.3) .. (0,0) .. controls (3.31,0.3) and (6.95,1.4) .. (10.93,3.29)   ;
\draw [color={rgb, 255:red, 74; green, 144; blue, 226 }  ,draw opacity=1 ]   (236.7,100.02) -- (218.94,112.76) ;
\draw [shift={(217.32,113.93)}, rotate = 324.33] [color={rgb, 255:red, 74; green, 144; blue, 226 }  ,draw opacity=1 ][line width=0.75]    (10.93,-3.29) .. controls (6.95,-1.4) and (3.31,-0.3) .. (0,0) .. controls (3.31,0.3) and (6.95,1.4) .. (10.93,3.29)   ;
\draw [color={rgb, 255:red, 74; green, 144; blue, 226 }  ,draw opacity=1 ]   (235.89,172.01) -- (217.32,158.46) ;
\draw [shift={(215.7,157.29)}, rotate = 36.11] [color={rgb, 255:red, 74; green, 144; blue, 226 }  ,draw opacity=1 ][line width=0.75]    (10.93,-3.29) .. controls (6.95,-1.4) and (3.31,-0.3) .. (0,0) .. controls (3.31,0.3) and (6.95,1.4) .. (10.93,3.29)   ;
\draw  [draw opacity=0] (143.12,98.21) .. controls (145.61,98.29) and (148.13,98.46) .. (150.65,98.73) .. controls (167.21,100.47) and (182.35,106.06) .. (195.16,114.44) -- (141.7,186.23) -- cycle ; \draw    (143.12,98.21) .. controls (145.61,98.29) and (148.13,98.46) .. (150.65,98.73) .. controls (167.21,100.47) and (182.35,106.06) .. (195.16,114.44) ;  
\draw  [draw opacity=0] (135.58,165.76) .. controls (125.5,164.66) and (115.33,161.98) .. (105.43,157.59) .. controls (91.22,151.29) and (79.17,142.17) .. (69.76,131.29) -- (140.24,77) -- cycle ; \draw    (135.58,165.76) .. controls (125.5,164.66) and (115.33,161.98) .. (105.43,157.59) .. controls (91.22,151.29) and (79.17,142.17) .. (69.76,131.29) ;  

\draw (395.19,128.09) node [anchor=north west][inner sep=0.75pt]  [font=\small,color={rgb, 255:red, 74; green, 144; blue, 226 }  ,opacity=1 ]  {$p'_{1}$};
\draw (294.46,132.03) node [anchor=north west][inner sep=0.75pt]  [font=\small,color={rgb, 255:red, 74; green, 144; blue, 226 }  ,opacity=1 ]  {$p'_{2}$};
\draw (337.71,197.61) node [anchor=north west][inner sep=0.75pt]  [font=\Large] [align=left] {SM};
\draw (365.95,8.96) node [anchor=north west][inner sep=0.75pt]    {$L$};
\draw (315.19,7.12) node [anchor=north west][inner sep=0.75pt]    {$L$};
\draw (295.46,55.18) node [anchor=north west][inner sep=0.75pt]    {$\mathcal{O}_{L^{2} L^{\dagger 2}}$};
\draw (374.91,77.24) node [anchor=north west][inner sep=0.75pt]    {$L^{\dagger }$};
\draw (331.23,86.12) node [anchor=north west][inner sep=0.75pt]    {$L^{\dagger }$};
\draw (384.2,178.25) node [anchor=north west][inner sep=0.75pt]    {$L$};
\draw (324.52,173.62) node [anchor=north west][inner sep=0.75pt]    {$L$};
\draw (363.75,241.68) node [anchor=north west][inner sep=0.75pt]    {$L^{\dagger }$};
\draw (322.16,237.92) node [anchor=north west][inner sep=0.75pt]    {$L^{\dagger }$};
\draw (296.18,21.91) node [anchor=north west][inner sep=0.75pt]  [font=\small,color={rgb, 255:red, 74; green, 144; blue, 226 }  ,opacity=1 ]  {$p_{1}$};
\draw (391.58,24.7) node [anchor=north west][inner sep=0.75pt]  [font=\small,color={rgb, 255:red, 74; green, 144; blue, 226 }  ,opacity=1 ]  {$p_{2}$};
\draw (293.79,235.53) node [anchor=north west][inner sep=0.75pt]  [font=\small,color={rgb, 255:red, 74; green, 144; blue, 226 }  ,opacity=1 ]  {$p_{3}$};
\draw (393.73,239.34) node [anchor=north west][inner sep=0.75pt]  [font=\small,color={rgb, 255:red, 74; green, 144; blue, 226 }  ,opacity=1 ]  {$p_{4}$};
\draw (473.66,124.04) node [anchor=north west][inner sep=0.75pt]  [font=\small,color={rgb, 255:red, 74; green, 144; blue, 226 }  ,opacity=1 ]  {$p'_{1}$};
\draw (574.46,123.05) node [anchor=north west][inner sep=0.75pt]  [font=\small,color={rgb, 255:red, 74; green, 144; blue, 226 }  ,opacity=1 ]  {$p'_{2}$};
\draw (521.13,55.44) node [anchor=north west][inner sep=0.75pt]  [font=\Large] [align=left] {SM};
\draw (502.25,240.89) node [anchor=north west][inner sep=0.75pt]    {$L^{\dagger }$};
\draw (547.57,241.86) node [anchor=north west][inner sep=0.75pt]    {$L^{\dagger }$};
\draw (547.52,191.2) node [anchor=north west][inner sep=0.75pt]    {$\mathcal{O}_{L^{2} L^{\dagger 2}}$};
\draw (498.05,175.09) node [anchor=north west][inner sep=0.75pt]    {$L$};
\draw (535.88,170.32) node [anchor=north west][inner sep=0.75pt]    {$L$};
\draw (487.6,69.77) node [anchor=north west][inner sep=0.75pt]    {$L^{\dagger }$};
\draw (546.13,80.15) node [anchor=north west][inner sep=0.75pt]    {$L^{\dagger }$};
\draw (507.39,5.47) node [anchor=north west][inner sep=0.75pt]    {$L$};
\draw (556.7,7.31) node [anchor=north west][inner sep=0.75pt]    {$L$};
\draw (471.07,226.95) node [anchor=north west][inner sep=0.75pt]  [font=\small,color={rgb, 255:red, 74; green, 144; blue, 226 }  ,opacity=1 ]  {$p_{3}$};
\draw (573.88,231.2) node [anchor=north west][inner sep=0.75pt]  [font=\small,color={rgb, 255:red, 74; green, 144; blue, 226 }  ,opacity=1 ]  {$p_{4}$};
\draw (480.41,17.74) node [anchor=north west][inner sep=0.75pt]  [font=\small,color={rgb, 255:red, 74; green, 144; blue, 226 }  ,opacity=1 ]  {$p_{1}$};
\draw (577.33,16.68) node [anchor=north west][inner sep=0.75pt]  [font=\small,color={rgb, 255:red, 74; green, 144; blue, 226 }  ,opacity=1 ]  {$p_{2}$};
\draw (128.04,75.39) node [anchor=north west][inner sep=0.75pt]  [font=\small,color={rgb, 255:red, 74; green, 144; blue, 226 }  ,opacity=1 ]  {$p'_{1}$};
\draw (128.85,171.1) node [anchor=north west][inner sep=0.75pt]  [font=\small,color={rgb, 255:red, 74; green, 144; blue, 226 }  ,opacity=1 ]  {$p'_{2}$};
\draw (186.28,122.82) node [anchor=north west][inner sep=0.75pt]  [font=\Large] [align=left] {SM};
\draw (29.93,81.05) node [anchor=north west][inner sep=0.75pt]    {$L$};
\draw (14.72,149.65) node [anchor=north west][inner sep=0.75pt]    {$L^{\dagger }$};
\draw (49.23,152.1) node [anchor=north west][inner sep=0.75pt]    {$\mathcal{O}_{L^{2} L^{\dagger 2}}$};
\draw (82.41,96.59) node [anchor=north west][inner sep=0.75pt]    {$L$};
\draw (86.58,132.47) node [anchor=north west][inner sep=0.75pt]    {$L^{\dagger }$};
\draw (176.2,89.12) node [anchor=north west][inner sep=0.75pt]    {$L^{\dagger }$};
\draw (170.41,145.68) node [anchor=north west][inner sep=0.75pt]    {$L$};
\draw (247.11,93.32) node [anchor=north west][inner sep=0.75pt]    {$L$};
\draw (243.21,161.92) node [anchor=north west][inner sep=0.75pt]    {$L^{\dagger }$};
\draw (13.39,97.48) node [anchor=north west][inner sep=0.75pt]  [font=\small,color={rgb, 255:red, 74; green, 144; blue, 226 }  ,opacity=1 ]  {$p_{1}$};
\draw (21.47,169.47) node [anchor=north west][inner sep=0.75pt]  [font=\small,color={rgb, 255:red, 74; green, 144; blue, 226 }  ,opacity=1 ]  {$p_{3}$};
\draw (234.62,81.94) node [anchor=north west][inner sep=0.75pt]  [font=\small,color={rgb, 255:red, 74; green, 144; blue, 226 }  ,opacity=1 ]  {$p_{2}$};
\draw (228.96,171.1) node [anchor=north west][inner sep=0.75pt]  [font=\small,color={rgb, 255:red, 74; green, 144; blue, 226 }  ,opacity=1 ]  {$p_{4}$};

\end{tikzpicture}

    \caption{The one-loop diagram when calculating the renormalization group contribution from the effective operator $\mc{O}_{L^2L^{\dagger 2}}$ to $\mc{O}_{L^2L^{\dagger 2}}$. The black arrows indicate the direction of the fermion lines, while the blue arrows represent the direction of momentum flow. }
    \label{fig:UCLLee}
\end{figure}
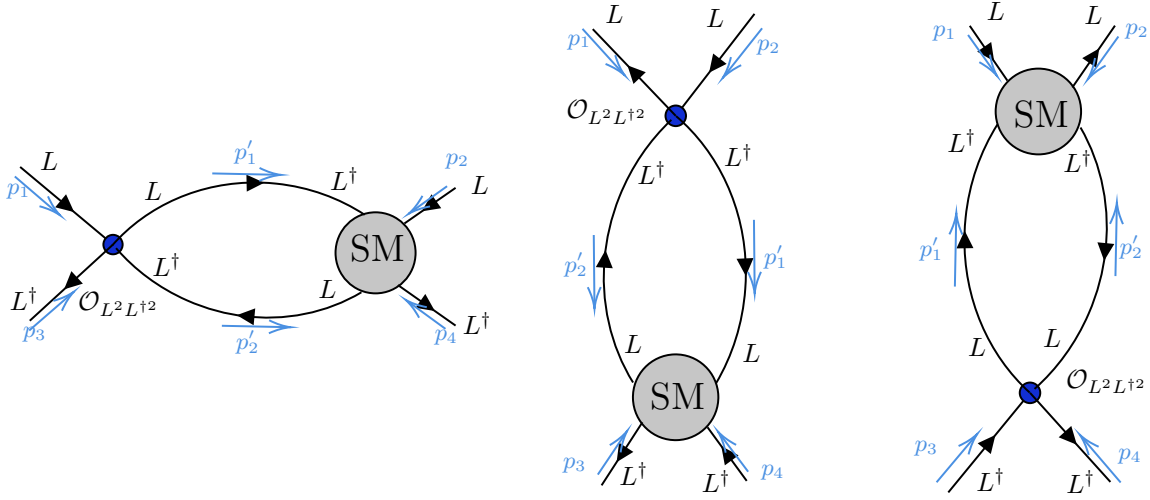
The Feynman diagram involved in the calculation is shown in Fig.~\ref{fig:UCLLee}. The Standard Model amplitude involved here is given by:
\begin{align}\begin{split}
    &\mc{A}_{\text{SM}}\left(L_{f_1i_1}(p_1),L_{f_2i_2}(p_2),L^{\dagger}{}^{i_3}_{f_3}(p_3),L^{\dagger}{}^{i_4}_{f_4}(p_4)\right)\\
    &\qquad=\left(2g_1^2y_l^2\delta^{i_1}_{i_3}\delta^{i_2}_{i_4}+g_2^2(\delta^{i_1}_{i_4}\delta^{i_2}_{i_3}-\frac12\delta^{i_1}_{i_3}\delta^{i_2}_{i_4})\right)\delta_{f_1f_3}\delta_{f_2f_4}\frac{\vev{12}[34]}{s_{13}} - (3\leftrightarrow 4)\ .
\end{split}\end{align}
Next, we relabel the particles according to fig.~\ref{fig:UCLLee} and then multiply the two amplitudes together to obtain the integrand. 
As shown in fig.~\ref{fig:UCLLee}, there are three possible product structures:
\begin{align}
\begin{split}
&\quad\tikzset{every picture/.style={line width=0.75pt}} 
\begin{tikzpicture}[x=0.75pt,y=0.75pt,yscale=-1,xscale=1]

\draw    (120.61,43) -- (157.49,73.51) ;
\draw [shift={(142.9,61.44)}, rotate = 219.6] [fill={rgb, 255:red, 0; green, 0; blue, 0 }  ][line width=0.08]  [draw opacity=0] (8.93,-4.29) -- (0,0) -- (8.93,4.29) -- cycle    ;
\draw  [fill={rgb, 255:red, 18; green, 47; blue, 210 }  ,fill opacity=1 ] (153.68,73.51) .. controls (153.68,71.4) and (155.38,69.69) .. (157.49,69.69) .. controls (159.6,69.69) and (161.31,71.4) .. (161.31,73.51) .. controls (161.31,75.62) and (159.6,77.33) .. (157.49,77.33) .. controls (155.38,77.33) and (153.68,75.62) .. (153.68,73.51) -- cycle ;
\draw    (157.49,73.51) -- (124.43,103.45) ;
\draw [shift={(137.25,91.84)}, rotate = 317.84] [fill={rgb, 255:red, 0; green, 0; blue, 0 }  ][line width=0.08]  [draw opacity=0] (8.93,-4.29) -- (0,0) -- (8.93,4.29) -- cycle    ;
\draw  [draw opacity=0] (157.45,73.56) .. controls (171.71,57.66) and (193.68,48.13) .. (217.64,49.22) -- (215.36,117.71) -- cycle ; \draw    (157.45,73.56) .. controls (171.14,58.3) and (191.94,48.9) .. (214.77,49.14) ; \draw [shift={(217.64,49.22)}, rotate = 177.83] [fill={rgb, 255:red, 0; green, 0; blue, 0 }  ][line width=0.08]  [draw opacity=0] (8.93,-4.29) -- (0,0) -- (8.93,4.29) -- cycle    ; 
\draw  [draw opacity=0] (255.95,92.38) .. controls (241.67,100.74) and (224.15,104.05) .. (206.51,101.23) -- (214.21,32.65) -- cycle ; \draw    (255.95,92.38) .. controls (242.46,100.28) and (226.06,103.67) .. (209.42,101.64) ; \draw [shift={(206.51,101.23)}, rotate = 4.26] [fill={rgb, 255:red, 0; green, 0; blue, 0 }  ][line width=0.08]  [draw opacity=0] (8.93,-4.29) -- (0,0) -- (8.93,4.29) -- cycle    ; 
\draw  [fill={rgb, 255:red, 198; green, 198; blue, 198 }  ,fill opacity=1 ] (245.88,76.76) .. controls (245.88,67.96) and (252.99,60.83) .. (261.77,60.83) .. controls (270.55,60.83) and (277.67,67.96) .. (277.67,76.76) .. controls (277.67,85.56) and (270.55,92.69) .. (261.77,92.69) .. controls (252.99,92.69) and (245.88,85.56) .. (245.88,76.76) -- cycle ;
\draw    (293.57,51.21) -- (273.22,65.23) ;
\draw [shift={(279.28,61.06)}, rotate = 325.44] [fill={rgb, 255:red, 0; green, 0; blue, 0 }  ][line width=0.08]  [draw opacity=0] (8.93,-4.29) -- (0,0) -- (8.93,4.29) -- cycle    ;
\draw    (270.68,89.44) -- (293.57,105.36) ;
\draw [shift={(286.23,100.26)}, rotate = 214.83] [fill={rgb, 255:red, 0; green, 0; blue, 0 }  ][line width=0.08]  [draw opacity=0] (8.93,-4.29) -- (0,0) -- (8.93,4.29) -- cycle    ;
\draw [color={rgb, 255:red, 74; green, 144; blue, 226 }  ,draw opacity=1 ]   (196.92,45.55) -- (222.26,46.08) ;
\draw [shift={(224.26,46.12)}, rotate = 181.2] [color={rgb, 255:red, 74; green, 144; blue, 226 }  ,draw opacity=1 ][line width=0.75]    (10.93,-3.29) .. controls (6.95,-1.4) and (3.31,-0.3) .. (0,0) .. controls (3.31,0.3) and (6.95,1.4) .. (10.93,3.29)   ;
\draw [color={rgb, 255:red, 74; green, 144; blue, 226 }  ,draw opacity=1 ]   (200.73,105.43) -- (226.07,105.96) ;
\draw [shift={(228.07,106)}, rotate = 181.2] [color={rgb, 255:red, 74; green, 144; blue, 226 }  ,draw opacity=1 ][line width=0.75]    (10.93,-3.29) .. controls (6.95,-1.4) and (3.31,-0.3) .. (0,0) .. controls (3.31,0.3) and (6.95,1.4) .. (10.93,3.29)   ;
\draw    (157.49,109.19) -- (157.49,86.34) ;
\draw [shift={(157.49,84.34)}, rotate = 90] [color={rgb, 255:red, 0; green, 0; blue, 0 }  ][line width=0.75]    (10.93,-3.29) .. controls (6.95,-1.4) and (3.31,-0.3) .. (0,0) .. controls (3.31,0.3) and (6.95,1.4) .. (10.93,3.29)   ;
\draw    (264.32,116.83) -- (264.32,98.45) ;
\draw [shift={(264.32,96.45)}, rotate = 90] [color={rgb, 255:red, 0; green, 0; blue, 0 }  ][line width=0.75]    (10.93,-3.29) .. controls (6.95,-1.4) and (3.31,-0.3) .. (0,0) .. controls (3.31,0.3) and (6.95,1.4) .. (10.93,3.29)   ;
\draw [color={rgb, 255:red, 74; green, 144; blue, 226 }  ,draw opacity=1 ]   (118.49,46.76) -- (134.98,60.75) ;
\draw [shift={(136.51,62.04)}, rotate = 220.32] [color={rgb, 255:red, 74; green, 144; blue, 226 }  ,draw opacity=1 ][line width=0.75]    (10.93,-3.29) .. controls (6.95,-1.4) and (3.31,-0.3) .. (0,0) .. controls (3.31,0.3) and (6.95,1.4) .. (10.93,3.29)   ;
\draw [color={rgb, 255:red, 74; green, 144; blue, 226 }  ,draw opacity=1 ]   (124.21,107.27) -- (140.5,93.29) ;
\draw [shift={(142.02,91.99)}, rotate = 139.35] [color={rgb, 255:red, 74; green, 144; blue, 226 }  ,draw opacity=1 ][line width=0.75]    (10.93,-3.29) .. controls (6.95,-1.4) and (3.31,-0.3) .. (0,0) .. controls (3.31,0.3) and (6.95,1.4) .. (10.93,3.29)   ;
\draw [color={rgb, 255:red, 74; green, 144; blue, 226 }  ,draw opacity=1 ]   (290.18,50.58) -- (276.55,60.25) ;
\draw [shift={(274.92,61.41)}, rotate = 324.64] [color={rgb, 255:red, 74; green, 144; blue, 226 }  ,draw opacity=1 ][line width=0.75]    (10.93,-3.29) .. controls (6.95,-1.4) and (3.31,-0.3) .. (0,0) .. controls (3.31,0.3) and (6.95,1.4) .. (10.93,3.29)   ;
\draw [color={rgb, 255:red, 74; green, 144; blue, 226 }  ,draw opacity=1 ]   (289.54,106.64) -- (275.27,96.34) ;
\draw [shift={(273.64,95.17)}, rotate = 35.8] [color={rgb, 255:red, 74; green, 144; blue, 226 }  ,draw opacity=1 ][line width=0.75]    (10.93,-3.29) .. controls (6.95,-1.4) and (3.31,-0.3) .. (0,0) .. controls (3.31,0.3) and (6.95,1.4) .. (10.93,3.29)   ;
\draw  [draw opacity=0] (216.55,49.17) .. controls (218.49,49.24) and (220.45,49.37) .. (222.42,49.57) .. controls (235.33,50.91) and (247.16,55.2) .. (257.18,61.62) -- (215.36,117.71) -- cycle ; \draw    (216.55,49.17) .. controls (218.49,49.24) and (220.45,49.37) .. (222.42,49.57) .. controls (235.33,50.91) and (247.16,55.2) .. (257.18,61.62) ;  
\draw  [draw opacity=0] (210.36,101.75) .. controls (202.48,100.88) and (194.53,98.8) .. (186.79,95.41) .. controls (175.68,90.53) and (166.24,83.49) .. (158.85,75.1) -- (214.21,32.65) -- cycle ; \draw    (210.36,101.75) .. controls (202.48,100.88) and (194.53,98.8) .. (186.79,95.41) .. controls (175.68,90.53) and (166.24,83.49) .. (158.85,75.1) ;  

\draw (202.9,29.71) node [anchor=north west][inner sep=0.75pt]  [font=\small,color={rgb, 255:red, 74; green, 144; blue, 226 }  ,opacity=1 ]  {$p'_{1}$};
\draw (203.54,104.25) node [anchor=north west][inner sep=0.75pt]  [font=\small,color={rgb, 255:red, 74; green, 144; blue, 226 }  ,opacity=1 ]  {$p'_{2}$};
\draw (249.23,112.18) node [anchor=north west][inner sep=0.75pt]  [font=\normalsize] [align=left] {SM};
\draw (126.06,34.12) node [anchor=north west][inner sep=0.75pt]    {$L$};
\draw (113.34,87.32) node [anchor=north west][inner sep=0.75pt]    {$L^{\dagger }$};
\draw (140.44,109.26) node [anchor=north west][inner sep=0.75pt]    {$\mathcal{O}_{L^{2} L^{\dagger 2}}$};
\draw (167.39,46.23) node [anchor=north west][inner sep=0.75pt]    {$L$};
\draw (169.93,73.95) node [anchor=north west][inner sep=0.75pt]    {$L^{\dagger }$};
\draw (240.51,40.18) node [anchor=north west][inner sep=0.75pt]    {$L^{\dagger }$};
\draw (236.7,84.45) node [anchor=north west][inner sep=0.75pt]    {$L$};
\draw (297.11,43.68) node [anchor=north west][inner sep=0.75pt]    {$L$};
\draw (293.29,96.88) node [anchor=north west][inner sep=0.75pt]    {$L^{\dagger }$};
\draw (112.61,46.92) node [anchor=north west][inner sep=0.75pt]  [font=\small,color={rgb, 255:red, 74; green, 144; blue, 226 }  ,opacity=1 ]  {$p_{1}$};
\draw (118.97,102.98) node [anchor=north west][inner sep=0.75pt]  [font=\small,color={rgb, 255:red, 74; green, 144; blue, 226 }  ,opacity=1 ]  {$p_{3}$};
\draw (286.84,34.81) node [anchor=north west][inner sep=0.75pt]  [font=\small,color={rgb, 255:red, 74; green, 144; blue, 226 }  ,opacity=1 ]  {$p_{2}$};
\draw (282.39,104.25) node [anchor=north west][inner sep=0.75pt]  [font=\small,color={rgb, 255:red, 74; green, 144; blue, 226 }  ,opacity=1 ]  {$p_{4}$};
\draw (20,65) node [anchor=north west][inner sep=0.75pt]    {$(\mathcal{A}_{L}\mathcal{A}_{R})^{(1)} \equiv$};
\end{tikzpicture} \\
&=-\mc{A}_{\text{eff}}\left(L_{f_1i_1}(p_1),L^{\dagger i_3}_{f_3}(p_3),L_{f'_{1}i'_{1}}(-p'_1),L^{\dagger i'_{2}}_{f'_{2}}(-p'_2)\right)\mc{A}_{\text{SM}}\left(L_{f'_{2}i'_{2}}(p'_2),L^{\dagger}{}^{i'_{1}}_{f'_{1}}(p'_1),L{}_{f_2i_2}(p_2),L^{\dagger i_4}{}_{f_4}(p_4)\right)\\
& = \mc{A}^{(1)}_1 \begin{pmatrix}
    \delta_{f_2f_4}C_{L^2_{[2]}L^{\dagger}{}^2_{[2]}}^{f_1f'f_3f'}, \delta_{f_2f_4}C_{L^2_{[1,1]}L^{\dagger}{}^2_{[1,1]}}^{f_1f'f_3f'}
    \end{pmatrix}\begin{pmatrix}
    3g_1^2y_l^2+\frac14g_2^2 & 3g_1^2y_l^2-\frac34g_2^2 \\
    g_1^2y_l^2-\frac14g_2^2 & g_1^2y_l^2+\frac34g_2^2 
    \end{pmatrix} \begin{pmatrix} T_1 \\ T_2 \end{pmatrix} \\
& \hspace{8em} + \mc{A}^{(1)}_2 \begin{pmatrix}
    C^{f_1f_2f_3f_4}_{L^2_{[2]}L^{\dagger}{}^2_{[2]}}, C^{f_1f_2f_3f_4}_{L^2_{[1,1]}L^{\dagger}{}^2_{[1,1]}}
    \end{pmatrix}\begin{pmatrix}
    2g_1^2y_l^2+g_2^2 & \frac32g_2^2 \\
    \frac12g_2^2 & 2g_1^2y_l^2
    \end{pmatrix} \begin{pmatrix} T_1 \\ T_2 \end{pmatrix} \ ,\\
&\qquad \text{with}\quad \mc{A}^{(1)}_1 = \frac{2\vev{11'}\vev{2'2}[32'][1'4]}{s_{24}} \ ,\qquad
\mc{A}^{(1)}_2 = \frac{2\vev{11'}\vev{2'2}[32'][1'4]}{s_{1'2}} \ .
\end{split}\label{eq:ALAR1}
\end{align}
The other two diagrams are related by conjugate, and yield the same integrand
\begin{align}
\begin{split}
&\quad\tikzset{every picture/.style={line width=0.75pt}} 
\begin{tikzpicture}[x=0.75pt,y=0.75pt,yscale=-1,xscale=1]

\draw    (123.55,49.28) -- (161.23,81.36) ;
\draw [shift={(146.19,68.56)}, rotate = 220.41] [fill={rgb, 255:red, 0; green, 0; blue, 0 }  ][line width=0.08]  [draw opacity=0] (8.93,-4.29) -- (0,0) -- (8.93,4.29) -- cycle    ;
\draw  [fill={rgb, 255:red, 18; green, 47; blue, 210 }  ,fill opacity=1 ] (157.33,81.36) .. controls (157.33,79.14) and (159.07,77.34) .. (161.23,77.34) .. controls (163.38,77.34) and (165.12,79.14) .. (165.12,81.36) .. controls (165.12,83.58) and (163.38,85.38) .. (161.23,85.38) .. controls (159.07,85.38) and (157.33,83.58) .. (157.33,81.36) -- cycle ;
\draw    (161.23,81.36) -- (127.44,112.83) ;
\draw [shift={(140.68,100.51)}, rotate = 317.02] [fill={rgb, 255:red, 0; green, 0; blue, 0 }  ][line width=0.08]  [draw opacity=0] (8.93,-4.29) -- (0,0) -- (8.93,4.29) -- cycle    ;
\draw  [draw opacity=0] (160.57,82.11) .. controls (175.09,65.01) and (197.76,54.7) .. (222.53,55.81) -- (220.35,127.83) -- cycle ; \draw    (160.57,82.11) .. controls (174.51,65.7) and (195.96,55.53) .. (219.57,55.73) ; \draw [shift={(222.53,55.81)}, rotate = 177.87] [fill={rgb, 255:red, 0; green, 0; blue, 0 }  ][line width=0.08]  [draw opacity=0] (8.93,-4.29) -- (0,0) -- (8.93,4.29) -- cycle    ; 
\draw  [draw opacity=0] (262.38,100.87) .. controls (247.8,109.84) and (229.82,113.46) .. (211.7,110.57) -- (219.17,38.4) -- cycle ; \draw    (262.38,100.87) .. controls (248.53,109.39) and (231.61,113.09) .. (214.42,110.95) ; \draw [shift={(211.7,110.57)}, rotate = 4.27] [fill={rgb, 255:red, 0; green, 0; blue, 0 }  ][line width=0.08]  [draw opacity=0] (8.93,-4.29) -- (0,0) -- (8.93,4.29) -- cycle    ; 
\draw  [fill={rgb, 255:red, 198; green, 198; blue, 198 }  ,fill opacity=1 ] (251.53,84.77) .. controls (251.53,75.53) and (258.8,68.03) .. (267.77,68.03) .. controls (276.74,68.03) and (284.01,75.53) .. (284.01,84.77) .. controls (284.01,94.02) and (276.74,101.52) .. (267.77,101.52) .. controls (258.8,101.52) and (251.53,94.02) .. (251.53,84.77) -- cycle ;
\draw    (300.25,57.92) -- (279.46,72.65) ;
\draw [shift={(285.78,68.18)}, rotate = 324.67] [fill={rgb, 255:red, 0; green, 0; blue, 0 }  ][line width=0.08]  [draw opacity=0] (8.93,-4.29) -- (0,0) -- (8.93,4.29) -- cycle    ;
\draw    (276.86,98.1) -- (300.25,114.84) ;
\draw [shift={(292.62,109.38)}, rotate = 215.6] [fill={rgb, 255:red, 0; green, 0; blue, 0 }  ][line width=0.08]  [draw opacity=0] (8.93,-4.29) -- (0,0) -- (8.93,4.29) -- cycle    ;
\draw [color={rgb, 255:red, 74; green, 144; blue, 226 }  ,draw opacity=1 ]   (201.5,51.96) -- (227.44,52.52) ;
\draw [shift={(229.44,52.56)}, rotate = 181.24] [color={rgb, 255:red, 74; green, 144; blue, 226 }  ,draw opacity=1 ][line width=0.75]    (10.93,-3.29) .. controls (6.95,-1.4) and (3.31,-0.3) .. (0,0) .. controls (3.31,0.3) and (6.95,1.4) .. (10.93,3.29)   ;
\draw [color={rgb, 255:red, 74; green, 144; blue, 226 }  ,draw opacity=1 ]   (205.4,114.91) -- (231.34,115.47) ;
\draw [shift={(233.34,115.51)}, rotate = 181.24] [color={rgb, 255:red, 74; green, 144; blue, 226 }  ,draw opacity=1 ][line width=0.75]    (10.93,-3.29) .. controls (6.95,-1.4) and (3.31,-0.3) .. (0,0) .. controls (3.31,0.3) and (6.95,1.4) .. (10.93,3.29)   ;
\draw    (161.23,118.86) -- (161.23,94.74) ;
\draw [shift={(161.23,92.74)}, rotate = 90] [color={rgb, 255:red, 0; green, 0; blue, 0 }  ][line width=0.75]    (10.93,-3.29) .. controls (6.95,-1.4) and (3.31,-0.3) .. (0,0) .. controls (3.31,0.3) and (6.95,1.4) .. (10.93,3.29)   ;
\draw    (270.37,126.9) -- (270.37,107.47) ;
\draw [shift={(270.37,105.47)}, rotate = 90] [color={rgb, 255:red, 0; green, 0; blue, 0 }  ][line width=0.75]    (10.93,-3.29) .. controls (6.95,-1.4) and (3.31,-0.3) .. (0,0) .. controls (3.31,0.3) and (6.95,1.4) .. (10.93,3.29)   ;
\draw [color={rgb, 255:red, 74; green, 144; blue, 226 }  ,draw opacity=1 ]   (121.38,53.23) -- (138.28,67.99) ;
\draw [shift={(139.79,69.3)}, rotate = 221.13] [color={rgb, 255:red, 74; green, 144; blue, 226 }  ,draw opacity=1 ][line width=0.75]    (10.93,-3.29) .. controls (6.95,-1.4) and (3.31,-0.3) .. (0,0) .. controls (3.31,0.3) and (6.95,1.4) .. (10.93,3.29)   ;
\draw [color={rgb, 255:red, 74; green, 144; blue, 226 }  ,draw opacity=1 ]   (127.23,116.85) -- (143.92,102.1) ;
\draw [shift={(145.42,100.78)}, rotate = 138.54] [color={rgb, 255:red, 74; green, 144; blue, 226 }  ,draw opacity=1 ][line width=0.75]    (10.93,-3.29) .. controls (6.95,-1.4) and (3.31,-0.3) .. (0,0) .. controls (3.31,0.3) and (6.95,1.4) .. (10.93,3.29)   ;
\draw [color={rgb, 255:red, 74; green, 144; blue, 226 }  ,draw opacity=1 ]   (296.79,57.25) -- (282.81,67.46) ;
\draw [shift={(281.19,68.63)}, rotate = 323.86] [color={rgb, 255:red, 74; green, 144; blue, 226 }  ,draw opacity=1 ][line width=0.75]    (10.93,-3.29) .. controls (6.95,-1.4) and (3.31,-0.3) .. (0,0) .. controls (3.31,0.3) and (6.95,1.4) .. (10.93,3.29)   ;
\draw [color={rgb, 255:red, 74; green, 144; blue, 226 }  ,draw opacity=1 ]   (296.14,116.18) -- (281.5,105.32) ;
\draw [shift={(279.89,104.13)}, rotate = 36.58] [color={rgb, 255:red, 74; green, 144; blue, 226 }  ,draw opacity=1 ][line width=0.75]    (10.93,-3.29) .. controls (6.95,-1.4) and (3.31,-0.3) .. (0,0) .. controls (3.31,0.3) and (6.95,1.4) .. (10.93,3.29)   ;
\draw  [draw opacity=0] (221.39,55.77) .. controls (223.43,55.83) and (225.48,55.97) .. (227.55,56.19) .. controls (241.06,57.64) and (253.42,62.33) .. (263.82,69.35) -- (220.35,127.83) -- cycle ; \draw    (221.39,55.77) .. controls (223.43,55.83) and (225.48,55.97) .. (227.55,56.19) .. controls (241.06,57.64) and (253.42,62.33) .. (263.82,69.35) ;  
\draw  [draw opacity=0] (215.7,111.1) .. controls (207.5,110.22) and (199.22,108.02) .. (191.17,104.38) .. controls (179.61,99.17) and (169.83,91.6) .. (162.23,82.57) -- (219.17,38.4) -- cycle ; \draw    (215.7,111.1) .. controls (207.5,110.22) and (199.22,108.02) .. (191.17,104.38) .. controls (179.61,99.17) and (169.83,91.6) .. (162.23,82.57) ;  

\draw (207.8,35.71) node [anchor=north west][inner sep=0.75pt]  [font=\small,color={rgb, 255:red, 74; green, 144; blue, 226 }  ,opacity=1 ]  {$p'_{1}$};
\draw (208.45,114.06) node [anchor=north west][inner sep=0.75pt]  [font=\small,color={rgb, 255:red, 74; green, 144; blue, 226 }  ,opacity=1 ]  {$p'_{2}$};
\draw (255.32,122.67) node [anchor=north west][inner sep=0.75pt]  [font=\normalsize] [align=left] {SM};
\draw (129.24,40.34) node [anchor=north west][inner sep=0.75pt]    {$L$};
\draw (117.19,96.54) node [anchor=north west][inner sep=0.75pt]    {$L$};
\draw (144.3,119.43) node [anchor=north west][inner sep=0.75pt]    {$\mathcal{O}_{L^{2} L^{\dagger 2}}$};
\draw (170.59,52.84) node [anchor=north west][inner sep=0.75pt]    {$L^{\dagger }$};
\draw (174.14,82.26) node [anchor=north west][inner sep=0.75pt]    {$L^{\dagger }$};
\draw (247.12,46.98) node [anchor=north west][inner sep=0.75pt]    {$L$};
\draw (242.28,93.25) node [anchor=north west][inner sep=0.75pt]    {$L$};
\draw (303.12,50.16) node [anchor=north west][inner sep=0.75pt]    {$L^{\dagger }$};
\draw (300.17,106.36) node [anchor=north west][inner sep=0.75pt]    {$L^{\dagger }$};
\draw (115.55,53.79) node [anchor=north west][inner sep=0.75pt]  [font=\small,color={rgb, 255:red, 74; green, 144; blue, 226 }  ,opacity=1 ]  {$p_{1}$};
\draw (122.04,112.72) node [anchor=north west][inner sep=0.75pt]  [font=\small,color={rgb, 255:red, 74; green, 144; blue, 226 }  ,opacity=1 ]  {$p_{2}$};
\draw (293.55,41.06) node [anchor=north west][inner sep=0.75pt]  [font=\small,color={rgb, 255:red, 74; green, 144; blue, 226 }  ,opacity=1 ]  {$p_{3}$};
\draw (289,114.06) node [anchor=north west][inner sep=0.75pt]  [font=\small,color={rgb, 255:red, 74; green, 144; blue, 226 }  ,opacity=1 ]  {$p_{4}$};
\draw (20,65) node [anchor=north west][inner sep=0.75pt]    {$(\mathcal{A}_{L}\mathcal{A}_{R})^{(2)} \equiv$};
\end{tikzpicture}\\
&=\frac12\mc{A}_{\text{eff}}\left(L_{f_1i_1}(p_1),L_{f_{2}i_{2}}(p_2),L^{\dagger i'_1}_{f'_1}(-p'_1),L^{\dagger i'_{2}}_{f'_{2}}(-p'_2)\right)\mc{A}_{\text{SM}}\left(L_{f'_{2}i'_{2}}(p'_2),L{}_{f'_1i'_1}(p'_1),L^{\dagger}{}^{i_{3}}_{f_{3}}(p_3),L^{\dagger i_4}{}_{f_4}(p_4)\right)\\
& = \mc{A}^{(2)} \left[(g_1^2y_l^2+\frac14g_2^2)C^{f_1f_2f_3f_4}_{L^2_{[2]}L^{\dagger}{}^2_{[2]}}T_1 + (g_1^2y_l^2-\frac34g_2^2)C^{f_1f_2f_3f_4}_{L^2_{[1,1]}L^{\dagger}{}^2_{[1,1]}}T_2\right] \\
&\qquad \text{with}\quad \mc{A}^{(2)} = 2\vev{12}[34]s_{34}\left(\frac{1}{s_{1'3}}+\frac{1}{s_{2'3}}\right) \ .
\end{split}\label{eq:ALAR2}
\end{align}
The negative sign in $(\mc{A}_L\mc{A}_R)^{(1)}$ arises from the fermion trace and the factor of $\frac12$ in $(\mc{A}_L\mc{A}_R)^{(2)}$ are symmetry factors due to the identical particles in the loop. 
To get the UV divergence, we compute the IR regulated phase space integral of them with eq.~\eqref{eq:b_master} applied on the kinematic factors $\mc{A}^{(1,2)}$. For example, the first integrand yields
\eq{
    &\frac{2}{\pi}\sum_{1'2'}\int {\rm d}\Phi_{1'2'}\oint_{\mathcal{C}_\infty} \frac{{\rm d}w}{w} \mc{A}^{(1)}_1 = \frac{2}{3}\vev{12}[34] \ ,\qquad
    \frac{2}{\pi}\sum_{1'2'}\int {\rm d}\Phi_{1'2'}\oint_{\mathcal{C}_\infty} \frac{{\rm d}w}{w} \mc{A}^{(1)}_2 = 3\vev{12}[34] \ ,\\
    &\Rightarrow\qquad b_{(13)} = \frac{2}{\pi}\sum_{1'2'}\int {\rm d}\Phi_{1'2'}\oint_{\mathcal{C}_\infty} \frac{{\rm d}w}{w} (\mc{A}_L\mc{A}_R)^{(1)}\\
    &\qquad\qquad\quad\ = \begin{pmatrix}
    \delta_{f_2f_4}C_{L^2_{[2]}L^{\dagger}{}^2_{[2]}}^{f_1f'f_3f'}, \delta_{f_2f_4}C_{L^2_{[1,1]}L^{\dagger}{}^2_{[1,1]}}^{f_1f'f_3f'}
    \end{pmatrix}\begin{pmatrix}
    g_1^2y_l^2+\frac1{12}g_2^2 & g_1^2y_l^2-\frac14g_2^2 \\
    \frac13g_1^2y_l^2-\frac1{12}g_2^2 & \frac13g_1^2y_l^2+\frac14g_2^2 
    \end{pmatrix} \begin{pmatrix}
    2\vev{12}[34]T_1 \\
    2\vev{12}[34]T_2
    \end{pmatrix} \\
    & \hspace{8em} + \begin{pmatrix}
    C^{f_1f_2f_3f_4}_{L^2_{[2]}L^{\dagger}{}^2_{[2]}}, C^{f_1f_2f_3f_4}_{L^2_{[1,1]}L^{\dagger}{}^2_{[1,1]}}
    \end{pmatrix}\begin{pmatrix}
    3g_1^2y_l^2+\frac32g_2^2 & \frac94g_2^2 \\
    \frac34g_2^2 & 3g_1^2y_l^2
    \end{pmatrix} \begin{pmatrix}
    2\vev{12}[34]T_1 \\
    2\vev{12}[34]T_2
    \end{pmatrix} \\
    &\qquad\qquad\quad\ \equiv 2\vev{12}[34](T_1\mc{K}^{bf}_{1;(13)} + T_2\mc{K}^{bf}_{2;(13)})\ .
}
The final result has exactly the same structure as the tree-level amplitudes in eq.~\eqref{eq:Fb:L4}, where we identify the $(13)$-channel contributions to the columns of the matrix $\mc{K}^{bf}_{j} \equiv \sum_i C_i\mc{K}^{bf}_{ij}$.
Sum over the 4 channels with the same diagram, which are simply permutations of the result:
\eq{
    b_{(13)} + b_{(14)} + b_{(23)} + b_{(24)} &= b_{(13)} - (1\leftrightarrow2) - (3\leftrightarrow4) + (1,2\leftrightarrow3,4) \\
    &= 2\vev{12}[34] \times \left(T_1\mc{Y}\left[\tiny{\young(12)},\tiny{\young(34)}\right]\mc{K}^{bf}_{1;(13)} + T_2 \mc{Y}\left[\tiny{\young(1,2)},\tiny{\young(3,4)}\right]\mc{K}^{bf}_{2;(13)}\right) \ .
}
Note that the permutations of fermions represented by the Young symmetrizers yield a minus sign. The kinematic factor $2\vev{12}[34]$ and the tensors $T_{1,2}$ have definite symmetries; thus, the Young symmetrizer ultimately acts on the coefficient structures with modified symmetries. For notational convenience, we make use of the symmetry \eqref{eq:frelation} to simplify and rename some of the symmetrized flavor tensors as
\eqs{
    \mc{Y}\left[\tiny{\young(12)},\tiny{\young(34)}\right]\delta_{f_2f_4}C_{L^2_{[2]}L^{\dagger}{}^2_{[2]}}^{f_1f'f_3f'} &\equiv 4\tilde{C}_{L^2_{[2]}L^{\dagger}{}^2_{[2]},1'}^{f_1f_2f_3f_4} \ ,&
    \mc{Y}\left[\tiny{\young(12)},\tiny{\young(34)}\right]\delta_{f_2f_4}C_{L^2_{[1,1]}L^{\dagger}{}^2_{[1,1]}}^{f_1f'f_3f'} &\equiv 4\tilde{C}_{L^2_{[2]}L^{\dagger}{}^2_{[2]},2'}^{f_1f_2f_3f_4} \ ,\label{eq:symC1}\\
    \mc{Y}\left[\tiny{\young(12)},\tiny{\young(34)}\right]C^{f_1f_2f_3f_4}_{L^2_{[2]}L^{\dagger}{}^2_{[2]}} &= 4C^{f_1f_2f_3f_4}_{L^2_{[2]}L^{\dagger}{}^2_{[2]}} \ ,&
    \mc{Y}\left[\tiny{\young(12)},\tiny{\young(34)}\right]C^{f_1f_2f_3f_4}_{L^2_{[1,1]}L^{\dagger}{}^2_{[1,1]}} &= 0 \ , \\
    \mc{Y}\left[\tiny{\young(1,2)},\tiny{\young(3,4)}\right]\delta_{f_2f_4}C_{L^2_{[2]}L^{\dagger}{}^2_{[2]}}^{f_1f'f_3f'} &\equiv 4\tilde{C}_{L^2_{[1,1]}L^{\dagger}{}^2_{[1,1]},1'}^{f_1f_2f_3f_4} \ ,&
    \mc{Y}\left[\tiny{\young(1,2)},\tiny{\young(3,4)}\right]\delta_{f_2f_4}C_{L^2_{[1,1]}L^{\dagger}{}^2_{[1,1]}}^{f_1f'f_3f'} &\equiv 4\tilde{C}_{L^2_{[1,1]}L^{\dagger}{}^2_{[1,1]},2'}^{f_1f_2f_3f_4} \ ,\\
    \mc{Y}\left[\tiny{\young(1,2)},\tiny{\young(3,4)}\right]C^{f_1f_2f_3f_4}_{L^2_{[2]}L^{\dagger}{}^2_{[2]}} &= 0 \ ,&
    \mc{Y}\left[\tiny{\young(1,2)},\tiny{\young(3,4)}\right]C^{f_1f_2f_3f_4}_{L^2_{[1,1]}L^{\dagger}{}^2_{[1,1]}} &= 4C^{f_1f_2f_3f_4}_{L^2_{[1,1]}L^{\dagger}{}^2_{[1,1]}} \ .\label{eq:symC2}
}
Note that the subscripts in the renamed result indicate the true symmetries of the flavor indices, which may be different from those of the contributing Wilson coefficient tensors on the left. The other two diagrams both contribute to the channel $b_{(12)}$, but vanish after the IR regulated phase space integration
\eq{
    b_{(12)} = \frac{2}{\pi}\sum_{1'2'}\int {\rm d}\Phi_{1'2'}\oint_{\mathcal{C}_\infty} \frac{{\rm d}w}{w} \mc{A}^{(2)} = 0 \ .
}
This can also be directly deduced, as the kinematic terms involving the loop momentum among them are either $1/t$ or $1/u$, which are precisely the infrared divergent terms. 

In summary, we have the UV contribution to the anomalous dimensions:
\eq{
    \gamma^{\rm UV}_1 &= \frac{1}{8\pi^2}\mc{Y}\left[\tiny{\young(12)},\tiny{\young(34)}\right]\mc{K}^{bf}_{1;(13)} \\
    &= \frac{1}{8\pi^2}\left[ 
    (12g_1^2y_l^2+6g_2^2)C_{L^2_{[2]}L^{\dagger}{}^2_{[2]}}^{f_1f_2f_3f_4}+
    (4g_1^2y_l^2+\frac13g_2^2)\tilde{C}_{L^2_{[2]}L^{\dagger}{}^2_{[2]},1'}^{f_1f_2f_3f_4}+
    (\frac43g_1^2y_l^2-\frac13g_2^2)\tilde{C}_{L^2_{[2]}L^{\dagger}{}^2_{[2]},2'}^{f_1f_2f_3f_4} \right] \\
    \gamma^{\rm UV}_2 &= \frac{1}{8\pi^2}\mc{Y}\left[\tiny{\young(1,2)},\tiny{\young(3,4)}\right]\mc{K}^{bf}_{2;(13)} \\
    &= \frac{1}{8\pi^2}\left[ 
    12g_1^2y_l^2 C_{L^2_{[1,1]}L^{\dagger}{}^2_{[1,1]}}^{f_1f_2f_3f_4}+
    (4g_1^2y_l^2-g_2^2) \tilde{C}_{L^2_{[1,1]}L^{\dagger}{}^2_{[1,1]},1'}^{f_1f_2f_3f_4}+
    (\frac43g_1^2y_l^2+g_2^2) \tilde{C}_{L^2_{[1,1]}L^{\dagger}{}^2_{[1,1]},2'}^{f_1f_2f_3f_4} \right]\ .
}

\subsection{Infrared Collinear Anomalous Dimensions in the Standard Model}
Although unitarity cuts provide an efficient way to extract ultraviolet divergences, they cannot be applied directly to determine collinear infrared divergences. This is because the coefficients of massless bubble integrals, corresponding to vanishing total momentum $(p_1'+p_2')^2=0$, are not accessible through standard unitarity cuts.
Nevertheless, collinear divergences are universal and depend only on the quantum numbers of the external fields. The one-loop collinear anomalous dimensions have been extensively studied and can be extracted from the corresponding tree-level amplitudes.
At one loop, the collinear IR divergence can be parametrized as
\begin{align}
    \mathcal{A}^{\text{1-loop}}_{\text{coll}}=-\frac{1}{(4\pi)^2}\sum_a\frac{\gamma^a_{\text{coll}}}{\epsilon}\mathcal{A}^{\text{tree}}.
\end{align}
We sum over all external legs' collinear factors and attach the tree amplitudes. Below are the collinear factors for all the SM fields,
\begin{align}
    \gamma^{H}_{\text{coll}}&=\gamma^{H^{\dagger}}_{\text{coll}}=-4g_1^2y_h^2-4C_2(2)g_2^2+{\rm Tr}\left[N_cY_u^{\dagger}Y_u+N_cY_d^{\dagger}Y_d+Y_e^{\dagger}Y_e\right],\\
    \gamma^{B}_{\text{coll}}&=\frac{2g_1^2}{3}\left(N_fN_c(2y_q^2+y_u^2+y_d^2)+N_f(2y_l^2+y_e^2)+y_h^2 \right),\\
    \gamma^{W}_{\text{coll}}&=-g_2^2\left(2\times\frac{11}{3}-\frac13\left(N_fN_c+N_f+\frac12\right) \right),\\
    \gamma^{G}_{\text{coll}}&=-g_3^2\left(3\times\frac{11}{3}-\frac43N_f \right),\\
    \left(\gamma^{l}_{\text{coll}}\right)_{pr}&=-3\left(y_l^2g_1^2+g_2^2C_2(2)\right)\delta_{pr}+\frac12\left[Y_e^{\dagger}Y_e\right]_{pr},\\
    \left(\gamma^{e}_{\text{coll}}\right)_{pr}&=-3y_e^2g_1^2\delta_{pr}+\left[Y_eY_e^{\dagger}\right]_{pr},\\
    \left(\gamma^q_{\text{coll}}\right)_{pr}&=-3(g_3^2C_2(N_c)+g_2^2C_2(2)+y_q^2g_1^2)\delta_{pr}-\frac12[Y_u^{\dagger}Y_u+Y_d^{\dagger}Y_d]_{pr},\\
    \left(\gamma^{u}_{\text{coll}}\right)_{pr}&=-3\left(y_u^2g_1^2+g_2^2C_2(2)\right)\delta_{pr}+\left[Y_uY_u^{\dagger}\right]_{pr},\\
    \left(\gamma^{d}_{\text{coll}}\right)_{pr}&=-3\left(y_d^2g_1^2+g_2^2C_2(2)\right)\delta_{pr}+\left[Y_dY_d^{\dagger}\right]_{pr},
\end{align}
where $C_2(N)=\frac{N^2-1}{2N}$. The subscript of fermion collinear factors indicates the contraction of flavor indices, as shown in the example below,
\begin{align}
C^{\; prst}_{L^2_{[2]}L^{\dagger}{}^2_{[2]}}\stackrel{\text{collinear IR}}{\longrightarrow} 2(\gamma^l_{\text{coll}})_{ss'}C^{\; prs't}_{L^2_{[2]}L^{\dagger}{}^2_{[2]}} +2C^{\; p'rst}_{L^2_{[2]}L^{\dagger}{}^2_{[2]}}(\gamma^l_{\text{coll}})_{p'p}.
\end{align}
Whether the flavor index is contracted with the first or second index of $\gamma_{\text{coll}}$ depends on whether the external leg corresponds to a particle or an antiparticle. The flavor indices in the superscript of the Wilson coefficient follow the same order as the fermions in the subscript.
 \begin{align}\begin{split}
&\dot{C}_{L^2_{[2]}L^{\dagger}{}^2_{[2]}}^{f_1f_2f_3f_4} \supset (12g_1^2y_l^2+3g_2^2)C_{L^2_{[2]}L^{\dagger}{}^2_{[2]}}^{f_1f_2f_3f_4}+(8g_1^2y_l^2+\frac23g_2^2)\tilde{C}_{L^2_{[2]}L^{\dagger}{}^2_{[2]},1'}^{f_1f_2f_3f_4}+(\frac83g_1^2y_l^2-\frac23g_2^2)\tilde{C}_{L^2_{[2]}L^{\dagger}{}^2_{[2]},2'}^{f_1f_2f_3f_4}\\
&\qquad\qquad\qquad + [Y_e^{\dagger}Y_e]_{f'f_1}C_{L^2_{[2]}L^{\dagger}{}^2_{[2]}}^{f'f_2f_3f_4}+ [Y_e^{\dagger}Y_e]_{f_3f'}C_{L^2_{[2]}L^{\dagger}{}^2_{[2]}}^{f_1f_2f'f_4},
\end{split}\\
\begin{split}
&\dot{C}_{L^2_{[1,1]}L^{\dagger}{}^2_{[1,1]}}^{f_1f_2f_3f_4} \supset (12g_1^2y_l^2-9g_2^2)C_{L^2_{[1,1]}L^{\dagger}{}^2_{[1,1]}}^{f_1f_2f_3f_4}+(8g_1^2y_l^2-2g_2^2)\tilde{C}_{L^2_{[1,1]}L^{\dagger}{}^2_{[1,1]},1'}^{f_1f_2f_3f_4}+(\frac83g_1^2y_l^2+2g_2^2)\tilde{C}_{L^2_{[1,1]}L^{\dagger}{}^2_{[1,1]},2'}^{f_1f_2f_3f_4}\\
&\qquad\qquad\qquad + [Y_e^{\dagger}Y_e]_{f'f_1}C_{L^2_{[1,1]}L^{\dagger}{}^2_{[1,1]}}^{f'f_2f_3f_4}+ [Y_e^{\dagger}Y_e]_{f_3f'}C_{L^2_{[1,1]}L^{\dagger}{}^2_{[1,1]}}^{f_1f_2f'f_4}.
\end{split}\end{align}
Where the $\tilde{C}$ tensors are defined in eqs.~(\ref{eq:symC1}-\ref{eq:symC2}) and the $\supset$ is used to indicate that there are other loop contributions in the full SMEFT RGE. As a concrete example of the RGE flavor components, we have
\eq{
    &\dot{C}_{L^2_{[2]}L^{\dagger}{}^2_{[2]}}^{1112} \supset (12g_1^2y_l^2+3g_2^2)C_{L^2_{[2]}L^{\dagger}{}^2_{[2]}}^{1112} 
    +\frac{1}{2}(8g_1^2y_l^2+\frac23g_2^2)C_{L^2_{[2]}L^{\dagger}{}^2_{[2]}}^{1f2f}+\frac{1}{2}(8g_1^2y_l^2-\frac23g_2^2)C_{L^2_{[1,1]}L^{\dagger}{}^2_{[1,1]}}^{1323}\\
    &\qquad\qquad\qquad + [Y_e^{\dagger}Y_e]_{f1}C_{L^2_{[2]}L^{\dagger}{}^2_{[2]}}^{f112} + \frac{1}{2}[Y_e^{\dagger}Y_e]_{1f}C_{L^2_{[2]}L^{\dagger}{}^2_{[2]}}^{11f2} + \frac{1}{2}[Y_e^{\dagger}Y_e]_{2f}C_{L^2_{[2]}L^{\dagger}{}^2_{[2]}}^{111f},
}
Here we derived the $\tilde{C}$ tensor as
\eq{
    \tilde{C}_{L^2_{[2]}L^{\dagger}{}^2_{[2]},1'}^{1112} &= \frac{1}{2}\left(\delta_{12}C_{L^2_{[2]}L^{\dagger}{}^2_{[2]}}^{1f1f} + \delta_{11}C_{L^2_{[2]}L^{\dagger}{}^2_{[2]}}^{1f2f} \right) = \frac{1}{2}C_{L^2_{[2]}L^{\dagger}{}^2_{[2]}}^{1f2f} \ ,\\
    \tilde{C}_{L^2_{[2]}L^{\dagger}{}^2_{[2]},2'}^{1112} &= \frac{1}{2}\left(\delta_{12}C_{L^2_{[1,1]}L^{\dagger}{}^2_{[1,1]}}^{1f1f} + \delta_{11}C_{L^2_{[1,1]}L^{\dagger}{}^2_{[1,1]}}^{1f2f} \right) = \frac{1}{2}C_{L^2_{[1,1]}L^{\dagger}{}^2_{[1,1]}}^{1323} \ .
}
where in the second tensor we sum over $f=1,2,3$ and only $f=3$ survives the anti-symmetry of $C_{L^2_{[1,1]}L^{\dagger}{}^2_{[1,1]}}$.

\section{The RGEs for the Four-Fermion Operators}
We provide the complete one-loop renormalization group equations for all four-fermion operators up to dimension-8 in the Standard Model Effective Field Theory.
The full results are contained in the accompanying \textbf{Mathematica} file \textbf{RGEresultfor4fermions.m}, stored as a hierarchical Association with four top-level keys:
\begin{center}
    \{"dim-6", "dim-7", "dim-8", "quadratic terms"\}.
\end{center}
These correspond respectively to the renormalization among dimension-6 operators, dimension-7 operators, dimension-8 operators, and the contributions from dimension-5, dimension-6 and dimension-7 operators to the running of dimension-8 operators. As a consistency check, our dimension-6 and dimension-7 results reproduce all previous calculations in Refs. \cite{Jenkins:2013wua,Alonso:2013hga,Liao:2016hru}. 
Furthermore, we have compared the full dimension-8 running with the recent independent computation of Ref. \cite{Boughezal:2024zqa}, and we find complete agreement for the parts where the same operator basis is used.

The second layer of keys in the Mathematica Association corresponds to the operator types, labeled according to the particle content introduced in Section \ref{sec:f-basis}.
For illustration, we display below the output for the renormalization of one particular dimension-6 operator type $L^2L^{\dagger 2}$:
\begin{align}\begin{split}
    \texttt{ In[1]:=}&\;\mathbf{RGEresult}=\mathbf{Import}[\texttt{"file path/RGEresultfor4fermions.m"}];\\
    \texttt{In[2]:=}&\;\mathbf{RGEresult}[\texttt{"dim-6"},\texttt{"L"}^2\texttt{"L}\dagger\texttt{"}^2]\\
    \texttt{Out[2]:=}&\;\Big\{\dot{C}^{\,prst}_{\{L^2_{y[2]}L\dagger^2_{y[2]},1 \}}\to -\frac23 g^2_1\delta_{ps}C^{\,rt}_{\{L\,H\,H\dagger\,L\dagger\,D,1 \}}+\frac{2}{3}g^2_2\delta_{ps}C^{\,rt}_{\{L\,H\,H\dagger\,L\dagger\,D,1 \}}-\frac43 g^2_1\delta_{ps}C^{\,rt}_{\{L\,H\,H\dagger\,L\dagger\,D,2 \}}-\\
    &\quad \frac83g^2_1\delta_{ps}C^{\,urut}_{\{dc\,L\,dc\dagger\,L\dagger,1 \}} -\frac83g^2_1\delta_{ps}C^{\,urut}_{\{ec\,L\,ec\dagger\,L\dagger,1 \}} -\frac83g^2_1\delta_{ps}C^{\,rutu}_{\{L\,Q\,L\dagger\,Q\dagger,1 \}}-\frac43g^2_1\delta_{ps}C^{\,rutu}_{\{L\,Q\,L\dagger\,Q\dagger,2 \}}+\\
    &\cdots
\end{split}\end{align}
In the expression above, the flavor indices carried by the Wilson coefficients follow exactly the ordering of the fermion fields in the operator label. For operators containing repeated fermionic fields, the Wilson coefficients must transform as irreducible flavor tensors of the corresponding symmetric group. We indicate this by attaching Young diagram labels on the subscript, in direct analogy with the construction of the p-basis and f-basis in section \ref{sec:f-basis}.

This symmetry assignment is crucial for renormalization.
Since the operators themselves form irreducible representations of the flavor-permutation group, the anomalous-dimension matrix must preserve the same Young symmetry. In practice, once the Wilson coefficients are projected onto some flavor-index symmetry, the resulting anomalous dimension inherit exactly the same symmetry as in eqs.~(\ref{eq:symC1}-\ref{eq:symC2}). 
All results presented below and in the \textbf{Mathematica} file follow this symmetry-projected convention.

In the subsections that follow, we list the complete one-loop RGEs for all dimension-8 four-fermion operators.
The various operator types can be efficiently located using Table \ref{tab:psi4}, which classifies all $\psi^4$ structures according to baryon and lepton number assignments.
Tables \ref{tab:psi2XH}, \ref{tab:psi2H2}, and \ref{tab:psi2F2} summarize the additional operators entering intermediate unitarity cuts that contribute to the renormalization of dimension-8 four-fermion operators.
Throughout, the superscripts $(d6)$ and $(d8)$ on Wilson coefficients indicate the mass dimension of the corresponding operators.

We now begin listing the explicit renormalization equations, starting with operators of type $Q^2Q^{\dagger 2}$.

\begin{table}[htbp]
\begin{align*}
\begin{array}{|c|c|c|c|c|c||c|c|}
\hline\hline
\multicolumn{8}{|c|}{\text{All types for }\psi^4} \\
\hline
\multicolumn{6}{|c|}{|\Delta B|=0, |\Delta L|=0} & \multicolumn{2}{|c|}{|\Delta B|=1, |\Delta L|=1}\\
\hline
Q^2Q^{\dagger 2} & \text{Tab. }\ref{tab:Q2Q2} & QQ^{\dagger }u_{\mathbb{C}}u_{\mathbb{C}}^{\dagger} & \text{Tab. }\ref{tab:Q2u2} & QQ^{\dagger }d_{\mathbb{C}}d_{\mathbb{C}}^{\dagger} & \text{Tab. }\ref{tab:Q2d2} & Q^3L & \text{Tab. }\ref{tab:LQ3}\\
u_{\mathbb{C}}{}^2u_{\mathbb{C}}^{\dagger 2} & \text{Tab. }\ref{tab:u2u2} & d_{\mathbb{C}}{}^2d_{\mathbb{C}}^{\dagger 2} & \text{Tab. }\ref{tab:d2d2} &  u_{\mathbb{C}}u_{\mathbb{C}}^{\dagger }d_{\mathbb{C}}d_{\mathbb{C}}^{\dagger} & \text{Tab. }\ref{tab:udud} & Q^{\dagger 2}u_{\mathbb{C}}e_{\mathbb{C}} & \text{Tab. }\ref{tab:Q2ue}\\
Q^2u_{\mathbb{C}}d_{\mathbb{C}} & \text{Tab. }\ref{tab:Q2ud} & QQ^{\dagger}LL^{\dagger} & \text{Tab. }\ref{tab:LLQQ} & QQ^{\dagger}e_{\mathbb{C}}e_{\mathbb{C}}^{\dagger} & \text{Tab. }\ref{tab:QQee} & Q^{\dagger}L^{\dagger}u_{\mathbb{C}}d_{\mathbb{C}} & \text{Tab. }\ref{tab:LQud} \\
u_{\mathbb{C}}u_{\mathbb{C}}^{\dagger}LL^{\dagger} & \text{Tab. }\ref{tab:uull} & u_{\mathbb{C}}u_{\mathbb{C}}^{\dagger}e_{\mathbb{C}}e_{\mathbb{C}}^{\dagger} & \text{Tab. }\ref{tab:uuee} &  d_{\mathbb{C}}d_{\mathbb{C}}^{\dagger}LL^{\dagger} & \text{Tab. }\ref{tab:ddll} & u_{\mathbb{C}}{}^2d_{\mathbb{C}}e_{\mathbb{C}} & \text{Tab. }\ref{tab:u2de} \\
d_{\mathbb{C}}d_{\mathbb{C}}^{\dagger}e_{\mathbb{C}}e_{\mathbb{C}}^{\dagger} & \text{Tab. }\ref{tab:ddee} & Qu_{\mathbb{C}}Le_{\mathbb{C}} & \text{Tab. }\ref{tab:LQue} & Q^{\dagger}d_{\mathbb{C}}^{\dagger}Le_{\mathbb{C}}^{\dagger} & \text{Tab. }\ref{tab:LQde} & & \\
L^2L^{\dagger 2} & \text{Tab. }\ref{tab:L2L2} & LL^{\dagger}e_{\mathbb{C}}e_{\mathbb{C}}^{\dagger} & \text{Tab. }\ref{tab:LLee} & e_{\mathbb{C}}{}^2e_{\mathbb{C}}^{\dagger 2} & \text{Tab. }\ref{tab:e2e2} & & \\
\hline
\end{array}
\end{align*}
\caption{Classification of all four-fermion operator types at dimension-6 and dimension-8, grouped according to their baryon- and lepton-number assignments. The $B$ and $L$ violating operators were not included in Ref.~\cite{Boughezal:2024zqa}.
The table also serves as an index for locating the explicit RGEs in Section 4.}
\label{tab:psi4}
\end{table}

\begin{table}[htbp]
\begin{align*}\small
\begin{array}{|c|c|c|}
\multicolumn{3}{c}{F\psi^2\phi D^n}\\
\hline\hline
\text{abbreviation} & \text{Wilson coefficient} & \text{operator}\\
\hline
C_{eB,1}^{(d6)} & C_ {B_L e_{\mathbb{C}}LH^{\dagger},1}^{\; pr} & iH^{\dagger i}B_{\rm L}^{\mu\nu}\left(e_{\mathbb{C} p}\sigma_{\mu \nu}L_r{}_{i}\right) \\
C_{eW,1}^{(d6)} & C_ {W_L e_{\mathbb{C}}LH^{\dagger},1}^{\; pr} & i\tau^I{}_{j}^{i}H^{\dagger j}W_{\rm L}^I{}^{\mu\nu}\left(e_{\mathbb{C} p}\sigma_{\mu \nu}L_r{}_{i}\right) \\
C_{uB,1}^{(d6)} & C_ {B_LQu_{\mathbb{C}}H,1}^{\; pr} & i\epsilon^{ij}H_jB_{\rm L}^{\mu\nu}\left(Q_p{}_{ai}\sigma_{\mu \nu}u_{\mathbb{C}}{}_r^{a}\right) \\
C_{uW,1}^{(d6)} & C_ {W_LQu_{\mathbb{C}}H,1}^{\; pr} & i\tau^I{}_{k}^{i}\epsilon^{jk}H_jW_{\rm L}^I{}^{\mu\nu}\left(Q_p{}_{ai}\sigma_{\mu \nu}u_{\mathbb{C}}{}_r^{a}\right) \\
C_{uG,1}^{(d6)} & C_ {G_LQu_{\mathbb{C}}H,1}^{\; pr} & i\lambda^A{}_{b}^{a}\epsilon^{ij}H_jG_{\rm L}^A{}^{\mu\nu}\left(Q_p{}_{ai}\sigma_{\mu \nu}u_{\mathbb{C}}{}_r^{b}\right) \\
C_{dB,1}^{(d6)} & C_ {B_Ld_{\mathbb{C}}QH^{\dagger},1}^{\; pr} & iH^{\dagger i}B_{\rm L}^{\mu\nu}\left(d_{\mathbb{C}}{}_p^{a}\sigma_{\mu \nu}Q_r{}_{ai}\right) \\
C_{dW,1}^{(d6)} & C_ {W_Ld_{\mathbb{C}}QH^{\dagger},1}^{\; pr} & i\tau^I{}_{j}^{i}H^{\dagger j}W_{\rm L}^I{}^{\mu\nu}\left(d_{\mathbb{C}}{}_p^{a}\sigma_{\mu \nu}Q_r{}_{ai}\right) \\
C_{dG,1}^{(d6)} & C_ {G_Ld_{\mathbb{C}}QH^{\dagger},1}^{\; pr} & i\lambda^A{}_{a}^{b}H^{\dagger i}G_{\rm L}^A{}^{\mu\nu}\left(d_{\mathbb{C}}{}_p^{a}\sigma_{\mu \nu}Q_r{}_{bi}\right) \\
\hline
C_{eB,1}^{(d8)} & C_ {B_L e_{\mathbb{C}}LH^{\dagger}D^2,1}^{\; pr} & iB_{\rm L}^{\lambda\mu}\left(D^{\nu}H^{\dagger i}\right)\left(e_{\mathbb{C} p}\sigma_{\mu \nu}\left(D_{\lambda}L_r{}_{i}\right)\right) \\
C_{eB,2}^{(d8)} & C_ {B_L e_{\mathbb{C}}LH^{\dagger}D^2,2}^{\; pr} & B_{\rm L}^{\mu\nu}\left(D_{\nu}H^{\dagger i}\right)\left(e_{\mathbb{C} p}\left(D_{\mu}L_r{}_{i}\right)\right) \\
C_{eB,3}^{(d8)} & C_ {B_R e_{\mathbb{C}}LH^{\dagger}D^2,1}^{\; pr} & iB_{\rm R}^{\lambda\mu}\left(D_{\lambda}D^{\nu}H^{\dagger i}\right)\left(e_{\mathbb{C} p}\sigma_{\mu \nu}L_r{}_{i}\right) \\
C_{eW,1}^{(d8)} & C_ {W_L e_{\mathbb{C}}LH^{\dagger}D^2,1}^{\; pr} & i\tau^I{}_{j}^{i}W_{\rm L}^I{}^{\lambda\mu}\left(D^{\nu}H^{\dagger j}\right)\left(e_{\mathbb{C} p}\sigma_{\mu \nu}\left(D_{\lambda}L_r{}_{i}\right)\right) \\
C_{eW,2}^{(d8)} & C_ {W_L e_{\mathbb{C}}LH^{\dagger}D^2,2}^{\; pr} & \tau^I{}_{j}^{i}W_{\rm L}^I{}^{\mu\nu}\left(D_{\nu}H^{\dagger j}\right)\left(e_{\mathbb{C} p}\left(D_{\mu}L_r{}_{i}\right)\right) \\
C_{eW,3}^{(d8)} & C_ {W_R e_{\mathbb{C}}LH^{\dagger}D^2,1}^{\; pr} & i\tau^I{}_{j}^{i}W_{\rm R}^I{}^{\lambda\mu}\left(D_{\lambda}D^{\nu}H^{\dagger j}\right)\left(e_{\mathbb{C} p}\sigma_{\mu \nu}L_r{}_{i}\right) \\
C_{uB,1}^{(d8)} & C_ {B_LQu_{\mathbb{C}}HD^2,1}^{\; pr} & i\epsilon^{ij}B_{\rm L}^{\lambda\mu}\left(D^{\nu}H_j\right)\left(Q_p{}_{ai}\sigma_{\mu \nu}\left(D_{\lambda}u_{\mathbb{C}}{}_r^{a}\right)\right) \\
C_{uB,2}^{(d8)} & C_ {B_LQu_{\mathbb{C}}HD^2,2}^{\; pr} & \epsilon^{ij}B_{\rm L}^{\mu\nu}\left(D_{\nu}H_j\right)\left(Q_p{}_{ai}\left(D_{\mu}u_{\mathbb{C}}{}_r^{a}\right)\right) \\
C_{uB,3}^{(d8)} & C_ {B_RQu_{\mathbb{C}}HD^2,1}^{\; pr} & i\epsilon^{ij}B_{\rm R}^{\lambda\mu}\left(D_{\lambda}D^{\nu}H_j\right)\left(Q_p{}_{ai}\sigma_{\mu \nu}u_{\mathbb{C}}{}_r^{a}\right) \\
C_{uW,1}^{(d8)} & C_ {W_LQu_{\mathbb{C}}HD^2,1}^{\; pr} & i\tau^I{}_{k}^{i}\epsilon^{jk}W_{\rm L}^I{}^{\lambda\mu}\left(D^{\nu}H_j\right)\left(Q_p{}_{ai}\sigma_{\mu \nu}\left(D_{\lambda}u_{\mathbb{C}}{}_r^{a}\right)\right) \\
C_{uW,2}^{(d8)} & C_ {W_LQu_{\mathbb{C}}HD^2,2}^{\; pr} & \tau^I{}_{k}^{i}\epsilon^{jk}W_{\rm L}^I{}^{\mu\nu}\left(D_{\nu}H_j\right)\left(Q_p{}_{ai}\left(D_{\mu}u_{\mathbb{C}}{}_r^{a}\right)\right) \\
C_{uW,3}^{(d8)} & C_ {W_RQu_{\mathbb{C}}HD^2,1}^{\; pr} & i\tau^I{}_{k}^{i}\epsilon^{jk}W_{\rm R}^I{}^{\lambda\mu}\left(D_{\lambda}D^{\nu}H_j\right)\left(Q_p{}_{ai}\sigma_{\mu \nu}u_{\mathbb{C}}{}_r^{a}\right) \\
C_{uG,1}^{(d8)} & C_ {G_LQu_{\mathbb{C}}HD^2,1}^{\; pr} & i\lambda^A{}_{b}^{a}\epsilon^{ij}G_{\rm L}^A{}^{\lambda\mu}\left(D^{\nu}H_j\right)\left(Q_p{}_{ai}\sigma_{\mu \nu}\left(D_{\lambda}u_{\mathbb{C}}{}_r^{b}\right)\right) \\
C_{uG,2}^{(d8)} & C_ {G_LQu_{\mathbb{C}}HD^2,2}^{\; pr} & \lambda^A{}_{b}^{a}\epsilon^{ij}G_{\rm L}^A{}^{\mu\nu}\left(D_{\nu}H_j\right)\left(Q_p{}_{ai}\left(D_{\mu}u_{\mathbb{C}}{}_r^{b}\right)\right) \\
C_{uG,3}^{(d8)} & C_ {G_RQu_{\mathbb{C}}HD^2,1}^{\; pr} & i\lambda^A{}_{b}^{a}\epsilon^{ij}G_{\rm R}^A{}^{\lambda\mu}\left(D_{\lambda}D^{\nu}H_j\right)\left(Q_p{}_{ai}\sigma_{\mu \nu}u_{\mathbb{C}}{}_r^{b}\right) \\
C_{dB,1}^{(d8)} & C_ {B_Ld_{\mathbb{C}}QH^{\dagger}D^2,1}^{\; pr} & iB_{\rm L}^{\lambda\mu}\left(D^{\nu}H^{\dagger i}\right)\left(d_{\mathbb{C}}{}_p^{a}\sigma_{\mu \nu}\left(D_{\lambda}Q_r{}_{ai}\right)\right) \\
C_{dB,2}^{(d8)} & C_ {B_Ld_{\mathbb{C}}QH^{\dagger}D^2,2}^{\; pr} & B_{\rm L}^{\mu\nu}\left(D_{\nu}H^{\dagger i}\right)\left(d_{\mathbb{C}}{}_p^{a}\left(D_{\mu}Q_r{}_{ai}\right)\right) \\
C_{dB,3}^{(d8)} & C_ {B_Rd_{\mathbb{C}}QH^{\dagger}D^2,1}^{\; pr} & iB_{\rm R}^{\lambda\mu}\left(D_{\lambda}D^{\nu}H^{\dagger i}\right)\left(d_{\mathbb{C}}{}_p^{a}\sigma_{\mu \nu}Q_r{}_{ai}\right) \\
C_{dW,1}^{(d8)} & C_ {W_Ld_{\mathbb{C}}QH^{\dagger}D^2,1}^{\; pr} & i\tau^I{}_{j}^{i}W_{\rm L}^I{}^{\lambda\mu}\left(D^{\nu}H^{\dagger j}\right)\left(d_{\mathbb{C}}{}_p^{a}\sigma_{\mu \nu}\left(D_{\lambda}Q_r{}_{ai}\right)\right) \\
C_{dW,2}^{(d8)} & C_ {W_Ld_{\mathbb{C}}QH^{\dagger}D^2,2}^{\; pr} & \tau^I{}_{j}^{i}W_{\rm L}^I{}^{\mu\nu}\left(D_{\nu}H^{\dagger j}\right)\left(d_{\mathbb{C}}{}_p^{a}\left(D_{\mu}Q_r{}_{ai}\right)\right) \\
C_{dW,3}^{(d8)} & C_ {W_Rd_{\mathbb{C}}QH^{\dagger}D^2,1}^{\; pr} & i\tau^I{}_{j}^{i}W_{\rm R}^I{}^{\lambda\mu}\left(D_{\lambda}D^{\nu}H^{\dagger j}\right)\left(d_{\mathbb{C}}{}_p^{a}\sigma_{\mu \nu}Q_r{}_{ai}\right) \\
C_{dG,1}^{(d8)} & C_ {G_Ld_{\mathbb{C}}QH^{\dagger}D^2,1}^{\; pr} & i\lambda^A{}_{a}^{b}G_{\rm L}^A{}^{\lambda\mu}\left(D^{\nu}H^{\dagger i}\right)\left(d_{\mathbb{C}}{}_p^{a}\sigma_{\mu \nu}\left(D_{\lambda}Q_r{}_{bi}\right)\right) \\
C_{dG,2}^{(d8)} & C_ {G_Ld_{\mathbb{C}}QH^{\dagger}D^2,2}^{\; pr} & \lambda^A{}_{a}^{b}G_{\rm L}^A{}^{\mu\nu}\left(D_{\nu}H^{\dagger i}\right)\left(d_{\mathbb{C}}{}_p^{a}\left(D_{\mu}Q_r{}_{bi}\right)\right) \\
C_{dG,3}^{(d8)} & C_ {G_Rd_{\mathbb{C}}QH^{\dagger}D^2,1}^{\; pr} & i\lambda^A{}_{a}^{b}G_{\rm R}^A{}^{\lambda\mu}\left(D_{\lambda}D^{\nu}H^{\dagger i}\right)\left(d_{\mathbb{C}}{}_p^{a}\sigma_{\mu \nu}Q_r{}_{bi}\right) \\
\hline
\end{array}
\end{align*}
\caption{$F\psi^2\phi D^n$-class operators containing both bosonic and fermionic fields that contribute to the one-loop renormalization of dimension-8 four-fermion operators through bosonic-loop unitarity cuts. The leftmost column contains the abbreviations for the Wilson coefficients in the case of $N_f=1$. While we adopt the definition of chiral basis for gauge boson fields: $\phi F_{{\rm L/R}}^{\mu\nu}\bar{\psi}\sigma_{\mu\nu}\psi=\frac12 \phi (F^{\mu\nu}\mp i\tilde{F}^{\mu\nu})\bar{\psi}\sigma_{\mu\nu}\psi $\ .}
\label{tab:psi2XH}
\end{table}

\begin{table}[htbp]
\begin{align*}\small
\begin{array}{|c|c|c|}
\multicolumn{3}{c}{\psi^2\phi^2 D^n}\\
\hline\hline
\text{abbreviation} & \text{Wilson coefficient} & \text{operator}\\
\hline
C_{Hl,1}^{(d5)} & C_ {L_{[2]}^2H_{[2]}^2,1}^{\; pr} & \left(L_p{}_{i}L_r{}_{j}\right)\epsilon^{ik}\epsilon^{jl}H_kH_l \\
\hline
C_{Hl,1}^{(d6)} & C_ {LL^{\dagger}HH^{\dagger}D,1}^{\; pr} & iH_j\left(D^{\mu}H^{\dagger i}\right)\left(L_p{}_{i}\sigma_{\mu}L^{\dagger}_r{}^{j}\right) \\
C_{Hl,2}^{(d6)} & C_ {LL^{\dagger}HH^{\dagger}D,2}^{\; pr} & iH_j\left(D^{\mu}H^{\dagger j}\right)\left(L_p{}_{i}\sigma_{\mu}L^{\dagger}_r{}^{i}\right) \\
C_{He,1}^{(d6)} & C_ {e_{\mathbb{C}}e_{\mathbb{C}}^{\dagger}HH^{\dagger}D,1}^{\; pr} & iH_i\left(D^{\mu}H^{\dagger i}\right)\left(e_{\mathbb{C} p}\sigma_{\mu}e^{\dagger}_{\mathbb{C} r}\right) \\
C_{Hq,1}^{(d6)} & C_ {QQ^{\dagger}HH^{\dagger}D,1}^{\; pr} & iH_j\left(D^{\mu}H^{\dagger i}\right)\left(Q_p{}_{ai}\sigma_{\mu}Q^{\dagger}_r{}^{aj}\right) \\
C_{Hq,2}^{(d6)} & C_ {QQ^{\dagger}HH^{\dagger}D,2}^{\; pr} & iH_j\left(D^{\mu}H^{\dagger j}\right)\left(Q_p{}_{ai}\sigma_{\mu}Q^{\dagger}_r{}^{ai}\right) \\
C_{Hu,1}^{(d6)} & C_ {u_{\mathbb{C}}u_{\mathbb{C}}^{\dagger}HH^{\dagger}D,1}^{\; pr} & iH_i\left(D^{\mu}H^{\dagger i}\right)\left(u_{\mathbb{C}}{}_p^{a}\sigma_{\mu}u_{\mathbb{C}}^{\dagger}{}_r{}_{a}\right) \\
C_{Hd,1}^{(d6)} & C_ {d_{\mathbb{C}}d_{\mathbb{C}}^{\dagger}HH^{\dagger}D,1}^{\; pr} & iH_i\left(D^{\mu}H^{\dagger i}\right)\left(d_{\mathbb{C}}{}_p^{a}\sigma_{\mu}d_{\mathbb{C}}^{\dagger}{}_r{}_{a}\right) \\
C_{Hud,1}^{(d6)} & C_ {d_{\mathbb{C}}u_{\mathbb{C}}^{\dagger}H^{\dagger 2}D,1}^{\; pr} & i\epsilon_{ij}H^{\dagger i}\left(D^{\mu}H^{\dagger j}\right)\left(d_{\mathbb{C}}{}_p^{a}\sigma_{\mu}u_{\mathbb{C}}^{\dagger}{}_r{}_{a}\right) \\
\hline
C_{Hl,1}^{(d7)} & C_ {L_{[2]}^2H_{[2]}^2D^2,1}^{\; pr} & \left(L_p{}_{i}L_r{}_{j}\right)\epsilon^{ik}\epsilon^{jl}\left(D_{\mu}H_k\right)\left(D^{\mu}H_l\right) \\
C_{Hl,2}^{(d7)} & C_ {L_{[2]}^2H_{[2]}^2D^2,2}^{\; pr} & i\epsilon^{ik}\epsilon^{jl}\left(D^{\mu}H_k\right)\left(D^{\nu}H_l\right)\left(L_p{}_{i}\sigma_{\mu \nu}L_r{}_{j}\right) \\
\hline
C_{Hl,1}^{(d8)} & C_ {LL^{\dagger}HH^{\dagger}D^3,1}^{\; pr} & iH_j\left(D_{\nu}D^{\mu}H^{\dagger i}\right)\left(L_p{}_{i}\sigma_{\mu}\left(D^{\nu}L^{\dagger}_r{}^{j}\right)\right) \\
C_{Hl,2}^{(d8)} & C_ {LL^{\dagger}HH^{\dagger}D^3,2}^{\; pr} & i\left(D_{\nu}H_j\right)\left(D^{\mu}H^{\dagger i}\right)\left(L_p{}_{i}\sigma_{\mu}\left(D^{\nu}L^{\dagger}_r{}^{j}\right)\right) \\
C_{Hl,3}^{(d8)} & C_ {LL^{\dagger}HH^{\dagger}D^3,3}^{\; pr} & iH_j\left(D_{\nu}D^{\mu}H^{\dagger j}\right)\left(L_p{}_{i}\sigma_{\mu}\left(D^{\nu}L^{\dagger}_r{}^{i}\right)\right) \\
C_{Hl,4}^{(d8)} & C_ {LL^{\dagger}HH^{\dagger}D^3,4}^{\; pr} & i\left(D_{\nu}H_j\right)\left(D^{\mu}H^{\dagger j}\right)\left(L_p{}_{i}\sigma_{\mu}\left(D^{\nu}L^{\dagger}_r{}^{i}\right)\right) \\
C_{He,1}^{(d8)} & C_ {e_{\mathbb{C}}e_{\mathbb{C}}^{\dagger}HH^{\dagger}D^2,1}^{\; pr} & iH_i\left(D_{\nu}D^{\mu}H^{\dagger i}\right)\left(e_{\mathbb{C} p}\sigma_{\mu}\left(D^{\nu}e^{\dagger}_{\mathbb{C} r}\right)\right) \\
C_{He,2}^{(d8)} & C_ {e_{\mathbb{C}}e_{\mathbb{C}}^{\dagger}HH^{\dagger}D^2,2}^{\; pr} & i\left(D_{\nu}H_i\right)\left(D^{\mu}H^{\dagger i}\right)\left(e_{\mathbb{C} p}\sigma_{\mu}\left(D^{\nu}e^{\dagger}_{\mathbb{C} r}\right)\right) \\
C_{Hq,1}^{(d8)} & C_ {QQ^{\dagger}HH^{\dagger}D^3,1}^{\; pr} & iH_j\left(D_{\nu}D^{\mu}H^{\dagger i}\right)\left(Q_p{}_{ai}\sigma_{\mu}\left(D^{\nu}Q^{\dagger}_r{}^{aj}\right)\right) \\
C_{Hq,2}^{(d8)} & C_ {QQ^{\dagger}HH^{\dagger}D^3,2}^{\; pr} & i\left(D_{\nu}H_j\right)\left(D^{\mu}H^{\dagger i}\right)\left(Q_p{}_{ai}\sigma_{\mu}\left(D^{\nu}Q^{\dagger}_r{}^{aj}\right)\right) \\
C_{Hq,3}^{(d8)} & C_ {QQ^{\dagger}HH^{\dagger}D^3,3}^{\; pr} & iH_j\left(D_{\nu}D^{\mu}H^{\dagger j}\right)\left(Q_p{}_{ai}\sigma_{\mu}\left(D^{\nu}Q^{\dagger}_r{}^{ai}\right)\right) \\
C_{Hq,4}^{(d8)} & C_ {QQ^{\dagger}HH^{\dagger}D^3,4}^{\; pr} & i\left(D_{\nu}H_j\right)\left(D^{\mu}H^{\dagger j}\right)\left(Q_p{}_{ai}\sigma_{\mu}\left(D^{\nu}Q^{\dagger}_r{}^{ai}\right)\right) \\
C_{Hu,1}^{(d8)} & C_ {u_{\mathbb{C}}u_{\mathbb{C}}^{\dagger}HH^{\dagger}D^3,1}^{\; pr} & iH_i\left(D_{\nu}D^{\mu}H^{\dagger i}\right)\left(u_{\mathbb{C}}{}_p^{a}\sigma_{\mu}\left(D^{\nu}u_{\mathbb{C}}^{\dagger}{}_r{}_{a}\right)\right) \\
C_{Hu,2}^{(d8)} & C_ {u_{\mathbb{C}}u_{\mathbb{C}}^{\dagger}HH^{\dagger}D^3,2}^{\; pr} & i\left(D_{\nu}H_i\right)\left(D^{\mu}H^{\dagger i}\right)\left(u_{\mathbb{C}}{}_p^{a}\sigma_{\mu}\left(D^{\nu}u_{\mathbb{C}}^{\dagger}{}_r{}_{a}\right)\right) \\
C_{Hd,1}^{(d8)} & C_ {d_{\mathbb{C}}d_{\mathbb{C}}^{\dagger}HH^{\dagger}D^3,1}^{\; pr} & iH_i\left(D_{\nu}D^{\mu}H^{\dagger i}\right)\left(d_{\mathbb{C}}{}_p^{a}\sigma_{\mu}\left(D^{\nu}d_{\mathbb{C}}^{\dagger}{}_r{}_{a}\right)\right) \\
C_{Hd,2}^{(d8)} & C_ {d_{\mathbb{C}}d_{\mathbb{C}}^{\dagger}HH^{\dagger}D^3,2}^{\; pr} & i\left(D_{\nu}H_i\right)\left(D^{\mu}H^{\dagger i}\right)\left(d_{\mathbb{C}}{}_p^{a}\sigma_{\mu}\left(D^{\nu}d_{\mathbb{C}}^{\dagger}{}_r{}_{a}\right)\right) \\
C_{Hud,1}^{(d8)} & C_ {d_{\mathbb{C}}u_{\mathbb{C}}^{\dagger}H^{\dagger 2}D^3,1}^{\; pr} & i\epsilon_{ij}H^{\dagger i}\left(D_{\nu}D^{\mu}H^{\dagger j}\right)\left(d_{\mathbb{C}}{}_p^{a}\sigma_{\mu}\left(D^{\nu}u_{\mathbb{C}}^{\dagger}{}_r{}_{a}\right)\right) \\
\hline
\end{array}
\end{align*}
\caption{$\psi^2\phi^2D^n$-class operators containing both bosonic and fermionic fields that contribute to the one-loop renormalization of dimension-8 four-fermion operators through bosonic-loop unitarity cuts. The leftmost column contains the abbreviations for the Wilson coefficients in the case of $N_f=1$.}
\label{tab:psi2H2}
\end{table}

\begin{table}[htbp]
\begin{align*}\small
\begin{array}{|c|c|c|}
\multicolumn{3}{c}{F^2\psi^2 D}\\
\hline\hline
\text{abbreviation} & \text{Wilson coefficient} & \text{operator}\\
\hline
C_{Bl,1}^{(d8)} & C_ {LL^{\dagger}B_LB_RD,1}^{\; pr} & iB_{\rm R}^{\lambda\mu}B_{\rm L}{}_{\lambda}{}^{\nu}\left(L_p{}_{i}\sigma_{\mu}\left(D_{\nu}L^{\dagger}_r{}^{i}\right)\right) \\
C_{Wl,1}^{(d8)} & C_ {LL^{\dagger}W_LW_RD,1}^{\; pr} & i\tau^K{}_{j}^{i}\epsilon^{IJK}W_{\rm R}^J{}^{\lambda\mu}W_{\rm L}^I{}_{\lambda}{}^{\nu}\left(L_p{}_{i}\sigma_{\mu}\left(D_{\nu}L^{\dagger}_r{}^{j}\right)\right) \\
C_{Wl,2}^{(d8)} & C_ {LL^{\dagger}W_LW_RD,2}^{\; pr} & iW_{\rm R}^I{}^{\lambda\mu}W_{\rm L}^I{}_{\lambda}{}^{\nu}\left(L_p{}_{i}\sigma_{\mu}\left(D_{\nu}L^{\dagger}_r{}^{i}\right)\right) \\
C_{BWl,1}^{(d8)} & C_ {LL^{\dagger}B_LW_RD,1}^{\; pr} & i\tau^I{}_{j}^{i}W_{\rm R}^I{}^{\lambda\mu}B_{\rm L}{}_{\lambda}{}^{\nu}\left(L_p{}_{i}\sigma_{\mu}\left(D_{\nu}L^{\dagger}_r{}^{j}\right)\right) \\
C_{Gl,1}^{(d8)} & C_ {LL^{\dagger}G_LG_RD,1}^{\; pr} & iG_{\rm R}^A{}^{\lambda\mu}G_{\rm L}^A{}_{\lambda}{}^{\nu}\left(L_p{}_{i}\sigma_{\mu}\left(D_{\nu}L^{\dagger}_r{}^{i}\right)\right) \\
C_{Be,1}^{(d8)} & C_ {e_{\mathbb{C}}e_{\mathbb{C}}^{\dagger}B_LB_RD,1}^{\; pr} & iB_{\rm R}^{\lambda\mu}B_{\rm L}{}_{\lambda}{}^{\nu}\left(e_{\mathbb{C} p}\sigma_{\mu}\left(D_{\nu}e^{\dagger}_{\mathbb{C} r}\right)\right) \\
C_{We,1}^{(d8)} & C_ {e_{\mathbb{C}}e_{\mathbb{C}}^{\dagger}W_LW_RD,1}^{\; pr} & iW_{\rm R}^I{}^{\lambda\mu}W_{\rm L}^I{}_{\lambda}{}^{\nu}\left(e_{\mathbb{C} p}\sigma_{\mu}\left(D_{\nu}e^{\dagger}_{\mathbb{C} r}\right)\right) \\
C_{Ge,1}^{(d8)} & C_ {e_{\mathbb{C}}e_{\mathbb{C}}^{\dagger}G_LG_RD,1}^{\; pr} & iG_{\rm R}^A{}^{\lambda\mu}G_{\rm L}^A{}_{\lambda}{}^{\nu}\left(e_{\mathbb{C} p}\sigma_{\mu}\left(D_{\nu}e^{\dagger}_{\mathbb{C} r}\right)\right) \\
C_{Bq,1}^{(d8)} & C_ {QQ^{\dagger}B_LB_RD,1}^{\; pr} & iB_{\rm R}^{\lambda\mu}B_{\rm L}{}_{\lambda}{}^{\nu}\left(Q_p{}_{ai}\sigma_{\mu}\left(D_{\nu}Q^{\dagger}_r{}^{ai}\right)\right) \\
C_{Wq,1}^{(d8)} & C_ {QQ^{\dagger}W_LW_RD,1}^{\; pr} & i\tau^K{}_{j}^{i}\epsilon^{IJK}W_{\rm R}^J{}^{\lambda\mu}W_{\rm L}^I{}_{\lambda}{}^{\nu}\left(Q_p{}_{ai}\sigma_{\mu}\left(D_{\nu}Q^{\dagger}_r{}^{aj}\right)\right) \\
C_{Wq,2}^{(d8)} & C_ {QQ^{\dagger}W_LW_RD,2}^{\; pr} & iW_{\rm R}^I{}^{\lambda\mu}W_{\rm L}^I{}_{\lambda}{}^{\nu}\left(Q_p{}_{ai}\sigma_{\mu}\left(D_{\nu}Q^{\dagger}_r{}^{ai}\right)\right) \\
C_{Gq,1}^{(d8)} & C_ {QQ^{\dagger}G_LG_RD,1}^{\; pr} & id^{ABC}\lambda^C{}_{b}^{a}G_{\rm R}^B{}^{\lambda\mu}G_{\rm L}^A{}_{\lambda}{}^{\nu}\left(Q_p{}_{ai}\sigma_{\mu}\left(D_{\nu}Q^{\dagger}_r{}^{bi}\right)\right) \\
C_{Gq,2}^{(d8)} & C_ {QQ^{\dagger}G_LG_RD,2}^{\; pr} & if^{ABC}\lambda^C{}_{b}^{a}G_{\rm R}^B{}^{\lambda\mu}G_{\rm L}^A{}_{\lambda}{}^{\nu}\left(Q_p{}_{ai}\sigma_{\mu}\left(D_{\nu}Q^{\dagger}_r{}^{bi}\right)\right) \\
C_{Gq,3}^{(d8)} & C_ {QQ^{\dagger}G_LG_RD,3}^{\; pr} & iG_{\rm R}^A{}^{\lambda\mu}G_{\rm L}^A{}_{\lambda}{}^{\nu}\left(Q_p{}_{ai}\sigma_{\mu}\left(D_{\nu}Q^{\dagger}_r{}^{ai}\right)\right) \\
C_{BWq,1}^{(d8)} & C_ {QQ^{\dagger}B_LW_RD,1}^{\; pr} & i\tau^I{}_{j}^{i}W_{\rm R}^I{}^{\lambda\mu}B_{\rm L}{}_{\lambda}{}^{\nu}\left(Q_p{}_{ai}\sigma_{\mu}\left(D_{\nu}Q^{\dagger}_r{}^{aj}\right)\right) \\
C_{BGq,1}^{(d8)} & C_ {QQ^{\dagger}B_LG_RD,1}^{\; pr} & i\lambda^A{}_{b}^{a}G_{\rm R}^A{}^{\lambda\mu}B_{\rm L}{}_{\lambda}{}^{\nu}\left(Q_p{}_{ai}\sigma_{\mu}\left(D_{\nu}Q^{\dagger}_r{}^{bi}\right)\right) \\
C_{WGq,1}^{(d8)} & C_ {QQ^{\dagger}W_LG_RD,1}^{\; pr} & i\lambda^A{}_{b}^{a}\tau^I{}_{j}^{i}G_{\rm R}^A{}^{\lambda\mu}W_{\rm L}^I{}_{\lambda}{}^{\nu}\left(Q_p{}_{ai}\sigma_{\mu}\left(D_{\nu}Q^{\dagger}_r{}^{bj}\right)\right) \\
C_{Bu,1}^{(d8)} & C_ {u_{\mathbb{C}}u_{\mathbb{C}}^{\dagger}B_LB_RD,1}^{\; pr} & iB_{\rm R}^{\lambda\mu}B_{\rm L}{}_{\lambda}{}^{\nu}\left(u_{\mathbb{C}}{}_p^{a}\sigma_{\mu}\left(D_{\nu}u_{\mathbb{C}}^{\dagger}{}_r{}_{a}\right)\right) \\
C_{Wu,1}^{(d8)} & C_ {u_{\mathbb{C}}u_{\mathbb{C}}^{\dagger}W_LW_RD,1}^{\; pr} & iW_{\rm R}^I{}^{\lambda\mu}W_{\rm L}^I{}_{\lambda}{}^{\nu}\left(u_{\mathbb{C}}{}_p^{a}\sigma_{\mu}\left(D_{\nu}u_{\mathbb{C}}^{\dagger}{}_r{}_{a}\right)\right) \\
C_{Gu,1}^{(d8)} & C_ {u_{\mathbb{C}}u_{\mathbb{C}}^{\dagger}G_LG_RD,1}^{\; pr} & id^{ABC}\lambda^C{}_{a}^{b}G_{\rm R}^B{}^{\lambda\mu}G_{\rm L}^A{}_{\lambda}{}^{\nu}\left(u_{\mathbb{C}}{}_p^{a}\sigma_{\mu}\left(D_{\nu}u_{\mathbb{C}}^{\dagger}{}_r{}_{b}\right)\right) \\
C_{Gu,2}^{(d8)} & C_ {u_{\mathbb{C}}u_{\mathbb{C}}^{\dagger}G_LG_RD,2}^{\; pr} & if^{ABC}\lambda^C{}_{a}^{b}G_{\rm R}^B{}^{\lambda\mu}G_{\rm L}^A{}_{\lambda}{}^{\nu}\left(u_{\mathbb{C}}{}_p^{a}\sigma_{\mu}\left(D_{\nu}u_{\mathbb{C}}^{\dagger}{}_r{}_{b}\right)\right) \\
C_{Gu,3}^{(d8)} & C_ {u_{\mathbb{C}}u_{\mathbb{C}}^{\dagger}G_LG_RD,3}^{\; pr} & iG_{\rm R}^A{}^{\lambda\mu}G_{\rm L}^A{}_{\lambda}{}^{\nu}\left(u_{\mathbb{C}}{}_p^{a}\sigma_{\mu}\left(D_{\nu}u_{\mathbb{C}}^{\dagger}{}_r{}_{a}\right)\right) \\
C_{BGu,1}^{(d8)} & C_ {u_{\mathbb{C}}u_{\mathbb{C}}^{\dagger}B_LG_RD,1}^{\; pr} & i\lambda^A{}_{a}^{b}G_{\rm R}^A{}^{\lambda\mu}B_{\rm L}{}_{\lambda}{}^{\nu}\left(u_{\mathbb{C}}{}_p^{a}\sigma_{\mu}\left(D_{\nu}u_{\mathbb{C}}^{\dagger}{}_r{}_{b}\right)\right) \\
C_{Bd,1}^{(d8)} & C_ {d_{\mathbb{C}}d_{\mathbb{C}}^{\dagger}B_LB_RD,1}^{\; pr} & iB_{\rm R}^{\lambda\mu}B_{\rm L}{}_{\lambda}{}^{\nu}\left(d_{\mathbb{C}}{}_p^{a}\sigma_{\mu}\left(D_{\nu}d_{\mathbb{C}}^{\dagger}{}_r{}_{a}\right)\right) \\
C_{Wd,1}^{(d8)} & C_ {d_{\mathbb{C}}d_{\mathbb{C}}^{\dagger}W_LW_RD,1}^{\; pr} & iW_{\rm R}^I{}^{\lambda\mu}W_{\rm L}^I{}_{\lambda}{}^{\nu}\left(d_{\mathbb{C}}{}_p^{a}\sigma_{\mu}\left(D_{\nu}d_{\mathbb{C}}^{\dagger}{}_r{}_{a}\right)\right) \\
C_{Gd,1}^{(d8)} & C_ {d_{\mathbb{C}}d_{\mathbb{C}}^{\dagger}G_LG_RD,1}^{\; pr} & id^{ABC}\lambda^C{}_{a}^{b}G_{\rm R}^B{}^{\lambda\mu}G_{\rm L}^A{}_{\lambda}{}^{\nu}\left(d_{\mathbb{C}}{}_p^{a}\sigma_{\mu}\left(D_{\nu}d_{\mathbb{C}}^{\dagger}{}_r{}_{b}\right)\right) \\
C_{Gd,2}^{(d8)} & C_ {d_{\mathbb{C}}d_{\mathbb{C}}^{\dagger}G_LG_RD,2}^{\; pr} & if^{ABC}\lambda^C{}_{a}^{b}G_{\rm R}^B{}^{\lambda\mu}G_{\rm L}^A{}_{\lambda}{}^{\nu}\left(d_{\mathbb{C}}{}_p^{a}\sigma_{\mu}\left(D_{\nu}d_{\mathbb{C}}^{\dagger}{}_r{}_{b}\right)\right) \\
C_{Gd,3}^{(d8)} & C_ {d_{\mathbb{C}}d_{\mathbb{C}}^{\dagger}G_LG_RD,3}^{\; pr} & iG_{\rm R}^A{}^{\lambda\mu}G_{\rm L}^A{}_{\lambda}{}^{\nu}\left(d_{\mathbb{C}}{}_p^{a}\sigma_{\mu}\left(D_{\nu}d_{\mathbb{C}}^{\dagger}{}_r{}_{a}\right)\right) \\
C_{BGd,1}^{(d8)} & C_ {d_{\mathbb{C}}d_{\mathbb{C}}^{\dagger}B_LG_RD,1}^{\; pr} & i\lambda^A{}_{a}^{b}G_{\rm R}^A{}^{\lambda\mu}B_{\rm L}{}_{\lambda}{}^{\nu}\left(d_{\mathbb{C}}{}_p^{a}\sigma_{\mu}\left(D_{\nu}d_{\mathbb{C}}^{\dagger}{}_r{}_{b}\right)\right) \\
\hline
\end{array}
\end{align*}
\caption{$\psi^2F^2D$-class operators containing both bosonic and fermionic fields that contribute to the one-loop renormalization of dimension-8 four-fermion operators through bosonic-loop unitarity cuts. The leftmost column contains the abbreviations for the Wilson coefficients in the case of $N_f=1$. While we adopt the definition of chiral basis for gauge boson fields as in eq.~\eqref{eq:FLFR}.}
\label{tab:psi2F2}
\end{table}

\subsection{$Q^2Q^{\dagger 2}$}
We first present the full RGEs for the dimension-6 $Q^2Q^{\dagger 2}$ operators. Throughout this section, Wilson coefficients are written in the f-basis.
The superscript flavor indices follow the ordering of fermion fields in the operator definition.
When identical fermion fields appear, the coefficients are projected onto irreducible
flavor tensors, labeled by Young-diagram symbols such as $[2]$, or $[1,1]$.
The anomalous dimensions preserve these symmetry assignments, so operators belonging to different Young sectors do not mix.
Each Wilson coefficient and operator is summarized in Table \ref{tab:Q2Q2}. Note that the superscripts of the Wilson coefficients represent flavor indices, and their order corresponds to the ordering of the fermion subscripts. The ordering follows the ascending helicity order, and for fermions with the same helicity, it follows the alphabetical sequence.
\begin{table}[htbp]
\begin{align*}
\begin{array}{|c|c|c|}
\hline\hline
\text{abbreviation} & \text{Wilson coefficient} & \text{operator}\\
\hline
C_{q,1}^{(d6)} & C_ {Q{}_ {[2]}^2 Q{}^{\dagger}{}_{[2]}^2,1}^{\; prst} & \left(Q{}_{pai}Q{}_{rbj}\right)(Q^{\dagger}{}_s^{ai} Q^{\dagger}{}_t^{bj}) \\
C_{q,2}^{(d6)} & C_ {Q{}_ {[2]}^2 Q{}^{\dagger}{}_{[2]}^2,2}^{\; prst} & \left(Q{}_{pai}Q{}_{rbj}\right)(Q^{\dagger}{}_s^{aj} Q^{\dagger}{}_t^{bi}) \\
 & C_ {Q{}_ {[1,1]}^2 Q{}^{\dagger}{}_{[1,1]}^2,1}^{\; prst} & \left(Q{}_{pai}Q{}_{rbj}\right)(Q^{\dagger}{}_s^{ai} Q^{\dagger}{}_t^{bj}) \\
 & C_ {Q{}_ {[1,1]}^2 Q{}^{\dagger}{}_{[1,1]}^2,2}^{\; prst} & \left(Q{}_{pai}Q{}_{rbj}\right)(Q^{\dagger}{}_s^{aj} Q^{\dagger}{}_t^{bi}) \\
\hline
C_{q,1}^{(d8)} & C_ {Q{}_ {[2]}^2 Q{}^{\dagger}{}_{[2]}^2 D^2,1}^{\; prst} & \left(Q{}_{pai} Q{}_{rbj}\right)(D^{\mu}Q^{\dagger}{}_s^{ai} D_{\mu}Q^{\dagger}{}_t^{bj}) \\
C_{q,2}^{(d8)} & C_ {Q{}_ {[2]}^2 Q{}^{\dagger}{}_{[2]}^2 D^2,2}^{\; prst} & i\left(Q{}_{pai} \sigma_{\mu\nu}Q{}_{rbj}\right)(D^{\mu}Q^{\dagger}{}_s^{ai} D^{\nu}Q^{\dagger}{}_t^{bj}) \\
C_{q,3}^{(d8)} & C_ {Q{}_ {[2]}^2 Q{}^{\dagger}{}_{[2]}^2 D^2,3}^{\; prst} & \left(Q{}_{pai} Q{}_{rbj}\right)(D^{\mu}Q^{\dagger}{}_s^{aj} D_{\mu}Q^{\dagger}{}_t^{bi}) \\
C_{q,4}^{(d8)} & C_ {Q{}_ {[2]}^2 Q{}^{\dagger}{}_{[2]}^2 D^2,4}^{\; prst} & i\left(Q{}_{pai} \sigma_{\mu\nu}Q{}_{rbj}\right)(D^{\mu}Q^{\dagger}{}_s^{aj} D^{\nu}Q^{\dagger}{}_t^{bi}) \\
 & C_ {Q{}_ {[1,1]}^2 Q{}^{\dagger}{}_{[1,1]}^2 D^2,1}^{\; prst} & \left(Q{}_{pai} Q{}_{rbj}\right)(D^{\mu}Q^{\dagger}{}_s^{ai} D_{\mu}Q^{\dagger}{}_t^{bj}) \\
 & C_ {Q{}_ {[1,1]}^2 Q{}^{\dagger}{}_{[1,1]}^2 D^2,2}^{\; prst} & i\left(Q{}_{pai} \sigma_{\mu\nu}Q{}_{rbj}\right)(D^{\mu}Q^{\dagger}{}_s^{ai} D^{\nu}Q^{\dagger}{}_t^{bj}) \\
 & C_ {Q{}_ {[1,1]}^2 Q{}^{\dagger}{}_{[1,1]}^2 D^2,3}^{\; prst} & \left(Q{}_{pai} Q{}_{rbj}\right)(D^{\mu}Q^{\dagger}{}_s^{aj} D_{\mu}Q^{\dagger}{}_t^{bi}) \\
 & C_ {Q{}_ {[1,1]}^2 Q{}^{\dagger}{}_{[1,1]}^2 D^2,4}^{\; prst} & i\left(Q{}_{pai} \sigma_{\mu\nu}Q{}_{rbj}\right)(D^{\mu}Q^{\dagger}{}_s^{aj} D^{\nu}Q^{\dagger}{}_t^{bi}) \\
\hline
\end{array}
\end{align*}
\caption{List of the $Q^2Q^{\dagger 2}$-type operators and the corresponding Wilson coefficients in dimension-6 and dimension-8. The leftmost column contains the abbreviations for the Wilson coefficients in the case of $N_f=1$.}
\label{tab:Q2Q2}
\end{table}
\begin{align}
\begin{split}\dot{C}_{Q_{[2]}^2 Q^{\dagger}{}_{[2]}^2,1}^{\; prst}&\supset \frac{1}{18} g_1^2 \delta _{ps} C_{Q H H^{\dagger}{} Q^{\dagger}{} D,1}^{\; rt}-\frac{1}{6} g_2^2 \delta _{ps} C_{Q H H^{\dagger}{} Q^{\dagger}{} D,1}^{\; rt}+\frac{1}{9} g_1^2 \delta _{ps} C_{Q H H^{\dagger}{} Q^{\dagger}{} D,2}^{\; rt}\\&+\frac{2}{9} g_1^2 \delta _{ps} C_{d_{\mathbb{C}}{} Q d_{\mathbb{C}}{}^{\dagger}{} Q^{\dagger}{},1}^{\; urut}+\frac{2}{27} g_1^2 \delta _{ps} C_{d_{\mathbb{C}}{} Q d_{\mathbb{C}}{}^{\dagger}{} Q^{\dagger}{},2}^{\; urut}+\frac{2}{9} g_3^2 \delta _{ps} C_{d_{\mathbb{C}}{} Q d_{\mathbb{C}}{}^{\dagger}{} Q^{\dagger}{},2}^{\; urut}\\&+\frac{2}{9} g_1^2 \delta _{ps} C_{e_{\mathbb{C}}{} Q e_{\mathbb{C}}{}^{\dagger}{} Q^{\dagger}{},1}^{\; urut}-\frac{2}{9} g_1^2 \delta _{ps} C_{L Q L^{\dagger}{} Q^{\dagger}{},1}^{\; urut}-\frac{1}{9} g_1^2 \delta _{ps} C_{L Q L^{\dagger}{} Q^{\dagger}{},2}^{\; urut}\\&-\frac{1}{3} g_2^2 \delta _{ps} C_{L Q L^{\dagger}{} Q^{\dagger}{},2}^{\; urut}-\frac{4}{27} g_1^2 \delta _{ps} C_{Q u_{\mathbb{C}}{} Q^{\dagger}{} u_{\mathbb{C}}{}^{\dagger}{},1}^{\; rutu}+\frac{2}{9} g_3^2 \delta _{ps} C_{Q u_{\mathbb{C}}{} Q^{\dagger}{} u_{\mathbb{C}}{}^{\dagger}{},1}^{\; rutu}\\&-\frac{4}{9} g_1^2 \delta _{ps} C_{Q u_{\mathbb{C}}{} Q^{\dagger}{} u_{\mathbb{C}}{}^{\dagger}{},2}^{\; rutu}-\left[Y_d\right]_{up} \left[Y_u\right]_{vr} C_{d_{\mathbb{C}}{}^{\dagger}{} Q^{\dagger}{}_{[2]}^2 u_{\mathbb{C}}{}^{\dagger}{},1}^{\; ustv}-\left[Y_d\right]_{up} \left[Y_u\right]_{vr} C_{d_{\mathbb{C}}{}^{\dagger}{} Q^{\dagger}{}_{[2]}^2 u_{\mathbb{C}}{}^{\dagger}{},2}^{\; ustv}\\&+\frac{1}{3} g_1^2 C_{Q_{[2]}^2 Q^{\dagger}{}_{[2]}^2,1}^{\; prst}-3 g_2^2 C_{Q_{[2]}^2 Q^{\dagger}{}_{[2]}^2,1}^{\; prst}-2 g_3^2 C_{Q_{[2]}^2 Q^{\dagger}{}_{[2]}^2,1}^{\; prst}\\&+\frac{14}{27} g_1^2 \delta _{ps} C_{Q_{[2]}^2 Q^{\dagger}{}_{[2]}^2,1}^{\; rutu}-\frac{2}{3} g_2^2 \delta _{ps} C_{Q_{[2]}^2 Q^{\dagger}{}_{[2]}^2,1}^{\; rutu}-\frac{4}{9} g_3^2 \delta _{ps} C_{Q_{[2]}^2 Q^{\dagger}{}_{[2]}^2,1}^{\; rutu}\\&+6 g_2^2 C_{Q_{[2]}^2 Q^{\dagger}{}_{[2]}^2,2}^{\; prst}+6 g_3^2 C_{Q_{[2]}^2 Q^{\dagger}{}_{[2]}^2,2}^{\; prst}+\frac{10}{27} g_1^2 \delta _{ps} C_{Q_{[2]}^2 Q^{\dagger}{}_{[2]}^2,2}^{\; rutu}\\&-2 g_2^2 \delta _{ps} C_{Q_{[2]}^2 Q^{\dagger}{}_{[2]}^2,2}^{\; rutu}-\frac{8}{9} g_3^2 \delta _{ps} C_{Q_{[2]}^2 Q^{\dagger}{}_{[2]}^2,2}^{\; rutu}+\frac{10}{27} g_1^2 \delta _{ps} C_{Q_{[1,1]}^2 Q^{\dagger}{}_{[1,1]}^2,1}^{\; rutu}\\&+\frac{2}{3} g_2^2 \delta _{ps} C_{Q_{[1,1]}^2 Q^{\dagger}{}_{[1,1]}^2,1}^{\; rutu}+\frac{4}{9} g_3^2 \delta _{ps} C_{Q_{[1,1]}^2 Q^{\dagger}{}_{[1,1]}^2,1}^{\; rutu}+\frac{2}{27} g_1^2 \delta _{ps} C_{Q_{[1,1]}^2 Q^{\dagger}{}_{[1,1]}^2,2}^{\; rutu}\\&-2 g_2^2 \delta _{ps} C_{Q_{[1,1]}^2 Q^{\dagger}{}_{[1,1]}^2,2}^{\; rutu}+\frac{8}{9} g_3^2 \delta _{ps} C_{Q_{[1,1]}^2 Q^{\dagger}{}_{[1,1]}^2,2}^{\; rutu}-\left[Y_d\right]_{up} \left[Y_d\right]_{us}^*C_{Q H H^{\dagger}{} Q^{\dagger}{} D,2}^{\; rt} \\&+\left[Y_d\right]_{uv} \left[Y_d\right]_{us}^*C_{Q_{[2]}^2 Q^{\dagger}{}_{[2]}^2,1}^{\; prtv} +\left[Y_d\right]_{up} \left[Y_d\right]_{uv}^*C_{Q_{[2]}^2 Q^{\dagger}{}_{[2]}^2,1}^{\; rvst} +\left[Y_d\right]_{up} \left[Y_d\right]_{vs}^*C_{d_{\mathbb{C}}{} Q d_{\mathbb{C}}{}^{\dagger}{} Q^{\dagger}{},1}^{\; vrut} \\&+\left[Y_u\right]_{up} \left[Y_u\right]_{us}^*C_{Q H H^{\dagger}{} Q^{\dagger}{} D,1}^{\; rt} +\left[Y_u\right]_{up} \left[Y_u\right]_{us}^*C_{Q H H^{\dagger}{} Q^{\dagger}{} D,2}^{\; rt} +\left[Y_u\right]_{uv} \left[Y_u\right]_{us}^*C_{Q_{[2]}^2 Q^{\dagger}{}_{[2]}^2,1}^{\; prtv} \\&+\left[Y_u\right]_{up} \left[Y_u\right]_{uv}^*C_{Q_{[2]}^2 Q^{\dagger}{}_{[2]}^2,1}^{\; rvst} +\left[Y_u\right]_{up} \left[Y_u\right]_{vs}^*C_{Q u_{\mathbb{C}}{} Q^{\dagger}{} u_{\mathbb{C}}{}^{\dagger}{},2}^{\; rvtu} -\left[Y_d\right]_{us}^* \left[Y_u\right]_{vt}^*C_{d_{\mathbb{C}}{} Q_{[2]}^2 u_{\mathbb{C}}{},1}^{\; uprv} \\&-\left[Y_d\right]_{us}^* \left[Y_u\right]_{vt}^*C_{d_{\mathbb{C}}{} Q_{[2]}^2 u_{\mathbb{C}}{},2}^{\; uprv} \end{split}\\

\begin{split}\dot{C}_{Q_{[2]}^2 Q^{\dagger}{}_{[2]}^2,2}^{\; prst}&\supset \frac{1}{3} g_2^2 \delta _{ps} C_{Q H H^{\dagger}{} Q^{\dagger}{} D,1}^{\; rt}-\frac{2}{3} g_3^2 \delta _{ps} C_{d_{\mathbb{C}}{} Q d_{\mathbb{C}}{}^{\dagger}{} Q^{\dagger}{},2}^{\; urut}+\frac{2}{3} g_2^2 \delta _{ps} C_{L Q L^{\dagger}{} Q^{\dagger}{},2}^{\; urut}\\&-\frac{2}{3} g_3^2 \delta _{ps} C_{Q u_{\mathbb{C}}{} Q^{\dagger}{} u_{\mathbb{C}}{}^{\dagger}{},1}^{\; rutu}+\left[Y_d\right]_{up} \left[Y_u\right]_{vr} C_{d_{\mathbb{C}}{}^{\dagger}{} Q^{\dagger}{}_{[2]}^2 u_{\mathbb{C}}{}^{\dagger}{},1}^{\; ustv}+\left[Y_d\right]_{up} \left[Y_u\right]_{vr} C_{d_{\mathbb{C}}{}^{\dagger}{} Q^{\dagger}{}_{[2]}^2 u_{\mathbb{C}}{}^{\dagger}{},2}^{\; ustv}\\&+6 g_2^2 C_{Q_{[2]}^2 Q^{\dagger}{}_{[2]}^2,1}^{\; prst}+6 g_3^2 C_{Q_{[2]}^2 Q^{\dagger}{}_{[2]}^2,1}^{\; prst}+\frac{4}{3} g_2^2 \delta _{ps} C_{Q_{[2]}^2 Q^{\dagger}{}_{[2]}^2,1}^{\; rutu}\\&+\frac{4}{3} g_3^2 \delta _{ps} C_{Q_{[2]}^2 Q^{\dagger}{}_{[2]}^2,1}^{\; rutu}+\frac{1}{3} g_1^2 C_{Q_{[2]}^2 Q^{\dagger}{}_{[2]}^2,2}^{\; prst}-3 g_2^2 C_{Q_{[2]}^2 Q^{\dagger}{}_{[2]}^2,2}^{\; prst}\\&-2 g_3^2 C_{Q_{[2]}^2 Q^{\dagger}{}_{[2]}^2,2}^{\; prst}+4 g_2^2 \delta _{ps} C_{Q_{[2]}^2 Q^{\dagger}{}_{[2]}^2,2}^{\; rutu}+\frac{8}{3} g_3^2 \delta _{ps} C_{Q_{[2]}^2 Q^{\dagger}{}_{[2]}^2,2}^{\; rutu}\\&-\frac{4}{3} g_2^2 \delta _{ps} C_{Q_{[1,1]}^2 Q^{\dagger}{}_{[1,1]}^2,1}^{\; rutu}-\frac{4}{3} g_3^2 \delta _{ps} C_{Q_{[1,1]}^2 Q^{\dagger}{}_{[1,1]}^2,1}^{\; rutu}+4 g_2^2 \delta _{ps} C_{Q_{[1,1]}^2 Q^{\dagger}{}_{[1,1]}^2,2}^{\; rutu}\\&-\frac{8}{3} g_3^2 \delta _{ps} C_{Q_{[1,1]}^2 Q^{\dagger}{}_{[1,1]}^2,2}^{\; rutu}-\left[Y_d\right]_{up} \left[Y_d\right]_{us}^*C_{Q H H^{\dagger}{} Q^{\dagger}{} D,1}^{\; rt} +\left[Y_d\right]_{uv} \left[Y_d\right]_{us}^*C_{Q_{[2]}^2 Q^{\dagger}{}_{[2]}^2,2}^{\; prtv} \\&+\left[Y_d\right]_{up} \left[Y_d\right]_{uv}^*C_{Q_{[2]}^2 Q^{\dagger}{}_{[2]}^2,2}^{\; rvst} +\left[Y_d\right]_{up} \left[Y_d\right]_{vs}^*C_{d_{\mathbb{C}}{} Q d_{\mathbb{C}}{}^{\dagger}{} Q^{\dagger}{},2}^{\; vrut} -\left[Y_u\right]_{up} \left[Y_u\right]_{us}^*C_{Q H H^{\dagger}{} Q^{\dagger}{} D,1}^{\; rt} \\&+\left[Y_u\right]_{uv} \left[Y_u\right]_{us}^*C_{Q_{[2]}^2 Q^{\dagger}{}_{[2]}^2,2}^{\; prtv} +\left[Y_u\right]_{up} \left[Y_u\right]_{uv}^*C_{Q_{[2]}^2 Q^{\dagger}{}_{[2]}^2,2}^{\; rvst} +\left[Y_u\right]_{up} \left[Y_u\right]_{vs}^*C_{Q u_{\mathbb{C}}{} Q^{\dagger}{} u_{\mathbb{C}}{}^{\dagger}{},1}^{\; rvtu} \\&+\left[Y_d\right]_{us}^* \left[Y_u\right]_{vt}^*C_{d_{\mathbb{C}}{} Q_{[2]}^2 u_{\mathbb{C}}{},1}^{\; uprv} +\left[Y_d\right]_{us}^* \left[Y_u\right]_{vt}^*C_{d_{\mathbb{C}}{} Q_{[2]}^2 u_{\mathbb{C}}{},2}^{\; uprv} \end{split}\\

\begin{split}\dot{C}_{Q_{[1,1]}^2 Q^{\dagger}{}_{[1,1]}^2,1}^{\; prst}&\supset \frac{1}{18} g_1^2 \delta _{ps} C_{Q H H^{\dagger}{} Q^{\dagger}{} D,1}^{\; rt}-\frac{1}{6} g_2^2 \delta _{ps} C_{Q H H^{\dagger}{} Q^{\dagger}{} D,1}^{\; rt}+\frac{1}{9} g_1^2 \delta _{ps} C_{Q H H^{\dagger}{} Q^{\dagger}{} D,2}^{\; rt}\\&+\frac{2}{9} g_1^2 \delta _{ps} C_{d_{\mathbb{C}}{} Q d_{\mathbb{C}}{}^{\dagger}{} Q^{\dagger}{},1}^{\; urut}+\frac{2}{27} g_1^2 \delta _{ps} C_{d_{\mathbb{C}}{} Q d_{\mathbb{C}}{}^{\dagger}{} Q^{\dagger}{},2}^{\; urut}+\frac{2}{9} g_3^2 \delta _{ps} C_{d_{\mathbb{C}}{} Q d_{\mathbb{C}}{}^{\dagger}{} Q^{\dagger}{},2}^{\; urut}\\&+\frac{2}{9} g_1^2 \delta _{ps} C_{e_{\mathbb{C}}{} Q e_{\mathbb{C}}{}^{\dagger}{} Q^{\dagger}{},1}^{\; urut}-\frac{2}{9} g_1^2 \delta _{ps} C_{L Q L^{\dagger}{} Q^{\dagger}{},1}^{\; urut}-\frac{1}{9} g_1^2 \delta _{ps} C_{L Q L^{\dagger}{} Q^{\dagger}{},2}^{\; urut}\\&-\frac{1}{3} g_2^2 \delta _{ps} C_{L Q L^{\dagger}{} Q^{\dagger}{},2}^{\; urut}-\frac{4}{27} g_1^2 \delta _{ps} C_{Q u_{\mathbb{C}}{} Q^{\dagger}{} u_{\mathbb{C}}{}^{\dagger}{},1}^{\; rutu}+\frac{2}{9} g_3^2 \delta _{ps} C_{Q u_{\mathbb{C}}{} Q^{\dagger}{} u_{\mathbb{C}}{}^{\dagger}{},1}^{\; rutu}\\&-\frac{4}{9} g_1^2 \delta _{ps} C_{Q u_{\mathbb{C}}{} Q^{\dagger}{} u_{\mathbb{C}}{}^{\dagger}{},2}^{\; rutu}+\left[Y_d\right]_{up} \left[Y_u\right]_{vr} C_{d_{\mathbb{C}}{}^{\dagger}{} Q^{\dagger}{}_{[1,1]}^2 u_{\mathbb{C}}{}^{\dagger}{},1}^{\; ustv}+\left[Y_d\right]_{up} \left[Y_u\right]_{vr} C_{d_{\mathbb{C}}{}^{\dagger}{} Q^{\dagger}{}_{[1,1]}^2 u_{\mathbb{C}}{}^{\dagger}{},2}^{\; ustv}\\&+\frac{14}{27} g_1^2 \delta _{ps} C_{Q_{[2]}^2 Q^{\dagger}{}_{[2]}^2,1}^{\; rutu}-\frac{2}{3} g_2^2 \delta _{ps} C_{Q_{[2]}^2 Q^{\dagger}{}_{[2]}^2,1}^{\; rutu}-\frac{4}{9} g_3^2 \delta _{ps} C_{Q_{[2]}^2 Q^{\dagger}{}_{[2]}^2,1}^{\; rutu}\\&+\frac{10}{27} g_1^2 \delta _{ps} C_{Q_{[2]}^2 Q^{\dagger}{}_{[2]}^2,2}^{\; rutu}-2 g_2^2 \delta _{ps} C_{Q_{[2]}^2 Q^{\dagger}{}_{[2]}^2,2}^{\; rutu}-\frac{8}{9} g_3^2 \delta _{ps} C_{Q_{[2]}^2 Q^{\dagger}{}_{[2]}^2,2}^{\; rutu}\\&+\frac{1}{3} g_1^2 C_{Q_{[1,1]}^2 Q^{\dagger}{}_{[1,1]}^2,1}^{\; prst}-3 g_2^2 C_{Q_{[1,1]}^2 Q^{\dagger}{}_{[1,1]}^2,1}^{\; prst}-2 g_3^2 C_{Q_{[1,1]}^2 Q^{\dagger}{}_{[1,1]}^2,1}^{\; prst}\\&+\frac{10}{27} g_1^2 \delta _{ps} C_{Q_{[1,1]}^2 Q^{\dagger}{}_{[1,1]}^2,1}^{\; rutu}+\frac{2}{3} g_2^2 \delta _{ps} C_{Q_{[1,1]}^2 Q^{\dagger}{}_{[1,1]}^2,1}^{\; rutu}+\frac{4}{9} g_3^2 \delta _{ps} C_{Q_{[1,1]}^2 Q^{\dagger}{}_{[1,1]}^2,1}^{\; rutu}\\&+6 g_2^2 C_{Q_{[1,1]}^2 Q^{\dagger}{}_{[1,1]}^2,2}^{\; prst}-6 g_3^2 C_{Q_{[1,1]}^2 Q^{\dagger}{}_{[1,1]}^2,2}^{\; prst}+\frac{2}{27} g_1^2 \delta _{ps} C_{Q_{[1,1]}^2 Q^{\dagger}{}_{[1,1]}^2,2}^{\; rutu}\\&-2 g_2^2 \delta _{ps} C_{Q_{[1,1]}^2 Q^{\dagger}{}_{[1,1]}^2,2}^{\; rutu}+\frac{8}{9} g_3^2 \delta _{ps} C_{Q_{[1,1]}^2 Q^{\dagger}{}_{[1,1]}^2,2}^{\; rutu}-\left[Y_d\right]_{up} \left[Y_d\right]_{us}^*C_{Q H H^{\dagger}{} Q^{\dagger}{} D,2}^{\; rt} \\&-\left[Y_d\right]_{uv} \left[Y_d\right]_{us}^*C_{Q_{[1,1]}^2 Q^{\dagger}{}_{[1,1]}^2,1}^{\; prtv} -\left[Y_d\right]_{up} \left[Y_d\right]_{uv}^*C_{Q_{[1,1]}^2 Q^{\dagger}{}_{[1,1]}^2,1}^{\; rvst} +\left[Y_d\right]_{up} \left[Y_d\right]_{vs}^*C_{d_{\mathbb{C}}{} Q d_{\mathbb{C}}{}^{\dagger}{} Q^{\dagger}{},1}^{\; vrut} \\&+\left[Y_u\right]_{up} \left[Y_u\right]_{us}^*C_{Q H H^{\dagger}{} Q^{\dagger}{} D,1}^{\; rt} +\left[Y_u\right]_{up} \left[Y_u\right]_{us}^*C_{Q H H^{\dagger}{} Q^{\dagger}{} D,2}^{\; rt} -\left[Y_u\right]_{uv} \left[Y_u\right]_{us}^*C_{Q_{[1,1]}^2 Q^{\dagger}{}_{[1,1]}^2,1}^{\; prtv} \\&-\left[Y_u\right]_{up} \left[Y_u\right]_{uv}^*C_{Q_{[1,1]}^2 Q^{\dagger}{}_{[1,1]}^2,1}^{\; rvst} +\left[Y_u\right]_{up} \left[Y_u\right]_{vs}^*C_{Q u_{\mathbb{C}}{} Q^{\dagger}{} u_{\mathbb{C}}{}^{\dagger}{},2}^{\; rvtu} +\left[Y_d\right]_{us}^* \left[Y_u\right]_{vt}^*C_{d_{\mathbb{C}}{} Q_{[1,1]}^2 u_{\mathbb{C}}{},1}^{\; uprv} \\&+\left[Y_d\right]_{us}^* \left[Y_u\right]_{vt}^*C_{d_{\mathbb{C}}{} Q_{[1,1]}^2 u_{\mathbb{C}}{},2}^{\; uprv} \end{split}\\

\begin{split}\dot{C}_{Q_{[1,1]}^2 Q^{\dagger}{}_{[1,1]}^2,2}^{\; prst}&\supset \frac{1}{3} g_2^2 \delta _{ps} C_{Q H H^{\dagger}{} Q^{\dagger}{} D,1}^{\; rt}+\frac{2}{3} g_3^2 \delta _{ps} C_{d_{\mathbb{C}}{} Q d_{\mathbb{C}}{}^{\dagger}{} Q^{\dagger}{},2}^{\; urut}+\frac{2}{3} g_2^2 \delta _{ps} C_{L Q L^{\dagger}{} Q^{\dagger}{},2}^{\; urut}\\&+\frac{2}{3} g_3^2 \delta _{ps} C_{Q u_{\mathbb{C}}{} Q^{\dagger}{} u_{\mathbb{C}}{}^{\dagger}{},1}^{\; rutu}-\left[Y_d\right]_{up} \left[Y_u\right]_{vr} C_{d_{\mathbb{C}}{}^{\dagger}{} Q^{\dagger}{}_{[1,1]}^2 u_{\mathbb{C}}{}^{\dagger}{},1}^{\; ustv}-\left[Y_d\right]_{up} \left[Y_u\right]_{vr} C_{d_{\mathbb{C}}{}^{\dagger}{} Q^{\dagger}{}_{[1,1]}^2 u_{\mathbb{C}}{}^{\dagger}{},2}^{\; ustv}\\&+\frac{4}{3} g_2^2 \delta _{ps} C_{Q_{[2]}^2 Q^{\dagger}{}_{[2]}^2,1}^{\; rutu}-\frac{4}{3} g_3^2 \delta _{ps} C_{Q_{[2]}^2 Q^{\dagger}{}_{[2]}^2,1}^{\; rutu}+4 g_2^2 \delta _{ps} C_{Q_{[2]}^2 Q^{\dagger}{}_{[2]}^2,2}^{\; rutu}\\&-\frac{8}{3} g_3^2 \delta _{ps} C_{Q_{[2]}^2 Q^{\dagger}{}_{[2]}^2,2}^{\; rutu}+6 g_2^2 C_{Q_{[1,1]}^2 Q^{\dagger}{}_{[1,1]}^2,1}^{\; prst}-6 g_3^2 C_{Q_{[1,1]}^2 Q^{\dagger}{}_{[1,1]}^2,1}^{\; prst}\\&-\frac{4}{3} g_2^2 \delta _{ps} C_{Q_{[1,1]}^2 Q^{\dagger}{}_{[1,1]}^2,1}^{\; rutu}+\frac{4}{3} g_3^2 \delta _{ps} C_{Q_{[1,1]}^2 Q^{\dagger}{}_{[1,1]}^2,1}^{\; rutu}+\frac{1}{3} g_1^2 C_{Q_{[1,1]}^2 Q^{\dagger}{}_{[1,1]}^2,2}^{\; prst}\\&-3 g_2^2 C_{Q_{[1,1]}^2 Q^{\dagger}{}_{[1,1]}^2,2}^{\; prst}-2 g_3^2 C_{Q_{[1,1]}^2 Q^{\dagger}{}_{[1,1]}^2,2}^{\; prst}+4 g_2^2 \delta _{ps} C_{Q_{[1,1]}^2 Q^{\dagger}{}_{[1,1]}^2,2}^{\; rutu}\\&+\frac{8}{3} g_3^2 \delta _{ps} C_{Q_{[1,1]}^2 Q^{\dagger}{}_{[1,1]}^2,2}^{\; rutu}-\left[Y_d\right]_{up} \left[Y_d\right]_{us}^*C_{Q H H^{\dagger}{} Q^{\dagger}{} D,1}^{\; rt} -\left[Y_d\right]_{uv} \left[Y_d\right]_{us}^*C_{Q_{[1,1]}^2 Q^{\dagger}{}_{[1,1]}^2,2}^{\; prtv} \\&-\left[Y_d\right]_{up} \left[Y_d\right]_{uv}^*C_{Q_{[1,1]}^2 Q^{\dagger}{}_{[1,1]}^2,2}^{\; rvst} -\left[Y_d\right]_{up} \left[Y_d\right]_{vs}^*C_{d_{\mathbb{C}}{} Q d_{\mathbb{C}}{}^{\dagger}{} Q^{\dagger}{},2}^{\; vrut} -\left[Y_u\right]_{up} \left[Y_u\right]_{us}^*C_{Q H H^{\dagger}{} Q^{\dagger}{} D,1}^{\; rt} \\&-\left[Y_u\right]_{uv} \left[Y_u\right]_{us}^*C_{Q_{[1,1]}^2 Q^{\dagger}{}_{[1,1]}^2,2}^{\; prtv} -\left[Y_u\right]_{up} \left[Y_u\right]_{uv}^*C_{Q_{[1,1]}^2 Q^{\dagger}{}_{[1,1]}^2,2}^{\; rvst} -\left[Y_u\right]_{up} \left[Y_u\right]_{vs}^*C_{Q u_{\mathbb{C}}{} Q^{\dagger}{} u_{\mathbb{C}}{}^{\dagger}{},1}^{\; rvtu} \\&-\left[Y_d\right]_{us}^* \left[Y_u\right]_{vt}^*C_{d_{\mathbb{C}}{} Q_{[1,1]}^2 u_{\mathbb{C}}{},1}^{\; uprv} -\left[Y_d\right]_{us}^* \left[Y_u\right]_{vt}^*C_{d_{\mathbb{C}}{} Q_{[1,1]}^2 u_{\mathbb{C}}{},2}^{\; uprv} \end{split}
\end{align}
The results presented above are written in the full-flavor form, with explicit flavor indices kept for all fermion fields.
This representation retains the complete flavor structure of the theory and is suitable for applications with arbitrary flavor assumptions.

For practical purposes, it is often useful to consider the simplified case of a single fermion generation.
We therefore also provide the corresponding results for $N_f=1$, obtained by setting all flavor indices equal and applying the appropriate symmetry projections.
A superscript asterisk on a Wilson coefficient indicates complex conjugation. Whenever a coefficient appears with a superscript *, the associated operator is understood to be the Hermitian conjugate of the original one.
In the remainder of this section, we present only the simplified results for $N_f=1$ for brevity.
The complete RGEs with full flavor dependence are available in the accompanying
\textbf{Mathematica} file.
\begin{align}
\begin{split}\dot{C}_{q,1}^{(d6)}&\supset \frac{1}{18} g_1^2 C_{Hq,1}^{(d6)}-\frac{1}{6} g_2^2 C_{Hq,1}^{(d6)}+\frac{1}{9} g_1^2 C_{Hq,2}^{(d6)}-\frac{2}{9} g_1^2 C_{lq,1}^{(d6)}-\frac{1}{9} g_1^2 C_{lq,2}^{(d6)}-\frac{1}{3} g_2^2 C_{lq,2}^{(d6)}\\&+\frac{23}{27} g_1^2 C_{q,1}^{(d6)}-\frac{11}{3} g_2^2 C_{q,1}^{(d6)}-\frac{22}{9} g_3^2 C_{q,1}^{(d6)}+\frac{10}{27} g_1^2 C_{q,2}^{(d6)}+4 g_2^2 C_{q,2}^{(d6)}+\frac{46}{9} g_3^2 C_{q,2}^{(d6)}\\&+\frac{2}{9} g_1^2 C_{qd,1}^{(d6)}+\frac{2}{27} g_1^2 C_{qd,2}^{(d6)}+\frac{2}{9} g_3^2 C_{qd,2}^{(d6)}+\frac{2}{9} g_1^2 C_{qe,1}^{(d6)}-\frac{4}{27} g_1^2 C_{qu,1}^{(d6)}+\frac{2}{9} g_3^2 C_{qu,1}^{(d6)}\\&-\frac{4}{9} g_1^2 C_{qu,2}^{(d6)}-Y_d Y_u C_{q^2ud,1}^{(d6)*}-Y_d Y_u C_{q^2ud,2}^{(d6)*}-Y_d C_{Hq,2}^{(d6)} Y_d{}^*+2 Y_d C_{q,1}^{(d6)} Y_d{}^*+Y_d C_{qd,1}^{(d6)} Y_d{}^*\\&+Y_u C_{Hq,1}^{(d6)} Y_u{}^*+Y_u C_{Hq,2}^{(d6)} Y_u{}^*+2 Y_u C_{q,1}^{(d6)} Y_u{}^*+Y_u C_{qu,2}^{(d6)} Y_u{}^*-C_{q^2ud,1}^{(d6)} Y_d{}^* Y_u{}^*-C_{q^2ud,2}^{(d6)} Y_d{}^* Y_u{}^*\end{split}\\

\begin{split}\dot{C}_{q,2}^{(d6)}&\supset \frac{1}{3} g_2^2 C_{Hq,1}^{(d6)}+\frac{2}{3} g_2^2 C_{lq,2}^{(d6)}+\frac{22}{3} g_2^2 C_{q,1}^{(d6)}+\frac{22}{3} g_3^2 C_{q,1}^{(d6)}+\frac{1}{3} g_1^2 C_{q,2}^{(d6)}+g_2^2 C_{q,2}^{(d6)}\\&+\frac{2}{3} g_3^2 C_{q,2}^{(d6)}-\frac{2}{3} g_3^2 C_{qd,2}^{(d6)}-\frac{2}{3} g_3^2 C_{qu,1}^{(d6)}+Y_d Y_u C_{q^2ud,1}^{(d6)*}+Y_d Y_u C_{q^2ud,2}^{(d6)*}-Y_d C_{Hq,1}^{(d6)} Y_d{}^*\\&+2 Y_d C_{q,2}^{(d6)} Y_d{}^*+Y_d C_{qd,2}^{(d6)} Y_d{}^*-Y_u C_{Hq,1}^{(d6)} Y_u{}^*+2 Y_u C_{q,2}^{(d6)} Y_u{}^*+Y_u C_{qu,1}^{(d6)} Y_u{}^*+C_{q^2ud,1}^{(d6)} Y_d{}^* Y_u{}^*\\&+C_{q^2ud,2}^{(d6)} Y_d{}^* Y_u{}^*\end{split}
\end{align}

\begin{align}
\begin{split}\dot{C}_{q,1}^{(d8)}&\supset \frac{5}{54} g_1 g_3 C_{BGq,1}^{(d8)}+\frac{5}{54} g_1 g_3 C_{BGq,1}^{(d8)*}-\frac{5}{108} g_1^2 C_{Bq,1}^{(d8)}+\frac{5}{36} g_1 g_2 C_{BWq,1}^{(d8)}+\frac{5}{36} g_1 g_2 C_{BWq,1}^{(d8)*}+\frac{25}{54} g_3^2 C_{Gq,1}^{(d8)}\\&-\frac{5}{12} i g_3^2 C_{Gq,2}^{(d8)}-\frac{20}{9} g_3^2 C_{Gq,3}^{(d8)}+\frac{1}{72} g_1^2 C_{Hq,1}^{(d8)}-\frac{1}{24} g_2^2 C_{Hq,1}^{(d8)}+\frac{1}{72} g_1^2 C_{Hq,2}^{(d8)}-\frac{1}{24} g_2^2 C_{Hq,2}^{(d8)}\\&+\frac{1}{36} g_1^2 C_{Hq,3}^{(d8)}+\frac{1}{36} g_1^2 C_{Hq,4}^{(d8)}-\frac{1}{12} g_1^2 C_{lq,1}^{(d8)}+\frac{5}{36} g_1^2 C_{lq,2}^{(d8)}-\frac{1}{24} g_1^2 C_{lq,3}^{(d8)}-\frac{1}{8} g_2^2 C_{lq,3}^{(d8)}\\&+\frac{5}{72} g_1^2 C_{lq,4}^{(d8)}+\frac{5}{24} g_2^2 C_{lq,4}^{(d8)}+\frac{97}{108} g_1^2 C_{q,1}^{(d8)}+\frac{1}{12} g_2^2 C_{q,1}^{(d8)}+\frac{101}{18} g_3^2 C_{q,1}^{(d8)}-\frac{25}{108} g_1^2 C_{q,2}^{(d8)}\\&+\frac{25}{4} g_2^2 C_{q,2}^{(d8)}+\frac{175}{18} g_3^2 C_{q,2}^{(d8)}+\frac{5}{36} g_1^2 C_{q,3}^{(d8)}+\frac{103}{12} g_2^2 C_{q,3}^{(d8)}+9 g_3^2 C_{q,3}^{(d8)}-\frac{5}{108} g_1^2 C_{q,4}^{(d8)}\\&+\frac{55}{12} g_2^2 C_{q,4}^{(d8)}-\frac{35}{9} g_3^2 C_{q,4}^{(d8)}+\frac{1}{12} g_1^2 C_{qd,1}^{(d8)}-\frac{5}{36} g_1^2 C_{qd,2}^{(d8)}+\frac{1}{36} g_1^2 C_{qd,3}^{(d8)}+\frac{1}{12} g_3^2 C_{qd,3}^{(d8)}\\&-\frac{5}{108} g_1^2 C_{qd,4}^{(d8)}-\frac{5}{36} g_3^2 C_{qd,4}^{(d8)}+\frac{1}{12} g_1^2 C_{qe,1}^{(d8)}-\frac{5}{36} g_1^2 C_{qe,2}^{(d8)}-\frac{1}{18} g_1^2 C_{qu,1}^{(d8)}+\frac{1}{12} g_3^2 C_{qu,1}^{(d8)}\\&+\frac{5}{54} g_1^2 C_{qu,2}^{(d8)}-\frac{5}{36} g_3^2 C_{qu,2}^{(d8)}-\frac{1}{6} g_1^2 C_{qu,3}^{(d8)}+\frac{5}{18} g_1^2 C_{qu,4}^{(d8)}+\frac{2}{3} Y_d Y_u C_{q^2ud,1}^{(d8)*}+\frac{2}{3} Y_d Y_u C_{q^2ud,2}^{(d8)*}\\&+\frac{4}{3} Y_d Y_u C_{q^2ud,3}^{(d8)*}+\frac{35}{18} g_2 g_3 C_{WGq,1}^{(d8)}+\frac{35}{18} g_2 g_3 C_{WGq,1}^{(d8)*}-\frac{5}{12} i g_2^2 C_{Wq,1}^{(d8)}-\frac{5}{4} g_2^2 C_{Wq,2}^{(d8)}-\frac{2}{3} Y_d C_{Hq,3}^{(d8)} Y_d{}^*\\&+\frac{1}{6} Y_d C_{Hq,4}^{(d8)} Y_d{}^*+2 Y_d C_{q,1}^{(d8)} Y_d{}^*+\frac{1}{6} Y_d C_{qd,1}^{(d8)} Y_d{}^*-\frac{5}{6} Y_d C_{qd,2}^{(d8)} Y_d{}^*-\frac{1}{6} Y_u C_{Hq,1}^{(d8)} Y_u{}^*+\frac{2}{3} Y_u C_{Hq,2}^{(d8)} Y_u{}^*\\&-\frac{1}{6} Y_u C_{Hq,3}^{(d8)} Y_u{}^*+\frac{2}{3} Y_u C_{Hq,4}^{(d8)} Y_u{}^*+2 Y_u C_{q,1}^{(d8)} Y_u{}^*+\frac{1}{6} Y_u C_{qu,3}^{(d8)} Y_u{}^*-\frac{5}{6} Y_u C_{qu,4}^{(d8)} Y_u{}^*+\frac{2}{3} C_{q^2ud,1}^{(d8)} Y_d{}^* Y_u{}^*\\&+\frac{2}{3} C_{q^2ud,2}^{(d8)} Y_d{}^* Y_u{}^*+\frac{4}{3} C_{q^2ud,3}^{(d8)} Y_d{}^* Y_u{}^*\end{split}\\

\begin{split}\dot{C}_{q,2}^{(d8)}&\supset \frac{1}{18} g_1 g_3 C_{BGq,1}^{(d8)}+\frac{1}{18} g_1 g_3 C_{BGq,1}^{(d8)*}-\frac{1}{36} g_1^2 C_{Bq,1}^{(d8)}+\frac{1}{12} g_1 g_2 C_{BWq,1}^{(d8)}+\frac{1}{12} g_1 g_2 C_{BWq,1}^{(d8)*}+\frac{5}{18} g_3^2 C_{Gq,1}^{(d8)}\\&-\frac{1}{4} i g_3^2 C_{Gq,2}^{(d8)}-\frac{4}{3} g_3^2 C_{Gq,3}^{(d8)}-\frac{1}{72} g_1^2 C_{Hq,1}^{(d8)}+\frac{1}{24} g_2^2 C_{Hq,1}^{(d8)}-\frac{1}{72} g_1^2 C_{Hq,2}^{(d8)}+\frac{1}{24} g_2^2 C_{Hq,2}^{(d8)}\\&-\frac{1}{36} g_1^2 C_{Hq,3}^{(d8)}-\frac{1}{36} g_1^2 C_{Hq,4}^{(d8)}+\frac{1}{12} g_1^2 C_{lq,1}^{(d8)}-\frac{5}{36} g_1^2 C_{lq,2}^{(d8)}+\frac{1}{24} g_1^2 C_{lq,3}^{(d8)}+\frac{1}{8} g_2^2 C_{lq,3}^{(d8)}\\&-\frac{5}{72} g_1^2 C_{lq,4}^{(d8)}-\frac{5}{24} g_2^2 C_{lq,4}^{(d8)}-\frac{7}{36} g_1^2 C_{q,1}^{(d8)}+\frac{17}{4} g_2^2 C_{q,1}^{(d8)}+\frac{37}{6} g_3^2 C_{q,1}^{(d8)}+\frac{37}{108} g_1^2 C_{q,2}^{(d8)}\\&+\frac{41}{12} g_2^2 C_{q,2}^{(d8)}+\frac{101}{18} g_3^2 C_{q,2}^{(d8)}-\frac{5}{36} g_1^2 C_{q,3}^{(d8)}+\frac{11}{4} g_2^2 C_{q,3}^{(d8)}+\frac{7}{3} g_3^2 C_{q,3}^{(d8)}+\frac{5}{108} g_1^2 C_{q,4}^{(d8)}\\&-\frac{5}{4} g_2^2 C_{q,4}^{(d8)}+\frac{5}{9} g_3^2 C_{q,4}^{(d8)}-\frac{1}{12} g_1^2 C_{qd,1}^{(d8)}+\frac{5}{36} g_1^2 C_{qd,2}^{(d8)}-\frac{1}{36} g_1^2 C_{qd,3}^{(d8)}-\frac{1}{12} g_3^2 C_{qd,3}^{(d8)}\\&+\frac{5}{108} g_1^2 C_{qd,4}^{(d8)}+\frac{5}{36} g_3^2 C_{qd,4}^{(d8)}-\frac{1}{12} g_1^2 C_{qe,1}^{(d8)}+\frac{5}{36} g_1^2 C_{qe,2}^{(d8)}+\frac{1}{18} g_1^2 C_{qu,1}^{(d8)}-\frac{1}{12} g_3^2 C_{qu,1}^{(d8)}\\&-\frac{5}{54} g_1^2 C_{qu,2}^{(d8)}+\frac{5}{36} g_3^2 C_{qu,2}^{(d8)}+\frac{1}{6} g_1^2 C_{qu,3}^{(d8)}-\frac{5}{18} g_1^2 C_{qu,4}^{(d8)}-\frac{5}{6} g_2 g_3 C_{WGq,1}^{(d8)}-\frac{5}{6} g_2 g_3 C_{WGq,1}^{(d8)*}\\&-\frac{1}{4} i g_2^2 C_{Wq,1}^{(d8)}-\frac{3}{4} g_2^2 C_{Wq,2}^{(d8)}+\frac{1}{2} Y_d C_{Hq,4}^{(d8)} Y_d{}^*+2 Y_d C_{q,2}^{(d8)} Y_d{}^*-\frac{1}{2} Y_d C_{qd,1}^{(d8)} Y_d{}^*+\frac{1}{2} Y_d C_{qd,2}^{(d8)} Y_d{}^*\\&-\frac{1}{2} Y_u C_{Hq,1}^{(d8)} Y_u{}^*-\frac{1}{2} Y_u C_{Hq,3}^{(d8)} Y_u{}^*+2 Y_u C_{q,2}^{(d8)} Y_u{}^*-\frac{1}{2} Y_u C_{qu,3}^{(d8)} Y_u{}^*+\frac{1}{2} Y_u C_{qu,4}^{(d8)} Y_u{}^*\end{split}\\

\begin{split}\dot{C}_{q,3}^{(d8)}&\supset -\frac{5}{18} g_1 g_3 C_{BGq,1}^{(d8)}-\frac{5}{18} g_1 g_3 C_{BGq,1}^{(d8)*}-\frac{5}{18} g_1 g_2 C_{BWq,1}^{(d8)}-\frac{5}{18} g_1 g_2 C_{BWq,1}^{(d8)*}-\frac{25}{18} g_3^2 C_{Gq,1}^{(d8)}\\&+\frac{5}{4} i g_3^2 C_{Gq,2}^{(d8)}+\frac{1}{12} g_2^2 C_{Hq,1}^{(d8)}+\frac{1}{12} g_2^2 C_{Hq,2}^{(d8)}+\frac{1}{4} g_2^2 C_{lq,3}^{(d8)}-\frac{5}{12} g_2^2 C_{lq,4}^{(d8)}+\frac{59}{6} g_2^2 C_{q,1}^{(d8)}\\&+\frac{59}{6} g_3^2 C_{q,1}^{(d8)}-\frac{5}{2} g_2^2 C_{q,2}^{(d8)}-\frac{5}{2} g_3^2 C_{q,2}^{(d8)}+\frac{19}{27} g_1^2 C_{q,3}^{(d8)}+\frac{11}{6} g_2^2 C_{q,3}^{(d8)}+\frac{61}{9} g_3^2 C_{q,3}^{(d8)}\\&-\frac{55}{6} g_2^2 C_{q,4}^{(d8)}+\frac{35}{3} g_3^2 C_{q,4}^{(d8)}-\frac{1}{4} g_3^2 C_{qd,3}^{(d8)}+\frac{5}{12} g_3^2 C_{qd,4}^{(d8)}-\frac{1}{4} g_3^2 C_{qu,1}^{(d8)}+\frac{5}{12} g_3^2 C_{qu,2}^{(d8)}\\&-\frac{2}{3} Y_d Y_u C_{q^2ud,1}^{(d8)*}-\frac{2}{3} Y_d Y_u C_{q^2ud,2}^{(d8)*}-\frac{4}{3} Y_d Y_u C_{q^2ud,3}^{(d8)*}-\frac{25}{18} g_2 g_3 C_{WGq,1}^{(d8)}-\frac{25}{18} g_2 g_3 C_{WGq,1}^{(d8)*}+\frac{5}{6} i g_2^2 C_{Wq,1}^{(d8)}\\&-\frac{2}{3} Y_d C_{Hq,1}^{(d8)} Y_d{}^*+\frac{1}{6} Y_d C_{Hq,2}^{(d8)} Y_d{}^*+2 Y_d C_{q,3}^{(d8)} Y_d{}^*+\frac{1}{6} Y_d C_{qd,3}^{(d8)} Y_d{}^*-\frac{5}{6} Y_d C_{qd,4}^{(d8)} Y_d{}^*+\frac{1}{6} Y_u C_{Hq,1}^{(d8)} Y_u{}^*\\&-\frac{2}{3} Y_u C_{Hq,2}^{(d8)} Y_u{}^*+2 Y_u C_{q,3}^{(d8)} Y_u{}^*+\frac{1}{6} Y_u C_{qu,1}^{(d8)} Y_u{}^*-\frac{5}{6} Y_u C_{qu,2}^{(d8)} Y_u{}^*-\frac{2}{3} C_{q^2ud,1}^{(d8)} Y_d{}^* Y_u{}^*-\frac{2}{3} C_{q^2ud,2}^{(d8)} Y_d{}^* Y_u{}^*\\&-\frac{4}{3} C_{q^2ud,3}^{(d8)} Y_d{}^* Y_u{}^*\end{split}\\

\begin{split}\dot{C}_{q,4}^{(d8)}&\supset \frac{1}{6} g_1 g_3 C_{BGq,1}^{(d8)}+\frac{1}{6} g_1 g_3 C_{BGq,1}^{(d8)*}-\frac{1}{6} g_1 g_2 C_{BWq,1}^{(d8)}-\frac{1}{6} g_1 g_2 C_{BWq,1}^{(d8)*}+\frac{5}{6} g_3^2 C_{Gq,1}^{(d8)}-\frac{3}{4} i g_3^2 C_{Gq,2}^{(d8)}\\&-\frac{1}{12} g_2^2 C_{Hq,1}^{(d8)}-\frac{1}{12} g_2^2 C_{Hq,2}^{(d8)}-\frac{1}{4} g_2^2 C_{lq,3}^{(d8)}+\frac{5}{12} g_2^2 C_{lq,4}^{(d8)}-\frac{5}{2} g_2^2 C_{q,1}^{(d8)}+\frac{5}{2} g_3^2 C_{q,1}^{(d8)}\\&-\frac{5}{6} g_2^2 C_{q,2}^{(d8)}+\frac{5}{6} g_3^2 C_{q,2}^{(d8)}-\frac{11}{2} g_2^2 C_{q,3}^{(d8)}+7 g_3^2 C_{q,3}^{(d8)}+\frac{1}{9} g_1^2 C_{q,4}^{(d8)}+\frac{11}{2} g_2^2 C_{q,4}^{(d8)}\\&+7 g_3^2 C_{q,4}^{(d8)}-\frac{1}{4} g_3^2 C_{qd,3}^{(d8)}+\frac{5}{12} g_3^2 C_{qd,4}^{(d8)}-\frac{1}{4} g_3^2 C_{qu,1}^{(d8)}+\frac{5}{12} g_3^2 C_{qu,2}^{(d8)}+\frac{1}{6} g_2 g_3 C_{WGq,1}^{(d8)}\\&+\frac{1}{6} g_2 g_3 C_{WGq,1}^{(d8)*}+\frac{1}{2} i g_2^2 C_{Wq,1}^{(d8)}+\frac{1}{2} Y_d C_{Hq,2}^{(d8)} Y_d{}^*+2 Y_d C_{q,4}^{(d8)} Y_d{}^*+\frac{1}{2} Y_d C_{qd,3}^{(d8)} Y_d{}^*-\frac{1}{2} Y_d C_{qd,4}^{(d8)} Y_d{}^*\\&+\frac{1}{2} Y_u C_{Hq,1}^{(d8)} Y_u{}^*+2 Y_u C_{q,4}^{(d8)} Y_u{}^*+\frac{1}{2} Y_u C_{qu,1}^{(d8)} Y_u{}^*-\frac{1}{2} Y_u C_{qu,2}^{(d8)} Y_u{}^*\end{split}
\end{align}
We now turn to the quadratic contributions, i.e. the dimension-8 running induced by double insertions of lower-dimensional operators.
\begin{align}
\begin{split}\dot{C}_{q,1}^{(d8)}&\supset \frac{2}{3} C_{Hq,1}^{(d6)} C_{Hq,2}^{(d6)}+\frac{2}{3} \left(C_{Hq,2}^{(d6)}\right){}^2+\frac{4}{3} \left(C_{lq,1}^{(d6)}\right){}^2+\frac{4}{3} C_{lq,1}^{(d6)} C_{lq,2}^{(d6)}+40 \left(C_{q,1}^{(d6)}\right){}^2+\frac{80}{3} C_{q,1}^{(d6)} C_{q,2}^{(d6)}\\&+\frac{64}{3} \left(C_{q,2}^{(d6)}\right){}^2+2 \left(C_{qd,1}^{(d6)}\right){}^2+\frac{4}{3} C_{qd,1}^{(d6)} C_{qd,2}^{(d6)}+\frac{2}{3} \left(C_{qe,1}^{(d6)}\right){}^2+\frac{4}{3} C_{qu,1}^{(d6)} C_{qu,2}^{(d6)}+2 \left(C_{qu,2}^{(d6)}\right){}^2\\&+C_{q^2ud,1}^{(d6)} C_{q^2ud,1}^{(d6)*}+C_{q^2ud,1}^{(d6)*} C_{q^2ud,2}^{(d6)}+C_{q^2ud,1}^{(d6)} C_{q^2ud,2}^{(d6)*}+C_{q^2ud,2}^{(d6)} C_{q^2ud,2}^{(d6)*}+8 C_{q^2ue,1}^{(d6)} C_{q^2ue,1}^{(d6)*}+9 C_{q^3l,1}^{(d6)} C_{q^3l,1}^{(d6)*}\end{split}\\

\begin{split}\dot{C}_{q,2}^{(d8)}&\supset -\frac{2}{3} C_{Hq,1}^{(d6)} C_{Hq,2}^{(d6)}-\frac{2}{3} \left(C_{Hq,2}^{(d6)}\right){}^2-\frac{4}{3} \left(C_{lq,1}^{(d6)}\right){}^2-\frac{4}{3} C_{lq,1}^{(d6)} C_{lq,2}^{(d6)}-\frac{56}{3} \left(C_{q,1}^{(d6)}\right){}^2\\&-\frac{80}{3} C_{q,1}^{(d6)} C_{q,2}^{(d6)}-\frac{16}{3} \left(C_{q,2}^{(d6)}\right){}^2-2 \left(C_{qd,1}^{(d6)}\right){}^2-\frac{4}{3} C_{qd,1}^{(d6)} C_{qd,2}^{(d6)}-\frac{2}{3} \left(C_{qe,1}^{(d6)}\right){}^2-\frac{4}{3} C_{qu,1}^{(d6)} C_{qu,2}^{(d6)}\\&-2 \left(C_{qu,2}^{(d6)}\right){}^2+\frac{1}{3} C_{q^2ud,1}^{(d6)} C_{q^2ud,1}^{(d6)*}-\frac{1}{3} C_{q^2ud,1}^{(d6)*} C_{q^2ud,2}^{(d6)}-\frac{1}{3} C_{q^2ud,1}^{(d6)} C_{q^2ud,2}^{(d6)*}+\frac{1}{3} C_{q^2ud,2}^{(d6)} C_{q^2ud,2}^{(d6)*}+\frac{1}{3} C_{q^3l,1}^{(d6)} C_{q^3l,1}^{(d6)*}\end{split}\\

\begin{split}\dot{C}_{q,3}^{(d8)}&\supset \frac{1}{3} \left(C_{Hq,1}^{(d6)}\right){}^2+\frac{2}{3} \left(C_{lq,2}^{(d6)}\right){}^2+\frac{128}{3} C_{q,1}^{(d6)} C_{q,2}^{(d6)}+\frac{40}{3} \left(C_{q,2}^{(d6)}\right){}^2+\frac{2}{3} \left(C_{qd,2}^{(d6)}\right){}^2+\frac{2}{3} \left(C_{qu,1}^{(d6)}\right){}^2\\&-C_{q^2ud,1}^{(d6)} C_{q^2ud,1}^{(d6)*}-C_{q^2ud,1}^{(d6)*} C_{q^2ud,2}^{(d6)}-C_{q^2ud,1}^{(d6)} C_{q^2ud,2}^{(d6)*}-C_{q^2ud,2}^{(d6)} C_{q^2ud,2}^{(d6)*}-8 C_{q^2ue,1}^{(d6)} C_{q^2ue,1}^{(d6)*}-9 C_{q^3l,1}^{(d6)} C_{q^3l,1}^{(d6)*}\end{split}\\

\begin{split}\dot{C}_{q,4}^{(d8)}&\supset -\frac{1}{3} \left(C_{Hq,1}^{(d6)}\right){}^2-\frac{2}{3} \left(C_{lq,2}^{(d6)}\right){}^2-\frac{8}{3} \left(C_{q,2}^{(d6)}\right){}^2+\frac{2}{3} \left(C_{qd,2}^{(d6)}\right){}^2+\frac{2}{3} \left(C_{qu,1}^{(d6)}\right){}^2\\&-\frac{1}{3} C_{q^2ud,1}^{(d6)} C_{q^2ud,1}^{(d6)*}+\frac{1}{3} C_{q^2ud,1}^{(d6)*} C_{q^2ud,2}^{(d6)}+\frac{1}{3} C_{q^2ud,1}^{(d6)} C_{q^2ud,2}^{(d6)*}-\frac{1}{3} C_{q^2ud,2}^{(d6)} C_{q^2ud,2}^{(d6)*}+\frac{1}{3} C_{q^3l,1}^{(d6)} C_{q^3l,1}^{(d6)*}\end{split}
\end{align}

\subsection{$QQ^{\dagger}u_{\mathbb{C}}u_{\mathbb{C}}^{\dagger}$}
We list below the RGEs for the dimension-8 type $QQ^{\dagger}u_{\mathbb{C}}u_{\mathbb{C}}^{\dagger}$. The corresponding operators and Wilson coefficients are defined in Table \ref{tab:Q2u2}.
\begin{table}[htbp]
\begin{align*}
\begin{array}{|c|c|c|}
\hline\hline
\text{abbreviation} & \text{Wilson coefficient} & \text{operator}\\
\hline
C_{qu,1}^{(d6)} & C_ {Qu_{\mathbb{C}}Q^{\dagger}u_{\mathbb{C}}^{\dagger},1}^{\; prst} & \left(Q{}_{pai}u_{\mathbb{C}}{}_{r}^a\right)(Q^{\dagger}{}_s^{ci} u_{\mathbb{C}}^{\dagger}{}_{tc}) \\
C_{qu,2}^{(d6)} & C_ {Qu_{\mathbb{C}}Q^{\dagger}u_{\mathbb{C}}^{\dagger},2}^{\; prst} & \left(Q{}_{pai}u_{\mathbb{C}}{}_{r}^b\right)(Q^{\dagger}{}_s^{ai} u_{\mathbb{C}}^{\dagger}{}_{tb}) \\
\hline
C_{qu,1}^{(d8)} & C_ {Qu_{\mathbb{C}}Q^{\dagger}u_{\mathbb{C}}^{\dagger}D^2,1}^{\; prst} & \left(Q{}_{pai}u_{\mathbb{C}}{}_{r}^a\right)(D^{\mu}Q^{\dagger}{}_s^{ci} D_{\mu}u_{\mathbb{C}}^{\dagger}{}_{tc}) \\
C_{qu,2}^{(d8)} & C_ {Qu_{\mathbb{C}}Q^{\dagger}u_{\mathbb{C}}^{\dagger}D^2,2}^{\; prst} & i\left(Q{}_{pai}\sigma_{\mu\nu}u_{\mathbb{C}}{}_{r}^a\right)(D^{\mu}Q^{\dagger}{}_s^{ci} D^{\nu}u_{\mathbb{C}}^{\dagger}{}_{tc}) \\
C_{qu,3}^{(d8)} & C_ {Qu_{\mathbb{C}}Q^{\dagger}u_{\mathbb{C}}^{\dagger}D^2,3}^{\; prst} & \left(Q{}_{pai}u_{\mathbb{C}}{}_{r}^b\right)(D^{\mu}Q^{\dagger}{}_s^{ai} D_{\mu}u_{\mathbb{C}}^{\dagger}{}_{tb}) \\
C_{qu,4}^{(d8)} & C_ {Qu_{\mathbb{C}}Q^{\dagger}u_{\mathbb{C}}^{\dagger}D^2,4}^{\; prst} & i\left(Q{}_{pai}\sigma_{\mu\nu}u_{\mathbb{C}}{}_{r}^b\right)(D^{\mu}Q^{\dagger}{}_s^{ai} D^{\nu}u_{\mathbb{C}}^{\dagger}{}_{tb}) \\
\hline
\end{array}
\end{align*}
\caption{List of the $QQ^{\dagger}u_{\mathbb{C}}u_{\mathbb{C}}^{\dagger}$-type operators and the corresponding Wilson coefficients in dimension-6 and dimension-8. The leftmost column contains the abbreviations for the Wilson coefficients in the case of $N_f=1$.}
\label{tab:Q2u2}
\end{table}

\begin{align}
\begin{split}\dot{C}_{qu,1}^{(d8)}&\supset -\frac{10}{9} g_1 g_3 C_{BGq,1}^{(d8)}-\frac{10}{9} g_1 g_3 C_{BGq,1}^{(d8)*}-\frac{5}{18} g_1 g_3 C_{BGu,1}^{(d8)}-\frac{5}{18} g_1 g_3 C_{BGu,1}^{(d8)*}-\frac{25}{18} g_3^2 C_{Gq,1}^{(d8)}\\&-\frac{5}{4} i g_3^2 C_{Gq,2}^{(d8)}-\frac{25}{18} g_3^2 C_{Gu,1}^{(d8)}+\frac{5}{4} i g_3^2 C_{Gu,2}^{(d8)}-2 Y_e Y_u C_{lequ,1}^{(d8)*}-\frac{2}{3} Y_e Y_u C_{lequ,2}^{(d8)*}-\frac{1}{2} g_3^2 C_{q,1}^{(d8)}\\&-\frac{5}{6} g_3^2 C_{q,2}^{(d8)}-g_3^2 C_{q,3}^{(d8)}-\frac{5}{3} g_3^2 C_{q,4}^{(d8)}+\frac{1}{4} g_3^2 C_{qd,3}^{(d8)}-\frac{5}{12} g_3^2 C_{qd,4}^{(d8)}-\frac{1}{2} g_1^2 C_{qu,1}^{(d8)}\\&+\frac{5}{2} g_2^2 C_{qu,1}^{(d8)}-\frac{61}{4} g_3^2 C_{qu,1}^{(d8)}+\frac{125}{54} g_1^2 C_{qu,2}^{(d8)}+\frac{5}{2} g_2^2 C_{qu,2}^{(d8)}-\frac{125}{36} g_3^2 C_{qu,2}^{(d8)}-\frac{28}{3} g_3^2 C_{qu,3}^{(d8)}\\&+\frac{10}{3} g_3^2 C_{qu,4}^{(d8)}-\frac{2}{3} Y_d Y_u C_{q^2ud,1}^{(d8)*}+\frac{10}{3} Y_d Y_u C_{q^2ud,2}^{(d8)*}-\frac{8}{3} Y_d Y_u C_{q^2ud,3}^{(d8)*}-\frac{1}{2} g_3^2 C_{u,1}^{(d8)}-\frac{5}{6} g_3^2 C_{u,2}^{(d8)}\\&-\frac{1}{4} g_3^2 C_{ud,3}^{(d8)}+\frac{5}{12} g_3^2 C_{ud,4}^{(d8)}+Y_d C_{qu,1}^{(d8)} Y_d{}^*+\frac{1}{6} Y_d C_{ud,3}^{(d8)} Y_d{}^*-\frac{5}{6} Y_d C_{ud,4}^{(d8)} Y_d{}^*+\frac{1}{3} Y_u C_{q,1}^{(d8)} Y_u{}^*\\&+\frac{5}{3} Y_u C_{q,2}^{(d8)} Y_u{}^*+\frac{2}{3} Y_u C_{q,3}^{(d8)} Y_u{}^*+\frac{10}{3} Y_u C_{q,4}^{(d8)} Y_u{}^*+15 Y_u C_{qu,1}^{(d8)} Y_u{}^*+\frac{13}{3} Y_u C_{qu,3}^{(d8)} Y_u{}^*+\frac{5}{3} Y_u C_{qu,4}^{(d8)} Y_u{}^*\\&+\frac{1}{3} Y_u C_{u,1}^{(d8)} Y_u{}^*+\frac{5}{3} Y_u C_{u,2}^{(d8)} Y_u{}^*-\frac{2}{3} C_{q^2ud,1}^{(d8)} Y_d{}^* Y_u{}^*+\frac{10}{3} C_{q^2ud,2}^{(d8)} Y_d{}^* Y_u{}^*-\frac{8}{3} C_{q^2ud,3}^{(d8)} Y_d{}^* Y_u{}^*-2 C_{lequ,1}^{(d8)} Y_e{}^* Y_u{}^*\\&-\frac{2}{3} C_{lequ,2}^{(d8)} Y_e{}^* Y_u{}^*\end{split}\\

\begin{split}\dot{C}_{qu,2}^{(d8)}&\supset -\frac{2}{3} g_1 g_3 C_{BGq,1}^{(d8)}-\frac{2}{3} g_1 g_3 C_{BGq,1}^{(d8)*}-\frac{1}{6} g_1 g_3 C_{BGu,1}^{(d8)}-\frac{1}{6} g_1 g_3 C_{BGu,1}^{(d8)*}-\frac{5}{6} g_3^2 C_{Gq,1}^{(d8)}\\&-\frac{3}{4} i g_3^2 C_{Gq,2}^{(d8)}-\frac{5}{6} g_3^2 C_{Gu,1}^{(d8)}+\frac{3}{4} i g_3^2 C_{Gu,2}^{(d8)}+\frac{1}{2} g_3^2 C_{q,1}^{(d8)}+\frac{5}{6} g_3^2 C_{q,2}^{(d8)}+g_3^2 C_{q,3}^{(d8)}\\&+\frac{5}{3} g_3^2 C_{q,4}^{(d8)}-\frac{1}{4} g_3^2 C_{qd,3}^{(d8)}+\frac{5}{12} g_3^2 C_{qd,4}^{(d8)}+\frac{25}{18} g_1^2 C_{qu,1}^{(d8)}+\frac{3}{2} g_2^2 C_{qu,1}^{(d8)}-\frac{25}{12} g_3^2 C_{qu,1}^{(d8)}\\&+\frac{17}{18} g_1^2 C_{qu,2}^{(d8)}+\frac{3}{2} g_2^2 C_{qu,2}^{(d8)}+\frac{79}{12} g_3^2 C_{qu,2}^{(d8)}+2 g_3^2 C_{qu,3}^{(d8)}+\frac{1}{2} g_3^2 C_{u,1}^{(d8)}+\frac{5}{6} g_3^2 C_{u,2}^{(d8)}\\&-\frac{5 g_1 Y_u C_{uB,3}^{(d8)*}}{3 \sqrt{2}}+\frac{1}{4} g_3^2 C_{ud,3}^{(d8)}-\frac{5}{12} g_3^2 C_{ud,4}^{(d8)}+\frac{2}{3} \sqrt{2} g_3 Y_u C_{uG,3}^{(d8)*}-\frac{3 g_2 Y_u C_{uW,3}^{(d8)*}}{\sqrt{2}}+Y_d C_{qu,2}^{(d8)} Y_d{}^*\\&-\frac{1}{2} Y_d C_{ud,3}^{(d8)} Y_d{}^*+\frac{1}{2} Y_d C_{ud,4}^{(d8)} Y_d{}^*-Y_u C_{q,1}^{(d8)} Y_u{}^*-Y_u C_{q,2}^{(d8)} Y_u{}^*-2 Y_u C_{q,3}^{(d8)} Y_u{}^*-2 Y_u C_{q,4}^{(d8)} Y_u{}^*\\&+3 Y_u C_{qu,2}^{(d8)} Y_u{}^*+Y_u C_{qu,3}^{(d8)} Y_u{}^*+Y_u C_{qu,4}^{(d8)} Y_u{}^*-Y_u C_{u,1}^{(d8)} Y_u{}^*-Y_u C_{u,2}^{(d8)} Y_u{}^*+\frac{5 g_1 C_{uB,3}^{(d8)} Y_u{}^*}{3 \sqrt{2}}\\&-\frac{2}{3} \sqrt{2} g_3 C_{uG,3}^{(d8)} Y_u{}^*+\frac{3 g_2 C_{uW,3}^{(d8)} Y_u{}^*}{\sqrt{2}}\end{split}\\

\begin{split}\dot{C}_{qu,3}^{(d8)}&\supset \frac{10}{27} g_1 g_3 C_{BGq,1}^{(d8)}+\frac{10}{27} g_1 g_3 C_{BGq,1}^{(d8)*}+\frac{5}{54} g_1 g_3 C_{BGu,1}^{(d8)}+\frac{5}{54} g_1 g_3 C_{BGu,1}^{(d8)*}-\frac{20}{27} g_1^2 C_{Bq,1}^{(d8)}-\frac{5}{108} g_1^2 C_{Bu,1}^{(d8)}\\&+\frac{1}{12} g_1^2 C_{eu,1}^{(d8)}-\frac{5}{36} g_1^2 C_{eu,2}^{(d8)}+\frac{25}{54} g_3^2 C_{Gq,1}^{(d8)}+\frac{5}{12} i g_3^2 C_{Gq,2}^{(d8)}-\frac{20}{9} g_3^2 C_{Gq,3}^{(d8)}+\frac{25}{54} g_3^2 C_{Gu,1}^{(d8)}\\&-\frac{5}{12} i g_3^2 C_{Gu,2}^{(d8)}-\frac{20}{9} g_3^2 C_{Gu,3}^{(d8)}-\frac{1}{18} g_1^2 C_{Hq,1}^{(d8)}-\frac{1}{18} g_1^2 C_{Hq,2}^{(d8)}-\frac{1}{9} g_1^2 C_{Hq,3}^{(d8)}-\frac{1}{9} g_1^2 C_{Hq,4}^{(d8)}\\&+\frac{1}{36} g_1^2 C_{Hu,1}^{(d8)}+\frac{1}{36} g_1^2 C_{Hu,2}^{(d8)}+\frac{1}{3} g_1^2 C_{lq,1}^{(d8)}-\frac{5}{9} g_1^2 C_{lq,2}^{(d8)}+\frac{1}{6} g_1^2 C_{lq,3}^{(d8)}-\frac{5}{18} g_1^2 C_{lq,4}^{(d8)}\\&-\frac{1}{12} g_1^2 C_{lu,1}^{(d8)}+\frac{5}{36} g_1^2 C_{lu,2}^{(d8)}-\frac{7}{9} g_1^2 C_{q,1}^{(d8)}+\frac{1}{6} g_3^2 C_{q,1}^{(d8)}+\frac{25}{27} g_1^2 C_{q,2}^{(d8)}+\frac{5}{18} g_3^2 C_{q,2}^{(d8)}\\&-\frac{5}{9} g_1^2 C_{q,3}^{(d8)}+\frac{1}{3} g_3^2 C_{q,3}^{(d8)}+\frac{5}{27} g_1^2 C_{q,4}^{(d8)}+\frac{5}{9} g_3^2 C_{q,4}^{(d8)}-\frac{1}{3} g_1^2 C_{qd,1}^{(d8)}+\frac{5}{9} g_1^2 C_{qd,2}^{(d8)}\\&-\frac{1}{9} g_1^2 C_{qd,3}^{(d8)}-\frac{1}{12} g_3^2 C_{qd,3}^{(d8)}+\frac{5}{27} g_1^2 C_{qd,4}^{(d8)}+\frac{5}{36} g_3^2 C_{qd,4}^{(d8)}-\frac{1}{3} g_1^2 C_{qe,1}^{(d8)}+\frac{5}{9} g_1^2 C_{qe,2}^{(d8)}\\&+\frac{1}{4} g_1^2 C_{qu,1}^{(d8)}-\frac{1}{4} g_3^2 C_{qu,1}^{(d8)}-\frac{5}{12} g_1^2 C_{qu,2}^{(d8)}+\frac{85}{12} g_3^2 C_{qu,2}^{(d8)}+\frac{1}{4} g_1^2 C_{qu,3}^{(d8)}+\frac{5}{2} g_2^2 C_{qu,3}^{(d8)}\\&+12 g_3^2 C_{qu,3}^{(d8)}+\frac{115}{108} g_1^2 C_{qu,4}^{(d8)}+\frac{5}{2} g_2^2 C_{qu,4}^{(d8)}+\frac{70}{9} g_3^2 C_{qu,4}^{(d8)}+2 Y_d Y_u C_{q^2ud,1}^{(d8)*}+\frac{2}{3} Y_d Y_u C_{q^2ud,2}^{(d8)*}\\&-\frac{4}{9} g_1^2 C_{u,1}^{(d8)}+\frac{1}{6} g_3^2 C_{u,1}^{(d8)}+\frac{10}{27} g_1^2 C_{u,2}^{(d8)}+\frac{5}{18} g_3^2 C_{u,2}^{(d8)}+\frac{1}{12} g_1^2 C_{ud,1}^{(d8)}-\frac{5}{36} g_1^2 C_{ud,2}^{(d8)}\\&+\frac{1}{36} g_1^2 C_{ud,3}^{(d8)}+\frac{1}{12} g_3^2 C_{ud,3}^{(d8)}-\frac{5}{108} g_1^2 C_{ud,4}^{(d8)}-\frac{5}{36} g_3^2 C_{ud,4}^{(d8)}-\frac{5}{4} g_2^2 C_{Wu,1}^{(d8)}-\frac{2}{3} Y_d C_{Hu,1}^{(d8)} Y_d{}^*\\&+\frac{1}{6} Y_d C_{Hu,2}^{(d8)} Y_d{}^*+Y_d C_{qu,3}^{(d8)} Y_d{}^*+\frac{1}{6} Y_d C_{ud,1}^{(d8)} Y_d{}^*-\frac{5}{6} Y_d C_{ud,2}^{(d8)} Y_d{}^*-\frac{1}{6} Y_u C_{Hq,1}^{(d8)} Y_u{}^*+\frac{2}{3} Y_u C_{Hq,2}^{(d8)} Y_u{}^*\\&-\frac{1}{3} Y_u C_{Hq,3}^{(d8)} Y_u{}^*+\frac{4}{3} Y_u C_{Hq,4}^{(d8)} Y_u{}^*-\frac{1}{6} Y_u C_{Hu,1}^{(d8)} Y_u{}^*+\frac{2}{3} Y_u C_{Hu,2}^{(d8)} Y_u{}^*+\frac{2}{3} Y_u C_{q,1}^{(d8)} Y_u{}^*-\frac{10}{3} Y_u C_{q,2}^{(d8)} Y_u{}^*\\&+\frac{1}{3} Y_u C_{q,3}^{(d8)} Y_u{}^*-\frac{5}{3} Y_u C_{q,4}^{(d8)} Y_u{}^*+\frac{1}{3} Y_u C_{qu,1}^{(d8)} Y_u{}^*+\frac{5}{3} Y_u C_{qu,2}^{(d8)} Y_u{}^*+3 Y_u C_{qu,3}^{(d8)} Y_u{}^*+\frac{1}{3} Y_u C_{u,1}^{(d8)} Y_u{}^*\\&-\frac{5}{3} Y_u C_{u,2}^{(d8)} Y_u{}^*+2 C_{q^2ud,1}^{(d8)} Y_d{}^* Y_u{}^*+\frac{2}{3} C_{q^2ud,2}^{(d8)} Y_d{}^* Y_u{}^*\end{split}\\

\begin{split}\dot{C}_{qu,4}^{(d8)}&\supset \frac{2}{9} g_1 g_3 C_{BGq,1}^{(d8)}+\frac{2}{9} g_1 g_3 C_{BGq,1}^{(d8)*}+\frac{1}{18} g_1 g_3 C_{BGu,1}^{(d8)}+\frac{1}{18} g_1 g_3 C_{BGu,1}^{(d8)*}-\frac{4}{9} g_1^2 C_{Bq,1}^{(d8)}-\frac{1}{36} g_1^2 C_{Bu,1}^{(d8)}\\&-\frac{1}{12} g_1^2 C_{eu,1}^{(d8)}+\frac{5}{36} g_1^2 C_{eu,2}^{(d8)}+\frac{5}{18} g_3^2 C_{Gq,1}^{(d8)}+\frac{1}{4} i g_3^2 C_{Gq,2}^{(d8)}-\frac{4}{3} g_3^2 C_{Gq,3}^{(d8)}+\frac{5}{18} g_3^2 C_{Gu,1}^{(d8)}\\&-\frac{1}{4} i g_3^2 C_{Gu,2}^{(d8)}-\frac{4}{3} g_3^2 C_{Gu,3}^{(d8)}+\frac{1}{18} g_1^2 C_{Hq,1}^{(d8)}+\frac{1}{18} g_1^2 C_{Hq,2}^{(d8)}+\frac{1}{9} g_1^2 C_{Hq,3}^{(d8)}+\frac{1}{9} g_1^2 C_{Hq,4}^{(d8)}\\&-\frac{1}{36} g_1^2 C_{Hu,1}^{(d8)}-\frac{1}{36} g_1^2 C_{Hu,2}^{(d8)}-\frac{1}{3} g_1^2 C_{lq,1}^{(d8)}+\frac{5}{9} g_1^2 C_{lq,2}^{(d8)}-\frac{1}{6} g_1^2 C_{lq,3}^{(d8)}+\frac{5}{18} g_1^2 C_{lq,4}^{(d8)}\\&+\frac{1}{12} g_1^2 C_{lu,1}^{(d8)}-\frac{5}{36} g_1^2 C_{lu,2}^{(d8)}+\frac{7}{9} g_1^2 C_{q,1}^{(d8)}-\frac{1}{6} g_3^2 C_{q,1}^{(d8)}-\frac{25}{27} g_1^2 C_{q,2}^{(d8)}-\frac{5}{18} g_3^2 C_{q,2}^{(d8)}\\&+\frac{5}{9} g_1^2 C_{q,3}^{(d8)}-\frac{1}{3} g_3^2 C_{q,3}^{(d8)}-\frac{5}{27} g_1^2 C_{q,4}^{(d8)}-\frac{5}{9} g_3^2 C_{q,4}^{(d8)}+\frac{1}{3} g_1^2 C_{qd,1}^{(d8)}-\frac{5}{9} g_1^2 C_{qd,2}^{(d8)}\\&+\frac{1}{9} g_1^2 C_{qd,3}^{(d8)}+\frac{1}{12} g_3^2 C_{qd,3}^{(d8)}-\frac{5}{27} g_1^2 C_{qd,4}^{(d8)}-\frac{5}{36} g_3^2 C_{qd,4}^{(d8)}+\frac{1}{3} g_1^2 C_{qe,1}^{(d8)}-\frac{5}{9} g_1^2 C_{qe,2}^{(d8)}\\&-\frac{1}{4} g_1^2 C_{qu,1}^{(d8)}+\frac{17}{4} g_3^2 C_{qu,1}^{(d8)}+\frac{5}{12} g_1^2 C_{qu,2}^{(d8)}-\frac{5}{12} g_3^2 C_{qu,2}^{(d8)}+\frac{23}{36} g_1^2 C_{qu,3}^{(d8)}+\frac{3}{2} g_2^2 C_{qu,3}^{(d8)}\\&+\frac{14}{3} g_3^2 C_{qu,3}^{(d8)}+\frac{79}{36} g_1^2 C_{qu,4}^{(d8)}+\frac{3}{2} g_2^2 C_{qu,4}^{(d8)}+\frac{16}{3} g_3^2 C_{qu,4}^{(d8)}+\frac{4}{9} g_1^2 C_{u,1}^{(d8)}-\frac{1}{6} g_3^2 C_{u,1}^{(d8)}\\&-\frac{10}{27} g_1^2 C_{u,2}^{(d8)}-\frac{5}{18} g_3^2 C_{u,2}^{(d8)}-\frac{1}{12} g_1^2 C_{ud,1}^{(d8)}+\frac{5}{36} g_1^2 C_{ud,2}^{(d8)}-\frac{1}{36} g_1^2 C_{ud,3}^{(d8)}-\frac{1}{12} g_3^2 C_{ud,3}^{(d8)}\\&+\frac{5}{108} g_1^2 C_{ud,4}^{(d8)}+\frac{5}{36} g_3^2 C_{ud,4}^{(d8)}-2 \sqrt{2} g_3 Y_u C_{uG,3}^{(d8)*}-\frac{3}{4} g_2^2 C_{Wu,1}^{(d8)}+\frac{1}{2} Y_d C_{Hu,2}^{(d8)} Y_d{}^*+Y_d C_{qu,4}^{(d8)} Y_d{}^*\\&-\frac{1}{2} Y_d C_{ud,1}^{(d8)} Y_d{}^*+\frac{1}{2} Y_d C_{ud,2}^{(d8)} Y_d{}^*-\frac{1}{2} Y_u C_{Hq,1}^{(d8)} Y_u{}^*-Y_u C_{Hq,3}^{(d8)} Y_u{}^*-\frac{1}{2} Y_u C_{Hu,1}^{(d8)} Y_u{}^*-2 Y_u C_{q,1}^{(d8)} Y_u{}^*\\&+2 Y_u C_{q,2}^{(d8)} Y_u{}^*-Y_u C_{q,3}^{(d8)} Y_u{}^*+Y_u C_{q,4}^{(d8)} Y_u{}^*+Y_u C_{qu,1}^{(d8)} Y_u{}^*+Y_u C_{qu,2}^{(d8)} Y_u{}^*+3 Y_u C_{qu,4}^{(d8)} Y_u{}^*\\&-Y_u C_{u,1}^{(d8)} Y_u{}^*+Y_u C_{u,2}^{(d8)} Y_u{}^*+2 \sqrt{2} g_3 C_{uG,3}^{(d8)} Y_u{}^*\end{split}
\end{align}
The following are the quadratic contributions.
\begin{align}
\begin{split}\dot{C}_{qu,1}^{(d8)}&\supset 4 C_{lequ,1}^{(d6)} C_{lequ,1}^{(d6)*}-2 C_{lequ,1}^{(d6)*} C_{lequ,2}^{(d6)}-2 C_{lequ,1}^{(d6)} C_{lequ,2}^{(d6)*}+C_{lequ,2}^{(d6)} C_{lequ,2}^{(d6)*}+\frac{4}{3} C_{q,1}^{(d6)} C_{qu,1}^{(d6)}+\frac{8}{3} C_{q,2}^{(d6)} C_{qu,1}^{(d6)}\\&+24 \left(C_{qu,1}^{(d6)}\right){}^2+\frac{56}{3} C_{qu,1}^{(d6)} C_{qu,2}^{(d6)}-\frac{2}{3} C_{qudl,1}^{(d6)} C_{qudl,1}^{(d6)*}+7 C_{q^2ud,1}^{(d6)} C_{q^2ud,1}^{(d6)*}-11 C_{q^2ud,1}^{(d6)*} C_{q^2ud,2}^{(d6)}-11 C_{q^2ud,1}^{(d6)} C_{q^2ud,2}^{(d6)*}\\&+16 C_{q^2ud,2}^{(d6)} C_{q^2ud,2}^{(d6)*}-\frac{8}{3} C_{q^2ue,1}^{(d6)} C_{q^2ue,1}^{(d6)*}+\frac{4}{3} C_{qu,1}^{(d6)} C_{u,1}^{(d6)}+\frac{2}{3} C_{qd,2}^{(d6)} C_{ud,2}^{(d6)}\end{split}\\

\begin{split}\dot{C}_{qu,2}^{(d8)}&\supset \frac{1}{3} C_{lequ,2}^{(d6)} C_{lequ,2}^{(d6)*}-\frac{4}{3} C_{q,1}^{(d6)} C_{qu,1}^{(d6)}-\frac{8}{3} C_{q,2}^{(d6)} C_{qu,1}^{(d6)}+\frac{8}{3} C_{qu,1}^{(d6)} C_{qu,2}^{(d6)}-\frac{2}{3} C_{qudl,1}^{(d6)} C_{qudl,1}^{(d6)*}+C_{q^2ud,1}^{(d6)} C_{q^2ud,1}^{(d6)*}\\&-\frac{1}{3} C_{q^2ud,1}^{(d6)*} C_{q^2ud,2}^{(d6)}-\frac{1}{3} C_{q^2ud,1}^{(d6)} C_{q^2ud,2}^{(d6)*}-\frac{8}{3} C_{q^2ue,1}^{(d6)} C_{q^2ue,1}^{(d6)*}-\frac{4}{3} C_{qu,1}^{(d6)} C_{u,1}^{(d6)}+\frac{8}{3} C_{uB,1}^{(d6)} C_{uB,1}^{(d6)*}-\frac{2}{3} C_{qd,2}^{(d6)} C_{ud,2}^{(d6)}\\&-\frac{16}{9} C_{uG,1}^{(d6)} C_{uG,1}^{(d6)*}+16 C_{uW,1}^{(d6)} C_{uW,1}^{(d6)*}\end{split}\\

\begin{split}\dot{C}_{qu,3}^{(d8)}&\supset \frac{1}{3} C_{Hq,1}^{(d6)} C_{Hu,1}^{(d6)}+\frac{2}{3} C_{Hq,2}^{(d6)} C_{Hu,1}^{(d6)}+\frac{4}{3} C_{lq,1}^{(d6)} C_{lu,1}^{(d6)}+\frac{2}{3} C_{lq,2}^{(d6)} C_{lu,1}^{(d6)}+\frac{2}{3} C_{eu,1}^{(d6)} C_{qe,1}^{(d6)}+\frac{8}{3} C_{q,1}^{(d6)} C_{qu,1}^{(d6)}\\&+\frac{4}{3} C_{q,2}^{(d6)} C_{qu,1}^{(d6)}+\frac{4}{3} \left(C_{qu,1}^{(d6)}\right){}^2+\frac{28}{3} C_{q,1}^{(d6)} C_{qu,2}^{(d6)}+\frac{20}{3} C_{q,2}^{(d6)} C_{qu,2}^{(d6)}+\frac{28}{3} \left(C_{qu,2}^{(d6)}\right){}^2+\frac{2}{3} C_{qudl,1}^{(d6)} C_{qudl,1}^{(d6)*}\\&+4 C_{q^2ud,1}^{(d6)} C_{q^2ud,1}^{(d6)*}-2 C_{q^2ud,1}^{(d6)*} C_{q^2ud,2}^{(d6)}-2 C_{q^2ud,1}^{(d6)} C_{q^2ud,2}^{(d6)*}+C_{q^2ud,2}^{(d6)} C_{q^2ud,2}^{(d6)*}+\frac{8}{3} C_{q^2ue,1}^{(d6)} C_{q^2ue,1}^{(d6)*}+\frac{4}{3} C_{qu,1}^{(d6)} C_{u,1}^{(d6)}\\&+\frac{16}{3} C_{qu,2}^{(d6)} C_{u,1}^{(d6)}+2 C_{qd,1}^{(d6)} C_{ud,1}^{(d6)}+\frac{2}{3} C_{qd,2}^{(d6)} C_{ud,1}^{(d6)}+\frac{2}{3} C_{qd,1}^{(d6)} C_{ud,2}^{(d6)}\end{split}\\

\begin{split}\dot{C}_{qu,4}^{(d8)}&\supset -\frac{1}{3} C_{Hq,1}^{(d6)} C_{Hu,1}^{(d6)}-\frac{2}{3} C_{Hq,2}^{(d6)} C_{Hu,1}^{(d6)}-\frac{4}{3} C_{lq,1}^{(d6)} C_{lu,1}^{(d6)}-\frac{2}{3} C_{lq,2}^{(d6)} C_{lu,1}^{(d6)}-\frac{2}{3} C_{eu,1}^{(d6)} C_{qe,1}^{(d6)}\\&-\frac{8}{3} C_{q,1}^{(d6)} C_{qu,1}^{(d6)}-\frac{4}{3} C_{q,2}^{(d6)} C_{qu,1}^{(d6)}+\frac{4}{3} \left(C_{qu,1}^{(d6)}\right){}^2-\frac{28}{3} C_{q,1}^{(d6)} C_{qu,2}^{(d6)}-\frac{20}{3} C_{q,2}^{(d6)} C_{qu,2}^{(d6)}+\frac{4}{3} \left(C_{qu,2}^{(d6)}\right){}^2\\&+\frac{2}{3} C_{qudl,1}^{(d6)} C_{qudl,1}^{(d6)*}+\frac{1}{3} C_{q^2ud,2}^{(d6)} C_{q^2ud,2}^{(d6)*}+\frac{8}{3} C_{q^2ue,1}^{(d6)} C_{q^2ue,1}^{(d6)*}-\frac{4}{3} C_{qu,1}^{(d6)} C_{u,1}^{(d6)}-\frac{16}{3} C_{qu,2}^{(d6)} C_{u,1}^{(d6)}-2 C_{qd,1}^{(d6)} C_{ud,1}^{(d6)}\\&-\frac{2}{3} C_{qd,2}^{(d6)} C_{ud,1}^{(d6)}-\frac{2}{3} C_{qd,1}^{(d6)} C_{ud,2}^{(d6)}+\frac{16}{3} C_{uG,1}^{(d6)} C_{uG,1}^{(d6)*}\end{split}
\end{align}

\subsection{$QQ^{\dagger}d_{\mathbb{C}}d_{\mathbb{C}}^{\dagger}$}
We list below the RGEs for the dimension-8 type $QQ^{\dagger}d_{\mathbb{C}}d_{\mathbb{C}}^{\dagger}$. The corresponding operators and Wilson coefficients are defined in Table \ref{tab:Q2d2}.
\begin{table}[htbp]
\begin{align*}
\begin{array}{|c|c|c|}
\hline\hline
\text{abbreviation} & \text{Wilson coefficient} & \text{operator}\\
\hline
C_{qd,1}^{(d6)} & C_ {d_{\mathbb{C}}Qd_{\mathbb{C}}^{\dagger}Q^{\dagger},2}^{\; prst} & \left(Q{}_{pai}d_{\mathbb{C}}{}_{r}^b\right)(Q^{\dagger}{}_s^{ai} d_{\mathbb{C}}^{\dagger}{}_{tb}) \\
C_{qd,2}^{(d6)} & C_ {d_{\mathbb{C}}Qd_{\mathbb{C}}^{\dagger}Q^{\dagger},1}^{\; prst} & \left(Q{}_{pai}d_{\mathbb{C}}{}_{r}^a\right)(Q^{\dagger}{}_s^{ci} d_{\mathbb{C}}^{\dagger}{}_{tc}) \\
\hline
C_{qd,1}^{(d8)} & C_ {d_{\mathbb{C}}Qd_{\mathbb{C}}^{\dagger}Q^{\dagger}D^2,1}^{\; prst} & \left(Q{}_{pai}d_{\mathbb{C}}{}_{r}^b\right)(D^{\mu}Q^{\dagger}{}_s^{ai} D_{\mu}d_{\mathbb{C}}^{\dagger}{}_{tb}) \\
C_{qd,2}^{(d8)} & C_ {d_{\mathbb{C}}Qd_{\mathbb{C}}^{\dagger}Q^{\dagger}D^2,2}^{\; prst} & i\left(Q{}_{pai}\sigma_{\mu\nu}d_{\mathbb{C}}{}_{r}^b\right)(D^{\mu}Q^{\dagger}{}_s^{ai} D^{\nu}d_{\mathbb{C}}^{\dagger}{}_{tb}) \\
C_{qd,3}^{(d8)} & C_ {d_{\mathbb{C}}Qd_{\mathbb{C}}^{\dagger}Q^{\dagger}D^2,3}^{\; prst} & \left(Q{}_{pai}d_{\mathbb{C}}{}_{r}^a\right)(D^{\mu}Q^{\dagger}{}_s^{ci} D_{\mu}d_{\mathbb{C}}^{\dagger}{}_{tc}) \\
C_{qd,4}^{(d8)} & C_ {d_{\mathbb{C}}Qd_{\mathbb{C}}^{\dagger}Q^{\dagger}D^2,4}^{\; prst} & i\left(Q{}_{pai}\sigma_{\mu\nu}d_{\mathbb{C}}{}_{r}^a\right)(D^{\mu}Q^{\dagger}{}_s^{ci} D^{\nu}d_{\mathbb{C}}^{\dagger}{}_{tc}) \\
\hline
\end{array}
\end{align*}
\caption{List of the $QQ^{\dagger}d_{\mathbb{C}}d_{\mathbb{C}}^{\dagger}$-type operators and the corresponding Wilson coefficients in dimension-6 and dimension-8. The leftmost column contains the abbreviations for the Wilson coefficients in the case of $N_f=1$.}
\label{tab:Q2d2}
\end{table}

\begin{align}
\begin{split}\dot{C}_{qd,1}^{(d8)}&\supset -\frac{5}{108} g_1^2 C_{Bd,1}^{(d8)}+\frac{5}{54} g_1 g_3 C_{BGd,1}^{(d8)}+\frac{5}{54} g_1 g_3 C_{BGd,1}^{(d8)*}-\frac{5}{27} g_1 g_3 C_{BGq,1}^{(d8)}-\frac{5}{27} g_1 g_3 C_{BGq,1}^{(d8)*}\\&-\frac{5}{27} g_1^2 C_{Bq,1}^{(d8)}+\frac{2}{9} g_1^2 C_{d,1}^{(d8)}+\frac{1}{6} g_3^2 C_{d,1}^{(d8)}-\frac{5}{27} g_1^2 C_{d,2}^{(d8)}+\frac{5}{18} g_3^2 C_{d,2}^{(d8)}+\frac{1}{12} g_1^2 C_{ed,1}^{(d8)}\\&-\frac{5}{36} g_1^2 C_{ed,2}^{(d8)}+\frac{25}{54} g_3^2 C_{Gd,1}^{(d8)}-\frac{5}{12} i g_3^2 C_{Gd,2}^{(d8)}-\frac{20}{9} g_3^2 C_{Gd,3}^{(d8)}+\frac{25}{54} g_3^2 C_{Gq,1}^{(d8)}+\frac{5}{12} i g_3^2 C_{Gq,2}^{(d8)}\\&-\frac{20}{9} g_3^2 C_{Gq,3}^{(d8)}+\frac{1}{36} g_1^2 C_{Hd,1}^{(d8)}+\frac{1}{36} g_1^2 C_{Hd,2}^{(d8)}+\frac{1}{36} g_1^2 C_{Hq,1}^{(d8)}+\frac{1}{36} g_1^2 C_{Hq,2}^{(d8)}+\frac{1}{18} g_1^2 C_{Hq,3}^{(d8)}\\&+\frac{1}{18} g_1^2 C_{Hq,4}^{(d8)}-\frac{1}{12} g_1^2 C_{ld,1}^{(d8)}+\frac{5}{36} g_1^2 C_{ld,2}^{(d8)}-\frac{1}{6} g_1^2 C_{lq,1}^{(d8)}+\frac{5}{18} g_1^2 C_{lq,2}^{(d8)}-\frac{1}{12} g_1^2 C_{lq,3}^{(d8)}\\&+\frac{5}{36} g_1^2 C_{lq,4}^{(d8)}+\frac{7}{18} g_1^2 C_{q,1}^{(d8)}+\frac{1}{6} g_3^2 C_{q,1}^{(d8)}-\frac{25}{54} g_1^2 C_{q,2}^{(d8)}+\frac{5}{18} g_3^2 C_{q,2}^{(d8)}+\frac{5}{18} g_1^2 C_{q,3}^{(d8)}\\&+\frac{1}{3} g_3^2 C_{q,3}^{(d8)}-\frac{5}{54} g_1^2 C_{q,4}^{(d8)}+\frac{5}{9} g_3^2 C_{q,4}^{(d8)}+\frac{7}{4} g_1^2 C_{qd,1}^{(d8)}+\frac{5}{2} g_2^2 C_{qd,1}^{(d8)}+12 g_3^2 C_{qd,1}^{(d8)}\\&-\frac{35}{108} g_1^2 C_{qd,2}^{(d8)}+\frac{5}{2} g_2^2 C_{qd,2}^{(d8)}+\frac{70}{9} g_3^2 C_{qd,2}^{(d8)}+\frac{1}{12} g_1^2 C_{qd,3}^{(d8)}-\frac{1}{4} g_3^2 C_{qd,3}^{(d8)}-\frac{5}{36} g_1^2 C_{qd,4}^{(d8)}\\&+\frac{85}{12} g_3^2 C_{qd,4}^{(d8)}+\frac{1}{6} g_1^2 C_{qe,1}^{(d8)}-\frac{5}{18} g_1^2 C_{qe,2}^{(d8)}-\frac{1}{9} g_1^2 C_{qu,1}^{(d8)}-\frac{1}{12} g_3^2 C_{qu,1}^{(d8)}+\frac{5}{27} g_1^2 C_{qu,2}^{(d8)}\\&+\frac{5}{36} g_3^2 C_{qu,2}^{(d8)}-\frac{1}{3} g_1^2 C_{qu,3}^{(d8)}+\frac{5}{9} g_1^2 C_{qu,4}^{(d8)}+2 Y_d Y_u C_{q^2ud,1}^{(d8)*}+\frac{2}{3} Y_d Y_u C_{q^2ud,2}^{(d8)*}-\frac{1}{6} g_1^2 C_{ud,1}^{(d8)}\\&+\frac{5}{18} g_1^2 C_{ud,2}^{(d8)}-\frac{1}{18} g_1^2 C_{ud,3}^{(d8)}+\frac{1}{12} g_3^2 C_{ud,3}^{(d8)}+\frac{5}{54} g_1^2 C_{ud,4}^{(d8)}-\frac{5}{36} g_3^2 C_{ud,4}^{(d8)}-\frac{5}{4} g_2^2 C_{Wd,1}^{(d8)}\\&+\frac{1}{3} Y_d C_{d,1}^{(d8)} Y_d{}^*-\frac{5}{3} Y_d C_{d,2}^{(d8)} Y_d{}^*-\frac{2}{3} Y_d C_{Hd,1}^{(d8)} Y_d{}^*+\frac{1}{6} Y_d C_{Hd,2}^{(d8)} Y_d{}^*-\frac{2}{3} Y_d C_{Hq,1}^{(d8)} Y_d{}^*+\frac{1}{6} Y_d C_{Hq,2}^{(d8)} Y_d{}^*\\&-\frac{4}{3} Y_d C_{Hq,3}^{(d8)} Y_d{}^*+\frac{1}{3} Y_d C_{Hq,4}^{(d8)} Y_d{}^*+\frac{2}{3} Y_d C_{q,1}^{(d8)} Y_d{}^*-\frac{10}{3} Y_d C_{q,2}^{(d8)} Y_d{}^*+\frac{1}{3} Y_d C_{q,3}^{(d8)} Y_d{}^*-\frac{5}{3} Y_d C_{q,4}^{(d8)} Y_d{}^*\\&+3 Y_d C_{qd,1}^{(d8)} Y_d{}^*+\frac{1}{3} Y_d C_{qd,3}^{(d8)} Y_d{}^*+\frac{5}{3} Y_d C_{qd,4}^{(d8)} Y_d{}^*-\frac{1}{6} Y_u C_{Hd,1}^{(d8)} Y_u{}^*+\frac{2}{3} Y_u C_{Hd,2}^{(d8)} Y_u{}^*+Y_u C_{qd,1}^{(d8)} Y_u{}^*\\&+\frac{1}{6} Y_u C_{ud,1}^{(d8)} Y_u{}^*-\frac{5}{6} Y_u C_{ud,2}^{(d8)} Y_u{}^*+2 C_{q^2ud,1}^{(d8)} Y_d{}^* Y_u{}^*+\frac{2}{3} C_{q^2ud,2}^{(d8)} Y_d{}^* Y_u{}^*\end{split}\\

\begin{split}\dot{C}_{qd,2}^{(d8)}&\supset -\frac{1}{36} g_1^2 C_{Bd,1}^{(d8)}+\frac{1}{18} g_1 g_3 C_{BGd,1}^{(d8)}+\frac{1}{18} g_1 g_3 C_{BGd,1}^{(d8)*}-\frac{1}{9} g_1 g_3 C_{BGq,1}^{(d8)}-\frac{1}{9} g_1 g_3 C_{BGq,1}^{(d8)*}\\&-\frac{1}{9} g_1^2 C_{Bq,1}^{(d8)}-\frac{2}{9} g_1^2 C_{d,1}^{(d8)}-\frac{1}{6} g_3^2 C_{d,1}^{(d8)}+\frac{5}{27} g_1^2 C_{d,2}^{(d8)}-\frac{5}{18} g_3^2 C_{d,2}^{(d8)}+2 \sqrt{2} g_3 Y_d C_{dG,3}^{(d8)*}\\&-\frac{1}{12} g_1^2 C_{ed,1}^{(d8)}+\frac{5}{36} g_1^2 C_{ed,2}^{(d8)}+\frac{5}{18} g_3^2 C_{Gd,1}^{(d8)}-\frac{1}{4} i g_3^2 C_{Gd,2}^{(d8)}-\frac{4}{3} g_3^2 C_{Gd,3}^{(d8)}+\frac{5}{18} g_3^2 C_{Gq,1}^{(d8)}\\&+\frac{1}{4} i g_3^2 C_{Gq,2}^{(d8)}-\frac{4}{3} g_3^2 C_{Gq,3}^{(d8)}-\frac{1}{36} g_1^2 C_{Hd,1}^{(d8)}-\frac{1}{36} g_1^2 C_{Hd,2}^{(d8)}-\frac{1}{36} g_1^2 C_{Hq,1}^{(d8)}-\frac{1}{36} g_1^2 C_{Hq,2}^{(d8)}\\&-\frac{1}{18} g_1^2 C_{Hq,3}^{(d8)}-\frac{1}{18} g_1^2 C_{Hq,4}^{(d8)}+\frac{1}{12} g_1^2 C_{ld,1}^{(d8)}-\frac{5}{36} g_1^2 C_{ld,2}^{(d8)}+\frac{1}{6} g_1^2 C_{lq,1}^{(d8)}-\frac{5}{18} g_1^2 C_{lq,2}^{(d8)}\\&+\frac{1}{12} g_1^2 C_{lq,3}^{(d8)}-\frac{5}{36} g_1^2 C_{lq,4}^{(d8)}-\frac{7}{18} g_1^2 C_{q,1}^{(d8)}-\frac{1}{6} g_3^2 C_{q,1}^{(d8)}+\frac{25}{54} g_1^2 C_{q,2}^{(d8)}-\frac{5}{18} g_3^2 C_{q,2}^{(d8)}\\&-\frac{5}{18} g_1^2 C_{q,3}^{(d8)}-\frac{1}{3} g_3^2 C_{q,3}^{(d8)}+\frac{5}{54} g_1^2 C_{q,4}^{(d8)}-\frac{5}{9} g_3^2 C_{q,4}^{(d8)}-\frac{7}{36} g_1^2 C_{qd,1}^{(d8)}+\frac{3}{2} g_2^2 C_{qd,1}^{(d8)}\\&+\frac{14}{3} g_3^2 C_{qd,1}^{(d8)}+\frac{25}{36} g_1^2 C_{qd,2}^{(d8)}+\frac{3}{2} g_2^2 C_{qd,2}^{(d8)}+\frac{16}{3} g_3^2 C_{qd,2}^{(d8)}-\frac{1}{12} g_1^2 C_{qd,3}^{(d8)}+\frac{17}{4} g_3^2 C_{qd,3}^{(d8)}\\&+\frac{5}{36} g_1^2 C_{qd,4}^{(d8)}-\frac{5}{12} g_3^2 C_{qd,4}^{(d8)}-\frac{1}{6} g_1^2 C_{qe,1}^{(d8)}+\frac{5}{18} g_1^2 C_{qe,2}^{(d8)}+\frac{1}{9} g_1^2 C_{qu,1}^{(d8)}+\frac{1}{12} g_3^2 C_{qu,1}^{(d8)}\\&-\frac{5}{27} g_1^2 C_{qu,2}^{(d8)}-\frac{5}{36} g_3^2 C_{qu,2}^{(d8)}+\frac{1}{3} g_1^2 C_{qu,3}^{(d8)}-\frac{5}{9} g_1^2 C_{qu,4}^{(d8)}+\frac{1}{6} g_1^2 C_{ud,1}^{(d8)}-\frac{5}{18} g_1^2 C_{ud,2}^{(d8)}\\&+\frac{1}{18} g_1^2 C_{ud,3}^{(d8)}-\frac{1}{12} g_3^2 C_{ud,3}^{(d8)}-\frac{5}{54} g_1^2 C_{ud,4}^{(d8)}+\frac{5}{36} g_3^2 C_{ud,4}^{(d8)}-\frac{3}{4} g_2^2 C_{Wd,1}^{(d8)}-Y_d C_{d,1}^{(d8)} Y_d{}^*\\&+Y_d C_{d,2}^{(d8)} Y_d{}^*-2 \sqrt{2} g_3 C_{dG,3}^{(d8)} Y_d{}^*+\frac{1}{2} Y_d C_{Hd,2}^{(d8)} Y_d{}^*+\frac{1}{2} Y_d C_{Hq,2}^{(d8)} Y_d{}^*+Y_d C_{Hq,4}^{(d8)} Y_d{}^*-2 Y_d C_{q,1}^{(d8)} Y_d{}^*\\&+2 Y_d C_{q,2}^{(d8)} Y_d{}^*-Y_d C_{q,3}^{(d8)} Y_d{}^*+Y_d C_{q,4}^{(d8)} Y_d{}^*+3 Y_d C_{qd,2}^{(d8)} Y_d{}^*+Y_d C_{qd,3}^{(d8)} Y_d{}^*+Y_d C_{qd,4}^{(d8)} Y_d{}^*\\&-\frac{1}{2} Y_u C_{Hd,1}^{(d8)} Y_u{}^*+Y_u C_{qd,2}^{(d8)} Y_u{}^*-\frac{1}{2} Y_u C_{ud,1}^{(d8)} Y_u{}^*+\frac{1}{2} Y_u C_{ud,2}^{(d8)} Y_u{}^*\end{split}\\

\begin{split}\dot{C}_{qd,3}^{(d8)}&\supset -\frac{5}{18} g_1 g_3 C_{BGd,1}^{(d8)}-\frac{5}{18} g_1 g_3 C_{BGd,1}^{(d8)*}+\frac{5}{9} g_1 g_3 C_{BGq,1}^{(d8)}+\frac{5}{9} g_1 g_3 C_{BGq,1}^{(d8)*}-\frac{1}{2} g_3^2 C_{d,1}^{(d8)}\\&-\frac{5}{6} g_3^2 C_{d,2}^{(d8)}-\frac{25}{18} g_3^2 C_{Gd,1}^{(d8)}+\frac{5}{4} i g_3^2 C_{Gd,2}^{(d8)}-\frac{25}{18} g_3^2 C_{Gq,1}^{(d8)}-\frac{5}{4} i g_3^2 C_{Gq,2}^{(d8)}-\frac{1}{2} g_3^2 C_{q,1}^{(d8)}\\&-\frac{5}{6} g_3^2 C_{q,2}^{(d8)}-g_3^2 C_{q,3}^{(d8)}-\frac{5}{3} g_3^2 C_{q,4}^{(d8)}-\frac{28}{3} g_3^2 C_{qd,1}^{(d8)}+\frac{10}{3} g_3^2 C_{qd,2}^{(d8)}+\frac{3}{2} g_1^2 C_{qd,3}^{(d8)}\\&+\frac{5}{2} g_2^2 C_{qd,3}^{(d8)}-\frac{61}{4} g_3^2 C_{qd,3}^{(d8)}+\frac{5}{54} g_1^2 C_{qd,4}^{(d8)}+\frac{5}{2} g_2^2 C_{qd,4}^{(d8)}-\frac{125}{36} g_3^2 C_{qd,4}^{(d8)}+\frac{1}{4} g_3^2 C_{qu,1}^{(d8)}\\&-\frac{5}{12} g_3^2 C_{qu,2}^{(d8)}-\frac{2}{3} Y_d Y_u C_{q^2ud,1}^{(d8)*}+\frac{10}{3} Y_d Y_u C_{q^2ud,2}^{(d8)*}-\frac{8}{3} Y_d Y_u C_{q^2ud,3}^{(d8)*}-\frac{1}{4} g_3^2 C_{ud,3}^{(d8)}+\frac{5}{12} g_3^2 C_{ud,4}^{(d8)}\\&+\frac{1}{3} Y_d C_{d,1}^{(d8)} Y_d{}^*+\frac{5}{3} Y_d C_{d,2}^{(d8)} Y_d{}^*+2 Y_e C_{ledq,1}^{(d8)*} Y_d{}^*+\frac{1}{3} Y_d C_{q,1}^{(d8)} Y_d{}^*+\frac{5}{3} Y_d C_{q,2}^{(d8)} Y_d{}^*+\frac{2}{3} Y_d C_{q,3}^{(d8)} Y_d{}^*\\&+\frac{10}{3} Y_d C_{q,4}^{(d8)} Y_d{}^*+\frac{13}{3} Y_d C_{qd,1}^{(d8)} Y_d{}^*+\frac{5}{3} Y_d C_{qd,2}^{(d8)} Y_d{}^*+15 Y_d C_{qd,3}^{(d8)} Y_d{}^*+2 Y_d C_{ledq,1}^{(d8)} Y_e{}^*+Y_u C_{qd,3}^{(d8)} Y_u{}^*\\&+\frac{1}{6} Y_u C_{ud,3}^{(d8)} Y_u{}^*-\frac{5}{6} Y_u C_{ud,4}^{(d8)} Y_u{}^*-\frac{2}{3} C_{q^2ud,1}^{(d8)} Y_d{}^* Y_u{}^*+\frac{10}{3} C_{q^2ud,2}^{(d8)} Y_d{}^* Y_u{}^*-\frac{8}{3} C_{q^2ud,3}^{(d8)} Y_d{}^* Y_u{}^*\end{split}\\

\begin{split}\dot{C}_{qd,4}^{(d8)}&\supset -\frac{1}{6} g_1 g_3 C_{BGd,1}^{(d8)}-\frac{1}{6} g_1 g_3 C_{BGd,1}^{(d8)*}+\frac{1}{3} g_1 g_3 C_{BGq,1}^{(d8)}+\frac{1}{3} g_1 g_3 C_{BGq,1}^{(d8)*}+\frac{1}{2} g_3^2 C_{d,1}^{(d8)}\\&+\frac{5}{6} g_3^2 C_{d,2}^{(d8)}-\frac{g_1 Y_d C_{dB,3}^{(d8)*}}{3 \sqrt{2}}-\frac{2}{3} \sqrt{2} g_3 Y_d C_{dG,3}^{(d8)*}-\frac{3 g_2 Y_d C_{dW,3}^{(d8)*}}{\sqrt{2}}-\frac{5}{6} g_3^2 C_{Gd,1}^{(d8)}+\frac{3}{4} i g_3^2 C_{Gd,2}^{(d8)}\\&-\frac{5}{6} g_3^2 C_{Gq,1}^{(d8)}-\frac{3}{4} i g_3^2 C_{Gq,2}^{(d8)}+\frac{1}{2} g_3^2 C_{q,1}^{(d8)}+\frac{5}{6} g_3^2 C_{q,2}^{(d8)}+g_3^2 C_{q,3}^{(d8)}+\frac{5}{3} g_3^2 C_{q,4}^{(d8)}\\&+2 g_3^2 C_{qd,1}^{(d8)}+\frac{1}{18} g_1^2 C_{qd,3}^{(d8)}+\frac{3}{2} g_2^2 C_{qd,3}^{(d8)}-\frac{25}{12} g_3^2 C_{qd,3}^{(d8)}+\frac{5}{18} g_1^2 C_{qd,4}^{(d8)}+\frac{3}{2} g_2^2 C_{qd,4}^{(d8)}\\&+\frac{79}{12} g_3^2 C_{qd,4}^{(d8)}-\frac{1}{4} g_3^2 C_{qu,1}^{(d8)}+\frac{5}{12} g_3^2 C_{qu,2}^{(d8)}+\frac{1}{4} g_3^2 C_{ud,3}^{(d8)}-\frac{5}{12} g_3^2 C_{ud,4}^{(d8)}-Y_d C_{d,1}^{(d8)} Y_d{}^*\\&-Y_d C_{d,2}^{(d8)} Y_d{}^*+\frac{g_1 C_{dB,3}^{(d8)} Y_d{}^*}{3 \sqrt{2}}+\frac{2}{3} \sqrt{2} g_3 C_{dG,3}^{(d8)} Y_d{}^*+\frac{3 g_2 C_{dW,3}^{(d8)} Y_d{}^*}{\sqrt{2}}-Y_d C_{q,1}^{(d8)} Y_d{}^*-Y_d C_{q,2}^{(d8)} Y_d{}^*\\&-2 Y_d C_{q,3}^{(d8)} Y_d{}^*-2 Y_d C_{q,4}^{(d8)} Y_d{}^*+Y_d C_{qd,1}^{(d8)} Y_d{}^*+Y_d C_{qd,2}^{(d8)} Y_d{}^*+3 Y_d C_{qd,4}^{(d8)} Y_d{}^*+Y_u C_{qd,4}^{(d8)} Y_u{}^*\\&-\frac{1}{2} Y_u C_{ud,3}^{(d8)} Y_u{}^*+\frac{1}{2} Y_u C_{ud,4}^{(d8)} Y_u{}^*\end{split}
\end{align}
The following are the quadratic contributions.
\begin{align}
\begin{split}\dot{C}_{qd,1}^{(d8)}&\supset \frac{1}{3} C_{Hd,1}^{(d6)} C_{Hq,1}^{(d6)}+\frac{2}{3} C_{Hd,1}^{(d6)} C_{Hq,2}^{(d6)}+\frac{4}{3} C_{ld,1}^{(d6)} C_{lq,1}^{(d6)}+\frac{2}{3} C_{ld,1}^{(d6)} C_{lq,2}^{(d6)}+\frac{16}{3} C_{d,1}^{(d6)} C_{qd,1}^{(d6)}+\frac{28}{3} C_{q,1}^{(d6)} C_{qd,1}^{(d6)}\\&+\frac{20}{3} C_{q,2}^{(d6)} C_{qd,1}^{(d6)}+\frac{28}{3} \left(C_{qd,1}^{(d6)}\right){}^2+\frac{4}{3} C_{d,1}^{(d6)} C_{qd,2}^{(d6)}+\frac{8}{3} C_{q,1}^{(d6)} C_{qd,2}^{(d6)}+\frac{4}{3} C_{q,2}^{(d6)} C_{qd,2}^{(d6)}+\frac{4}{3} \left(C_{qd,2}^{(d6)}\right){}^2\\&+\frac{2}{3} C_{ed,1}^{(d6)} C_{qe,1}^{(d6)}+\frac{2}{3} C_{qudl,1}^{(d6)} C_{qudl,1}^{(d6)*}+4 C_{q^2ud,1}^{(d6)} C_{q^2ud,1}^{(d6)*}-2 C_{q^2ud,1}^{(d6)*} C_{q^2ud,2}^{(d6)}-2 C_{q^2ud,1}^{(d6)} C_{q^2ud,2}^{(d6)*}+C_{q^2ud,2}^{(d6)} C_{q^2ud,2}^{(d6)*}\\&+\frac{2}{3} C_{qu,1}^{(d6)} C_{ud,1}^{(d6)}+2 C_{qu,2}^{(d6)} C_{ud,1}^{(d6)}+\frac{2}{3} C_{qu,2}^{(d6)} C_{ud,2}^{(d6)}\end{split}\\

\begin{split}\dot{C}_{qd,2}^{(d8)}&\supset \frac{16}{3} C_{dG,1}^{(d6)} C_{dG,1}^{(d6)*}-\frac{1}{3} C_{Hd,1}^{(d6)} C_{Hq,1}^{(d6)}-\frac{2}{3} C_{Hd,1}^{(d6)} C_{Hq,2}^{(d6)}-\frac{4}{3} C_{ld,1}^{(d6)} C_{lq,1}^{(d6)}-\frac{2}{3} C_{ld,1}^{(d6)} C_{lq,2}^{(d6)}-\frac{16}{3} C_{d,1}^{(d6)} C_{qd,1}^{(d6)}\\&-\frac{28}{3} C_{q,1}^{(d6)} C_{qd,1}^{(d6)}-\frac{20}{3} C_{q,2}^{(d6)} C_{qd,1}^{(d6)}+\frac{4}{3} \left(C_{qd,1}^{(d6)}\right){}^2-\frac{4}{3} C_{d,1}^{(d6)} C_{qd,2}^{(d6)}-\frac{8}{3} C_{q,1}^{(d6)} C_{qd,2}^{(d6)}-\frac{4}{3} C_{q,2}^{(d6)} C_{qd,2}^{(d6)}\\&+\frac{4}{3} \left(C_{qd,2}^{(d6)}\right){}^2-\frac{2}{3} C_{ed,1}^{(d6)} C_{qe,1}^{(d6)}+\frac{2}{3} C_{qudl,1}^{(d6)} C_{qudl,1}^{(d6)*}+\frac{1}{3} C_{q^2ud,2}^{(d6)} C_{q^2ud,2}^{(d6)*}-\frac{2}{3} C_{qu,1}^{(d6)} C_{ud,1}^{(d6)}-2 C_{qu,2}^{(d6)} C_{ud,1}^{(d6)}\\&-\frac{2}{3} C_{qu,2}^{(d6)} C_{ud,2}^{(d6)}\end{split}\\

\begin{split}\dot{C}_{qd,3}^{(d8)}&\supset 4 C_{ledq,1}^{(d6)} C_{ledq,1}^{(d6)*}+\frac{4}{3} C_{d,1}^{(d6)} C_{qd,2}^{(d6)}+\frac{4}{3} C_{q,1}^{(d6)} C_{qd,2}^{(d6)}+\frac{8}{3} C_{q,2}^{(d6)} C_{qd,2}^{(d6)}+\frac{56}{3} C_{qd,1}^{(d6)} C_{qd,2}^{(d6)}+24 \left(C_{qd,2}^{(d6)}\right){}^2\\&-\frac{2}{3} C_{qudl,1}^{(d6)} C_{qudl,1}^{(d6)*}+7 C_{q^2ud,1}^{(d6)} C_{q^2ud,1}^{(d6)*}-11 C_{q^2ud,1}^{(d6)*} C_{q^2ud,2}^{(d6)}-11 C_{q^2ud,1}^{(d6)} C_{q^2ud,2}^{(d6)*}+16 C_{q^2ud,2}^{(d6)} C_{q^2ud,2}^{(d6)*}+\frac{2}{3} C_{qu,1}^{(d6)} C_{ud,2}^{(d6)}\end{split}\\

\begin{split}\dot{C}_{qd,4}^{(d8)}&\supset \frac{8}{3} C_{dB,1}^{(d6)} C_{dB,1}^{(d6)*}-\frac{16}{9} C_{dG,1}^{(d6)} C_{dG,1}^{(d6)*}+8 C_{dW,1}^{(d6)} C_{dW,1}^{(d6)*}-\frac{4}{3} C_{d,1}^{(d6)} C_{qd,2}^{(d6)}-\frac{4}{3} C_{q,1}^{(d6)} C_{qd,2}^{(d6)}-\frac{8}{3} C_{q,2}^{(d6)} C_{qd,2}^{(d6)}\\&+\frac{8}{3} C_{qd,1}^{(d6)} C_{qd,2}^{(d6)}-\frac{2}{3} C_{qudl,1}^{(d6)} C_{qudl,1}^{(d6)*}+C_{q^2ud,1}^{(d6)} C_{q^2ud,1}^{(d6)*}-\frac{1}{3} C_{q^2ud,1}^{(d6)*} C_{q^2ud,2}^{(d6)}-\frac{1}{3} C_{q^2ud,1}^{(d6)} C_{q^2ud,2}^{(d6)*}-\frac{2}{3} C_{qu,1}^{(d6)} C_{ud,2}^{(d6)}\end{split}
\end{align}

\subsection{$u_{\mathbb{C}}^2u_{\mathbb{C}}^{\dagger 2}$}
We list below the RGEs for the dimension-8 type $u_{\mathbb{C}}^2u_{\mathbb{C}}^{\dagger 2}$. The corresponding operators and Wilson coefficients are defined in Table \ref{tab:u2u2}.
\begin{table}[htbp]
\begin{align*}
\begin{array}{|c|c|c|}
\hline\hline
\text{abbreviation} & \text{Wilson coefficient} & \text{operator}\\
\hline
C_{u,1}^{(d6)} & C_{u_{\mathbb{C}}{}_{[2]}^2 u_{\mathbb{C}}{}^{\dagger}{}_{[2]}^2,1}^{\; prst} & \left(u_{\mathbb{C}}{}_{p}^{a}u_{\mathbb{C}}{}_{r}^{b}\right)(u_{\mathbb{C}}^{\dagger}{}_{sa} u_{\mathbb{C}}^{\dagger}{}_{tb}) \\
 & C_ {u_{\mathbb{C}}{}_ {[1,1]}^2 u_{\mathbb{C}}{}^{\dagger}{}_{[1,1]}^2,1}^{\; prst}& \left(u_{\mathbb{C}}{}_{p}^{a}u_{\mathbb{C}}{}_{r}^{b}\right)(u_{\mathbb{C}}^{\dagger}{}_{sa} u_{\mathbb{C}}^{\dagger}{}_{tb}) \\
\hline
C_{u,1}^{(d8)} & C_{u_{\mathbb{C}}{}_{[2]}^2 u_{\mathbb{C}}{}^{\dagger}{}_{[2]}^2D^2,1}^{\; prst}  & \left(u_{\mathbb{C}}{}_{p}^{a}u_{\mathbb{C}}{}_{r}^{b}\right)(D_{\mu}u_{\mathbb{C}}^{\dagger}{}_{sa}D^{\mu} u_{\mathbb{C}}^{\dagger}{}_{tb}) \\
C_{u,2}^{(d8)} & C_{u_{\mathbb{C}}{}_{[2]}^2 u_{\mathbb{C}}{}^{\dagger}{}_{[2]}^2D^2,2}^{\; prst} & i\left(u_{\mathbb{C}}{}_{p}^{a}\sigma_{\mu\nu}u_{\mathbb{C}}{}_{r}^{b}\right)(D^{\mu}u_{\mathbb{C}}^{\dagger}{}_{sa}D^{\nu} u_{\mathbb{C}}^{\dagger}{}_{tb}) \\
 & C_{u_{\mathbb{C}}{}_{[1,1]}^2 u_{\mathbb{C}}{}^{\dagger}{}_{[1,1]}^2D^2,1}^{\; prst}  & \left(u_{\mathbb{C}}{}_{p}^{a}u_{\mathbb{C}}{}_{r}^{b}\right)(D_{\mu}u_{\mathbb{C}}^{\dagger}{}_{sa}D^{\mu} u_{\mathbb{C}}^{\dagger}{}_{tb}) \\
 & C_{u_{\mathbb{C}}{}_{[1,1]}^2 u_{\mathbb{C}}{}^{\dagger}{}_{[1,1]}^2D^2,2}^{\; prst} & i\left(u_{\mathbb{C}}{}_{p}^{a}\sigma_{\mu\nu}u_{\mathbb{C}}{}_{r}^{b}\right)(D^{\mu}u_{\mathbb{C}}^{\dagger}{}_{sa}D^{\nu} u_{\mathbb{C}}^{\dagger}{}_{tb}) \\
\hline
\end{array}
\end{align*}
\caption{List of the $u_{\mathbb{C}}^2u_{\mathbb{C}}^{\dagger 2}$-type operators and the corresponding Wilson coefficients in dimension-6 and dimension-8. The leftmost column contains the abbreviations for the Wilson coefficients in the case of $N_f=1$.}
\label{tab:u2u2}
\end{table}

\begin{align}
\begin{split}\dot{C}_{u,1}^{(d8)}&\supset -\frac{20}{27} g_1 g_3 C_{BGu,1}^{(d8)}-\frac{20}{27} g_1 g_3 C_{BGu,1}^{(d8)*}-\frac{20}{27} g_1^2 C_{Bu,1}^{(d8)}-\frac{1}{3} g_1^2 C_{eu,1}^{(d8)}+\frac{5}{9} g_1^2 C_{eu,2}^{(d8)}\\&-\frac{25}{27} g_3^2 C_{Gu,1}^{(d8)}-\frac{5}{6} i g_3^2 C_{Gu,2}^{(d8)}-\frac{20}{9} g_3^2 C_{Gu,3}^{(d8)}-\frac{1}{9} g_1^2 C_{Hu,1}^{(d8)}-\frac{1}{9} g_1^2 C_{Hu,2}^{(d8)}+\frac{1}{3} g_1^2 C_{lu,1}^{(d8)}\\&-\frac{5}{9} g_1^2 C_{lu,2}^{(d8)}-\frac{1}{9} g_1^2 C_{qu,1}^{(d8)}-\frac{1}{3} g_3^2 C_{qu,1}^{(d8)}+\frac{5}{27} g_1^2 C_{qu,2}^{(d8)}+\frac{5}{9} g_3^2 C_{qu,2}^{(d8)}-\frac{1}{3} g_1^2 C_{qu,3}^{(d8)}\\&+\frac{5}{9} g_1^2 C_{qu,4}^{(d8)}+\frac{352}{27} g_1^2 C_{u,1}^{(d8)}+\frac{139}{9} g_3^2 C_{u,1}^{(d8)}-\frac{40}{27} g_1^2 C_{u,2}^{(d8)}+\frac{65}{9} g_3^2 C_{u,2}^{(d8)}-\frac{1}{3} g_1^2 C_{ud,1}^{(d8)}\\&+\frac{5}{9} g_1^2 C_{ud,2}^{(d8)}-\frac{1}{9} g_1^2 C_{ud,3}^{(d8)}+\frac{1}{6} g_3^2 C_{ud,3}^{(d8)}+\frac{5}{27} g_1^2 C_{ud,4}^{(d8)}-\frac{5}{18} g_3^2 C_{ud,4}^{(d8)}-\frac{1}{3} Y_u C_{Hu,1}^{(d8)} Y_u{}^*\\&+\frac{4}{3} Y_u C_{Hu,2}^{(d8)} Y_u{}^*+\frac{1}{3} Y_u C_{qu,1}^{(d8)} Y_u{}^*-\frac{5}{3} Y_u C_{qu,2}^{(d8)} Y_u{}^*+\frac{1}{3} Y_u C_{qu,3}^{(d8)} Y_u{}^*-\frac{5}{3} Y_u C_{qu,4}^{(d8)} Y_u{}^*+4 Y_u C_{u,1}^{(d8)} Y_u{}^*\end{split}\\

\begin{split}\dot{C}_{u,2}^{(d8)}&\supset \frac{8}{9} g_1 g_3 C_{BGu,1}^{(d8)}+\frac{8}{9} g_1 g_3 C_{BGu,1}^{(d8)*}-\frac{4}{9} g_1^2 C_{Bu,1}^{(d8)}+\frac{1}{3} g_1^2 C_{eu,1}^{(d8)}-\frac{5}{9} g_1^2 C_{eu,2}^{(d8)}+\frac{10}{9} g_3^2 C_{Gu,1}^{(d8)}\\&+i g_3^2 C_{Gu,2}^{(d8)}-\frac{4}{3} g_3^2 C_{Gu,3}^{(d8)}+\frac{1}{9} g_1^2 C_{Hu,1}^{(d8)}+\frac{1}{9} g_1^2 C_{Hu,2}^{(d8)}-\frac{1}{3} g_1^2 C_{lu,1}^{(d8)}+\frac{5}{9} g_1^2 C_{lu,2}^{(d8)}\\&+\frac{1}{9} g_1^2 C_{qu,1}^{(d8)}-\frac{2}{3} g_3^2 C_{qu,1}^{(d8)}-\frac{5}{27} g_1^2 C_{qu,2}^{(d8)}+\frac{10}{9} g_3^2 C_{qu,2}^{(d8)}+\frac{1}{3} g_1^2 C_{qu,3}^{(d8)}-\frac{5}{9} g_1^2 C_{qu,4}^{(d8)}\\&-\frac{16}{9} g_1^2 C_{u,1}^{(d8)}+\frac{26}{3} g_3^2 C_{u,1}^{(d8)}+\frac{88}{27} g_1^2 C_{u,2}^{(d8)}+\frac{58}{9} g_3^2 C_{u,2}^{(d8)}+\frac{1}{3} g_1^2 C_{ud,1}^{(d8)}-\frac{5}{9} g_1^2 C_{ud,2}^{(d8)}\\&+\frac{1}{9} g_1^2 C_{ud,3}^{(d8)}+\frac{1}{3} g_3^2 C_{ud,3}^{(d8)}-\frac{5}{27} g_1^2 C_{ud,4}^{(d8)}-\frac{5}{9} g_3^2 C_{ud,4}^{(d8)}-Y_u C_{Hu,1}^{(d8)} Y_u{}^*+Y_u C_{qu,1}^{(d8)} Y_u{}^*\\&-Y_u C_{qu,2}^{(d8)} Y_u{}^*-Y_u C_{qu,3}^{(d8)} Y_u{}^*+Y_u C_{qu,4}^{(d8)} Y_u{}^*+4 Y_u C_{u,2}^{(d8)} Y_u{}^*\end{split}
\end{align}
The following are the quadratic contributions.
\begin{align}
\begin{split}\dot{C}_{u,1}^{(d8)}&\supset \frac{2}{3} \left(C_{eu,1}^{(d6)}\right){}^2+\frac{2}{3} \left(C_{Hu,1}^{(d6)}\right){}^2+\frac{4}{3} \left(C_{lu,1}^{(d6)}\right){}^2+\frac{4}{3} \left(C_{qu,1}^{(d6)}\right){}^2+\frac{8}{3} C_{qu,1}^{(d6)} C_{qu,2}^{(d6)}+4 \left(C_{qu,2}^{(d6)}\right){}^2\\&+32 \left(C_{u,1}^{(d6)}\right){}^2+2 \left(C_{ud,1}^{(d6)}\right){}^2+\frac{4}{3} C_{ud,1}^{(d6)} C_{ud,2}^{(d6)}+\frac{2}{3} \left(C_{ud,2}^{(d6)}\right){}^2\end{split}\\

\begin{split}\dot{C}_{u,2}^{(d8)}&\supset -\frac{2}{3} \left(C_{eu,1}^{(d6)}\right){}^2-\frac{2}{3} \left(C_{Hu,1}^{(d6)}\right){}^2-\frac{4}{3} \left(C_{lu,1}^{(d6)}\right){}^2+\frac{4}{3} \left(C_{qu,1}^{(d6)}\right){}^2-\frac{8}{3} C_{qu,1}^{(d6)} C_{qu,2}^{(d6)}\\&-4 \left(C_{qu,2}^{(d6)}\right){}^2+\frac{2}{3} C_{u^2de,1}^{(d6)} C_{u^2de,1}^{(d6)*}-\frac{32}{3} \left(C_{u,1}^{(d6)}\right){}^2-2 \left(C_{ud,1}^{(d6)}\right){}^2-\frac{4}{3} C_{ud,1}^{(d6)} C_{ud,2}^{(d6)}+\frac{2}{3} \left(C_{ud,2}^{(d6)}\right){}^2\end{split}
\end{align}

\subsection{$d_{\mathbb{C}}^2d_{\mathbb{C}}^{\dagger 2}$}
We list below the RGEs for the dimension-8 type $d_{\mathbb{C}}^2d_{\mathbb{C}}^{\dagger 2}$. The corresponding operators and Wilson coefficients are defined in Table \ref{tab:d2d2}.
\begin{table}[htbp]
\begin{align*}
\begin{array}{|c|c|c|}
\hline\hline
\text{abbreviation} & \text{Wilson coefficient} & \text{operator}\\
\hline
C_{d,1}^{(d6)} & C_{d_{\mathbb{C}}{}_{[2]}^2 d_{\mathbb{C}}{}^{\dagger}{}_{[2]}^2,1}^{\; prst} & \left(d_{\mathbb{C}}{}_{p}^{a}d_{\mathbb{C}}{}_{r}^{b}\right)(d_{\mathbb{C}}^{\dagger}{}_{sa} d_{\mathbb{C}}^{\dagger}{}_{tb}) \\
 & C_ {d_{\mathbb{C}}{}_ {[1,1]}^2 d_{\mathbb{C}}{}^{\dagger}{}_{[1,1]}^2,1}^{\; prst}& \left(d_{\mathbb{C}}{}_{p}^{a}d_{\mathbb{C}}{}_{r}^{b}\right)(d_{\mathbb{C}}^{\dagger}{}_{sa} d_{\mathbb{C}}^{\dagger}{}_{tb}) \\
\hline
C_{d,1}^{(d8)} & C_{d_{\mathbb{C}}{}_{[2]}^2 d_{\mathbb{C}}{}^{\dagger}{}_{[2]}^2D^2,1}^{\; prst}  & \left(d_{\mathbb{C}}{}_{p}^{a}d_{\mathbb{C}}{}_{r}^{b}\right)(D_{\mu}d_{\mathbb{C}}^{\dagger}{}_{sa}D^{\mu} d_{\mathbb{C}}^{\dagger}{}_{tb}) \\
C_{d,2}^{(d8)} & C_{d_{\mathbb{C}}{}_{[2]}^2 d_{\mathbb{C}}{}^{\dagger}{}_{[2]}^2D^2,2}^{\; prst} & i\left(d_{\mathbb{C}}{}_{p}^{a}\sigma_{\mu\nu}d_{\mathbb{C}}{}_{r}^{b}\right)(D^{\mu}d_{\mathbb{C}}^{\dagger}{}_{sa}D^{\nu} d_{\mathbb{C}}^{\dagger}{}_{tb}) \\
 & C_{d_{\mathbb{C}}{}_{[1,1]}^2 d_{\mathbb{C}}{}^{\dagger}{}_{[1,1]}^2D^2,1}^{\; prst}  & \left(d_{\mathbb{C}}{}_{p}^{a}d_{\mathbb{C}}{}_{r}^{b}\right)(D_{\mu}d_{\mathbb{C}}^{\dagger}{}_{sa}D^{\mu} d_{\mathbb{C}}^{\dagger}{}_{tb}) \\
 & C_{d_{\mathbb{C}}{}_{[1,1]}^2 d_{\mathbb{C}}{}^{\dagger}{}_{[1,1]}^2D^2,2}^{\; prst} & i\left(d_{\mathbb{C}}{}_{p}^{a}\sigma_{\mu\nu}d_{\mathbb{C}}{}_{r}^{b}\right)(D^{\mu}d_{\mathbb{C}}^{\dagger}{}_{sa}D^{\nu} d_{\mathbb{C}}^{\dagger}{}_{tb}) \\
\hline
\end{array}
\end{align*}
\caption{List of the $d_{\mathbb{C}}^2d_{\mathbb{C}}^{\dagger 2}$-type operators and the corresponding Wilson coefficients in dimension-6 and dimension-8. The leftmost column contains the abbreviations for the Wilson coefficients in the case of $N_f=1$.}
\label{tab:d2d2}
\end{table}

\begin{align}
\begin{split}\dot{C}_{d,1}^{(d8)}&\supset -\frac{5}{27} g_1^2 C_{Bd,1}^{(d8)}+\frac{10}{27} g_1 g_3 C_{BGd,1}^{(d8)}+\frac{10}{27} g_1 g_3 C_{BGd,1}^{(d8)*}+\frac{88}{27} g_1^2 C_{d,1}^{(d8)}+\frac{139}{9} g_3^2 C_{d,1}^{(d8)}\\&-\frac{10}{27} g_1^2 C_{d,2}^{(d8)}+\frac{65}{9} g_3^2 C_{d,2}^{(d8)}+\frac{1}{6} g_1^2 C_{ed,1}^{(d8)}-\frac{5}{18} g_1^2 C_{ed,2}^{(d8)}-\frac{25}{27} g_3^2 C_{Gd,1}^{(d8)}-\frac{5}{6} i g_3^2 C_{Gd,2}^{(d8)}\\&-\frac{20}{9} g_3^2 C_{Gd,3}^{(d8)}+\frac{1}{18} g_1^2 C_{Hd,1}^{(d8)}+\frac{1}{18} g_1^2 C_{Hd,2}^{(d8)}-\frac{1}{6} g_1^2 C_{ld,1}^{(d8)}+\frac{5}{18} g_1^2 C_{ld,2}^{(d8)}+\frac{1}{6} g_1^2 C_{qd,1}^{(d8)}\\&-\frac{5}{18} g_1^2 C_{qd,2}^{(d8)}+\frac{1}{18} g_1^2 C_{qd,3}^{(d8)}-\frac{1}{3} g_3^2 C_{qd,3}^{(d8)}-\frac{5}{54} g_1^2 C_{qd,4}^{(d8)}+\frac{5}{9} g_3^2 C_{qd,4}^{(d8)}-\frac{1}{3} g_1^2 C_{ud,1}^{(d8)}\\&+\frac{5}{9} g_1^2 C_{ud,2}^{(d8)}-\frac{1}{9} g_1^2 C_{ud,3}^{(d8)}+\frac{1}{6} g_3^2 C_{ud,3}^{(d8)}+\frac{5}{27} g_1^2 C_{ud,4}^{(d8)}-\frac{5}{18} g_3^2 C_{ud,4}^{(d8)}+4 Y_d C_{d,1}^{(d8)} Y_d{}^*\\&-\frac{4}{3} Y_d C_{Hd,1}^{(d8)} Y_d{}^*+\frac{1}{3} Y_d C_{Hd,2}^{(d8)} Y_d{}^*+\frac{1}{3} Y_d C_{qd,1}^{(d8)} Y_d{}^*-\frac{5}{3} Y_d C_{qd,2}^{(d8)} Y_d{}^*+\frac{1}{3} Y_d C_{qd,3}^{(d8)} Y_d{}^*-\frac{5}{3} Y_d C_{qd,4}^{(d8)} Y_d{}^*\end{split}\\

\begin{split}\dot{C}_{d,2}^{(d8)}&\supset -\frac{1}{9} g_1^2 C_{Bd,1}^{(d8)}-\frac{4}{9} g_1 g_3 C_{BGd,1}^{(d8)}-\frac{4}{9} g_1 g_3 C_{BGd,1}^{(d8)*}-\frac{4}{9} g_1^2 C_{d,1}^{(d8)}+\frac{26}{3} g_3^2 C_{d,1}^{(d8)}\\&+\frac{22}{27} g_1^2 C_{d,2}^{(d8)}+\frac{58}{9} g_3^2 C_{d,2}^{(d8)}-\frac{1}{6} g_1^2 C_{ed,1}^{(d8)}+\frac{5}{18} g_1^2 C_{ed,2}^{(d8)}+\frac{10}{9} g_3^2 C_{Gd,1}^{(d8)}+i g_3^2 C_{Gd,2}^{(d8)}\\&-\frac{4}{3} g_3^2 C_{Gd,3}^{(d8)}-\frac{1}{18} g_1^2 C_{Hd,1}^{(d8)}-\frac{1}{18} g_1^2 C_{Hd,2}^{(d8)}+\frac{1}{6} g_1^2 C_{ld,1}^{(d8)}-\frac{5}{18} g_1^2 C_{ld,2}^{(d8)}-\frac{1}{6} g_1^2 C_{qd,1}^{(d8)}\\&+\frac{5}{18} g_1^2 C_{qd,2}^{(d8)}-\frac{1}{18} g_1^2 C_{qd,3}^{(d8)}-\frac{2}{3} g_3^2 C_{qd,3}^{(d8)}+\frac{5}{54} g_1^2 C_{qd,4}^{(d8)}+\frac{10}{9} g_3^2 C_{qd,4}^{(d8)}+\frac{1}{3} g_1^2 C_{ud,1}^{(d8)}\\&-\frac{5}{9} g_1^2 C_{ud,2}^{(d8)}+\frac{1}{9} g_1^2 C_{ud,3}^{(d8)}+\frac{1}{3} g_3^2 C_{ud,3}^{(d8)}-\frac{5}{27} g_1^2 C_{ud,4}^{(d8)}-\frac{5}{9} g_3^2 C_{ud,4}^{(d8)}+4 Y_d C_{d,2}^{(d8)} Y_d{}^*\\&+Y_d C_{Hd,2}^{(d8)} Y_d{}^*-Y_d C_{qd,1}^{(d8)} Y_d{}^*+Y_d C_{qd,2}^{(d8)} Y_d{}^*+Y_d C_{qd,3}^{(d8)} Y_d{}^*-Y_d C_{qd,4}^{(d8)} Y_d{}^*\end{split}
\end{align}
The following are the quadratic contributions.
\begin{align}
\begin{split}\dot{C}_{d,1}^{(d8)}&\supset 32 \left(C_{d,1}^{(d6)}\right){}^2+\frac{2}{3} \left(C_{ed,1}^{(d6)}\right){}^2+\frac{2}{3} \left(C_{Hd,1}^{(d6)}\right){}^2+\frac{4}{3} \left(C_{ld,1}^{(d6)}\right){}^2+4 \left(C_{qd,1}^{(d6)}\right){}^2+\frac{8}{3} C_{qd,1}^{(d6)} C_{qd,2}^{(d6)}\\&+\frac{4}{3} \left(C_{qd,2}^{(d6)}\right){}^2+2 \left(C_{ud,1}^{(d6)}\right){}^2+\frac{4}{3} C_{ud,1}^{(d6)} C_{ud,2}^{(d6)}+\frac{2}{3} \left(C_{ud,2}^{(d6)}\right){}^2\end{split}\\

\begin{split}\dot{C}_{d,2}^{(d8)}&\supset -\frac{32}{3} \left(C_{d,1}^{(d6)}\right){}^2-\frac{2}{3} \left(C_{ed,1}^{(d6)}\right){}^2-\frac{2}{3} \left(C_{Hd,1}^{(d6)}\right){}^2-\frac{4}{3} \left(C_{ld,1}^{(d6)}\right){}^2-4 \left(C_{qd,1}^{(d6)}\right){}^2\\&-\frac{8}{3} C_{qd,1}^{(d6)} C_{qd,2}^{(d6)}+\frac{4}{3} \left(C_{qd,2}^{(d6)}\right){}^2-2 \left(C_{ud,1}^{(d6)}\right){}^2-\frac{4}{3} C_{ud,1}^{(d6)} C_{ud,2}^{(d6)}+\frac{2}{3} \left(C_{ud,2}^{(d6)}\right){}^2\end{split}
\end{align}

\subsection{$u_{\mathbb{C}}u_{\mathbb{C}}^{\dagger}d_{\mathbb{C}}d_{\mathbb{C}}^{\dagger}$}
We list below the RGEs for the dimension-8 type $u_{\mathbb{C}}u_{\mathbb{C}}^{\dagger}d_{\mathbb{C}}d_{\mathbb{C}}^{\dagger}$. The corresponding operators and Wilson coefficients are defined in Table \ref{tab:udud}.
\begin{table}[htbp]
\begin{align*}
\begin{array}{|c|c|c|}
\hline\hline
\text{abbreviation} & \text{Wilson coefficient} & \text{operator}\\
\hline
C_{ud,1}^{(d6)} & C_{d_{\mathbb{C}} u_{\mathbb{C}}d_{\mathbb{C}}^{\dagger}u_{\mathbb{C}}^{\dagger}  ,1}^{\; prst} & \left(d_{\mathbb{C}}{}_{p}^{a}u_{\mathbb{C}}{}_{r}^{b}\right)(d_{\mathbb{C}}^{\dagger}{}_{sa} u_{\mathbb{C}}^{\dagger}{}_{tb}) \\
C_{ud,2}^{(d6)} & C_{d_{\mathbb{C}} u_{\mathbb{C}}d_{\mathbb{C}}^{\dagger}u_{\mathbb{C}}^{\dagger}  ,2}^{\; prst} & \left(d_{\mathbb{C}}{}_{p}^{a}u_{\mathbb{C}}{}_{r}^{b}\right)(d_{\mathbb{C}}^{\dagger}{}_{sb} u_{\mathbb{C}}^{\dagger}{}_{ta}) \\
\hline
C_{ud,1}^{(d8)} & C_{d_{\mathbb{C}} u_{\mathbb{C}} d_{\mathbb{C}}^{\dagger}u_{\mathbb{C}}^{\dagger}D^2,1}^{\; prst}  & \left(d_{\mathbb{C}}{}_{p}^{a}u_{\mathbb{C}}{}_{r}^{b}\right)(D_{\mu}d_{\mathbb{C}}^{\dagger}{}_{sa}D^{\mu} u_{\mathbb{C}}^{\dagger}{}_{tb}) \\
C_{ud,2}^{(d8)} & C_{d_{\mathbb{C}} u_{\mathbb{C}} d_{\mathbb{C}}^{\dagger}u_{\mathbb{C}}^{\dagger}D^2,2}^{\; prst}  & i\left(d_{\mathbb{C}}{}_{p}^{a}\sigma_{\mu\nu}u_{\mathbb{C}}{}_{r}^{b}\right)(D^{\mu}d_{\mathbb{C}}^{\dagger}{}_{sa}D^{\nu} u_{\mathbb{C}}^{\dagger}{}_{tb}) \\
C_{ud,3}^{(d8)} & C_{d_{\mathbb{C}} u_{\mathbb{C}} d_{\mathbb{C}}^{\dagger}u_{\mathbb{C}}^{\dagger}D^2,3}^{\; prst}  & \left(d_{\mathbb{C}}{}_{p}^{a}u_{\mathbb{C}}{}_{r}^{b}\right)(D_{\mu}d_{\mathbb{C}}^{\dagger}{}_{sb}D^{\mu} u_{\mathbb{C}}^{\dagger}{}_{ta}) \\
C_{ud,4}^{(d8)} & C_{d_{\mathbb{C}} u_{\mathbb{C}} d_{\mathbb{C}}^{\dagger}u_{\mathbb{C}}^{\dagger}D^2,4}^{\; prst}  & i\left(d_{\mathbb{C}}{}_{p}^{a}\sigma_{\mu\nu}u_{\mathbb{C}}{}_{r}^{b}\right)(D^{\mu}d_{\mathbb{C}}^{\dagger}{}_{sb}D^{\nu} u_{\mathbb{C}}^{\dagger}{}_{ta}) \\
\hline
\end{array}
\end{align*}
\caption{List of the $u_{\mathbb{C}}u_{\mathbb{C}}^{\dagger}d_{\mathbb{C}}d_{\mathbb{C}}^{\dagger}$-type operators and the corresponding Wilson coefficients in dimension-6 and dimension-8. The leftmost column contains the abbreviations for the Wilson coefficients in the case of $N_f=1$.}
\label{tab:udud}
\end{table}

\begin{align}
\begin{split}\dot{C}_{ud,1}^{(d8)}&\supset -\frac{20}{27} g_1^2 C_{Bd,1}^{(d8)}+\frac{10}{27} g_1 g_3 C_{BGd,1}^{(d8)}+\frac{10}{27} g_1 g_3 C_{BGd,1}^{(d8)*}-\frac{5}{27} g_1 g_3 C_{BGu,1}^{(d8)}-\frac{5}{27} g_1 g_3 C_{BGu,1}^{(d8)*}\\&-\frac{5}{27} g_1^2 C_{Bu,1}^{(d8)}-\frac{8}{9} g_1^2 C_{d,1}^{(d8)}-\frac{1}{6} g_3^2 C_{d,1}^{(d8)}+\frac{20}{27} g_1^2 C_{d,2}^{(d8)}-\frac{5}{18} g_3^2 C_{d,2}^{(d8)}-\frac{1}{3} g_1^2 C_{ed,1}^{(d8)}\\&+\frac{5}{9} g_1^2 C_{ed,2}^{(d8)}+\frac{1}{6} g_1^2 C_{eu,1}^{(d8)}-\frac{5}{18} g_1^2 C_{eu,2}^{(d8)}+\frac{25}{54} g_3^2 C_{Gd,1}^{(d8)}+\frac{5}{12} i g_3^2 C_{Gd,2}^{(d8)}-\frac{20}{9} g_3^2 C_{Gd,3}^{(d8)}\\&+\frac{25}{54} g_3^2 C_{Gu,1}^{(d8)}+\frac{5}{12} i g_3^2 C_{Gu,2}^{(d8)}-\frac{20}{9} g_3^2 C_{Gu,3}^{(d8)}-\frac{1}{9} g_1^2 C_{Hd,1}^{(d8)}-\frac{1}{9} g_1^2 C_{Hd,2}^{(d8)}+\frac{1}{18} g_1^2 C_{Hu,1}^{(d8)}\\&+\frac{1}{18} g_1^2 C_{Hu,2}^{(d8)}+\frac{1}{3} g_1^2 C_{ld,1}^{(d8)}-\frac{5}{9} g_1^2 C_{ld,2}^{(d8)}-\frac{1}{6} g_1^2 C_{lu,1}^{(d8)}+\frac{5}{18} g_1^2 C_{lu,2}^{(d8)}-\frac{1}{3} g_1^2 C_{qd,1}^{(d8)}\\&+\frac{5}{9} g_1^2 C_{qd,2}^{(d8)}-\frac{1}{9} g_1^2 C_{qd,3}^{(d8)}+\frac{1}{6} g_3^2 C_{qd,3}^{(d8)}+\frac{5}{27} g_1^2 C_{qd,4}^{(d8)}-\frac{5}{18} g_3^2 C_{qd,4}^{(d8)}+\frac{1}{18} g_1^2 C_{qu,1}^{(d8)}\\&+\frac{1}{6} g_3^2 C_{qu,1}^{(d8)}-\frac{5}{54} g_1^2 C_{qu,2}^{(d8)}-\frac{5}{18} g_3^2 C_{qu,2}^{(d8)}+\frac{1}{6} g_1^2 C_{qu,3}^{(d8)}-\frac{5}{18} g_1^2 C_{qu,4}^{(d8)}+\frac{4}{3} Y_d Y_u C_{q^2ud,1}^{(d8)*}\\&+\frac{4}{3} Y_d Y_u C_{q^2ud,2}^{(d8)*}+\frac{8}{3} Y_d Y_u C_{q^2ud,3}^{(d8)*}-\frac{8}{9} g_1^2 C_{u,1}^{(d8)}-\frac{1}{6} g_3^2 C_{u,1}^{(d8)}+\frac{20}{27} g_1^2 C_{u,2}^{(d8)}-\frac{5}{18} g_3^2 C_{u,2}^{(d8)}\\&-\frac{79}{54} g_1^2 C_{ud,1}^{(d8)}+\frac{52}{9} g_3^2 C_{ud,1}^{(d8)}+\frac{35}{18} g_1^2 C_{ud,2}^{(d8)}+10 g_3^2 C_{ud,2}^{(d8)}+\frac{5}{18} g_1^2 C_{ud,3}^{(d8)}+\frac{55}{6} g_3^2 C_{ud,3}^{(d8)}\\&-\frac{25}{54} g_1^2 C_{ud,4}^{(d8)}+\frac{65}{18} g_3^2 C_{ud,4}^{(d8)}-\frac{4}{3} Y_d C_{Hu,1}^{(d8)} Y_d{}^*+\frac{1}{3} Y_d C_{Hu,2}^{(d8)} Y_d{}^*+\frac{1}{3} Y_d C_{qu,3}^{(d8)} Y_d{}^*-\frac{5}{3} Y_d C_{qu,4}^{(d8)} Y_d{}^*\\&+2 Y_d C_{ud,1}^{(d8)} Y_d{}^*-\frac{1}{3} Y_u C_{Hd,1}^{(d8)} Y_u{}^*+\frac{4}{3} Y_u C_{Hd,2}^{(d8)} Y_u{}^*+\frac{1}{3} Y_u C_{qd,1}^{(d8)} Y_u{}^*-\frac{5}{3} Y_u C_{qd,2}^{(d8)} Y_u{}^*+2 Y_u C_{ud,1}^{(d8)} Y_u{}^*\\&+\frac{4}{3} C_{q^2ud,1}^{(d8)} Y_d{}^* Y_u{}^*+\frac{4}{3} C_{q^2ud,2}^{(d8)} Y_d{}^* Y_u{}^*+\frac{8}{3} C_{q^2ud,3}^{(d8)} Y_d{}^* Y_u{}^*\end{split}\\

\begin{split}\dot{C}_{ud,2}^{(d8)}&\supset -\frac{4}{9} g_1^2 C_{Bd,1}^{(d8)}+\frac{2}{9} g_1 g_3 C_{BGd,1}^{(d8)}+\frac{2}{9} g_1 g_3 C_{BGd,1}^{(d8)*}-\frac{1}{9} g_1 g_3 C_{BGu,1}^{(d8)}-\frac{1}{9} g_1 g_3 C_{BGu,1}^{(d8)*}\\&-\frac{1}{9} g_1^2 C_{Bu,1}^{(d8)}+\frac{8}{9} g_1^2 C_{d,1}^{(d8)}+\frac{1}{6} g_3^2 C_{d,1}^{(d8)}-\frac{20}{27} g_1^2 C_{d,2}^{(d8)}+\frac{5}{18} g_3^2 C_{d,2}^{(d8)}+\frac{1}{3} g_1^2 C_{ed,1}^{(d8)}\\&-\frac{5}{9} g_1^2 C_{ed,2}^{(d8)}-\frac{1}{6} g_1^2 C_{eu,1}^{(d8)}+\frac{5}{18} g_1^2 C_{eu,2}^{(d8)}+\frac{5}{18} g_3^2 C_{Gd,1}^{(d8)}+\frac{1}{4} i g_3^2 C_{Gd,2}^{(d8)}-\frac{4}{3} g_3^2 C_{Gd,3}^{(d8)}\\&+\frac{5}{18} g_3^2 C_{Gu,1}^{(d8)}+\frac{1}{4} i g_3^2 C_{Gu,2}^{(d8)}-\frac{4}{3} g_3^2 C_{Gu,3}^{(d8)}+\frac{1}{9} g_1^2 C_{Hd,1}^{(d8)}+\frac{1}{9} g_1^2 C_{Hd,2}^{(d8)}-\frac{1}{18} g_1^2 C_{Hu,1}^{(d8)}\\&-\frac{1}{18} g_1^2 C_{Hu,2}^{(d8)}-\frac{1}{3} g_1^2 C_{ld,1}^{(d8)}+\frac{5}{9} g_1^2 C_{ld,2}^{(d8)}+\frac{1}{6} g_1^2 C_{lu,1}^{(d8)}-\frac{5}{18} g_1^2 C_{lu,2}^{(d8)}+\frac{1}{3} g_1^2 C_{qd,1}^{(d8)}\\&-\frac{5}{9} g_1^2 C_{qd,2}^{(d8)}+\frac{1}{9} g_1^2 C_{qd,3}^{(d8)}-\frac{1}{6} g_3^2 C_{qd,3}^{(d8)}-\frac{5}{27} g_1^2 C_{qd,4}^{(d8)}+\frac{5}{18} g_3^2 C_{qd,4}^{(d8)}-\frac{1}{18} g_1^2 C_{qu,1}^{(d8)}\\&-\frac{1}{6} g_3^2 C_{qu,1}^{(d8)}+\frac{5}{54} g_1^2 C_{qu,2}^{(d8)}+\frac{5}{18} g_3^2 C_{qu,2}^{(d8)}-\frac{1}{6} g_1^2 C_{qu,3}^{(d8)}+\frac{5}{18} g_1^2 C_{qu,4}^{(d8)}+\frac{8}{9} g_1^2 C_{u,1}^{(d8)}\\&+\frac{1}{6} g_3^2 C_{u,1}^{(d8)}-\frac{20}{27} g_1^2 C_{u,2}^{(d8)}+\frac{5}{18} g_3^2 C_{u,2}^{(d8)}+\frac{7}{6} g_1^2 C_{ud,1}^{(d8)}+6 g_3^2 C_{ud,1}^{(d8)}+\frac{5}{2} g_1^2 C_{ud,2}^{(d8)}\\&+\frac{16}{3} g_3^2 C_{ud,2}^{(d8)}-\frac{5}{18} g_1^2 C_{ud,3}^{(d8)}+\frac{13}{6} g_3^2 C_{ud,3}^{(d8)}+\frac{25}{54} g_1^2 C_{ud,4}^{(d8)}-\frac{5}{18} g_3^2 C_{ud,4}^{(d8)}+Y_d C_{Hu,2}^{(d8)} Y_d{}^*\\&-Y_d C_{qu,3}^{(d8)} Y_d{}^*+Y_d C_{qu,4}^{(d8)} Y_d{}^*+2 Y_d C_{ud,2}^{(d8)} Y_d{}^*-Y_u C_{Hd,1}^{(d8)} Y_u{}^*-Y_u C_{qd,1}^{(d8)} Y_u{}^*+Y_u C_{qd,2}^{(d8)} Y_u{}^*\\&+2 Y_u C_{ud,2}^{(d8)} Y_u{}^*\end{split}\\

\begin{split}\dot{C}_{ud,3}^{(d8)}&\supset -\frac{10}{9} g_1 g_3 C_{BGd,1}^{(d8)}-\frac{10}{9} g_1 g_3 C_{BGd,1}^{(d8)*}+\frac{5}{9} g_1 g_3 C_{BGu,1}^{(d8)}+\frac{5}{9} g_1 g_3 C_{BGu,1}^{(d8)*}+\frac{1}{2} g_3^2 C_{d,1}^{(d8)}\\&+\frac{5}{6} g_3^2 C_{d,2}^{(d8)}-\frac{25}{18} g_3^2 C_{Gd,1}^{(d8)}-\frac{5}{4} i g_3^2 C_{Gd,2}^{(d8)}-\frac{25}{18} g_3^2 C_{Gu,1}^{(d8)}-\frac{5}{4} i g_3^2 C_{Gu,2}^{(d8)}-\frac{1}{2} g_3^2 C_{qd,3}^{(d8)}\\&+\frac{5}{6} g_3^2 C_{qd,4}^{(d8)}-\frac{1}{2} g_3^2 C_{qu,1}^{(d8)}+\frac{5}{6} g_3^2 C_{qu,2}^{(d8)}-\frac{4}{3} Y_d Y_u C_{q^2ud,1}^{(d8)*}-\frac{4}{3} Y_d Y_u C_{q^2ud,2}^{(d8)*}-\frac{8}{3} Y_d Y_u C_{q^2ud,3}^{(d8)*}\\&+\frac{1}{2} g_3^2 C_{u,1}^{(d8)}+\frac{5}{6} g_3^2 C_{u,2}^{(d8)}+\frac{28}{3} g_3^2 C_{ud,1}^{(d8)}-\frac{10}{3} g_3^2 C_{ud,2}^{(d8)}-\frac{62}{27} g_1^2 C_{ud,3}^{(d8)}+\frac{113}{18} g_3^2 C_{ud,3}^{(d8)}\\&+\frac{10}{3} g_1^2 C_{ud,4}^{(d8)}-\frac{65}{6} g_3^2 C_{ud,4}^{(d8)}+Y_u C_{Hud,1}^{(d8)} Y_d{}^*+\frac{1}{3} Y_d C_{qu,1}^{(d8)} Y_d{}^*-\frac{5}{3} Y_d C_{qu,2}^{(d8)} Y_d{}^*+2 Y_d C_{ud,3}^{(d8)} Y_d{}^*\\&+Y_d C_{Hud,1}^{(d8)*} Y_u{}^*+\frac{1}{3} Y_u C_{qd,3}^{(d8)} Y_u{}^*-\frac{5}{3} Y_u C_{qd,4}^{(d8)} Y_u{}^*+2 Y_u C_{ud,3}^{(d8)} Y_u{}^*-\frac{4}{3} C_{q^2ud,1}^{(d8)} Y_d{}^* Y_u{}^*-\frac{4}{3} C_{q^2ud,2}^{(d8)} Y_d{}^* Y_u{}^*\\&-\frac{8}{3} C_{q^2ud,3}^{(d8)} Y_d{}^* Y_u{}^*\end{split}\\

\begin{split}\dot{C}_{ud,4}^{(d8)}&\supset -\frac{2}{3} g_1 g_3 C_{BGd,1}^{(d8)}-\frac{2}{3} g_1 g_3 C_{BGd,1}^{(d8)*}+\frac{1}{3} g_1 g_3 C_{BGu,1}^{(d8)}+\frac{1}{3} g_1 g_3 C_{BGu,1}^{(d8)*}-\frac{1}{2} g_3^2 C_{d,1}^{(d8)}\\&-\frac{5}{6} g_3^2 C_{d,2}^{(d8)}-\frac{5}{6} g_3^2 C_{Gd,1}^{(d8)}-\frac{3}{4} i g_3^2 C_{Gd,2}^{(d8)}-\frac{5}{6} g_3^2 C_{Gu,1}^{(d8)}-\frac{3}{4} i g_3^2 C_{Gu,2}^{(d8)}+\frac{1}{2} g_3^2 C_{qd,3}^{(d8)}\\&-\frac{5}{6} g_3^2 C_{qd,4}^{(d8)}+\frac{1}{2} g_3^2 C_{qu,1}^{(d8)}-\frac{5}{6} g_3^2 C_{qu,2}^{(d8)}-\frac{1}{2} g_3^2 C_{u,1}^{(d8)}-\frac{5}{6} g_3^2 C_{u,2}^{(d8)}-2 g_3^2 C_{ud,1}^{(d8)}\\&+2 g_1^2 C_{ud,3}^{(d8)}-\frac{13}{2} g_3^2 C_{ud,3}^{(d8)}+\frac{10}{9} g_1^2 C_{ud,4}^{(d8)}+\frac{37}{6} g_3^2 C_{ud,4}^{(d8)}+Y_u C_{Hud,1}^{(d8)} Y_d{}^*-Y_d C_{qu,1}^{(d8)} Y_d{}^*\\&+Y_d C_{qu,2}^{(d8)} Y_d{}^*+2 Y_d C_{ud,4}^{(d8)} Y_d{}^*+Y_d C_{Hud,1}^{(d8)*} Y_u{}^*-Y_u C_{qd,3}^{(d8)} Y_u{}^*+Y_u C_{qd,4}^{(d8)} Y_u{}^*+2 Y_u C_{ud,4}^{(d8)} Y_u{}^*\end{split}
\end{align}
The following are the quadratic contributions.
\begin{align}
\begin{split}\dot{C}_{ud,1}^{(d8)}&\supset \frac{2}{3} C_{ed,1}^{(d6)} C_{eu,1}^{(d6)}+\frac{2}{3} C_{Hd,1}^{(d6)} C_{Hu,1}^{(d6)}+\frac{4}{3} C_{ld,1}^{(d6)} C_{lu,1}^{(d6)}+\frac{4}{3} C_{qd,1}^{(d6)} C_{qu,1}^{(d6)}+4 C_{qd,1}^{(d6)} C_{qu,2}^{(d6)}+\frac{4}{3} C_{qd,2}^{(d6)} C_{qu,2}^{(d6)}\\&+8 C_{qudl,1}^{(d6)} C_{qudl,1}^{(d6)*}+2 C_{q^2ud,1}^{(d6)} C_{q^2ud,1}^{(d6)*}+2 C_{q^2ud,1}^{(d6)*} C_{q^2ud,2}^{(d6)}+2 C_{q^2ud,1}^{(d6)} C_{q^2ud,2}^{(d6)*}+2 C_{q^2ud,2}^{(d6)} C_{q^2ud,2}^{(d6)*}+9 C_{u^2de,1}^{(d6)} C_{u^2de,1}^{(d6)*}\\&+\frac{16}{3} C_{d,1}^{(d6)} C_{ud,1}^{(d6)}+\frac{16}{3} C_{u,1}^{(d6)} C_{ud,1}^{(d6)}+\frac{28}{3} \left(C_{ud,1}^{(d6)}\right){}^2+\frac{4}{3} C_{d,1}^{(d6)} C_{ud,2}^{(d6)}+\frac{4}{3} C_{u,1}^{(d6)} C_{ud,2}^{(d6)}+8 \left(C_{ud,2}^{(d6)}\right){}^2\end{split}\\

\begin{split}\dot{C}_{ud,2}^{(d8)}&\supset -\frac{2}{3} C_{ed,1}^{(d6)} C_{eu,1}^{(d6)}-\frac{2}{3} C_{Hd,1}^{(d6)} C_{Hu,1}^{(d6)}-\frac{4}{3} C_{ld,1}^{(d6)} C_{lu,1}^{(d6)}-\frac{4}{3} C_{qd,1}^{(d6)} C_{qu,1}^{(d6)}-4 C_{qd,1}^{(d6)} C_{qu,2}^{(d6)}\\&-\frac{4}{3} C_{qd,2}^{(d6)} C_{qu,2}^{(d6)}+\frac{2}{3} C_{q^2ud,1}^{(d6)} C_{q^2ud,1}^{(d6)*}-\frac{2}{3} C_{q^2ud,1}^{(d6)*} C_{q^2ud,2}^{(d6)}-\frac{2}{3} C_{q^2ud,1}^{(d6)} C_{q^2ud,2}^{(d6)*}+\frac{2}{3} C_{q^2ud,2}^{(d6)} C_{q^2ud,2}^{(d6)*}+\frac{1}{3} C_{u^2de,1}^{(d6)} C_{u^2de,1}^{(d6)*}\\&-\frac{16}{3} C_{d,1}^{(d6)} C_{ud,1}^{(d6)}-\frac{16}{3} C_{u,1}^{(d6)} C_{ud,1}^{(d6)}+\frac{4}{3} \left(C_{ud,1}^{(d6)}\right){}^2-\frac{4}{3} C_{d,1}^{(d6)} C_{ud,2}^{(d6)}-\frac{4}{3} C_{u,1}^{(d6)} C_{ud,2}^{(d6)}\end{split}\\

\begin{split}\dot{C}_{ud,3}^{(d8)}&\supset \frac{4}{3} C_{Hud,1}^{(d6)} C_{Hud,1}^{(d6)*}+\frac{4}{3} C_{qd,2}^{(d6)} C_{qu,1}^{(d6)}-8 C_{qudl,1}^{(d6)} C_{qudl,1}^{(d6)*}-2 C_{q^2ud,1}^{(d6)} C_{q^2ud,1}^{(d6)*}-2 C_{q^2ud,1}^{(d6)*} C_{q^2ud,2}^{(d6)}-2 C_{q^2ud,1}^{(d6)} C_{q^2ud,2}^{(d6)*}\\&-2 C_{q^2ud,2}^{(d6)} C_{q^2ud,2}^{(d6)*}-9 C_{u^2de,1}^{(d6)} C_{u^2de,1}^{(d6)*}+\frac{4}{3} C_{d,1}^{(d6)} C_{ud,2}^{(d6)}+\frac{4}{3} C_{u,1}^{(d6)} C_{ud,2}^{(d6)}+\frac{56}{3} C_{ud,1}^{(d6)} C_{ud,2}^{(d6)}+4 \left(C_{ud,2}^{(d6)}\right){}^2\end{split}\\

\begin{split}\dot{C}_{ud,4}^{(d8)}&\supset \frac{4}{3} C_{Hud,1}^{(d6)} C_{Hud,1}^{(d6)*}-\frac{4}{3} C_{qd,2}^{(d6)} C_{qu,1}^{(d6)}+\frac{2}{3} C_{q^2ud,1}^{(d6)} C_{q^2ud,1}^{(d6)*}-\frac{2}{3} C_{q^2ud,1}^{(d6)*} C_{q^2ud,2}^{(d6)}-\frac{2}{3} C_{q^2ud,1}^{(d6)} C_{q^2ud,2}^{(d6)*}+\frac{2}{3} C_{q^2ud,2}^{(d6)} C_{q^2ud,2}^{(d6)*}\\&-\frac{1}{3} C_{u^2de,1}^{(d6)} C_{u^2de,1}^{(d6)*}-\frac{4}{3} C_{d,1}^{(d6)} C_{ud,2}^{(d6)}-\frac{4}{3} C_{u,1}^{(d6)} C_{ud,2}^{(d6)}+\frac{8}{3} C_{ud,1}^{(d6)} C_{ud,2}^{(d6)}+4 \left(C_{ud,2}^{(d6)}\right){}^2\end{split}
\end{align}

\subsection{$Q^2 u_{\mathbb{C}}d_{\mathbb{C}}$ }
We list below the RGEs for the dimension-8 type $Q^2 u_{\mathbb{C}}d_{\mathbb{C}}$. The corresponding operators and Wilson coefficients are defined in Table \ref{tab:Q2ud}.
\begin{table}[htbp]
\begin{align*}
\begin{array}{|c|c|c|}
\hline\hline
\text{abbreviation} & \text{Wilson coefficient} & \text{operator}\\
\hline
C_{q^2ud,1}^{(d6)} & C_ {d_{\mathbb{C}}Q^2_{[2]}u_{\mathbb{C}},1}^{\; prst} & \epsilon^{ij}\left(d_{\mathbb{C}}{}^a_p Q_{rbi}\right)\left(Q_{saj}u_{\mathbb{C}}{}_t^b\right) \\
C_{q^2ud,2}^{(d6)} & C_ {d_{\mathbb{C}}Q^2_{[2]}u_{\mathbb{C}},2}^{\; prst} & \epsilon^{ij}\left(d_{\mathbb{C}}{}^a_p Q_{saj}\right)\left(Q_{rbi}u_{\mathbb{C}}{}_t^b\right) \\
 & C_ {d_{\mathbb{C}}Q^2_{[1,1]}u_{\mathbb{C}},1}^{\; prst} & \epsilon^{ij}\left(d_{\mathbb{C}}{}^a_p Q_{rbi}\right)\left(Q_{saj}u_{\mathbb{C}}{}_t^b\right) \\
 & C_ {d_{\mathbb{C}}Q^2_{[1,1]}u_{\mathbb{C}},2}^{\; prst} & \epsilon^{ij}\left(d_{\mathbb{C}}{}^a_p Q_{saj}\right)\left(Q_{rbi}u_{\mathbb{C}}{}_t^b\right) \\
\hline
C_{q^2ud,1}^{(d8)} & C_ {d_{\mathbb{C}}Q^2_{[2]}u_{\mathbb{C}}D^2,1}^{\; prst} & \epsilon^{ij}\left(d_{\mathbb{C}}{}^a_p Q_{rbi}\right)\left(D^{\mu}Q_{saj}D_{\mu}u_{\mathbb{C}}{}_t^b\right) \\
C_{q^2ud,2}^{(d8)} & C_ {d_{\mathbb{C}}Q^2_{[2]}u_{\mathbb{C}}D^2,2}^{\; prst} & \epsilon^{ij}\left(d_{\mathbb{C}}{}^a_p Q_{saj}\right)\left(D^{\mu}Q_{rbi}D_{\mu}u_{\mathbb{C}}{}_t^b\right) \\
C_{q^2ud,3}^{(d8)} & C_ {d_{\mathbb{C}}Q^2_{[2]}u_{\mathbb{C}}D^2,3}^{\; prst} & i\epsilon^{ij}\left(d_{\mathbb{C}}{}^a_p\sigma_{\mu\nu} Q_{rbi}\right)\left(D^{\mu}Q_{saj}D^{\nu}u_{\mathbb{C}}{}_t^b\right) \\
 & C_ {d_{\mathbb{C}}Q^2_{[1,1]}u_{\mathbb{C}}D^2,1}^{\; prst} & \epsilon^{ij}\left(d_{\mathbb{C}}{}^a_p Q_{rbi}\right)\left(D^{\mu}Q_{saj}D_{\mu}u_{\mathbb{C}}{}_t^b\right) \\
 & C_ {d_{\mathbb{C}}Q^2_{[1,1]}u_{\mathbb{C}}D^2,2}^{\; prst} & \epsilon^{ij}\left(d_{\mathbb{C}}{}^a_p Q_{saj}\right)\left(D^{\mu}Q_{rbi}D_{\mu}u_{\mathbb{C}}{}_t^b\right) \\
 & C_ {d_{\mathbb{C}}Q^2_{[1,1]}u_{\mathbb{C}}D^2,3}^{\; prst} & i\epsilon^{ij}\left(d_{\mathbb{C}}{}^a_p\sigma_{\mu\nu} Q_{rbi}\right)\left(D^{\mu}Q_{saj}D^{\nu}u_{\mathbb{C}}{}_t^b\right) \\
\hline
\end{array}
\end{align*}
\caption{List of the $Q^2 u_{\mathbb{C}}d_{\mathbb{C}}$-type operators and the corresponding Wilson coefficients in dimension-6 and dimension-8. The leftmost column contains the abbreviations for the Wilson coefficients in the case of $N_f=1$.}
\label{tab:Q2ud}
\end{table}

\begin{align}
\begin{split}\dot{C}_{q^2ud,1}^{(d8)}&\supset \frac{g_1 Y_u C_{dB,1}^{(d8)}}{2 \sqrt{2}}+\frac{5 g_1 Y_u C_{dB,2}^{(d8)}}{6 \sqrt{2}}-\frac{1}{3} \sqrt{2} g_3 Y_u C_{dG,2}^{(d8)}-\frac{3 g_2 Y_u C_{dW,1}^{(d8)}}{2 \sqrt{2}}+\frac{3 g_2 Y_u C_{dW,2}^{(d8)}}{2 \sqrt{2}}+2 Y_d Y_u C_{qd,1}^{(d8)}\\&+2 Y_d Y_u C_{qu,3}^{(d8)}+\frac{2}{3} g_1^2 C_{q^2ud,1}^{(d8)}+3 g_2^2 C_{q^2ud,1}^{(d8)}+16 g_3^2 C_{q^2ud,1}^{(d8)}+\frac{1}{6} g_1^2 C_{q^2ud,2}^{(d8)}+\frac{3}{2} g_2^2 C_{q^2ud,2}^{(d8)}\\&+2 g_3^2 C_{q^2ud,2}^{(d8)}-\frac{5}{18} g_1^2 C_{q^2ud,3}^{(d8)}-\frac{3}{2} g_2^2 C_{q^2ud,3}^{(d8)}+\frac{8}{3} g_3^2 C_{q^2ud,3}^{(d8)}-\frac{g_1 Y_d C_{uB,1}^{(d8)}}{2 \sqrt{2}}+\frac{g_1 Y_d C_{uB,2}^{(d8)}}{6 \sqrt{2}}\\&+\frac{1}{3} \sqrt{2} g_3 Y_d C_{uG,2}^{(d8)}+\frac{3 g_2 Y_d C_{uW,1}^{(d8)}}{2 \sqrt{2}}+\frac{3 g_2 Y_d C_{uW,2}^{(d8)}}{2 \sqrt{2}}+2 Y_d C_{q^2ud,1}^{(d8)} Y_d{}^*+2 Y_u C_{q^2ud,1}^{(d8)} Y_u{}^*\end{split}\\

\begin{split}\dot{C}_{q^2ud,2}^{(d8)}&\supset \frac{g_1 Y_u C_{dB,1}^{(d8)}}{6 \sqrt{2}}-\frac{5 g_1 Y_u C_{dB,2}^{(d8)}}{6 \sqrt{2}}+\frac{1}{3} \sqrt{2} g_3 Y_u C_{dG,2}^{(d8)}-\frac{g_2 Y_u C_{dW,1}^{(d8)}}{2 \sqrt{2}}-\frac{3 g_2 Y_u C_{dW,2}^{(d8)}}{2 \sqrt{2}}+2 Y_e Y_u C_{ledq,1}^{(d8)*}\\&+4 Y_d Y_u C_{q,1}^{(d8)}-4 Y_d Y_u C_{q,3}^{(d8)}+2 Y_d Y_u C_{qd,1}^{(d8)}+4 Y_d Y_u C_{qd,3}^{(d8)}+4 Y_d Y_u C_{qu,1}^{(d8)}+2 Y_d Y_u C_{qu,3}^{(d8)}\\&-\frac{1}{2} g_1^2 C_{q^2ud,1}^{(d8)}-\frac{3}{2} g_2^2 C_{q^2ud,1}^{(d8)}-18 g_3^2 C_{q^2ud,2}^{(d8)}-\frac{5}{6} g_1^2 C_{q^2ud,3}^{(d8)}-\frac{9}{2} g_2^2 C_{q^2ud,3}^{(d8)}+8 g_3^2 C_{q^2ud,3}^{(d8)}\\&-\frac{g_1 Y_d C_{uB,1}^{(d8)}}{6 \sqrt{2}}-\frac{g_1 Y_d C_{uB,2}^{(d8)}}{6 \sqrt{2}}+2 Y_d Y_u C_{ud,1}^{(d8)}-2 Y_d Y_u C_{ud,3}^{(d8)}-\frac{1}{3} \sqrt{2} g_3 Y_d C_{uG,2}^{(d8)}+\frac{g_2 Y_d C_{uW,1}^{(d8)}}{2 \sqrt{2}}\\&-\frac{3 g_2 Y_d C_{uW,2}^{(d8)}}{2 \sqrt{2}}+\frac{22}{3} Y_d C_{q^2ud,2}^{(d8)} Y_d{}^*-4 Y_d C_{q^2ud,3}^{(d8)} Y_d{}^*-2 Y_d C_{lequ,1}^{(d8)} Y_e{}^*-\frac{2}{3} Y_d C_{lequ,2}^{(d8)} Y_e{}^*+\frac{22}{3} Y_u C_{q^2ud,2}^{(d8)} Y_u{}^*\\&-4 Y_u C_{q^2ud,3}^{(d8)} Y_u{}^*\end{split}\\

\begin{split}\dot{C}_{q^2ud,3}^{(d8)}&\supset \frac{g_1 Y_u C_{dB,1}^{(d8)}}{2 \sqrt{2}}-\frac{5 g_1 Y_u C_{dB,2}^{(d8)}}{6 \sqrt{2}}+\frac{4}{3} \sqrt{2} g_3 Y_u C_{dG,2}^{(d8)}-\frac{3 g_2 Y_u C_{dW,1}^{(d8)}}{2 \sqrt{2}}-\frac{3 g_2 Y_u C_{dW,2}^{(d8)}}{2 \sqrt{2}}+4 Y_d Y_u C_{q,1}^{(d8)}\\&-4 Y_d Y_u C_{q,3}^{(d8)}-\frac{2}{3} g_1^2 C_{q^2ud,1}^{(d8)}-3 g_2^2 C_{q^2ud,1}^{(d8)}-\frac{2}{3} g_1^2 C_{q^2ud,2}^{(d8)}-3 g_2^2 C_{q^2ud,2}^{(d8)}-8 g_3^2 C_{q^2ud,2}^{(d8)}\\&-\frac{1}{18} g_1^2 C_{q^2ud,3}^{(d8)}-\frac{3}{2} g_2^2 C_{q^2ud,3}^{(d8)}+\frac{16}{3} g_3^2 C_{q^2ud,3}^{(d8)}-\frac{g_1 Y_d C_{uB,1}^{(d8)}}{2 \sqrt{2}}-\frac{g_1 Y_d C_{uB,2}^{(d8)}}{6 \sqrt{2}}+2 Y_d Y_u C_{ud,1}^{(d8)}\\&-2 Y_d Y_u C_{ud,3}^{(d8)}-\frac{4}{3} \sqrt{2} g_3 Y_d C_{uG,2}^{(d8)}+\frac{3 g_2 Y_d C_{uW,1}^{(d8)}}{2 \sqrt{2}}-\frac{3 g_2 Y_d C_{uW,2}^{(d8)}}{2 \sqrt{2}}+2 Y_d C_{q^2ud,3}^{(d8)} Y_d{}^*+2 Y_u C_{q^2ud,3}^{(d8)} Y_u{}^*\end{split}
\end{align}
The following are the quadratic contributions.
\begin{align}
\begin{split}\dot{C}_{q^2ud,1}^{(d8)}&\supset 4 C_{qd,1}^{(d6)} C_{q^2ud,1}^{(d6)}+4 C_{qu,2}^{(d6)} C_{q^2ud,1}^{(d6)}-2 C_{qd,1}^{(d6)} C_{q^2ud,2}^{(d6)}-2 C_{qu,2}^{(d6)} C_{q^2ud,2}^{(d6)}\end{split}\\

\begin{split}\dot{C}_{q^2ud,2}^{(d8)}&\supset -4 C_{ledq,1}^{(d6)*} C_{lequ,1}^{(d6)}+2 C_{ledq,1}^{(d6)*} C_{lequ,2}^{(d6)}-4 C_{q,1}^{(d6)} C_{q^2ud,1}^{(d6)}+4 C_{q,2}^{(d6)} C_{q^2ud,1}^{(d6)}-2 C_{qd,1}^{(d6)} C_{q^2ud,1}^{(d6)}\\&-10 C_{qd,2}^{(d6)} C_{q^2ud,1}^{(d6)}-10 C_{qu,1}^{(d6)} C_{q^2ud,1}^{(d6)}-2 C_{qu,2}^{(d6)} C_{q^2ud,1}^{(d6)}-4 C_{q,1}^{(d6)} C_{q^2ud,2}^{(d6)}+4 C_{q,2}^{(d6)} C_{q^2ud,2}^{(d6)}+4 C_{qd,1}^{(d6)} C_{q^2ud,2}^{(d6)}\\&+14 C_{qd,2}^{(d6)} C_{q^2ud,2}^{(d6)}+14 C_{qu,1}^{(d6)} C_{q^2ud,2}^{(d6)}+4 C_{qu,2}^{(d6)} C_{q^2ud,2}^{(d6)}-12 C_{qudl,1}^{(d6)} C_{q^3l,1}^{(d6)}-12 C_{q^2ue,1}^{(d6)*} C_{u^2de,1}^{(d6)}-2 C_{q^2ud,1}^{(d6)} C_{ud,1}^{(d6)}\\&-2 C_{q^2ud,2}^{(d6)} C_{ud,1}^{(d6)}+2 C_{q^2ud,1}^{(d6)} C_{ud,2}^{(d6)}+2 C_{q^2ud,2}^{(d6)} C_{ud,2}^{(d6)}\end{split}\\

\begin{split}\dot{C}_{q^2ud,3}^{(d8)}&\supset -4 C_{q,1}^{(d6)} C_{q^2ud,1}^{(d6)}+4 C_{q,2}^{(d6)} C_{q^2ud,1}^{(d6)}-4 C_{q,1}^{(d6)} C_{q^2ud,2}^{(d6)}+4 C_{q,2}^{(d6)} C_{q^2ud,2}^{(d6)}-12 C_{qudl,1}^{(d6)} C_{q^3l,1}^{(d6)}\\&-12 C_{q^2ue,1}^{(d6)*} C_{u^2de,1}^{(d6)}-2 C_{q^2ud,1}^{(d6)} C_{ud,1}^{(d6)}-2 C_{q^2ud,2}^{(d6)} C_{ud,1}^{(d6)}+2 C_{q^2ud,1}^{(d6)} C_{ud,2}^{(d6)}+2 C_{q^2ud,2}^{(d6)} C_{ud,2}^{(d6)}\end{split}
\end{align}

\subsection{$Q^3L$}
We list below the RGEs for the dimension-8 type $Q^3L$. The corresponding operators and Wilson coefficients are defined in Table \ref{tab:LQ3}.
\begin{table}[htbp]
\begin{align*}
\begin{array}{|c|c|c|}
\hline\hline
\text{abbreviation} & \text{Wilson coefficient} & \text{operator}\\
\hline
C_{q^3l,1}^{(d6)} & C_{LQ^3_{[3]}  ,1}^{\; prst} & \epsilon^{abc}\epsilon^{ik}\epsilon^{jl}(L_{pi}Q_{raj})(Q_{sbk}Q_{tcl}) \\
 & C_{LQ^3_{[2,1]}  ,1}^{\; prst} & \epsilon^{abc}\epsilon^{ik}\epsilon^{jl}(L_{pi}Q_{raj})(Q_{sbk}Q_{tcl}) \\
 & C_{LQ^3_{[1,1,1]}  ,1}^{\; prst} & \epsilon^{abc}\epsilon^{ik}\epsilon^{jl}(L_{pi}Q_{raj})(Q_{sbk}Q_{tcl}) \\
\hline
C_{q^3l,1}^{(d8)} & C_{LQ^3_{[3]} D^2 ,1}^{\; prst} & \epsilon^{abc}\epsilon^{ik}\epsilon^{jl}(L_{pi}Q_{raj})(D^{\mu}Q_{sbk}D_{\mu}Q_{tcl}) \\
 & C_{LQ^3_{[2,1]} D^2 ,1}^{\; prst} & \epsilon^{abc}\epsilon^{ik}\epsilon^{jl}(L_{pi}Q_{raj})(D^{\mu}Q_{sbk}D_{\mu}Q_{tcl}) \\
 & C_{LQ^3_{[2,1]} D^2 ,2}^{\; prst} & \epsilon^{abc}\epsilon^{ij}\epsilon^{kl}(L_{pi}Q_{raj})(D^{\mu}Q_{sbk}D_{\mu}Q_{tcl}) \\
 & C_{LQ^3_{[1,1,1]} D^2 ,1}^{\; prst} & \epsilon^{abc}\epsilon^{ik}\epsilon^{jl}(L_{pi}Q_{raj})(D^{\mu}Q_{sbk}D_{\mu}Q_{tcl}) \\
\hline
\end{array}
\end{align*}
\caption{List of the $Q^3L$-type operators and the corresponding Wilson coefficients in dimension-6 and dimension-8. The leftmost column contains the abbreviations for the Wilson coefficients in the case of $N_f=1$.}
\label{tab:LQ3}
\end{table}

\begin{align}
\begin{split}\dot{C}_{q^3l,1}^{(d8)}&\supset -4 Y_d Y_u C_{qudl,1}^{(d8)*}-4 Y_e Y_u C_{q^2ue,1}^{(d8)*}+\frac{2}{9} g_1^2 C_{q^3l,1}^{(d8)}+\frac{8}{3} g_3^2 C_{q^3l,1}^{(d8)}+\frac{3}{2} Y_d C_{q^3l,1}^{(d8)} Y_d{}^*\\&+\frac{1}{2} Y_e C_{q^3l,1}^{(d8)} Y_e{}^*+\frac{3}{2} Y_u C_{q^3l,1}^{(d8)} Y_u{}^*\end{split}
\end{align}
The following are the quadratic contributions.
\begin{align}
\begin{split}\dot{C}_{q^3l,1}^{(d8)}&\supset 4 C_{qudl,1}^{(d6)*} C_{q^2ud,1}^{(d6)}+4 C_{qudl,1}^{(d6)*} C_{q^2ud,2}^{(d6)}-4 C_{lequ,1}^{(d6)} C_{q^2ue,1}^{(d6)*}-4 C_{lequ,2}^{(d6)} C_{q^2ue,1}^{(d6)*}+6 C_{lq,1}^{(d6)} C_{q^3l,1}^{(d6)}-6 C_{lq,2}^{(d6)} C_{q^3l,1}^{(d6)}\\&+12 C_{q,1}^{(d6)} C_{q^3l,1}^{(d6)}-12 C_{q,2}^{(d6)} C_{q^3l,1}^{(d6)}\end{split}
\end{align}

\subsection{$Q^{\dagger 2} u_{\mathbb{C}}e_{\mathbb{C}}$}
We list below the RGEs for the dimension-8 type $Q^{\dagger 2} u_{\mathbb{C}}e_{\mathbb{C}}$. The corresponding operators and Wilson coefficients are defined in Table \ref{tab:Q2ue}.
\begin{table}[htbp]
\begin{align*}
\begin{array}{|c|c|c|}
\hline\hline
\text{abbreviation} & \text{Wilson coefficient} & \text{operator}\\
\hline
C_{q^2ue,1}^{(d6)} & C_{e_{\mathbb{C}}u_{\mathbb{C}}Q^{\dagger}{}^2_{[2]}  ,1}^{\; prst} & \epsilon_{abc}\epsilon_{ij}\left(e_{\mathbb{C} p}u_{\mathbb{C}}{}_r^{a}\right)\left(Q^{\dagger}_s{}^{bi}Q^{\dagger}_t{}^{cj}\right) \\
\hline
C_{q^2ue,1}^{(d8)} & C_{e_{\mathbb{C}}u_{\mathbb{C}}Q^{\dagger}{}^2_{[2]} D^2 ,1}^{\; prst} & \epsilon_{abc}\epsilon_{ij}\left(e_{\mathbb{C} p}u_{\mathbb{C}}{}_r^{a}\right)\left(\left(D_{\mu}Q^{\dagger}_s{}^{bi}\right)\left(D^{\mu}Q^{\dagger}_t{}^{cj}\right)\right) \\
 & C_{e_{\mathbb{C}}u_{\mathbb{C}}Q^{\dagger}{}^2_{[1,1]} D^2 ,1}^{\; prst} & i\epsilon_{abc}\epsilon_{ij}\left(e_{\mathbb{C} p}\sigma_{\mu \nu}u_{\mathbb{C}}{}_r^{a}\right)\left(\left(D^{\mu}Q^{\dagger}_s{}^{bi}\right)\left(D^{\nu}Q^{\dagger}_t{}^{cj}\right)\right) \\
\hline
\end{array}
\end{align*}
\caption{List of the $Q^{\dagger 2} u_{\mathbb{C}}e_{\mathbb{C}}$-type operators and the corresponding Wilson coefficients in dimension-6 and dimension-8. The leftmost column contains the abbreviations for the Wilson coefficients in the case of $N_f=1$.}
\label{tab:Q2ue}
\end{table}

\begin{align}
\begin{split}\dot{C}_{q^2ue,1}^{(d8)}&\supset -\frac{187}{54} g_1^2 C_{q^2ue,1}^{(d8)}-\frac{9}{2} g_2^2 C_{q^2ue,1}^{(d8)}+\frac{4}{9} g_3^2 C_{q^2ue,1}^{(d8)}-\frac{4}{3} Y_e Y_u C_{q^3l,1}^{(d8)*}-\frac{1}{6} Y_e C_{qudl,1}^{(d8)} Y_d{}^*\\&+\frac{5}{6} Y_e C_{qudl,2}^{(d8)} Y_d{}^*+Y_d C_{q^2ue,1}^{(d8)} Y_d{}^*+Y_e C_{q^2ue,1}^{(d8)} Y_e{}^*+\frac{5}{3} Y_u C_{q^2ue,1}^{(d8)} Y_u{}^*-\frac{4}{3} C_{u^2de,1}^{(d8)} Y_d{}^* Y_u{}^*\end{split}
\end{align}
The following are the quadratic contributions.
\begin{align}
\begin{split}\dot{C}_{q^2ue,1}^{(d8)}&\supset -\frac{2}{3} C_{ledq,1}^{(d6)} C_{qudl,1}^{(d6)}+4 C_{eu,1}^{(d6)} C_{q^2ue,1}^{(d6)}+8 C_{q,1}^{(d6)} C_{q^2ue,1}^{(d6)}-8 C_{q,2}^{(d6)} C_{q^2ue,1}^{(d6)}+\frac{4}{3} C_{qe,1}^{(d6)} C_{q^2ue,1}^{(d6)}\\&-\frac{4}{3} C_{qu,1}^{(d6)} C_{q^2ue,1}^{(d6)}+\frac{4}{3} C_{qu,2}^{(d6)} C_{q^2ue,1}^{(d6)}-3 C_{lequ,1}^{(d6)} C_{q^3l,1}^{(d6)*}-3 C_{lequ,2}^{(d6)} C_{q^3l,1}^{(d6)*}+3 C_{q^2ud,1}^{(d6)*} C_{u^2de,1}^{(d6)}+3 C_{q^2ud,2}^{(d6)*} C_{u^2de,1}^{(d6)}\end{split}
\end{align}

\subsection{$Q^{\dagger}L^{\dagger} u_{\mathbb{C}}d_{\mathbb{C}}$}
We list below the RGEs for the dimension-8 type $Q^{\dagger}L^{\dagger} u_{\mathbb{C}}d_{\mathbb{C}}$. The corresponding operators and Wilson coefficients are defined in Table \ref{tab:LQud}.
\begin{table}[htbp]
\begin{align*}
\begin{array}{|c|c|c|}
\hline\hline
\text{abbreviation} & \text{Wilson coefficient} & \text{operator}\\
\hline
C_{qudl,1}^{(d6)} & C_{d_{\mathbb{C}}u_{\mathbb{C}}L^{\dagger}Q^{\dagger}  ,1}^{\; prst} & \epsilon_{abc}\epsilon_{ij}\left(d_{\mathbb{C}}{}_p^{a}u_{\mathbb{C}}{}_r^{b}\right)\left(L^{\dagger}_s{}^{i}Q^{\dagger}_t{}^{cj}\right) \\
\hline
C_{qudl,1}^{(d8)} & C_{d_{\mathbb{C}}u_{\mathbb{C}}L^{\dagger}Q^{\dagger} D^2 ,1}^{\; prst} & \epsilon_{abc}\epsilon_{ij}\left(d_{\mathbb{C}}{}_p^{a}u_{\mathbb{C}}{}_r^{b}\right)\left(\left(D_{\mu}L^{\dagger}_s{}^{i}\right)\left(D^{\mu}Q^{\dagger}_t{}^{cj}\right)\right) \\
C_{qudl,2}^{(d8)} & C_{d_{\mathbb{C}}u_{\mathbb{C}}L^{\dagger}Q^{\dagger} D^2 ,1}^{\; prst} & i\epsilon_{abc}\epsilon_{ij}\left(d_{\mathbb{C}}{}_p^{a}\sigma_{\mu \nu}u_{\mathbb{C}}{}_r^{b}\right)\left(\left(D^{\mu}L^{\dagger}_s{}^{i}\right)\left(D^{\nu}Q^{\dagger}_t{}^{cj}\right)\right) \\
\hline
\end{array}
\end{align*}
\caption{List of the $Q^{\dagger}L^{\dagger} u_{\mathbb{C}}d_{\mathbb{C}}$-type operators and the corresponding Wilson coefficients in dimension-6 and dimension-8. The leftmost column contains the abbreviations for the Wilson coefficients in the case of $N_f=1$.}
\label{tab:LQud}
\end{table}

\begin{align}
\begin{split}\dot{C}_{qudl,1}^{(d8)}&\supset -\frac{79}{54} g_1^2 C_{qudl,1}^{(d8)}-\frac{9}{2} g_2^2 C_{qudl,1}^{(d8)}+\frac{4}{9} g_3^2 C_{qudl,1}^{(d8)}-\frac{20}{9} g_1^2 C_{qudl,2}^{(d8)}-\frac{8}{3} Y_d Y_u C_{q^3l,1}^{(d8)*}\\&+\frac{4}{3} Y_d C_{qudl,1}^{(d8)} Y_d{}^*-\frac{5}{6} Y_d C_{qudl,2}^{(d8)} Y_d{}^*+\frac{1}{2} Y_e C_{qudl,1}^{(d8)} Y_e{}^*-\frac{1}{3} Y_d C_{q^2ue,1}^{(d8)} Y_e{}^*+\frac{4}{3} Y_u C_{qudl,1}^{(d8)} Y_u{}^*+\frac{5}{6} Y_u C_{qudl,2}^{(d8)} Y_u{}^*\\&-\frac{4}{3} C_{u^2de,1}^{(d8)} Y_e{}^* Y_u{}^*\end{split}\\

\begin{split}\dot{C}_{qudl,2}^{(d8)}&\supset -\frac{4}{3} g_1^2 C_{qudl,1}^{(d8)}+\frac{5}{6} g_1^2 C_{qudl,2}^{(d8)}+\frac{3}{2} g_2^2 C_{qudl,2}^{(d8)}+4 g_3^2 C_{qudl,2}^{(d8)}-\frac{1}{2} Y_d C_{qudl,1}^{(d8)} Y_d{}^*\\&+Y_d C_{qudl,2}^{(d8)} Y_d{}^*+\frac{1}{2} Y_e C_{qudl,2}^{(d8)} Y_e{}^*+Y_d C_{q^2ue,1}^{(d8)} Y_e{}^*+\frac{1}{2} Y_u C_{qudl,1}^{(d8)} Y_u{}^*+Y_u C_{qudl,2}^{(d8)} Y_u{}^*\end{split}
\end{align}
The following are the quadratic contributions.
\begin{align}
\begin{split}\dot{C}_{qudl,1}^{(d8)}&\supset \frac{2}{3} C_{ld,1}^{(d6)} C_{qudl,1}^{(d6)}+4 C_{lq,1}^{(d6)} C_{qudl,1}^{(d6)}-4 C_{lq,2}^{(d6)} C_{qudl,1}^{(d6)}+\frac{2}{3} C_{lu,1}^{(d6)} C_{qudl,1}^{(d6)}+\frac{2}{3} C_{qd,1}^{(d6)} C_{qudl,1}^{(d6)}-\frac{2}{3} C_{qd,2}^{(d6)} C_{qudl,1}^{(d6)}\\&-\frac{2}{3} C_{qu,1}^{(d6)} C_{qudl,1}^{(d6)}+\frac{2}{3} C_{qu,2}^{(d6)} C_{qudl,1}^{(d6)}-\frac{4}{3} C_{ledq,1}^{(d6)*} C_{q^2ue,1}^{(d6)}+6 C_{q^2ud,1}^{(d6)} C_{q^3l,1}^{(d6)*}+6 C_{q^2ud,2}^{(d6)} C_{q^3l,1}^{(d6)*}-3 C_{lequ,1}^{(d6)*} C_{u^2de,1}^{(d6)}\\&-3 C_{lequ,2}^{(d6)*} C_{u^2de,1}^{(d6)}+4 C_{qudl,1}^{(d6)} C_{ud,1}^{(d6)}-4 C_{qudl,1}^{(d6)} C_{ud,2}^{(d6)}\end{split}\\

\begin{split}\dot{C}_{qudl,2}^{(d8)}&\supset -\frac{2}{3} C_{ld,1}^{(d6)} C_{qudl,1}^{(d6)}+\frac{2}{3} C_{lu,1}^{(d6)} C_{qudl,1}^{(d6)}+\frac{2}{3} C_{qd,1}^{(d6)} C_{qudl,1}^{(d6)}-\frac{2}{3} C_{qd,2}^{(d6)} C_{qudl,1}^{(d6)}+\frac{2}{3} C_{qu,1}^{(d6)} C_{qudl,1}^{(d6)}\\&-\frac{2}{3} C_{qu,2}^{(d6)} C_{qudl,1}^{(d6)}+\frac{4}{3} C_{ledq,1}^{(d6)*} C_{q^2ue,1}^{(d6)}+\frac{1}{3} C_{lequ,1}^{(d6)*} C_{u^2de,1}^{(d6)}-\frac{1}{3} C_{lequ,2}^{(d6)*} C_{u^2de,1}^{(d6)}\end{split}
\end{align}

\subsection{$u_{\mathbb{C}}^2 d_{\mathbb{C}} e_{\mathbb{C}}$}
We list below the RGEs for the dimension-8 type $u_{\mathbb{C}}^2 d_{\mathbb{C}} e_{\mathbb{C}}$. The corresponding operators and Wilson coefficients are defined in Table \ref{tab:u2de}.
\begin{table}[htbp]
\begin{align*}
\begin{array}{|c|c|c|}
\hline\hline
\text{abbreviation} & \text{Wilson coefficient} & \text{operator}\\
\hline
C_{u^2de,1}^{(d6)} & C_{d_{\mathbb{C}}e_{\mathbb{C}}u_{\mathbb{C}}{}^2_{[2]}  ,1}^{\; prst} & \epsilon_{abc}(d_{\mathbb{C}}{}_{p}^a u_{\mathbb{C}}{}_{s}^b)(e_{\mathbb{C}}{}_{r}u_{\mathbb{C}}{}_{t}^c) \\
 & C_{d_{\mathbb{C}}e_{\mathbb{C}}u_{\mathbb{C}}{}^2_{[1,1]}  ,1}^{\; prst} & \epsilon_{abc}(d_{\mathbb{C}}{}_{p}^a e_{\mathbb{C}}{}_{r})(u_{\mathbb{C}}{}_{s}^b u_{\mathbb{C}}{}_{t}^c) \\
\hline
C_{u^2de,1}^{(d8)} & C_{d_{\mathbb{C}}e_{\mathbb{C}}u_{\mathbb{C}}{}^2_{[2]} D^2  ,1}^{\; prst} & \epsilon_{abc}(d_{\mathbb{C}}{}_{p}^a u_{\mathbb{C}}{}_{s}^b)(D^{\mu}e_{\mathbb{C}}{}_{r}D_{\mu}u_{\mathbb{C}}{}_{t}^c) \\
 & C_{d_{\mathbb{C}}e_{\mathbb{C}}u_{\mathbb{C}}{}^2_{[1,1]}D^2  ,1}^{\; prst} & \epsilon_{abc}(d_{\mathbb{C}}{}_{p}^a e_{\mathbb{C}}{}_{r})(D^{\mu}u_{\mathbb{C}}{}_{s}^b D_{\mu}u_{\mathbb{C}}{}_{t}^c) \\
  & C_{d_{\mathbb{C}}e_{\mathbb{C}}u_{\mathbb{C}}{}^2_{[1,1]} D^2  ,2}^{\; prst} & \epsilon_{abc}(d_{\mathbb{C}}{}_{p}^a u_{\mathbb{C}}{}_{s}^b)(D^{\mu}e_{\mathbb{C}}{}_{r}D_{\mu}u_{\mathbb{C}}{}_{t}^c) \\
\hline
\end{array}
\end{align*}
\caption{List of the $u_{\mathbb{C}}^2d_{\mathbb{C}}e_{\mathbb{C}}$-type operators and the corresponding Wilson coefficients in dimension-6 and dimension-8. The leftmost column contains the abbreviations for the Wilson coefficients in the case of $N_f=1$.}
\label{tab:u2de}
\end{table}

\begin{align}
\begin{split}\dot{C}_{u^2de,1}^{(d8)}&\supset -4 Y_e Y_u C_{qudl,1}^{(d8)}-8 Y_d Y_u C_{q^2ue,1}^{(d8)}+\frac{2}{9} g_1^2 C_{u^2de,1}^{(d8)}+\frac{8}{3} g_3^2 C_{u^2de,1}^{(d8)}+Y_d C_{u^2de,1}^{(d8)} Y_d{}^*\\&+Y_e C_{u^2de,1}^{(d8)} Y_e{}^*+2 Y_u C_{u^2de,1}^{(d8)} Y_u{}^*\end{split}
\end{align}
The following are the quadratic contributions.
\begin{align}
\begin{split}\dot{C}_{u^2de,1}^{(d8)}&\supset -4 C_{lequ,1}^{(d6)} C_{qudl,1}^{(d6)}-4 C_{lequ,2}^{(d6)} C_{qudl,1}^{(d6)}+8 C_{q^2ud,1}^{(d6)} C_{q^2ue,1}^{(d6)}+8 C_{q^2ud,2}^{(d6)} C_{q^2ue,1}^{(d6)}+6 C_{eu,1}^{(d6)} C_{u^2de,1}^{(d6)}\\&+6 C_{u^2de,1}^{(d6)} C_{ud,1}^{(d6)}-6 C_{u^2de,1}^{(d6)} C_{ud,2}^{(d6)}\end{split}
\end{align}

\subsection{$LL^{\dagger}QQ^{\dagger}$}
We list below the RGEs for the dimension-8 type $LL^{\dagger}QQ^{\dagger}$. The corresponding operators and Wilson coefficients are defined in Table \ref{tab:LLQQ}.
\begin{table}[htbp]
\begin{align*}
\begin{array}{|c|c|c|}
\hline\hline
\text{abbreviation} & \text{Wilson coefficient} & \text{operator}\\
\hline
C_{lq,1}^{(d6)} & C_{LQL^{\dagger}Q^{\dagger} ,1}^{\; prst} & (L_{pi}Q_{raj})(L^{\dagger}{}_s^i Q^{\dagger}{}_t^{aj}) \\
C_{lq,2}^{(d6)} & C_{LQL^{\dagger}Q^{\dagger} ,2}^{\; prst} & (L_{pi}Q_{raj})(L^{\dagger}{}_s^j Q^{\dagger}{}_t^{ai}) \\
\hline
C_{lq,1}^{(d8)} & C_{LQL^{\dagger}Q^{\dagger} D^2 ,1}^{\; prst} & (L_{pi}Q_{raj})(D^{\mu}L^{\dagger}{}_s^i D_{\mu}Q^{\dagger}{}_t^{aj}) \\
C_{lq,2}^{(d8)} & C_{LQL^{\dagger}Q^{\dagger} D^2 ,2}^{\; prst} & i(L_{pi}\sigma_{\mu\nu}Q_{raj})(D^{\mu}L^{\dagger}{}_s^i D^{\nu}Q^{\dagger}{}_t^{aj}) \\
C_{lq,3}^{(d8)} & C_{LQL^{\dagger}Q^{\dagger} D^2 ,3}^{\; prst} & (L_{pi}Q_{raj})(D^{\mu}L^{\dagger}{}_s^j D_{\mu}Q^{\dagger}{}_t^{ai}) \\
C_{lq,4}^{(d8)} & C_{LQL^{\dagger}Q^{\dagger} D^2 ,4}^{\; prst} & i(L_{pi}\sigma_{\mu\nu}Q_{raj})(D^{\mu}L^{\dagger}{}_s^j D^{\nu}Q^{\dagger}{}_t^{ai}) \\
\hline
\end{array}
\end{align*}
\caption{List of the $LL^{\dagger}QQ^{\dagger}$-type operators and the corresponding Wilson coefficients in dimension-6 and dimension-8. The leftmost column contains the abbreviations for the Wilson coefficients in the case of $N_f=1$.}
\label{tab:LLQQ}
\end{table}

\begin{align}
\begin{split}\dot{C}_{lq,1}^{(d8)}&\supset -\frac{5}{108} g_1^2 C_{Bl,1}^{(d8)}-\frac{5}{12} g_1^2 C_{Bq,1}^{(d8)}+\frac{5}{36} g_1 g_2 C_{BWl,1}^{(d8)}+\frac{5}{36} g_1 g_2 C_{BWl,1}^{(d8)*}-\frac{5}{12} g_1 g_2 C_{BWq,1}^{(d8)}\\&-\frac{5}{12} g_1 g_2 C_{BWq,1}^{(d8)*}-\frac{20}{9} g_3^2 C_{Gl,1}^{(d8)}+\frac{1}{72} g_1^2 C_{Hl,1}^{(d8)}-\frac{1}{24} g_2^2 C_{Hl,1}^{(d8)}+\frac{1}{72} g_1^2 C_{Hl,2}^{(d8)}-\frac{1}{24} g_2^2 C_{Hl,2}^{(d8)}\\&+\frac{1}{36} g_1^2 C_{Hl,3}^{(d8)}+\frac{1}{36} g_1^2 C_{Hl,4}^{(d8)}-\frac{1}{24} g_1^2 C_{Hq,1}^{(d8)}-\frac{1}{24} g_2^2 C_{Hq,1}^{(d8)}-\frac{1}{24} g_1^2 C_{Hq,2}^{(d8)}-\frac{1}{24} g_2^2 C_{Hq,2}^{(d8)}\\&-\frac{1}{12} g_1^2 C_{Hq,3}^{(d8)}-\frac{1}{12} g_1^2 C_{Hq,4}^{(d8)}-\frac{1}{4} g_1^2 C_{l,1}^{(d8)}-\frac{1}{4} g_2^2 C_{l,1}^{(d8)}+\frac{5}{36} g_1^2 C_{l,2}^{(d8)}-\frac{5}{12} g_2^2 C_{l,2}^{(d8)}\\&+\frac{1}{12} g_1^2 C_{ld,1}^{(d8)}-\frac{5}{36} g_1^2 C_{ld,2}^{(d8)}+\frac{1}{12} g_1^2 C_{le,1}^{(d8)}-\frac{5}{36} g_1^2 C_{le,2}^{(d8)}-\frac{2}{3} Y_e Y_u C_{lequ,1}^{(d8)*}-\frac{2}{3} Y_e Y_u C_{lequ,2}^{(d8)*}\\&-\frac{4}{3} Y_e Y_u C_{lequ,3}^{(d8)*}-\frac{8}{27} g_1^2 C_{lq,1}^{(d8)}+\frac{1}{3} g_2^2 C_{lq,1}^{(d8)}+\frac{40}{9} g_3^2 C_{lq,1}^{(d8)}+\frac{25}{27} g_1^2 C_{lq,2}^{(d8)}+\frac{20}{3} g_2^2 C_{lq,2}^{(d8)}\\&+\frac{40}{9} g_3^2 C_{lq,2}^{(d8)}+\frac{1}{6} g_1^2 C_{lq,3}^{(d8)}+\frac{53}{6} g_2^2 C_{lq,3}^{(d8)}-\frac{5}{18} g_1^2 C_{lq,4}^{(d8)}+\frac{25}{6} g_2^2 C_{lq,4}^{(d8)}-\frac{1}{6} g_1^2 C_{lu,1}^{(d8)}\\&+\frac{5}{18} g_1^2 C_{lu,2}^{(d8)}-\frac{7}{12} g_1^2 C_{q,1}^{(d8)}-\frac{1}{4} g_2^2 C_{q,1}^{(d8)}+\frac{25}{36} g_1^2 C_{q,2}^{(d8)}-\frac{5}{12} g_2^2 C_{q,2}^{(d8)}-\frac{5}{12} g_1^2 C_{q,3}^{(d8)}\\&-\frac{3}{4} g_2^2 C_{q,3}^{(d8)}+\frac{5}{36} g_1^2 C_{q,4}^{(d8)}+\frac{5}{4} g_2^2 C_{q,4}^{(d8)}-\frac{1}{4} g_1^2 C_{qd,1}^{(d8)}+\frac{5}{12} g_1^2 C_{qd,2}^{(d8)}-\frac{1}{12} g_1^2 C_{qd,3}^{(d8)}\\&+\frac{5}{36} g_1^2 C_{qd,4}^{(d8)}-\frac{1}{4} g_1^2 C_{qe,1}^{(d8)}+\frac{5}{12} g_1^2 C_{qe,2}^{(d8)}+\frac{1}{6} g_1^2 C_{qu,1}^{(d8)}-\frac{5}{18} g_1^2 C_{qu,2}^{(d8)}+\frac{1}{2} g_1^2 C_{qu,3}^{(d8)}\\&-\frac{5}{6} g_1^2 C_{qu,4}^{(d8)}-\frac{5}{12} i g_2^2 C_{Wl,1}^{(d8)}-\frac{5}{4} g_2^2 C_{Wl,2}^{(d8)}-\frac{5}{12} i g_2^2 C_{Wq,1}^{(d8)}-\frac{5}{4} g_2^2 C_{Wq,2}^{(d8)}-\frac{2}{3} Y_d C_{Hl,3}^{(d8)} Y_d{}^*\\&+\frac{1}{6} Y_d C_{Hl,4}^{(d8)} Y_d{}^*+\frac{1}{6} Y_d C_{ld,1}^{(d8)} Y_d{}^*-\frac{5}{6} Y_d C_{ld,2}^{(d8)} Y_d{}^*+Y_d C_{lq,1}^{(d8)} Y_d{}^*-\frac{2}{3} Y_e C_{Hq,3}^{(d8)} Y_e{}^*+\frac{1}{6} Y_e C_{Hq,4}^{(d8)} Y_e{}^*\\&+Y_e C_{lq,1}^{(d8)} Y_e{}^*+\frac{1}{6} Y_e C_{qe,1}^{(d8)} Y_e{}^*-\frac{5}{6} Y_e C_{qe,2}^{(d8)} Y_e{}^*-\frac{1}{6} Y_u C_{Hl,1}^{(d8)} Y_u{}^*+\frac{2}{3} Y_u C_{Hl,2}^{(d8)} Y_u{}^*-\frac{1}{6} Y_u C_{Hl,3}^{(d8)} Y_u{}^*\\&+\frac{2}{3} Y_u C_{Hl,4}^{(d8)} Y_u{}^*+Y_u C_{lq,1}^{(d8)} Y_u{}^*+\frac{1}{6} Y_u C_{lu,1}^{(d8)} Y_u{}^*-\frac{5}{6} Y_u C_{lu,2}^{(d8)} Y_u{}^*-\frac{2}{3} C_{lequ,1}^{(d8)} Y_e{}^* Y_u{}^*-\frac{2}{3} C_{lequ,2}^{(d8)} Y_e{}^* Y_u{}^*\\&-\frac{4}{3} C_{lequ,3}^{(d8)} Y_e{}^* Y_u{}^*\end{split}\\

\begin{split}\dot{C}_{lq,2}^{(d8)}&\supset -\frac{1}{36} g_1^2 C_{Bl,1}^{(d8)}-\frac{1}{4} g_1^2 C_{Bq,1}^{(d8)}+\frac{1}{12} g_1 g_2 C_{BWl,1}^{(d8)}+\frac{1}{12} g_1 g_2 C_{BWl,1}^{(d8)*}-\frac{1}{4} g_1 g_2 C_{BWq,1}^{(d8)}\\&-\frac{1}{4} g_1 g_2 C_{BWq,1}^{(d8)*}-\frac{4}{3} g_3^2 C_{Gl,1}^{(d8)}-\frac{1}{72} g_1^2 C_{Hl,1}^{(d8)}+\frac{1}{24} g_2^2 C_{Hl,1}^{(d8)}-\frac{1}{72} g_1^2 C_{Hl,2}^{(d8)}+\frac{1}{24} g_2^2 C_{Hl,2}^{(d8)}\\&-\frac{1}{36} g_1^2 C_{Hl,3}^{(d8)}-\frac{1}{36} g_1^2 C_{Hl,4}^{(d8)}+\frac{1}{24} g_1^2 C_{Hq,1}^{(d8)}+\frac{1}{24} g_2^2 C_{Hq,1}^{(d8)}+\frac{1}{24} g_1^2 C_{Hq,2}^{(d8)}+\frac{1}{24} g_2^2 C_{Hq,2}^{(d8)}\\&+\frac{1}{12} g_1^2 C_{Hq,3}^{(d8)}+\frac{1}{12} g_1^2 C_{Hq,4}^{(d8)}+\frac{1}{4} g_1^2 C_{l,1}^{(d8)}+\frac{1}{4} g_2^2 C_{l,1}^{(d8)}-\frac{5}{36} g_1^2 C_{l,2}^{(d8)}+\frac{5}{12} g_2^2 C_{l,2}^{(d8)}\\&-\frac{1}{12} g_1^2 C_{ld,1}^{(d8)}+\frac{5}{36} g_1^2 C_{ld,2}^{(d8)}-\frac{1}{12} g_1^2 C_{le,1}^{(d8)}+\frac{5}{36} g_1^2 C_{le,2}^{(d8)}+\frac{5}{9} g_1^2 C_{lq,1}^{(d8)}+4 g_2^2 C_{lq,1}^{(d8)}\\&+\frac{8}{3} g_3^2 C_{lq,1}^{(d8)}+\frac{10}{9} g_1^2 C_{lq,2}^{(d8)}+3 g_2^2 C_{lq,2}^{(d8)}+\frac{8}{3} g_3^2 C_{lq,2}^{(d8)}-\frac{1}{6} g_1^2 C_{lq,3}^{(d8)}+\frac{5}{2} g_2^2 C_{lq,3}^{(d8)}\\&+\frac{5}{18} g_1^2 C_{lq,4}^{(d8)}-\frac{5}{6} g_2^2 C_{lq,4}^{(d8)}+\frac{1}{6} g_1^2 C_{lu,1}^{(d8)}-\frac{5}{18} g_1^2 C_{lu,2}^{(d8)}+\frac{7}{12} g_1^2 C_{q,1}^{(d8)}+\frac{1}{4} g_2^2 C_{q,1}^{(d8)}\\&-\frac{25}{36} g_1^2 C_{q,2}^{(d8)}+\frac{5}{12} g_2^2 C_{q,2}^{(d8)}+\frac{5}{12} g_1^2 C_{q,3}^{(d8)}+\frac{3}{4} g_2^2 C_{q,3}^{(d8)}-\frac{5}{36} g_1^2 C_{q,4}^{(d8)}-\frac{5}{4} g_2^2 C_{q,4}^{(d8)}\\&+\frac{1}{4} g_1^2 C_{qd,1}^{(d8)}-\frac{5}{12} g_1^2 C_{qd,2}^{(d8)}+\frac{1}{12} g_1^2 C_{qd,3}^{(d8)}-\frac{5}{36} g_1^2 C_{qd,4}^{(d8)}+\frac{1}{4} g_1^2 C_{qe,1}^{(d8)}-\frac{5}{12} g_1^2 C_{qe,2}^{(d8)}\\&-\frac{1}{6} g_1^2 C_{qu,1}^{(d8)}+\frac{5}{18} g_1^2 C_{qu,2}^{(d8)}-\frac{1}{2} g_1^2 C_{qu,3}^{(d8)}+\frac{5}{6} g_1^2 C_{qu,4}^{(d8)}-\frac{1}{4} i g_2^2 C_{Wl,1}^{(d8)}-\frac{3}{4} g_2^2 C_{Wl,2}^{(d8)}\\&-\frac{1}{4} i g_2^2 C_{Wq,1}^{(d8)}-\frac{3}{4} g_2^2 C_{Wq,2}^{(d8)}+\frac{1}{2} Y_d C_{Hl,4}^{(d8)} Y_d{}^*-\frac{1}{2} Y_d C_{ld,1}^{(d8)} Y_d{}^*+\frac{1}{2} Y_d C_{ld,2}^{(d8)} Y_d{}^*+Y_d C_{lq,2}^{(d8)} Y_d{}^*\\&+\frac{1}{2} Y_e C_{Hq,4}^{(d8)} Y_e{}^*+Y_e C_{lq,2}^{(d8)} Y_e{}^*-\frac{1}{2} Y_e C_{qe,1}^{(d8)} Y_e{}^*+\frac{1}{2} Y_e C_{qe,2}^{(d8)} Y_e{}^*-\frac{1}{2} Y_u C_{Hl,1}^{(d8)} Y_u{}^*-\frac{1}{2} Y_u C_{Hl,3}^{(d8)} Y_u{}^*\\&+Y_u C_{lq,2}^{(d8)} Y_u{}^*-\frac{1}{2} Y_u C_{lu,1}^{(d8)} Y_u{}^*+\frac{1}{2} Y_u C_{lu,2}^{(d8)} Y_u{}^*\end{split}\\

\begin{split}\dot{C}_{lq,3}^{(d8)}&\supset -\frac{5}{18} g_1 g_2 C_{BWl,1}^{(d8)}-\frac{5}{18} g_1 g_2 C_{BWl,1}^{(d8)*}+\frac{5}{6} g_1 g_2 C_{BWq,1}^{(d8)}+\frac{5}{6} g_1 g_2 C_{BWq,1}^{(d8)*}+\frac{1}{12} g_2^2 C_{Hl,1}^{(d8)}\\&+\frac{1}{12} g_2^2 C_{Hl,2}^{(d8)}+\frac{1}{12} g_2^2 C_{Hq,1}^{(d8)}+\frac{1}{12} g_2^2 C_{Hq,2}^{(d8)}+\frac{1}{2} g_2^2 C_{l,1}^{(d8)}+\frac{5}{6} g_2^2 C_{l,2}^{(d8)}+\frac{2}{3} Y_e Y_u C_{lequ,1}^{(d8)*}\\&+\frac{2}{3} Y_e Y_u C_{lequ,2}^{(d8)*}+\frac{4}{3} Y_e Y_u C_{lequ,3}^{(d8)*}+\frac{28}{3} g_2^2 C_{lq,1}^{(d8)}-\frac{10}{3} g_2^2 C_{lq,2}^{(d8)}-\frac{17}{27} g_1^2 C_{lq,3}^{(d8)}+\frac{4}{3} g_2^2 C_{lq,3}^{(d8)}\\&+\frac{40}{9} g_3^2 C_{lq,3}^{(d8)}+\frac{40}{27} g_1^2 C_{lq,4}^{(d8)}-\frac{25}{3} g_2^2 C_{lq,4}^{(d8)}+\frac{40}{9} g_3^2 C_{lq,4}^{(d8)}+\frac{1}{2} g_2^2 C_{q,1}^{(d8)}+\frac{5}{6} g_2^2 C_{q,2}^{(d8)}\\&+\frac{3}{2} g_2^2 C_{q,3}^{(d8)}-\frac{5}{2} g_2^2 C_{q,4}^{(d8)}+\frac{5}{6} i g_2^2 C_{Wl,1}^{(d8)}+\frac{5}{6} i g_2^2 C_{Wq,1}^{(d8)}-\frac{2}{3} Y_d C_{Hl,1}^{(d8)} Y_d{}^*+\frac{1}{6} Y_d C_{Hl,2}^{(d8)} Y_d{}^*\\&+\frac{1}{6} Y_e C_{ledq,1}^{(d8)*} Y_d{}^*-\frac{5}{6} Y_e C_{ledq,2}^{(d8)*} Y_d{}^*+Y_d C_{lq,3}^{(d8)} Y_d{}^*-\frac{2}{3} Y_e C_{Hq,1}^{(d8)} Y_e{}^*+\frac{1}{6} Y_e C_{Hq,2}^{(d8)} Y_e{}^*+\frac{1}{6} Y_d C_{ledq,1}^{(d8)} Y_e{}^*\\&-\frac{5}{6} Y_d C_{ledq,2}^{(d8)} Y_e{}^*+Y_e C_{lq,3}^{(d8)} Y_e{}^*+\frac{1}{6} Y_u C_{Hl,1}^{(d8)} Y_u{}^*-\frac{2}{3} Y_u C_{Hl,2}^{(d8)} Y_u{}^*+Y_u C_{lq,3}^{(d8)} Y_u{}^*+\frac{2}{3} C_{lequ,1}^{(d8)} Y_e{}^* Y_u{}^*\\&+\frac{2}{3} C_{lequ,2}^{(d8)} Y_e{}^* Y_u{}^*+\frac{4}{3} C_{lequ,3}^{(d8)} Y_e{}^* Y_u{}^*\end{split}\\

\begin{split}\dot{C}_{lq,4}^{(d8)}&\supset -\frac{1}{6} g_1 g_2 C_{BWl,1}^{(d8)}-\frac{1}{6} g_1 g_2 C_{BWl,1}^{(d8)*}+\frac{1}{2} g_1 g_2 C_{BWq,1}^{(d8)}+\frac{1}{2} g_1 g_2 C_{BWq,1}^{(d8)*}-\frac{1}{12} g_2^2 C_{Hl,1}^{(d8)}\\&-\frac{1}{12} g_2^2 C_{Hl,2}^{(d8)}-\frac{1}{12} g_2^2 C_{Hq,1}^{(d8)}-\frac{1}{12} g_2^2 C_{Hq,2}^{(d8)}-\frac{1}{2} g_2^2 C_{l,1}^{(d8)}-\frac{5}{6} g_2^2 C_{l,2}^{(d8)}-2 g_2^2 C_{lq,1}^{(d8)}\\&+\frac{8}{9} g_1^2 C_{lq,3}^{(d8)}-5 g_2^2 C_{lq,3}^{(d8)}+\frac{8}{3} g_3^2 C_{lq,3}^{(d8)}+\frac{5}{9} g_1^2 C_{lq,4}^{(d8)}+\frac{14}{3} g_2^2 C_{lq,4}^{(d8)}+\frac{8}{3} g_3^2 C_{lq,4}^{(d8)}\\&-\frac{1}{2} g_2^2 C_{q,1}^{(d8)}-\frac{5}{6} g_2^2 C_{q,2}^{(d8)}-\frac{3}{2} g_2^2 C_{q,3}^{(d8)}+\frac{5}{2} g_2^2 C_{q,4}^{(d8)}+\frac{1}{2} i g_2^2 C_{Wl,1}^{(d8)}+\frac{1}{2} i g_2^2 C_{Wq,1}^{(d8)}\\&+\frac{1}{2} Y_d C_{Hl,2}^{(d8)} Y_d{}^*+\frac{1}{2} Y_e C_{ledq,1}^{(d8)*} Y_d{}^*-\frac{1}{2} Y_e C_{ledq,2}^{(d8)*} Y_d{}^*+Y_d C_{lq,4}^{(d8)} Y_d{}^*+\frac{1}{2} Y_e C_{Hq,2}^{(d8)} Y_e{}^*+\frac{1}{2} Y_d C_{ledq,1}^{(d8)} Y_e{}^*\\&-\frac{1}{2} Y_d C_{ledq,2}^{(d8)} Y_e{}^*+Y_e C_{lq,4}^{(d8)} Y_e{}^*+\frac{1}{2} Y_u C_{Hl,1}^{(d8)} Y_u{}^*+Y_u C_{lq,4}^{(d8)} Y_u{}^*\end{split}
\end{align}
The following are the quadratic contributions.
\begin{align}
\begin{split}\dot{C}_{lq,1}^{(d8)}&\supset \frac{1}{3} C_{Hl,2}^{(d6)} C_{Hq,1}^{(d6)}+\frac{1}{3} C_{Hl,1}^{(d6)} C_{Hq,2}^{(d6)}+\frac{2}{3} C_{Hl,2}^{(d6)} C_{Hq,2}^{(d6)}+C_{lequ,1}^{(d6)} C_{lequ,1}^{(d6)*}+C_{lequ,1}^{(d6)*} C_{lequ,2}^{(d6)}+C_{lequ,1}^{(d6)} C_{lequ,2}^{(d6)*}\\&+C_{lequ,2}^{(d6)} C_{lequ,2}^{(d6)*}+4 C_{l,1}^{(d6)} C_{lq,1}^{(d6)}+\frac{28}{3} \left(C_{lq,1}^{(d6)}\right){}^2+\frac{4}{3} C_{l,1}^{(d6)} C_{lq,2}^{(d6)}+8 \left(C_{lq,2}^{(d6)}\right){}^2+\frac{28}{3} C_{lq,1}^{(d6)} C_{q,1}^{(d6)}\\&+4 C_{lq,2}^{(d6)} C_{q,1}^{(d6)}+\frac{20}{3} C_{lq,1}^{(d6)} C_{q,2}^{(d6)}+\frac{4}{3} C_{lq,2}^{(d6)} C_{q,2}^{(d6)}+2 C_{ld,1}^{(d6)} C_{qd,1}^{(d6)}+\frac{2}{3} C_{ld,1}^{(d6)} C_{qd,2}^{(d6)}+\frac{2}{3} C_{le,1}^{(d6)} C_{qe,1}^{(d6)}\\&+\frac{2}{3} C_{lu,1}^{(d6)} C_{qu,1}^{(d6)}+2 C_{lu,1}^{(d6)} C_{qu,2}^{(d6)}+8 C_{qudl,1}^{(d6)} C_{qudl,1}^{(d6)*}+18 C_{q^3l,1}^{(d6)} C_{q^3l,1}^{(d6)*}\end{split}\\

\begin{split}\dot{C}_{lq,2}^{(d8)}&\supset -\frac{1}{3} C_{Hl,2}^{(d6)} C_{Hq,1}^{(d6)}-\frac{1}{3} C_{Hl,1}^{(d6)} C_{Hq,2}^{(d6)}-\frac{2}{3} C_{Hl,2}^{(d6)} C_{Hq,2}^{(d6)}+\frac{1}{3} C_{lequ,1}^{(d6)} C_{lequ,1}^{(d6)*}-\frac{1}{3} C_{lequ,1}^{(d6)*} C_{lequ,2}^{(d6)}\\&-\frac{1}{3} C_{lequ,1}^{(d6)} C_{lequ,2}^{(d6)*}+\frac{1}{3} C_{lequ,2}^{(d6)} C_{lequ,2}^{(d6)*}-4 C_{l,1}^{(d6)} C_{lq,1}^{(d6)}+\frac{4}{3} \left(C_{lq,1}^{(d6)}\right){}^2-\frac{4}{3} C_{l,1}^{(d6)} C_{lq,2}^{(d6)}-\frac{28}{3} C_{lq,1}^{(d6)} C_{q,1}^{(d6)}\\&-4 C_{lq,2}^{(d6)} C_{q,1}^{(d6)}-\frac{20}{3} C_{lq,1}^{(d6)} C_{q,2}^{(d6)}-\frac{4}{3} C_{lq,2}^{(d6)} C_{q,2}^{(d6)}-2 C_{ld,1}^{(d6)} C_{qd,1}^{(d6)}-\frac{2}{3} C_{ld,1}^{(d6)} C_{qd,2}^{(d6)}-\frac{2}{3} C_{le,1}^{(d6)} C_{qe,1}^{(d6)}\\&-\frac{2}{3} C_{lu,1}^{(d6)} C_{qu,1}^{(d6)}-2 C_{lu,1}^{(d6)} C_{qu,2}^{(d6)}+\frac{2}{3} C_{q^3l,1}^{(d6)} C_{q^3l,1}^{(d6)*}\end{split}\\

\begin{split}\dot{C}_{lq,3}^{(d8)}&\supset \frac{1}{3} C_{Hl,1}^{(d6)} C_{Hq,1}^{(d6)}+\frac{2}{3} C_{ledq,1}^{(d6)} C_{ledq,1}^{(d6)*}-C_{lequ,1}^{(d6)} C_{lequ,1}^{(d6)*}-C_{lequ,1}^{(d6)*} C_{lequ,2}^{(d6)}-C_{lequ,1}^{(d6)} C_{lequ,2}^{(d6)*}-C_{lequ,2}^{(d6)} C_{lequ,2}^{(d6)*}\\&+\frac{4}{3} C_{l,1}^{(d6)} C_{lq,2}^{(d6)}+\frac{56}{3} C_{lq,1}^{(d6)} C_{lq,2}^{(d6)}+\frac{8}{3} \left(C_{lq,2}^{(d6)}\right){}^2+\frac{4}{3} C_{lq,2}^{(d6)} C_{q,1}^{(d6)}+4 C_{lq,2}^{(d6)} C_{q,2}^{(d6)}-8 C_{qudl,1}^{(d6)} C_{qudl,1}^{(d6)*}\\&-18 C_{q^3l,1}^{(d6)} C_{q^3l,1}^{(d6)*}\end{split}\\

\begin{split}\dot{C}_{lq,4}^{(d8)}&\supset -\frac{1}{3} C_{Hl,1}^{(d6)} C_{Hq,1}^{(d6)}+\frac{2}{3} C_{ledq,1}^{(d6)} C_{ledq,1}^{(d6)*}-\frac{1}{3} C_{lequ,1}^{(d6)} C_{lequ,1}^{(d6)*}+\frac{1}{3} C_{lequ,1}^{(d6)*} C_{lequ,2}^{(d6)}+\frac{1}{3} C_{lequ,1}^{(d6)} C_{lequ,2}^{(d6)*}\\&-\frac{1}{3} C_{lequ,2}^{(d6)} C_{lequ,2}^{(d6)*}-\frac{4}{3} C_{l,1}^{(d6)} C_{lq,2}^{(d6)}+\frac{8}{3} C_{lq,1}^{(d6)} C_{lq,2}^{(d6)}+\frac{8}{3} \left(C_{lq,2}^{(d6)}\right){}^2-\frac{4}{3} C_{lq,2}^{(d6)} C_{q,1}^{(d6)}-4 C_{lq,2}^{(d6)} C_{q,2}^{(d6)}\\&+\frac{2}{3} C_{q^3l,1}^{(d6)} C_{q^3l,1}^{(d6)*}\end{split}
\end{align}

\subsection{$QQ^{\dagger}e_{\mathbb{C}}e_{\mathbb{C}}^{\dagger}$}
We list below the RGEs for the dimension-8 type $QQ^{\dagger}e_{\mathbb{C}}e_{\mathbb{C}}^{\dagger}$. The corresponding operators and Wilson coefficients are defined in Table \ref{tab:QQee}.
\begin{table}[htbp]
\begin{align*}
\begin{array}{|c|c|c|}
\hline\hline
\text{abbreviation} & \text{Wilson coefficient} & \text{operator}\\
\hline
C_{qe,1}^{(d6)} & C_{e_{\mathbb{C}}Qe_{\mathbb{C}}^{\dagger}Q^{\dagger} ,1}^{\; prst} & (e_{\mathbb{C}}{}_{p}Q_{rai})(e_{\mathbb{C}}^{\dagger}{}_s Q^{\dagger}{}_t^{ai}) \\
\hline
C_{qe,1}^{(d8)} & C_{e_{\mathbb{C}}Qe_{\mathbb{C}}^{\dagger}Q^{\dagger} D^2 ,1}^{\; prst} & (e_{\mathbb{C}}{}_{p}Q_{rai})(D^{\mu}e_{\mathbb{C}}^{\dagger}{}_s D_{\mu}Q^{\dagger}{}_t^{ai}) \\
C_{qe,2}^{(d8)} & C_{e_{\mathbb{C}}Qe_{\mathbb{C}}^{\dagger}Q^{\dagger} D^2 ,2}^{\; prst} & i(e_{\mathbb{C}}{}_{p}\sigma_{\mu\nu}Q_{rai})(D^{\mu}e_{\mathbb{C}}^{\dagger}{}_s D^{\nu}Q^{\dagger}{}_t^{ai}) \\
\hline
\end{array}
\end{align*}
\caption{List of the $QQ^{\dagger}e_{\mathbb{C}}e_{\mathbb{C}}^{\dagger}$-type operators and the corresponding Wilson coefficients in dimension-6 and dimension-8. The leftmost column contains the abbreviations for the Wilson coefficients in the case of $N_f=1$.}
\label{tab:QQee}
\end{table}

\begin{align}
\begin{split}\dot{C}_{qe,1}^{(d8)}&\supset -\frac{5}{108} g_1^2 C_{Be,1}^{(d8)}-\frac{5}{3} g_1^2 C_{Bq,1}^{(d8)}+\frac{1}{3} g_1^2 C_{e,1}^{(d8)}+\frac{1}{12} g_1^2 C_{ed,1}^{(d8)}-\frac{5}{36} g_1^2 C_{ed,2}^{(d8)}\\&-\frac{1}{6} g_1^2 C_{eu,1}^{(d8)}+\frac{5}{18} g_1^2 C_{eu,2}^{(d8)}-\frac{20}{9} g_3^2 C_{Ge,1}^{(d8)}+\frac{1}{36} g_1^2 C_{He,1}^{(d8)}+\frac{1}{36} g_1^2 C_{He,2}^{(d8)}+\frac{1}{12} g_1^2 C_{Hq,1}^{(d8)}\\&+\frac{1}{12} g_1^2 C_{Hq,2}^{(d8)}+\frac{1}{6} g_1^2 C_{Hq,3}^{(d8)}+\frac{1}{6} g_1^2 C_{Hq,4}^{(d8)}-\frac{1}{12} g_1^2 C_{le,1}^{(d8)}+\frac{5}{36} g_1^2 C_{le,2}^{(d8)}-\frac{2}{3} Y_e Y_u C_{lequ,1}^{(d8)*}\\&-2 Y_e Y_u C_{lequ,2}^{(d8)*}+\frac{4}{3} Y_e Y_u C_{lequ,3}^{(d8)*}-\frac{1}{2} g_1^2 C_{lq,1}^{(d8)}+\frac{5}{6} g_1^2 C_{lq,2}^{(d8)}-\frac{1}{4} g_1^2 C_{lq,3}^{(d8)}+\frac{5}{12} g_1^2 C_{lq,4}^{(d8)}\\&+\frac{7}{6} g_1^2 C_{q,1}^{(d8)}-\frac{25}{18} g_1^2 C_{q,2}^{(d8)}+\frac{5}{6} g_1^2 C_{q,3}^{(d8)}-\frac{5}{18} g_1^2 C_{q,4}^{(d8)}+\frac{1}{2} g_1^2 C_{qd,1}^{(d8)}-\frac{5}{6} g_1^2 C_{qd,2}^{(d8)}\\&+\frac{1}{6} g_1^2 C_{qd,3}^{(d8)}-\frac{5}{18} g_1^2 C_{qd,4}^{(d8)}+\frac{769}{108} g_1^2 C_{qe,1}^{(d8)}+\frac{5}{2} g_2^2 C_{qe,1}^{(d8)}+\frac{40}{9} g_3^2 C_{qe,1}^{(d8)}+\frac{145}{108} g_1^2 C_{qe,2}^{(d8)}\\&+\frac{5}{2} g_2^2 C_{qe,2}^{(d8)}+\frac{40}{9} g_3^2 C_{qe,2}^{(d8)}-\frac{1}{3} g_1^2 C_{qu,1}^{(d8)}+\frac{5}{9} g_1^2 C_{qu,2}^{(d8)}-g_1^2 C_{qu,3}^{(d8)}+\frac{5}{3} g_1^2 C_{qu,4}^{(d8)}\\&-\frac{5}{4} g_2^2 C_{We,1}^{(d8)}+\frac{1}{6} Y_d C_{ed,1}^{(d8)} Y_d{}^*-\frac{5}{6} Y_d C_{ed,2}^{(d8)} Y_d{}^*-\frac{2}{3} Y_d C_{He,1}^{(d8)} Y_d{}^*+\frac{1}{6} Y_d C_{He,2}^{(d8)} Y_d{}^*+\frac{1}{6} Y_e C_{ledq,1}^{(d8)*} Y_d{}^*\\&+\frac{5}{6} Y_e C_{ledq,2}^{(d8)*} Y_d{}^*+Y_d C_{qe,1}^{(d8)} Y_d{}^*-\frac{2}{3} Y_e C_{Hq,1}^{(d8)} Y_e{}^*+\frac{1}{6} Y_e C_{Hq,2}^{(d8)} Y_e{}^*-\frac{4}{3} Y_e C_{Hq,3}^{(d8)} Y_e{}^*+\frac{1}{3} Y_e C_{Hq,4}^{(d8)} Y_e{}^*\\&+\frac{1}{6} Y_d C_{ledq,1}^{(d8)} Y_e{}^*+\frac{5}{6} Y_d C_{ledq,2}^{(d8)} Y_e{}^*+\frac{1}{3} Y_e C_{lq,1}^{(d8)} Y_e{}^*-\frac{5}{3} Y_e C_{lq,2}^{(d8)} Y_e{}^*+\frac{1}{6} Y_e C_{lq,3}^{(d8)} Y_e{}^*-\frac{5}{6} Y_e C_{lq,4}^{(d8)} Y_e{}^*\\&+2 Y_e C_{qe,1}^{(d8)} Y_e{}^*+\frac{1}{6} Y_u C_{eu,1}^{(d8)} Y_u{}^*-\frac{5}{6} Y_u C_{eu,2}^{(d8)} Y_u{}^*-\frac{1}{6} Y_u C_{He,1}^{(d8)} Y_u{}^*+\frac{2}{3} Y_u C_{He,2}^{(d8)} Y_u{}^*+Y_u C_{qe,1}^{(d8)} Y_u{}^*\\&-\frac{2}{3} C_{lequ,1}^{(d8)} Y_e{}^* Y_u{}^*-2 C_{lequ,2}^{(d8)} Y_e{}^* Y_u{}^*+\frac{4}{3} C_{lequ,3}^{(d8)} Y_e{}^* Y_u{}^*\end{split}\\

\begin{split}\dot{C}_{qe,2}^{(d8)}&\supset -\frac{1}{36} g_1^2 C_{Be,1}^{(d8)}-g_1^2 C_{Bq,1}^{(d8)}-\frac{1}{3} g_1^2 C_{e,1}^{(d8)}-\frac{1}{12} g_1^2 C_{ed,1}^{(d8)}+\frac{5}{36} g_1^2 C_{ed,2}^{(d8)}\\&+\frac{1}{6} g_1^2 C_{eu,1}^{(d8)}-\frac{5}{18} g_1^2 C_{eu,2}^{(d8)}-\frac{4}{3} g_3^2 C_{Ge,1}^{(d8)}-\frac{1}{36} g_1^2 C_{He,1}^{(d8)}-\frac{1}{36} g_1^2 C_{He,2}^{(d8)}-\frac{1}{12} g_1^2 C_{Hq,1}^{(d8)}\\&-\frac{1}{12} g_1^2 C_{Hq,2}^{(d8)}-\frac{1}{6} g_1^2 C_{Hq,3}^{(d8)}-\frac{1}{6} g_1^2 C_{Hq,4}^{(d8)}+\frac{1}{12} g_1^2 C_{le,1}^{(d8)}-\frac{5}{36} g_1^2 C_{le,2}^{(d8)}+\frac{1}{2} g_1^2 C_{lq,1}^{(d8)}\\&-\frac{5}{6} g_1^2 C_{lq,2}^{(d8)}+\frac{1}{4} g_1^2 C_{lq,3}^{(d8)}-\frac{5}{12} g_1^2 C_{lq,4}^{(d8)}-\frac{7}{6} g_1^2 C_{q,1}^{(d8)}+\frac{25}{18} g_1^2 C_{q,2}^{(d8)}-\frac{5}{6} g_1^2 C_{q,3}^{(d8)}\\&+\frac{5}{18} g_1^2 C_{q,4}^{(d8)}-\frac{1}{2} g_1^2 C_{qd,1}^{(d8)}+\frac{5}{6} g_1^2 C_{qd,2}^{(d8)}-\frac{1}{6} g_1^2 C_{qd,3}^{(d8)}+\frac{5}{18} g_1^2 C_{qd,4}^{(d8)}+\frac{29}{36} g_1^2 C_{qe,1}^{(d8)}\\&+\frac{3}{2} g_2^2 C_{qe,1}^{(d8)}+\frac{8}{3} g_3^2 C_{qe,1}^{(d8)}+\frac{109}{36} g_1^2 C_{qe,2}^{(d8)}+\frac{3}{2} g_2^2 C_{qe,2}^{(d8)}+\frac{8}{3} g_3^2 C_{qe,2}^{(d8)}+\frac{1}{3} g_1^2 C_{qu,1}^{(d8)}\\&-\frac{5}{9} g_1^2 C_{qu,2}^{(d8)}+g_1^2 C_{qu,3}^{(d8)}-\frac{5}{3} g_1^2 C_{qu,4}^{(d8)}-\frac{3}{4} g_2^2 C_{We,1}^{(d8)}-\frac{1}{2} Y_d C_{ed,1}^{(d8)} Y_d{}^*+\frac{1}{2} Y_d C_{ed,2}^{(d8)} Y_d{}^*\\&+\frac{1}{2} Y_d C_{He,2}^{(d8)} Y_d{}^*+\frac{1}{2} Y_e C_{ledq,1}^{(d8)*} Y_d{}^*+\frac{1}{2} Y_e C_{ledq,2}^{(d8)*} Y_d{}^*+Y_d C_{qe,2}^{(d8)} Y_d{}^*+\frac{1}{2} Y_e C_{Hq,2}^{(d8)} Y_e{}^*+Y_e C_{Hq,4}^{(d8)} Y_e{}^*\\&+\frac{1}{2} Y_d C_{ledq,1}^{(d8)} Y_e{}^*+\frac{1}{2} Y_d C_{ledq,2}^{(d8)} Y_e{}^*-Y_e C_{lq,1}^{(d8)} Y_e{}^*+Y_e C_{lq,2}^{(d8)} Y_e{}^*-\frac{1}{2} Y_e C_{lq,3}^{(d8)} Y_e{}^*+\frac{1}{2} Y_e C_{lq,4}^{(d8)} Y_e{}^*\\&+2 Y_e C_{qe,2}^{(d8)} Y_e{}^*-\frac{1}{2} Y_u C_{eu,1}^{(d8)} Y_u{}^*+\frac{1}{2} Y_u C_{eu,2}^{(d8)} Y_u{}^*-\frac{1}{2} Y_u C_{He,1}^{(d8)} Y_u{}^*+Y_u C_{qe,2}^{(d8)} Y_u{}^*\end{split}
\end{align}
The following are the quadratic contributions.
\begin{align}
\begin{split}\dot{C}_{qe,1}^{(d8)}&\supset \frac{1}{3} C_{He,1}^{(d6)} C_{Hq,1}^{(d6)}+\frac{2}{3} C_{He,1}^{(d6)} C_{Hq,2}^{(d6)}+\frac{2}{3} C_{ledq,1}^{(d6)} C_{ledq,1}^{(d6)*}+C_{lequ,1}^{(d6)} C_{lequ,1}^{(d6)*}-2 C_{lequ,1}^{(d6)*} C_{lequ,2}^{(d6)}-2 C_{lequ,1}^{(d6)} C_{lequ,2}^{(d6)*}\\&+4 C_{lequ,2}^{(d6)} C_{lequ,2}^{(d6)*}+\frac{4}{3} C_{le,1}^{(d6)} C_{lq,1}^{(d6)}+\frac{2}{3} C_{le,1}^{(d6)} C_{lq,2}^{(d6)}+2 C_{ed,1}^{(d6)} C_{qd,1}^{(d6)}+\frac{2}{3} C_{ed,1}^{(d6)} C_{qd,2}^{(d6)}+\frac{8}{3} C_{e,1}^{(d6)} C_{qe,1}^{(d6)}\\&+\frac{28}{3} C_{q,1}^{(d6)} C_{qe,1}^{(d6)}+\frac{20}{3} C_{q,2}^{(d6)} C_{qe,1}^{(d6)}+\frac{28}{3} \left(C_{qe,1}^{(d6)}\right){}^2+\frac{2}{3} C_{eu,1}^{(d6)} C_{qu,1}^{(d6)}+2 C_{eu,1}^{(d6)} C_{qu,2}^{(d6)}+\frac{16}{3} C_{q^2ue,1}^{(d6)} C_{q^2ue,1}^{(d6)*}\end{split}\\

\begin{split}\dot{C}_{qe,2}^{(d8)}&\supset -\frac{1}{3} C_{He,1}^{(d6)} C_{Hq,1}^{(d6)}-\frac{2}{3} C_{He,1}^{(d6)} C_{Hq,2}^{(d6)}+\frac{2}{3} C_{ledq,1}^{(d6)} C_{ledq,1}^{(d6)*}+\frac{1}{3} C_{lequ,1}^{(d6)} C_{lequ,1}^{(d6)*}-\frac{4}{3} C_{le,1}^{(d6)} C_{lq,1}^{(d6)}\\&-\frac{2}{3} C_{le,1}^{(d6)} C_{lq,2}^{(d6)}-2 C_{ed,1}^{(d6)} C_{qd,1}^{(d6)}-\frac{2}{3} C_{ed,1}^{(d6)} C_{qd,2}^{(d6)}-\frac{8}{3} C_{e,1}^{(d6)} C_{qe,1}^{(d6)}-\frac{28}{3} C_{q,1}^{(d6)} C_{qe,1}^{(d6)}-\frac{20}{3} C_{q,2}^{(d6)} C_{qe,1}^{(d6)}\\&+\frac{4}{3} \left(C_{qe,1}^{(d6)}\right){}^2-\frac{2}{3} C_{eu,1}^{(d6)} C_{qu,1}^{(d6)}-2 C_{eu,1}^{(d6)} C_{qu,2}^{(d6)}+\frac{16}{3} C_{q^2ue,1}^{(d6)} C_{q^2ue,1}^{(d6)*}\end{split}
\end{align}

\subsection{$LL^{\dagger}u_{\mathbb{C}}u_{\mathbb{C}}^{\dagger}$}
We list below the RGEs for the dimension-8 type $LL^{\dagger}u_{\mathbb{C}}u_{\mathbb{C}}^{\dagger}$. The corresponding operators and Wilson coefficients are defined in Table \ref{tab:uull}.
\begin{table}[htbp]
\begin{align*}
\begin{array}{|c|c|c|}
\hline\hline
\text{abbreviation} & \text{Wilson coefficient} & \text{operator}\\
\hline
C_{lu,1}^{(d6)} & C_{Lu_{\mathbb{C}}L^{\dagger}u_{\mathbb{C}}^{\dagger} ,1}^{\; prst} & (L_{pi}u_{\mathbb{C}}{}_r^a)(L^{\dagger}{}_s^i u_{\mathbb{C}}^{\dagger}{}_{ta}) \\
\hline
C_{lu,1}^{(d8)} & C_{Lu_{\mathbb{C}}L^{\dagger}u_{\mathbb{C}}^{\dagger} D^2 ,1}^{\; prst} & (L_{pi}u_{\mathbb{C}}{}_r^a)(D^{\mu}L^{\dagger}{}_s^i D_{\mu}u_{\mathbb{C}}^{\dagger}{}_{ta}) \\
C_{lu,2}^{(d8)} & C_{Lu_{\mathbb{C}}L^{\dagger}u_{\mathbb{C}}^{\dagger} D^2 ,2}^{\; prst} & i(L_{pi}\sigma_{\mu\nu}u_{\mathbb{C}}{}_r^a)(D^{\mu}L^{\dagger}{}_s^i D^{\nu}u_{\mathbb{C}}^{\dagger}{}_{ta}) \\
\hline
\end{array}
\end{align*}
\caption{List of the $LL^{\dagger}u_{\mathbb{C}}u_{\mathbb{C}}^{\dagger}$-type operators and the corresponding Wilson coefficients in dimension-6 and dimension-8. The leftmost column contains the abbreviations for the Wilson coefficients in the case of $N_f=1$.}
\label{tab:uull}
\end{table}

\begin{align}
\begin{split}\dot{C}_{lu,1}^{(d8)}&\supset -\frac{20}{27} g_1^2 C_{Bl,1}^{(d8)}-\frac{5}{12} g_1^2 C_{Bu,1}^{(d8)}-\frac{1}{4} g_1^2 C_{eu,1}^{(d8)}+\frac{5}{12} g_1^2 C_{eu,2}^{(d8)}-\frac{20}{9} g_3^2 C_{Gl,1}^{(d8)}\\&-\frac{1}{18} g_1^2 C_{Hl,1}^{(d8)}-\frac{1}{18} g_1^2 C_{Hl,2}^{(d8)}-\frac{1}{9} g_1^2 C_{Hl,3}^{(d8)}-\frac{1}{9} g_1^2 C_{Hl,4}^{(d8)}-\frac{1}{12} g_1^2 C_{Hu,1}^{(d8)}-\frac{1}{12} g_1^2 C_{Hu,2}^{(d8)}\\&+g_1^2 C_{l,1}^{(d8)}-\frac{5}{9} g_1^2 C_{l,2}^{(d8)}-\frac{1}{3} g_1^2 C_{ld,1}^{(d8)}+\frac{5}{9} g_1^2 C_{ld,2}^{(d8)}-\frac{1}{3} g_1^2 C_{le,1}^{(d8)}+\frac{5}{9} g_1^2 C_{le,2}^{(d8)}\\&-\frac{2}{3} Y_e Y_u C_{lequ,1}^{(d8)*}-2 Y_e Y_u C_{lequ,2}^{(d8)*}+\frac{4}{3} Y_e Y_u C_{lequ,3}^{(d8)*}-\frac{1}{3} g_1^2 C_{lq,1}^{(d8)}+\frac{5}{9} g_1^2 C_{lq,2}^{(d8)}-\frac{1}{6} g_1^2 C_{lq,3}^{(d8)}\\&+\frac{5}{18} g_1^2 C_{lq,4}^{(d8)}+\frac{1021}{108} g_1^2 C_{lu,1}^{(d8)}+\frac{5}{2} g_2^2 C_{lu,1}^{(d8)}+\frac{40}{9} g_3^2 C_{lu,1}^{(d8)}-\frac{155}{108} g_1^2 C_{lu,2}^{(d8)}+\frac{5}{2} g_2^2 C_{lu,2}^{(d8)}\\&+\frac{40}{9} g_3^2 C_{lu,2}^{(d8)}-\frac{1}{12} g_1^2 C_{qu,1}^{(d8)}+\frac{5}{36} g_1^2 C_{qu,2}^{(d8)}-\frac{1}{4} g_1^2 C_{qu,3}^{(d8)}+\frac{5}{12} g_1^2 C_{qu,4}^{(d8)}+\frac{4}{3} g_1^2 C_{u,1}^{(d8)}\\&-\frac{10}{9} g_1^2 C_{u,2}^{(d8)}-\frac{1}{4} g_1^2 C_{ud,1}^{(d8)}+\frac{5}{12} g_1^2 C_{ud,2}^{(d8)}-\frac{1}{12} g_1^2 C_{ud,3}^{(d8)}+\frac{5}{36} g_1^2 C_{ud,4}^{(d8)}-\frac{5}{4} g_2^2 C_{Wu,1}^{(d8)}\\&+\frac{1}{6} Y_e C_{eu,1}^{(d8)} Y_e{}^*-\frac{5}{6} Y_e C_{eu,2}^{(d8)} Y_e{}^*-\frac{2}{3} Y_e C_{Hu,1}^{(d8)} Y_e{}^*+\frac{1}{6} Y_e C_{Hu,2}^{(d8)} Y_e{}^*+Y_e C_{lu,1}^{(d8)} Y_e{}^*-\frac{1}{6} Y_u C_{Hl,1}^{(d8)} Y_u{}^*\\&+\frac{2}{3} Y_u C_{Hl,2}^{(d8)} Y_u{}^*-\frac{1}{3} Y_u C_{Hl,3}^{(d8)} Y_u{}^*+\frac{4}{3} Y_u C_{Hl,4}^{(d8)} Y_u{}^*+\frac{1}{3} Y_u C_{lq,1}^{(d8)} Y_u{}^*-\frac{5}{3} Y_u C_{lq,2}^{(d8)} Y_u{}^*+\frac{1}{6} Y_u C_{lq,3}^{(d8)} Y_u{}^*\\&-\frac{5}{6} Y_u C_{lq,4}^{(d8)} Y_u{}^*+2 Y_u C_{lu,1}^{(d8)} Y_u{}^*-\frac{2}{3} C_{lequ,1}^{(d8)} Y_e{}^* Y_u{}^*-2 C_{lequ,2}^{(d8)} Y_e{}^* Y_u{}^*+\frac{4}{3} C_{lequ,3}^{(d8)} Y_e{}^* Y_u{}^*\end{split}\\

\begin{split}\dot{C}_{lu,2}^{(d8)}&\supset -\frac{4}{9} g_1^2 C_{Bl,1}^{(d8)}-\frac{1}{4} g_1^2 C_{Bu,1}^{(d8)}+\frac{1}{4} g_1^2 C_{eu,1}^{(d8)}-\frac{5}{12} g_1^2 C_{eu,2}^{(d8)}-\frac{4}{3} g_3^2 C_{Gl,1}^{(d8)}\\&+\frac{1}{18} g_1^2 C_{Hl,1}^{(d8)}+\frac{1}{18} g_1^2 C_{Hl,2}^{(d8)}+\frac{1}{9} g_1^2 C_{Hl,3}^{(d8)}+\frac{1}{9} g_1^2 C_{Hl,4}^{(d8)}+\frac{1}{12} g_1^2 C_{Hu,1}^{(d8)}+\frac{1}{12} g_1^2 C_{Hu,2}^{(d8)}\\&-g_1^2 C_{l,1}^{(d8)}+\frac{5}{9} g_1^2 C_{l,2}^{(d8)}+\frac{1}{3} g_1^2 C_{ld,1}^{(d8)}-\frac{5}{9} g_1^2 C_{ld,2}^{(d8)}+\frac{1}{3} g_1^2 C_{le,1}^{(d8)}-\frac{5}{9} g_1^2 C_{le,2}^{(d8)}\\&+\frac{1}{3} g_1^2 C_{lq,1}^{(d8)}-\frac{5}{9} g_1^2 C_{lq,2}^{(d8)}+\frac{1}{6} g_1^2 C_{lq,3}^{(d8)}-\frac{5}{18} g_1^2 C_{lq,4}^{(d8)}-\frac{31}{36} g_1^2 C_{lu,1}^{(d8)}+\frac{3}{2} g_2^2 C_{lu,1}^{(d8)}\\&+\frac{8}{3} g_3^2 C_{lu,1}^{(d8)}+\frac{35}{12} g_1^2 C_{lu,2}^{(d8)}+\frac{3}{2} g_2^2 C_{lu,2}^{(d8)}+\frac{8}{3} g_3^2 C_{lu,2}^{(d8)}+\frac{1}{12} g_1^2 C_{qu,1}^{(d8)}-\frac{5}{36} g_1^2 C_{qu,2}^{(d8)}\\&+\frac{1}{4} g_1^2 C_{qu,3}^{(d8)}-\frac{5}{12} g_1^2 C_{qu,4}^{(d8)}-\frac{4}{3} g_1^2 C_{u,1}^{(d8)}+\frac{10}{9} g_1^2 C_{u,2}^{(d8)}+\frac{1}{4} g_1^2 C_{ud,1}^{(d8)}-\frac{5}{12} g_1^2 C_{ud,2}^{(d8)}\\&+\frac{1}{12} g_1^2 C_{ud,3}^{(d8)}-\frac{5}{36} g_1^2 C_{ud,4}^{(d8)}-\frac{3}{4} g_2^2 C_{Wu,1}^{(d8)}-\frac{1}{2} Y_e C_{eu,1}^{(d8)} Y_e{}^*+\frac{1}{2} Y_e C_{eu,2}^{(d8)} Y_e{}^*+\frac{1}{2} Y_e C_{Hu,2}^{(d8)} Y_e{}^*\\&+Y_e C_{lu,2}^{(d8)} Y_e{}^*-\frac{1}{2} Y_u C_{Hl,1}^{(d8)} Y_u{}^*-Y_u C_{Hl,3}^{(d8)} Y_u{}^*-Y_u C_{lq,1}^{(d8)} Y_u{}^*+Y_u C_{lq,2}^{(d8)} Y_u{}^*-\frac{1}{2} Y_u C_{lq,3}^{(d8)} Y_u{}^*\\&+\frac{1}{2} Y_u C_{lq,4}^{(d8)} Y_u{}^*+2 Y_u C_{lu,2}^{(d8)} Y_u{}^*\end{split}
\end{align}
The following are the quadratic contributions.
\begin{align}
\begin{split}\dot{C}_{lu,1}^{(d8)}&\supset \frac{1}{3} C_{Hl,1}^{(d6)} C_{Hu,1}^{(d6)}+\frac{2}{3} C_{Hl,2}^{(d6)} C_{Hu,1}^{(d6)}+\frac{2}{3} C_{eu,1}^{(d6)} C_{le,1}^{(d6)}+C_{lequ,1}^{(d6)} C_{lequ,1}^{(d6)*}-2 C_{lequ,1}^{(d6)*} C_{lequ,2}^{(d6)}-2 C_{lequ,1}^{(d6)} C_{lequ,2}^{(d6)*}\\&+4 C_{lequ,2}^{(d6)} C_{lequ,2}^{(d6)*}+4 C_{l,1}^{(d6)} C_{lu,1}^{(d6)}+\frac{28}{3} \left(C_{lu,1}^{(d6)}\right){}^2+\frac{4}{3} C_{lq,1}^{(d6)} C_{qu,1}^{(d6)}+\frac{2}{3} C_{lq,2}^{(d6)} C_{qu,1}^{(d6)}+4 C_{lq,1}^{(d6)} C_{qu,2}^{(d6)}\\&+2 C_{lq,2}^{(d6)} C_{qu,2}^{(d6)}+\frac{4}{3} C_{qudl,1}^{(d6)} C_{qudl,1}^{(d6)*}+\frac{16}{3} C_{lu,1}^{(d6)} C_{u,1}^{(d6)}+2 C_{ld,1}^{(d6)} C_{ud,1}^{(d6)}+\frac{2}{3} C_{ld,1}^{(d6)} C_{ud,2}^{(d6)}\end{split}\\

\begin{split}\dot{C}_{lu,2}^{(d8)}&\supset -\frac{1}{3} C_{Hl,1}^{(d6)} C_{Hu,1}^{(d6)}-\frac{2}{3} C_{Hl,2}^{(d6)} C_{Hu,1}^{(d6)}-\frac{2}{3} C_{eu,1}^{(d6)} C_{le,1}^{(d6)}+\frac{1}{3} C_{lequ,1}^{(d6)} C_{lequ,1}^{(d6)*}-4 C_{l,1}^{(d6)} C_{lu,1}^{(d6)}\\&+\frac{4}{3} \left(C_{lu,1}^{(d6)}\right){}^2-\frac{4}{3} C_{lq,1}^{(d6)} C_{qu,1}^{(d6)}-\frac{2}{3} C_{lq,2}^{(d6)} C_{qu,1}^{(d6)}-4 C_{lq,1}^{(d6)} C_{qu,2}^{(d6)}-2 C_{lq,2}^{(d6)} C_{qu,2}^{(d6)}+\frac{4}{3} C_{qudl,1}^{(d6)} C_{qudl,1}^{(d6)*}\\&-\frac{16}{3} C_{lu,1}^{(d6)} C_{u,1}^{(d6)}-2 C_{ld,1}^{(d6)} C_{ud,1}^{(d6)}-\frac{2}{3} C_{ld,1}^{(d6)} C_{ud,2}^{(d6)}\end{split}
\end{align}

\subsection{$LL^{\dagger}d_{\mathbb{C}}d_{\mathbb{C}}^{\dagger}$}
We list below the RGEs for the dimension-8 type $LL^{\dagger}d_{\mathbb{C}}d_{\mathbb{C}}^{\dagger}$. The corresponding operators and Wilson coefficients are defined in Table \ref{tab:ddll}.
\begin{table}[htbp]
\begin{align*}
\begin{array}{|c|c|c|}
\hline\hline
\text{abbreviation} & \text{Wilson coefficient} & \text{operator}\\
\hline
C_{ld,1}^{(d6)} & C_{d_{\mathbb{C}}Ld_{\mathbb{C}}^{\dagger}L^{\dagger} ,1}^{\; prst} & (d_{\mathbb{C}}{}_p^a L_{ri})(d_{\mathbb{C}}^{\dagger}{}_{sa} L^{\dagger}{}_t^i) \\
\hline
C_{ld,1}^{(d8)} & C_{d_{\mathbb{C}}Ld_{\mathbb{C}}^{\dagger}L^{\dagger} D^2 ,1}^{\; prst} & (d_{\mathbb{C}}{}_p^a L_{ri})(D^{\mu}d_{\mathbb{C}}^{\dagger}{}_{sa} D_{\mu}L^{\dagger}{}_t^i) \\
C_{ld,2}^{(d8)} & C_{d_{\mathbb{C}}Ld_{\mathbb{C}}^{\dagger}L^{\dagger} D^2 ,2}^{\; prst} & i(d_{\mathbb{C}}{}_p^a\sigma_{\mu\nu} L_{ri})(D^{\mu}d_{\mathbb{C}}^{\dagger}{}_{sa} D^{\nu}L^{\dagger}{}_t^i) \\
\hline
\end{array}
\end{align*}
\caption{List of the $LL^{\dagger}d_{\mathbb{C}}d_{\mathbb{C}}^{\dagger}$-type operators and the corresponding Wilson coefficients in dimension-6 and dimension-8. The leftmost column contains the abbreviations for the Wilson coefficients in the case of $N_f=1$.}
\label{tab:ddll}
\end{table}

\begin{align}
\begin{split}\dot{C}_{ld,1}^{(d8)}&\supset -\frac{5}{12} g_1^2 C_{Bd,1}^{(d8)}-\frac{5}{27} g_1^2 C_{Bl,1}^{(d8)}-\frac{2}{3} g_1^2 C_{d,1}^{(d8)}+\frac{5}{9} g_1^2 C_{d,2}^{(d8)}-\frac{1}{4} g_1^2 C_{ed,1}^{(d8)}\\&+\frac{5}{12} g_1^2 C_{ed,2}^{(d8)}-\frac{20}{9} g_3^2 C_{Gl,1}^{(d8)}-\frac{1}{12} g_1^2 C_{Hd,1}^{(d8)}-\frac{1}{12} g_1^2 C_{Hd,2}^{(d8)}+\frac{1}{36} g_1^2 C_{Hl,1}^{(d8)}+\frac{1}{36} g_1^2 C_{Hl,2}^{(d8)}\\&+\frac{1}{18} g_1^2 C_{Hl,3}^{(d8)}+\frac{1}{18} g_1^2 C_{Hl,4}^{(d8)}-\frac{1}{2} g_1^2 C_{l,1}^{(d8)}+\frac{5}{18} g_1^2 C_{l,2}^{(d8)}-\frac{161}{108} g_1^2 C_{ld,1}^{(d8)}+\frac{5}{2} g_2^2 C_{ld,1}^{(d8)}\\&+\frac{40}{9} g_3^2 C_{ld,1}^{(d8)}+\frac{175}{108} g_1^2 C_{ld,2}^{(d8)}+\frac{5}{2} g_2^2 C_{ld,2}^{(d8)}+\frac{40}{9} g_3^2 C_{ld,2}^{(d8)}+\frac{1}{6} g_1^2 C_{le,1}^{(d8)}-\frac{5}{18} g_1^2 C_{le,2}^{(d8)}\\&+\frac{1}{6} g_1^2 C_{lq,1}^{(d8)}-\frac{5}{18} g_1^2 C_{lq,2}^{(d8)}+\frac{1}{12} g_1^2 C_{lq,3}^{(d8)}-\frac{5}{36} g_1^2 C_{lq,4}^{(d8)}-\frac{1}{3} g_1^2 C_{lu,1}^{(d8)}+\frac{5}{9} g_1^2 C_{lu,2}^{(d8)}\\&-\frac{1}{4} g_1^2 C_{qd,1}^{(d8)}+\frac{5}{12} g_1^2 C_{qd,2}^{(d8)}-\frac{1}{12} g_1^2 C_{qd,3}^{(d8)}+\frac{5}{36} g_1^2 C_{qd,4}^{(d8)}+\frac{1}{2} g_1^2 C_{ud,1}^{(d8)}-\frac{5}{6} g_1^2 C_{ud,2}^{(d8)}\\&+\frac{1}{6} g_1^2 C_{ud,3}^{(d8)}-\frac{5}{18} g_1^2 C_{ud,4}^{(d8)}-\frac{5}{4} g_2^2 C_{Wd,1}^{(d8)}-\frac{2}{3} Y_d C_{Hl,1}^{(d8)} Y_d{}^*+\frac{1}{6} Y_d C_{Hl,2}^{(d8)} Y_d{}^*-\frac{4}{3} Y_d C_{Hl,3}^{(d8)} Y_d{}^*\\&+\frac{1}{3} Y_d C_{Hl,4}^{(d8)} Y_d{}^*+2 Y_d C_{ld,1}^{(d8)} Y_d{}^*+\frac{1}{6} Y_e C_{ledq,1}^{(d8)*} Y_d{}^*+\frac{5}{6} Y_e C_{ledq,2}^{(d8)*} Y_d{}^*+\frac{1}{3} Y_d C_{lq,1}^{(d8)} Y_d{}^*-\frac{5}{3} Y_d C_{lq,2}^{(d8)} Y_d{}^*\\&+\frac{1}{6} Y_d C_{lq,3}^{(d8)} Y_d{}^*-\frac{5}{6} Y_d C_{lq,4}^{(d8)} Y_d{}^*+\frac{1}{6} Y_e C_{ed,1}^{(d8)} Y_e{}^*-\frac{5}{6} Y_e C_{ed,2}^{(d8)} Y_e{}^*-\frac{2}{3} Y_e C_{Hd,1}^{(d8)} Y_e{}^*+\frac{1}{6} Y_e C_{Hd,2}^{(d8)} Y_e{}^*\\&+Y_e C_{ld,1}^{(d8)} Y_e{}^*+\frac{1}{6} Y_d C_{ledq,1}^{(d8)} Y_e{}^*+\frac{5}{6} Y_d C_{ledq,2}^{(d8)} Y_e{}^*\end{split}\\

\begin{split}\dot{C}_{ld,2}^{(d8)}&\supset -\frac{1}{4} g_1^2 C_{Bd,1}^{(d8)}-\frac{1}{9} g_1^2 C_{Bl,1}^{(d8)}+\frac{2}{3} g_1^2 C_{d,1}^{(d8)}-\frac{5}{9} g_1^2 C_{d,2}^{(d8)}+\frac{1}{4} g_1^2 C_{ed,1}^{(d8)}\\&-\frac{5}{12} g_1^2 C_{ed,2}^{(d8)}-\frac{4}{3} g_3^2 C_{Gl,1}^{(d8)}+\frac{1}{12} g_1^2 C_{Hd,1}^{(d8)}+\frac{1}{12} g_1^2 C_{Hd,2}^{(d8)}-\frac{1}{36} g_1^2 C_{Hl,1}^{(d8)}-\frac{1}{36} g_1^2 C_{Hl,2}^{(d8)}\\&-\frac{1}{18} g_1^2 C_{Hl,3}^{(d8)}-\frac{1}{18} g_1^2 C_{Hl,4}^{(d8)}+\frac{1}{2} g_1^2 C_{l,1}^{(d8)}-\frac{5}{18} g_1^2 C_{l,2}^{(d8)}+\frac{35}{36} g_1^2 C_{ld,1}^{(d8)}+\frac{3}{2} g_2^2 C_{ld,1}^{(d8)}\\&+\frac{8}{3} g_3^2 C_{ld,1}^{(d8)}+\frac{17}{12} g_1^2 C_{ld,2}^{(d8)}+\frac{3}{2} g_2^2 C_{ld,2}^{(d8)}+\frac{8}{3} g_3^2 C_{ld,2}^{(d8)}-\frac{1}{6} g_1^2 C_{le,1}^{(d8)}+\frac{5}{18} g_1^2 C_{le,2}^{(d8)}\\&-\frac{1}{6} g_1^2 C_{lq,1}^{(d8)}+\frac{5}{18} g_1^2 C_{lq,2}^{(d8)}-\frac{1}{12} g_1^2 C_{lq,3}^{(d8)}+\frac{5}{36} g_1^2 C_{lq,4}^{(d8)}+\frac{1}{3} g_1^2 C_{lu,1}^{(d8)}-\frac{5}{9} g_1^2 C_{lu,2}^{(d8)}\\&+\frac{1}{4} g_1^2 C_{qd,1}^{(d8)}-\frac{5}{12} g_1^2 C_{qd,2}^{(d8)}+\frac{1}{12} g_1^2 C_{qd,3}^{(d8)}-\frac{5}{36} g_1^2 C_{qd,4}^{(d8)}-\frac{1}{2} g_1^2 C_{ud,1}^{(d8)}+\frac{5}{6} g_1^2 C_{ud,2}^{(d8)}\\&-\frac{1}{6} g_1^2 C_{ud,3}^{(d8)}+\frac{5}{18} g_1^2 C_{ud,4}^{(d8)}-\frac{3}{4} g_2^2 C_{Wd,1}^{(d8)}+\frac{1}{2} Y_d C_{Hl,2}^{(d8)} Y_d{}^*+Y_d C_{Hl,4}^{(d8)} Y_d{}^*+2 Y_d C_{ld,2}^{(d8)} Y_d{}^*\\&+\frac{1}{2} Y_e C_{ledq,1}^{(d8)*} Y_d{}^*+\frac{1}{2} Y_e C_{ledq,2}^{(d8)*} Y_d{}^*-Y_d C_{lq,1}^{(d8)} Y_d{}^*+Y_d C_{lq,2}^{(d8)} Y_d{}^*-\frac{1}{2} Y_d C_{lq,3}^{(d8)} Y_d{}^*+\frac{1}{2} Y_d C_{lq,4}^{(d8)} Y_d{}^*\\&-\frac{1}{2} Y_e C_{ed,1}^{(d8)} Y_e{}^*+\frac{1}{2} Y_e C_{ed,2}^{(d8)} Y_e{}^*+\frac{1}{2} Y_e C_{Hd,2}^{(d8)} Y_e{}^*+Y_e C_{ld,2}^{(d8)} Y_e{}^*+\frac{1}{2} Y_d C_{ledq,1}^{(d8)} Y_e{}^*+\frac{1}{2} Y_d C_{ledq,2}^{(d8)} Y_e{}^*\end{split}
\end{align}
The following are the quadratic contributions.
\begin{align}
\begin{split}\dot{C}_{ld,1}^{(d8)}&\supset \frac{1}{3} C_{Hd,1}^{(d6)} C_{Hl,1}^{(d6)}+\frac{2}{3} C_{Hd,1}^{(d6)} C_{Hl,2}^{(d6)}+\frac{16}{3} C_{d,1}^{(d6)} C_{ld,1}^{(d6)}+4 C_{l,1}^{(d6)} C_{ld,1}^{(d6)}+\frac{28}{3} \left(C_{ld,1}^{(d6)}\right){}^2+\frac{2}{3} C_{ed,1}^{(d6)} C_{le,1}^{(d6)}\\&+\frac{2}{3} C_{ledq,1}^{(d6)} C_{ledq,1}^{(d6)*}+4 C_{lq,1}^{(d6)} C_{qd,1}^{(d6)}+2 C_{lq,2}^{(d6)} C_{qd,1}^{(d6)}+\frac{4}{3} C_{lq,1}^{(d6)} C_{qd,2}^{(d6)}+\frac{2}{3} C_{lq,2}^{(d6)} C_{qd,2}^{(d6)}+\frac{4}{3} C_{qudl,1}^{(d6)} C_{qudl,1}^{(d6)*}\\&+2 C_{lu,1}^{(d6)} C_{ud,1}^{(d6)}+\frac{2}{3} C_{lu,1}^{(d6)} C_{ud,2}^{(d6)}\end{split}\\

\begin{split}\dot{C}_{ld,2}^{(d8)}&\supset -\frac{1}{3} C_{Hd,1}^{(d6)} C_{Hl,1}^{(d6)}-\frac{2}{3} C_{Hd,1}^{(d6)} C_{Hl,2}^{(d6)}-\frac{16}{3} C_{d,1}^{(d6)} C_{ld,1}^{(d6)}-4 C_{l,1}^{(d6)} C_{ld,1}^{(d6)}+\frac{4}{3} \left(C_{ld,1}^{(d6)}\right){}^2\\&-\frac{2}{3} C_{ed,1}^{(d6)} C_{le,1}^{(d6)}+\frac{2}{3} C_{ledq,1}^{(d6)} C_{ledq,1}^{(d6)*}-4 C_{lq,1}^{(d6)} C_{qd,1}^{(d6)}-2 C_{lq,2}^{(d6)} C_{qd,1}^{(d6)}-\frac{4}{3} C_{lq,1}^{(d6)} C_{qd,2}^{(d6)}-\frac{2}{3} C_{lq,2}^{(d6)} C_{qd,2}^{(d6)}\\&+\frac{4}{3} C_{qudl,1}^{(d6)} C_{qudl,1}^{(d6)*}-2 C_{lu,1}^{(d6)} C_{ud,1}^{(d6)}-\frac{2}{3} C_{lu,1}^{(d6)} C_{ud,2}^{(d6)}\end{split}
\end{align}

\subsection{$u_{\mathbb{C}}u_{\mathbb{C}}^{\dagger}e_{\mathbb{C}}e_{\mathbb{C}}^{\dagger}$}
We list below the RGEs for the dimension-8 type $u_{\mathbb{C}}u_{\mathbb{C}}^{\dagger}e_{\mathbb{C}}e_{\mathbb{C}}^{\dagger}$. The corresponding operators and Wilson coefficients are defined in Table \ref{tab:uuee}.
\begin{table}[htbp]
\begin{align*}
\begin{array}{|c|c|c|}
\hline\hline
\text{abbreviation} & \text{Wilson coefficient} & \text{operator}\\
\hline
C_{eu,1}^{(d6)} & C_{e_{\mathbb{C}}u_{\mathbb{C}}e_{\mathbb{C}}^{\dagger}u_{\mathbb{C}}^{\dagger} ,1}^{\; prst} & (e_{\mathbb{C}}{}_{p}u_{\mathbb{C}}{}_{r}^a)(e_{\mathbb{C}}^{\dagger}{}_s u_{\mathbb{C}}^{\dagger}{}_{ta}) \\
\hline
C_{eu,1}^{(d8)} & C_{e_{\mathbb{C}}u_{\mathbb{C}}e_{\mathbb{C}}^{\dagger}u_{\mathbb{C}}^{\dagger} D^2 ,1}^{\; prst} & (e_{\mathbb{C}}{}_{p}u_{\mathbb{C}}{}_{r}^a)(D^{\mu}e_{\mathbb{C}}^{\dagger}{}_s D_{\mu}u_{\mathbb{C}}^{\dagger}{}_{ta}) \\
C_{eu,2}^{(d8)} & C_{e_{\mathbb{C}}u_{\mathbb{C}}e_{\mathbb{C}}^{\dagger}u_{\mathbb{C}}^{\dagger} D^2 ,2}^{\; prst} & i(e_{\mathbb{C}}{}_{p}\sigma_{\mu\nu}u_{\mathbb{C}}{}_{r}^a)(D^{\mu}e_{\mathbb{C}}^{\dagger}{}_s D^{\nu}u_{\mathbb{C}}^{\dagger}{}_{ta}) \\
\hline
\end{array}
\end{align*}
\caption{List of the $u_{\mathbb{C}}u_{\mathbb{C}}^{\dagger}e_{\mathbb{C}}e_{\mathbb{C}}^{\dagger}$-type operators and the corresponding Wilson coefficients in dimension-6 and dimension-8. The leftmost column contains the abbreviations for the Wilson coefficients in the case of $N_f=1$.}
\label{tab:uuee}
\end{table}

\begin{align}
\begin{split}\dot{C}_{eu,1}^{(d8)}&\supset -\frac{20}{27} g_1^2 C_{Be,1}^{(d8)}-\frac{5}{3} g_1^2 C_{Bu,1}^{(d8)}-\frac{4}{3} g_1^2 C_{e,1}^{(d8)}-\frac{1}{3} g_1^2 C_{ed,1}^{(d8)}+\frac{5}{9} g_1^2 C_{ed,2}^{(d8)}\\&-\frac{349}{54} g_1^2 C_{eu,1}^{(d8)}+\frac{40}{9} g_3^2 C_{eu,1}^{(d8)}+\frac{395}{54} g_1^2 C_{eu,2}^{(d8)}+\frac{40}{9} g_3^2 C_{eu,2}^{(d8)}-\frac{20}{9} g_3^2 C_{Ge,1}^{(d8)}-\frac{1}{9} g_1^2 C_{He,1}^{(d8)}\\&-\frac{1}{9} g_1^2 C_{He,2}^{(d8)}+\frac{1}{6} g_1^2 C_{Hu,1}^{(d8)}+\frac{1}{6} g_1^2 C_{Hu,2}^{(d8)}+\frac{1}{3} g_1^2 C_{le,1}^{(d8)}-\frac{5}{9} g_1^2 C_{le,2}^{(d8)}-\frac{4}{3} Y_e Y_u C_{lequ,1}^{(d8)*}\\&-\frac{4}{3} Y_e Y_u C_{lequ,2}^{(d8)*}-\frac{8}{3} Y_e Y_u C_{lequ,3}^{(d8)*}-\frac{1}{2} g_1^2 C_{lu,1}^{(d8)}+\frac{5}{6} g_1^2 C_{lu,2}^{(d8)}-\frac{1}{3} g_1^2 C_{qe,1}^{(d8)}+\frac{5}{9} g_1^2 C_{qe,2}^{(d8)}\\&+\frac{1}{6} g_1^2 C_{qu,1}^{(d8)}-\frac{5}{18} g_1^2 C_{qu,2}^{(d8)}+\frac{1}{2} g_1^2 C_{qu,3}^{(d8)}-\frac{5}{6} g_1^2 C_{qu,4}^{(d8)}-\frac{8}{3} g_1^2 C_{u,1}^{(d8)}+\frac{20}{9} g_1^2 C_{u,2}^{(d8)}\\&+\frac{1}{2} g_1^2 C_{ud,1}^{(d8)}-\frac{5}{6} g_1^2 C_{ud,2}^{(d8)}+\frac{1}{6} g_1^2 C_{ud,3}^{(d8)}-\frac{5}{18} g_1^2 C_{ud,4}^{(d8)}+2 Y_e C_{eu,1}^{(d8)} Y_e{}^*-\frac{4}{3} Y_e C_{Hu,1}^{(d8)} Y_e{}^*\\&+\frac{1}{3} Y_e C_{Hu,2}^{(d8)} Y_e{}^*+\frac{1}{3} Y_e C_{lu,1}^{(d8)} Y_e{}^*-\frac{5}{3} Y_e C_{lu,2}^{(d8)} Y_e{}^*+2 Y_u C_{eu,1}^{(d8)} Y_u{}^*-\frac{1}{3} Y_u C_{He,1}^{(d8)} Y_u{}^*+\frac{4}{3} Y_u C_{He,2}^{(d8)} Y_u{}^*\\&+\frac{1}{3} Y_u C_{qe,1}^{(d8)} Y_u{}^*-\frac{5}{3} Y_u C_{qe,2}^{(d8)} Y_u{}^*-\frac{4}{3} C_{lequ,1}^{(d8)} Y_e{}^* Y_u{}^*-\frac{4}{3} C_{lequ,2}^{(d8)} Y_e{}^* Y_u{}^*-\frac{8}{3} C_{lequ,3}^{(d8)} Y_e{}^* Y_u{}^*\end{split}\\

\begin{split}\dot{C}_{eu,2}^{(d8)}&\supset -\frac{4}{9} g_1^2 C_{Be,1}^{(d8)}-g_1^2 C_{Bu,1}^{(d8)}+\frac{4}{3} g_1^2 C_{e,1}^{(d8)}+\frac{1}{3} g_1^2 C_{ed,1}^{(d8)}-\frac{5}{9} g_1^2 C_{ed,2}^{(d8)}\\&+\frac{79}{18} g_1^2 C_{eu,1}^{(d8)}+\frac{8}{3} g_3^2 C_{eu,1}^{(d8)}+\frac{29}{6} g_1^2 C_{eu,2}^{(d8)}+\frac{8}{3} g_3^2 C_{eu,2}^{(d8)}-\frac{4}{3} g_3^2 C_{Ge,1}^{(d8)}+\frac{1}{9} g_1^2 C_{He,1}^{(d8)}\\&+\frac{1}{9} g_1^2 C_{He,2}^{(d8)}-\frac{1}{6} g_1^2 C_{Hu,1}^{(d8)}-\frac{1}{6} g_1^2 C_{Hu,2}^{(d8)}-\frac{1}{3} g_1^2 C_{le,1}^{(d8)}+\frac{5}{9} g_1^2 C_{le,2}^{(d8)}+\frac{1}{2} g_1^2 C_{lu,1}^{(d8)}\\&-\frac{5}{6} g_1^2 C_{lu,2}^{(d8)}+\frac{1}{3} g_1^2 C_{qe,1}^{(d8)}-\frac{5}{9} g_1^2 C_{qe,2}^{(d8)}-\frac{1}{6} g_1^2 C_{qu,1}^{(d8)}+\frac{5}{18} g_1^2 C_{qu,2}^{(d8)}-\frac{1}{2} g_1^2 C_{qu,3}^{(d8)}\\&+\frac{5}{6} g_1^2 C_{qu,4}^{(d8)}+\frac{8}{3} g_1^2 C_{u,1}^{(d8)}-\frac{20}{9} g_1^2 C_{u,2}^{(d8)}-\frac{1}{2} g_1^2 C_{ud,1}^{(d8)}+\frac{5}{6} g_1^2 C_{ud,2}^{(d8)}-\frac{1}{6} g_1^2 C_{ud,3}^{(d8)}\\&+\frac{5}{18} g_1^2 C_{ud,4}^{(d8)}+2 Y_e C_{eu,2}^{(d8)} Y_e{}^*+Y_e C_{Hu,2}^{(d8)} Y_e{}^*-Y_e C_{lu,1}^{(d8)} Y_e{}^*+Y_e C_{lu,2}^{(d8)} Y_e{}^*+2 Y_u C_{eu,2}^{(d8)} Y_u{}^*\\&-Y_u C_{He,1}^{(d8)} Y_u{}^*-Y_u C_{qe,1}^{(d8)} Y_u{}^*+Y_u C_{qe,2}^{(d8)} Y_u{}^*\end{split}
\end{align}
The following are the quadratic contributions.
\begin{align}
\begin{split}\dot{C}_{eu,1}^{(d8)}&\supset \frac{8}{3} C_{e,1}^{(d6)} C_{eu,1}^{(d6)}+\frac{28}{3} \left(C_{eu,1}^{(d6)}\right){}^2+\frac{2}{3} C_{He,1}^{(d6)} C_{Hu,1}^{(d6)}+2 C_{lequ,1}^{(d6)} C_{lequ,1}^{(d6)*}+2 C_{lequ,1}^{(d6)*} C_{lequ,2}^{(d6)}+2 C_{lequ,1}^{(d6)} C_{lequ,2}^{(d6)*}\\&+2 C_{lequ,2}^{(d6)} C_{lequ,2}^{(d6)*}+\frac{4}{3} C_{le,1}^{(d6)} C_{lu,1}^{(d6)}+\frac{4}{3} C_{qe,1}^{(d6)} C_{qu,1}^{(d6)}+4 C_{qe,1}^{(d6)} C_{qu,2}^{(d6)}+32 C_{q^2ue,1}^{(d6)} C_{q^2ue,1}^{(d6)*}+18 C_{u^2de,1}^{(d6)} C_{u^2de,1}^{(d6)*}\\&+\frac{16}{3} C_{eu,1}^{(d6)} C_{u,1}^{(d6)}+2 C_{ed,1}^{(d6)} C_{ud,1}^{(d6)}+\frac{2}{3} C_{ed,1}^{(d6)} C_{ud,2}^{(d6)}\end{split}\\

\begin{split}\dot{C}_{eu,2}^{(d8)}&\supset -\frac{8}{3} C_{e,1}^{(d6)} C_{eu,1}^{(d6)}+\frac{4}{3} \left(C_{eu,1}^{(d6)}\right){}^2-\frac{2}{3} C_{He,1}^{(d6)} C_{Hu,1}^{(d6)}+\frac{2}{3} C_{lequ,1}^{(d6)} C_{lequ,1}^{(d6)*}-\frac{2}{3} C_{lequ,1}^{(d6)*} C_{lequ,2}^{(d6)}\\&-\frac{2}{3} C_{lequ,1}^{(d6)} C_{lequ,2}^{(d6)*}+\frac{2}{3} C_{lequ,2}^{(d6)} C_{lequ,2}^{(d6)*}-\frac{4}{3} C_{le,1}^{(d6)} C_{lu,1}^{(d6)}-\frac{4}{3} C_{qe,1}^{(d6)} C_{qu,1}^{(d6)}-4 C_{qe,1}^{(d6)} C_{qu,2}^{(d6)}+\frac{2}{3} C_{u^2de,1}^{(d6)} C_{u^2de,1}^{(d6)*}\\&-\frac{16}{3} C_{eu,1}^{(d6)} C_{u,1}^{(d6)}-2 C_{ed,1}^{(d6)} C_{ud,1}^{(d6)}-\frac{2}{3} C_{ed,1}^{(d6)} C_{ud,2}^{(d6)}\end{split}
\end{align}

\subsection{$d_{\mathbb{C}}d_{\mathbb{C}}^{\dagger}e_{\mathbb{C}}e_{\mathbb{C}}^{\dagger}$}
We list below the RGEs for the dimension-8 type $d_{\mathbb{C}}d_{\mathbb{C}}^{\dagger}e_{\mathbb{C}}e_{\mathbb{C}}^{\dagger}$. The corresponding operators and Wilson coefficients are defined in Table \ref{tab:ddee}.
\begin{table}[htbp]
\begin{align*}
\begin{array}{|c|c|c|}
\hline\hline
\text{abbreviation} & \text{Wilson coefficient} & \text{operator}\\
\hline
C_{ed,1}^{(d6)} & C_{d_{\mathbb{C}}e_{\mathbb{C}}d_{\mathbb{C}}^{\dagger}e_{\mathbb{C}}^{\dagger} ,1}^{\; prst} & (d_{\mathbb{C}}{}_p^a e_{\mathbb{C}}{}_{r})(d_{\mathbb{C}}^{\dagger}{}_{sa} e_{\mathbb{C}}^{\dagger}{}_t) \\
\hline
C_{ed,1}^{(d8)} & C_{d_{\mathbb{C}}e_{\mathbb{C}}d_{\mathbb{C}}^{\dagger}e_{\mathbb{C}}^{\dagger} D^2 ,1}^{\; prst} & (d_{\mathbb{C}}{}_p^a e_{\mathbb{C}}{}_{r})(D^{\mu}d_{\mathbb{C}}^{\dagger}{}_{sa} D_{\mu}e_{\mathbb{C}}^{\dagger}{}_t) \\
C_{ed,2}^{(d8)} & C_{d_{\mathbb{C}}e_{\mathbb{C}}d_{\mathbb{C}}^{\dagger}e_{\mathbb{C}}^{\dagger} D^2 ,2}^{\; prst} & i(d_{\mathbb{C}}{}_p^a \sigma_{\mu\nu}e_{\mathbb{C}}{}_{r})(D^{\mu}d_{\mathbb{C}}^{\dagger}{}_{sa} D^{\nu}e_{\mathbb{C}}^{\dagger}{}_t) \\
\hline
\end{array}
\end{align*}
\caption{List of the $e_{\mathbb{C}}e_{\mathbb{C}}^{\dagger}d_{\mathbb{C}}d_{\mathbb{C}}^{\dagger}$-type operators and the corresponding Wilson coefficients in dimension-6 and dimension-8. The leftmost column contains the abbreviations for the Wilson coefficients in the case of $N_f=1$.}
\label{tab:ddee}
\end{table}

\begin{align}
\begin{split}\dot{C}_{ed,1}^{(d8)}&\supset -\frac{5}{3} g_1^2 C_{Bd,1}^{(d8)}-\frac{5}{27} g_1^2 C_{Be,1}^{(d8)}+\frac{4}{3} g_1^2 C_{d,1}^{(d8)}-\frac{10}{9} g_1^2 C_{d,2}^{(d8)}+\frac{2}{3} g_1^2 C_{e,1}^{(d8)}\\&+\frac{286}{27} g_1^2 C_{ed,1}^{(d8)}+\frac{40}{9} g_3^2 C_{ed,1}^{(d8)}+\frac{10}{27} g_1^2 C_{ed,2}^{(d8)}+\frac{40}{9} g_3^2 C_{ed,2}^{(d8)}-\frac{1}{3} g_1^2 C_{eu,1}^{(d8)}+\frac{5}{9} g_1^2 C_{eu,2}^{(d8)}\\&-\frac{20}{9} g_3^2 C_{Ge,1}^{(d8)}+\frac{1}{6} g_1^2 C_{Hd,1}^{(d8)}+\frac{1}{6} g_1^2 C_{Hd,2}^{(d8)}+\frac{1}{18} g_1^2 C_{He,1}^{(d8)}+\frac{1}{18} g_1^2 C_{He,2}^{(d8)}-\frac{1}{2} g_1^2 C_{ld,1}^{(d8)}\\&+\frac{5}{6} g_1^2 C_{ld,2}^{(d8)}-\frac{1}{6} g_1^2 C_{le,1}^{(d8)}+\frac{5}{18} g_1^2 C_{le,2}^{(d8)}+\frac{1}{2} g_1^2 C_{qd,1}^{(d8)}-\frac{5}{6} g_1^2 C_{qd,2}^{(d8)}+\frac{1}{6} g_1^2 C_{qd,3}^{(d8)}\\&-\frac{5}{18} g_1^2 C_{qd,4}^{(d8)}+\frac{1}{6} g_1^2 C_{qe,1}^{(d8)}-\frac{5}{18} g_1^2 C_{qe,2}^{(d8)}-g_1^2 C_{ud,1}^{(d8)}+\frac{5}{3} g_1^2 C_{ud,2}^{(d8)}-\frac{1}{3} g_1^2 C_{ud,3}^{(d8)}\\&+\frac{5}{9} g_1^2 C_{ud,4}^{(d8)}+2 Y_d C_{ed,1}^{(d8)} Y_d{}^*-\frac{4}{3} Y_d C_{He,1}^{(d8)} Y_d{}^*+\frac{1}{3} Y_d C_{He,2}^{(d8)} Y_d{}^*+\frac{1}{3} Y_e C_{ledq,1}^{(d8)*} Y_d{}^*-\frac{5}{3} Y_e C_{ledq,2}^{(d8)*} Y_d{}^*\\&+\frac{1}{3} Y_d C_{qe,1}^{(d8)} Y_d{}^*-\frac{5}{3} Y_d C_{qe,2}^{(d8)} Y_d{}^*+2 Y_e C_{ed,1}^{(d8)} Y_e{}^*-\frac{4}{3} Y_e C_{Hd,1}^{(d8)} Y_e{}^*+\frac{1}{3} Y_e C_{Hd,2}^{(d8)} Y_e{}^*+\frac{1}{3} Y_e C_{ld,1}^{(d8)} Y_e{}^*\\&-\frac{5}{3} Y_e C_{ld,2}^{(d8)} Y_e{}^*+\frac{1}{3} Y_d C_{ledq,1}^{(d8)} Y_e{}^*-\frac{5}{3} Y_d C_{ledq,2}^{(d8)} Y_e{}^*\end{split}\\

\begin{split}\dot{C}_{ed,2}^{(d8)}&\supset -g_1^2 C_{Bd,1}^{(d8)}-\frac{1}{9} g_1^2 C_{Be,1}^{(d8)}-\frac{4}{3} g_1^2 C_{d,1}^{(d8)}+\frac{10}{9} g_1^2 C_{d,2}^{(d8)}-\frac{2}{3} g_1^2 C_{e,1}^{(d8)}\\&+\frac{2}{9} g_1^2 C_{ed,1}^{(d8)}+\frac{8}{3} g_3^2 C_{ed,1}^{(d8)}+\frac{10}{3} g_1^2 C_{ed,2}^{(d8)}+\frac{8}{3} g_3^2 C_{ed,2}^{(d8)}+\frac{1}{3} g_1^2 C_{eu,1}^{(d8)}-\frac{5}{9} g_1^2 C_{eu,2}^{(d8)}\\&-\frac{4}{3} g_3^2 C_{Ge,1}^{(d8)}-\frac{1}{6} g_1^2 C_{Hd,1}^{(d8)}-\frac{1}{6} g_1^2 C_{Hd,2}^{(d8)}-\frac{1}{18} g_1^2 C_{He,1}^{(d8)}-\frac{1}{18} g_1^2 C_{He,2}^{(d8)}+\frac{1}{2} g_1^2 C_{ld,1}^{(d8)}\\&-\frac{5}{6} g_1^2 C_{ld,2}^{(d8)}+\frac{1}{6} g_1^2 C_{le,1}^{(d8)}-\frac{5}{18} g_1^2 C_{le,2}^{(d8)}-\frac{1}{2} g_1^2 C_{qd,1}^{(d8)}+\frac{5}{6} g_1^2 C_{qd,2}^{(d8)}-\frac{1}{6} g_1^2 C_{qd,3}^{(d8)}\\&+\frac{5}{18} g_1^2 C_{qd,4}^{(d8)}-\frac{1}{6} g_1^2 C_{qe,1}^{(d8)}+\frac{5}{18} g_1^2 C_{qe,2}^{(d8)}+g_1^2 C_{ud,1}^{(d8)}-\frac{5}{3} g_1^2 C_{ud,2}^{(d8)}+\frac{1}{3} g_1^2 C_{ud,3}^{(d8)}\\&-\frac{5}{9} g_1^2 C_{ud,4}^{(d8)}+2 Y_d C_{ed,2}^{(d8)} Y_d{}^*+Y_d C_{He,2}^{(d8)} Y_d{}^*+Y_e C_{ledq,1}^{(d8)*} Y_d{}^*-Y_e C_{ledq,2}^{(d8)*} Y_d{}^*-Y_d C_{qe,1}^{(d8)} Y_d{}^*\\&+Y_d C_{qe,2}^{(d8)} Y_d{}^*+2 Y_e C_{ed,2}^{(d8)} Y_e{}^*+Y_e C_{Hd,2}^{(d8)} Y_e{}^*-Y_e C_{ld,1}^{(d8)} Y_e{}^*+Y_e C_{ld,2}^{(d8)} Y_e{}^*+Y_d C_{ledq,1}^{(d8)} Y_e{}^*\\&-Y_d C_{ledq,2}^{(d8)} Y_e{}^*\end{split}
\end{align}
The following are the quadratic contributions.
\begin{align}
\begin{split}\dot{C}_{ed,1}^{(d8)}&\supset \frac{16}{3} C_{d,1}^{(d6)} C_{ed,1}^{(d6)}+\frac{8}{3} C_{e,1}^{(d6)} C_{ed,1}^{(d6)}+\frac{28}{3} \left(C_{ed,1}^{(d6)}\right){}^2+\frac{2}{3} C_{Hd,1}^{(d6)} C_{He,1}^{(d6)}+\frac{4}{3} C_{ld,1}^{(d6)} C_{le,1}^{(d6)}+\frac{4}{3} C_{ledq,1}^{(d6)} C_{ledq,1}^{(d6)*}\\&+4 C_{qd,1}^{(d6)} C_{qe,1}^{(d6)}+\frac{4}{3} C_{qd,2}^{(d6)} C_{qe,1}^{(d6)}+2 C_{eu,1}^{(d6)} C_{ud,1}^{(d6)}+\frac{2}{3} C_{eu,1}^{(d6)} C_{ud,2}^{(d6)}\end{split}\\

\begin{split}\dot{C}_{ed,2}^{(d8)}&\supset -\frac{16}{3} C_{d,1}^{(d6)} C_{ed,1}^{(d6)}-\frac{8}{3} C_{e,1}^{(d6)} C_{ed,1}^{(d6)}+\frac{4}{3} \left(C_{ed,1}^{(d6)}\right){}^2-\frac{2}{3} C_{Hd,1}^{(d6)} C_{He,1}^{(d6)}-\frac{4}{3} C_{ld,1}^{(d6)} C_{le,1}^{(d6)}\\&+\frac{4}{3} C_{ledq,1}^{(d6)} C_{ledq,1}^{(d6)*}-4 C_{qd,1}^{(d6)} C_{qe,1}^{(d6)}-\frac{4}{3} C_{qd,2}^{(d6)} C_{qe,1}^{(d6)}+\frac{4}{3} C_{u^2de,1}^{(d6)} C_{u^2de,1}^{(d6)*}-2 C_{eu,1}^{(d6)} C_{ud,1}^{(d6)}-\frac{2}{3} C_{eu,1}^{(d6)} C_{ud,2}^{(d6)}\end{split}
\end{align}

\subsection{$LQ u_{\mathbb{C}} e_{\mathbb{C}} $}
We list below the RGEs for the dimension-8 type $LQ u_{\mathbb{C}} e_{\mathbb{C}} $. The corresponding operators and Wilson coefficients are defined in Table \ref{tab:LQue}.
\begin{table}[htbp]
\begin{align*}
\begin{array}{|c|c|c|}
\hline\hline
\text{abbreviation} & \text{Wilson coefficient} & \text{operator}\\
\hline
C_{lequ,1}^{(d6)} & C_{e_{\mathbb{C}}LQu_{\mathbb{C}} ,1}^{\; prst} & \epsilon^{ij}(e_{\mathbb{C}}{}_p L_{ri})(Q_{saj} u_{\mathbb{C}}{}_t^a) \\
C_{lequ,2}^{(d6)} & C_{e_{\mathbb{C}}LQu_{\mathbb{C}} ,2}^{\; prst} & \epsilon^{ij}(e_{\mathbb{C}}{}_p Q_{saj})(L_{ri} u_{\mathbb{C}}{}_t^a) \\
\hline
C_{lequ,1}^{(d8)} & C_{e_{\mathbb{C}}LQu_{\mathbb{C}} D^2 ,1}^{\; prst} & \epsilon^{ij}(e_{\mathbb{C}}{}_p L_{ri})(D^{\mu}Q_{saj} D_{\mu}u_{\mathbb{C}}{}_t^a) \\
C_{lequ,2}^{(d8)} & C_{e_{\mathbb{C}}LQu_{\mathbb{C}} D^2 ,2}^{\; prst} & \epsilon^{ij}(e_{\mathbb{C}}{}_p Q_{saj})(D^{\mu}L_{ri} D_{\mu}u_{\mathbb{C}}{}_t^a) \\
C_{lequ,3}^{(d8)} & C_{e_{\mathbb{C}}LQu_{\mathbb{C}} D^2 ,3}^{\; prst} & i\epsilon^{ij}(e_{\mathbb{C}}{}_p \sigma_{\mu\nu}L_{ri})(D^{\mu}Q_{saj} D^{\nu}u_{\mathbb{C}}{}_t^a) \\
\hline
\end{array}
\end{align*}
\caption{List of the $LQu_{\mathbb{C}}e_{\mathbb{C}}$-type operators and the corresponding Wilson coefficients in dimension-6 and dimension-8. The leftmost column contains the abbreviations for the Wilson coefficients in the case of $N_f=1$.}
\label{tab:LQue}
\end{table}

\begin{align}
\begin{split}\dot{C}_{lequ,1}^{(d8)}&\supset \frac{g_1 Y_u C_{eB,1}^{(d8)}}{3 \sqrt{2}}-\frac{g_2 Y_u C_{eW,1}^{(d8)}}{\sqrt{2}}-2 Y_e Y_u C_{le,1}^{(d8)}-4 Y_d Y_u C_{ledq,1}^{(d8)}-\frac{8}{3} g_1^2 C_{lequ,1}^{(d8)}+3 g_2^2 C_{lequ,1}^{(d8)}\\&-8 g_3^2 C_{lequ,1}^{(d8)}-\frac{17}{18} g_1^2 C_{lequ,2}^{(d8)}+\frac{3}{2} g_2^2 C_{lequ,2}^{(d8)}-\frac{16}{3} g_3^2 C_{lequ,2}^{(d8)}-\frac{5}{2} g_1^2 C_{lequ,3}^{(d8)}-\frac{3}{2} g_2^2 C_{lequ,3}^{(d8)}\\&-6 Y_e Y_u C_{qu,1}^{(d8)}-2 Y_e Y_u C_{qu,3}^{(d8)}-\frac{g_1 Y_e C_{uB,1}^{(d8)}}{3 \sqrt{2}}+\frac{g_2 Y_e C_{uW,1}^{(d8)}}{\sqrt{2}}+\frac{1}{2} Y_d C_{lequ,1}^{(d8)} Y_d{}^*-\frac{16}{3} Y_e C_{q^2ud,2}^{(d8)} Y_d{}^*\\&+4 Y_e C_{q^2ud,3}^{(d8)} Y_d{}^*+\frac{7}{2} Y_e C_{lequ,1}^{(d8)} Y_e{}^*+\frac{2}{3} Y_e C_{lequ,2}^{(d8)} Y_e{}^*+\frac{15}{2} Y_u C_{lequ,1}^{(d8)} Y_u{}^*+2 Y_u C_{lequ,2}^{(d8)} Y_u{}^*\end{split}\\

\begin{split}\dot{C}_{lequ,2}^{(d8)}&\supset -\frac{g_1 Y_u C_{eB,1}^{(d8)}}{\sqrt{2}}-2 Y_e Y_u C_{eu,1}^{(d8)}+\frac{3 g_2 Y_u C_{eW,1}^{(d8)}}{\sqrt{2}}-\frac{1}{2} g_1^2 C_{lequ,1}^{(d8)}-\frac{3}{2} g_2^2 C_{lequ,1}^{(d8)}\\&+\frac{20}{3} g_1^2 C_{lequ,2}^{(d8)}+8 g_3^2 C_{lequ,2}^{(d8)}-\frac{15}{2} g_1^2 C_{lequ,3}^{(d8)}-\frac{9}{2} g_2^2 C_{lequ,3}^{(d8)}-2 Y_e Y_u C_{lq,1}^{(d8)}+2 Y_e Y_u C_{lq,3}^{(d8)}\\&-2 Y_e Y_u C_{lu,1}^{(d8)}-2 Y_e Y_u C_{qe,1}^{(d8)}+\frac{g_1 Y_e C_{uB,1}^{(d8)}}{\sqrt{2}}-\frac{3 g_2 Y_e C_{uW,1}^{(d8)}}{\sqrt{2}}+\frac{1}{2} Y_d C_{lequ,2}^{(d8)} Y_d{}^*+\frac{3}{2} Y_e C_{lequ,2}^{(d8)} Y_e{}^*\\&+\frac{3}{2} Y_u C_{lequ,2}^{(d8)} Y_u{}^*\end{split}\\

\begin{split}\dot{C}_{lequ,3}^{(d8)}&\supset -\frac{g_1 Y_u C_{eB,1}^{(d8)}}{2 \sqrt{2}}+\frac{5 g_1 Y_u C_{eB,2}^{(d8)}}{6 \sqrt{2}}-2 Y_e Y_u C_{eu,1}^{(d8)}+\frac{3 g_2 Y_u C_{eW,1}^{(d8)}}{2 \sqrt{2}}+\frac{3 g_2 Y_u C_{eW,2}^{(d8)}}{2 \sqrt{2}}\\&-4 g_1^2 C_{lequ,1}^{(d8)}-3 g_2^2 C_{lequ,1}^{(d8)}-\frac{16}{9} g_1^2 C_{lequ,2}^{(d8)}-3 g_2^2 C_{lequ,2}^{(d8)}+\frac{8}{3} g_3^2 C_{lequ,2}^{(d8)}-\frac{41}{18} g_1^2 C_{lequ,3}^{(d8)}\\&-\frac{3}{2} g_2^2 C_{lequ,3}^{(d8)}+\frac{8}{3} g_3^2 C_{lequ,3}^{(d8)}-2 Y_e Y_u C_{lq,1}^{(d8)}+2 Y_e Y_u C_{lq,3}^{(d8)}+\frac{g_1 Y_e C_{uB,1}^{(d8)}}{2 \sqrt{2}}+\frac{3 g_1 Y_e C_{uB,2}^{(d8)}}{2 \sqrt{2}}\\&-\frac{3 g_2 Y_e C_{uW,1}^{(d8)}}{2 \sqrt{2}}+\frac{3 g_2 Y_e C_{uW,2}^{(d8)}}{2 \sqrt{2}}+\frac{1}{2} Y_d C_{lequ,3}^{(d8)} Y_d{}^*+\frac{3}{2} Y_e C_{lequ,3}^{(d8)} Y_e{}^*+\frac{3}{2} Y_u C_{lequ,3}^{(d8)} Y_u{}^*\end{split}
\end{align}
The following are the quadratic contributions.
\begin{align}
\begin{split}\dot{C}_{lequ,1}^{(d8)}&\supset 4 C_{le,1}^{(d6)} C_{lequ,1}^{(d6)}-2 C_{le,1}^{(d6)} C_{lequ,2}^{(d6)}+12 C_{lequ,1}^{(d6)} C_{qu,1}^{(d6)}-6 C_{lequ,2}^{(d6)} C_{qu,1}^{(d6)}+4 C_{lequ,1}^{(d6)} C_{qu,2}^{(d6)}-2 C_{lequ,2}^{(d6)} C_{qu,2}^{(d6)}\\&+10 C_{ledq,1}^{(d6)} C_{q^2ud,1}^{(d6)}-14 C_{ledq,1}^{(d6)} C_{q^2ud,2}^{(d6)}\end{split}\\

\begin{split}\dot{C}_{lequ,2}^{(d8)}&\supset -2 C_{eu,1}^{(d6)} C_{lequ,1}^{(d6)}-2 C_{eu,1}^{(d6)} C_{lequ,2}^{(d6)}-2 C_{lequ,1}^{(d6)} C_{lq,1}^{(d6)}-2 C_{lequ,2}^{(d6)} C_{lq,1}^{(d6)}+2 C_{lequ,1}^{(d6)} C_{lq,2}^{(d6)}\\&+2 C_{lequ,2}^{(d6)} C_{lq,2}^{(d6)}-2 C_{lequ,1}^{(d6)} C_{lu,1}^{(d6)}+4 C_{lequ,2}^{(d6)} C_{lu,1}^{(d6)}-2 C_{lequ,1}^{(d6)} C_{qe,1}^{(d6)}+4 C_{lequ,2}^{(d6)} C_{qe,1}^{(d6)}+24 C_{q^2ue,1}^{(d6)} C_{q^3l,1}^{(d6)}\\&+12 C_{qudl,1}^{(d6)*} C_{u^2de,1}^{(d6)}\end{split}\\

\begin{split}\dot{C}_{lequ,3}^{(d8)}&\supset -2 C_{eu,1}^{(d6)} C_{lequ,1}^{(d6)}-2 C_{eu,1}^{(d6)} C_{lequ,2}^{(d6)}-2 C_{lequ,1}^{(d6)} C_{lq,1}^{(d6)}-2 C_{lequ,2}^{(d6)} C_{lq,1}^{(d6)}+2 C_{lequ,1}^{(d6)} C_{lq,2}^{(d6)}\\&+2 C_{lequ,2}^{(d6)} C_{lq,2}^{(d6)}+24 C_{q^2ue,1}^{(d6)} C_{q^3l,1}^{(d6)}+12 C_{qudl,1}^{(d6)*} C_{u^2de,1}^{(d6)}\end{split}
\end{align}

\subsection{$Q^{\dagger} L d_{\mathbb{C}}^{\dagger} e_{\mathbb{C}} $}
We list below the RGEs for the dimension-8 type $Q^{\dagger} L d_{\mathbb{C}}^{\dagger} e_{\mathbb{C}} $. The corresponding operators and Wilson coefficients are defined in Table \ref{tab:LQde}.
\begin{table}[htbp]
\begin{align*}
\begin{array}{|c|c|c|}
\hline\hline
\text{abbreviation} & \text{Wilson coefficient} & \text{operator}\\
\hline
C_{ledq,1}^{(d6)} & C_{e_{\mathbb{C}}Ld_{\mathbb{C}}^{\dagger}Q^{\dagger} ,1}^{\; prst} & \left(d_{\mathbb{C}}^{\dagger}{}_s{}_{a}Q^{\dagger}_t{}^{ai}\right)\left(e_{\mathbb{C} p}L_r{}_{i}\right) \\
\hline
C_{ledq,1}^{(d8)} & C_{e_{\mathbb{C}}Ld_{\mathbb{C}}^{\dagger}Q^{\dagger} D^2 ,1}^{\; prst} & \left(e_{\mathbb{C} p}L_r{}_{i}\right)\left(\left(D_{\mu}d_{\mathbb{C}}^{\dagger}{}_s{}_{a}\right)\left(D^{\mu}Q^{\dagger}_t{}^{ai}\right)\right) \\
C_{ledq,2}^{(d8)} & C_{e_{\mathbb{C}}Ld_{\mathbb{C}}^{\dagger}Q^{\dagger} D^2 ,2}^{\; prst} & i\left(e_{\mathbb{C} p}\sigma_{\mu \nu}L_r{}_{i}\right)\left(\left(D^{\mu}d_{\mathbb{C}}^{\dagger}{}_s{}_{a}\right)\left(D^{\nu}Q^{\dagger}_t{}^{ai}\right)\right) \\
\hline
\end{array}
\end{align*}
\caption{List of the $Q^{\dagger}d_{\mathbb{C}}^{\dagger}Le_{\mathbb{C}}$-type operators and the corresponding Wilson coefficients in dimension-6 and dimension-8. The leftmost column contains the abbreviations for the Wilson coefficients in the case of $N_f=1$.}
\label{tab:LQde}
\end{table}

\begin{align}
\begin{split}\dot{C}_{ledq,1}^{(d8)}&\supset -\frac{11}{6} g_1^2 C_{ledq,1}^{(d8)}+\frac{5}{2} g_2^2 C_{ledq,1}^{(d8)}-8 g_3^2 C_{ledq,1}^{(d8)}+\frac{5}{6} g_1^2 C_{ledq,2}^{(d8)}+\frac{5}{2} g_2^2 C_{ledq,2}^{(d8)}\\&+\frac{16}{3} Y_e Y_u C_{q^2ud,2}^{(d8)*}-4 Y_e Y_u C_{q^2ud,3}^{(d8)*}+\frac{1}{6} Y_e C_{ed,1}^{(d8)} Y_d{}^*+\frac{5}{6} Y_e C_{ed,2}^{(d8)} Y_d{}^*+\frac{1}{6} Y_e C_{ld,1}^{(d8)} Y_d{}^*+\frac{5}{6} Y_e C_{ld,2}^{(d8)} Y_d{}^*\\&+2 Y_e C_{le,1}^{(d8)} Y_d{}^*+\frac{15}{2} Y_d C_{ledq,1}^{(d8)} Y_d{}^*+\frac{1}{6} Y_e C_{lq,1}^{(d8)} Y_d{}^*+\frac{5}{6} Y_e C_{lq,2}^{(d8)} Y_d{}^*+\frac{1}{3} Y_e C_{lq,3}^{(d8)} Y_d{}^*+\frac{5}{3} Y_e C_{lq,4}^{(d8)} Y_d{}^*\\&+2 Y_e C_{qd,1}^{(d8)} Y_d{}^*+6 Y_e C_{qd,3}^{(d8)} Y_d{}^*+\frac{1}{6} Y_e C_{qe,1}^{(d8)} Y_d{}^*+\frac{5}{6} Y_e C_{qe,2}^{(d8)} Y_d{}^*+\frac{7}{2} Y_e C_{ledq,1}^{(d8)} Y_e{}^*+\frac{1}{2} Y_u C_{ledq,1}^{(d8)} Y_u{}^*\\&-4 C_{lequ,1}^{(d8)} Y_d{}^* Y_u{}^*-\frac{4}{3} C_{lequ,2}^{(d8)} Y_d{}^* Y_u{}^*\end{split}\\

\begin{split}\dot{C}_{ledq,2}^{(d8)}&\supset -\frac{3 g_1 Y_e C_{dB,3}^{(d8)*}}{\sqrt{2}}-\frac{3 g_2 Y_e C_{dW,3}^{(d8)*}}{\sqrt{2}}+\frac{1}{2} g_1^2 C_{ledq,1}^{(d8)}+\frac{3}{2} g_2^2 C_{ledq,1}^{(d8)}+\frac{25}{18} g_1^2 C_{ledq,2}^{(d8)}\\&+\frac{3}{2} g_2^2 C_{ledq,2}^{(d8)}+\frac{8}{3} g_3^2 C_{ledq,2}^{(d8)}+\frac{g_1 C_{eB,3}^{(d8)} Y_d{}^*}{3 \sqrt{2}}-\frac{1}{2} Y_e C_{ed,1}^{(d8)} Y_d{}^*-\frac{1}{2} Y_e C_{ed,2}^{(d8)} Y_d{}^*+\frac{3 g_2 C_{eW,3}^{(d8)} Y_d{}^*}{\sqrt{2}}\\&+\frac{1}{2} Y_e C_{ld,1}^{(d8)} Y_d{}^*+\frac{1}{2} Y_e C_{ld,2}^{(d8)} Y_d{}^*+\frac{3}{2} Y_d C_{ledq,2}^{(d8)} Y_d{}^*-\frac{1}{2} Y_e C_{lq,1}^{(d8)} Y_d{}^*-\frac{1}{2} Y_e C_{lq,2}^{(d8)} Y_d{}^*-Y_e C_{lq,3}^{(d8)} Y_d{}^*\\&-Y_e C_{lq,4}^{(d8)} Y_d{}^*+\frac{1}{2} Y_e C_{qe,1}^{(d8)} Y_d{}^*+\frac{1}{2} Y_e C_{qe,2}^{(d8)} Y_d{}^*+\frac{3}{2} Y_e C_{ledq,2}^{(d8)} Y_e{}^*+\frac{1}{2} Y_u C_{ledq,2}^{(d8)} Y_u{}^*\end{split}
\end{align}
The following are the quadratic contributions.
\begin{align}
\begin{split}\dot{C}_{ledq,1}^{(d8)}&\supset \frac{2}{3} C_{ed,1}^{(d6)} C_{ledq,1}^{(d6)}+\frac{2}{3} C_{ld,1}^{(d6)} C_{ledq,1}^{(d6)}+4 C_{le,1}^{(d6)} C_{ledq,1}^{(d6)}+\frac{2}{3} C_{ledq,1}^{(d6)} C_{lq,1}^{(d6)}+\frac{4}{3} C_{ledq,1}^{(d6)} C_{lq,2}^{(d6)}+4 C_{ledq,1}^{(d6)} C_{qd,1}^{(d6)}\\&+12 C_{ledq,1}^{(d6)} C_{qd,2}^{(d6)}+\frac{2}{3} C_{ledq,1}^{(d6)} C_{qe,1}^{(d6)}+10 C_{lequ,1}^{(d6)} C_{q^2ud,1}^{(d6)*}-5 C_{lequ,2}^{(d6)} C_{q^2ud,1}^{(d6)*}-14 C_{lequ,1}^{(d6)} C_{q^2ud,2}^{(d6)*}+7 C_{lequ,2}^{(d6)} C_{q^2ud,2}^{(d6)*}\\&-\frac{8}{3} C_{qudl,1}^{(d6)*} C_{q^2ue,1}^{(d6)}\end{split}\\

\begin{split}\dot{C}_{ledq,2}^{(d8)}&\supset \frac{8}{3} C_{dB,1}^{(d6)*} C_{eB,1}^{(d6)}+8 C_{dW,1}^{(d6)*} C_{eW,1}^{(d6)}-\frac{2}{3} C_{ed,1}^{(d6)} C_{ledq,1}^{(d6)}+\frac{2}{3} C_{ld,1}^{(d6)} C_{ledq,1}^{(d6)}-\frac{2}{3} C_{ledq,1}^{(d6)} C_{lq,1}^{(d6)}-\frac{4}{3} C_{ledq,1}^{(d6)} C_{lq,2}^{(d6)}\\&+\frac{2}{3} C_{ledq,1}^{(d6)} C_{qe,1}^{(d6)}-C_{lequ,2}^{(d6)} C_{q^2ud,1}^{(d6)*}+\frac{1}{3} C_{lequ,2}^{(d6)} C_{q^2ud,2}^{(d6)*}-\frac{8}{3} C_{qudl,1}^{(d6)*} C_{q^2ue,1}^{(d6)}\end{split}
\end{align}

\subsection{$L^2L^{\dagger 2}$}
We list below the RGEs for the dimension-8 type $L^2L^{\dagger 2}$. The corresponding operators and Wilson coefficients are defined in Table \ref{tab:L2L2}.
\begin{table}[htbp]
\begin{align*}
\begin{array}{|c|c|c|}
\hline\hline
\text{abbreviation} & \text{Wilson coefficient} & \text{operator}\\
\hline
C_{l,1}^{(d6)} & C_{L_{[2]}^2L^{\dagger}{}_{[2]}^2 ,1}^{\; prst} & (L_{pi}L_{rj})(L^{\dagger}{}_s^i L^{\dagger}{}_t^j) \\
 & C_{L_{[1,1]}^2L^{\dagger}{}_{[1,1]}^2 ,1}^{\; prst} & (L_{pi}L_{rj})(L^{\dagger}{}_s^i L^{\dagger}{}_t^j) \\
\hline
C_{l,1}^{(d8)} & C_{L_{[2]}^2L^{\dagger}{}_{[2]}^2 D^2 ,1}^{\; prst} & (L_{pi}L_{rj})(D^{\mu}L^{\dagger}{}_s^i D_{\mu}L^{\dagger}{}_t^j) \\
C_{l,2}^{(d8)} & C_{L_{[2]}^2L^{\dagger}{}_{[2]}^2 D^2 ,2}^{\; prst} & i(L_{pi}\sigma_{\mu\nu}L_{rj})(D^{\mu}L^{\dagger}{}_s^i D^{\nu}L^{\dagger}{}_t^j) \\
 & C_{L_{[1,1]}^2L^{\dagger}{}_{[1,1]}^2 D^2 ,1}^{\; prst} & (L_{pi}L_{rj})(D^{\mu}L^{\dagger}{}_s^i D_{\mu}L^{\dagger}{}_t^j) \\
 & C_{L_{[1,1]}^2L^{\dagger}{}_{[1,1]}^2 D^2 ,2}^{\; prst} & i(L_{pi}\sigma_{\mu\nu}L_{rj})(D^{\mu}L^{\dagger}{}_s^i D^{\nu}L^{\dagger}{}_t^j) \\
\hline
\end{array}
\end{align*}
\caption{List of the $L^2L^{\dagger 2}$-type operators and the corresponding Wilson coefficients in dimension-6 and dimension-8. The leftmost column contains the abbreviations for the Wilson coefficients in the case of $N_f=1$.}
\label{tab:L2L2}
\end{table}

\begin{align}
\begin{split}\dot{C}_{l,1}^{(d8)}&\supset -\frac{5}{12} g_1^2 C_{Bl,1}^{(d8)}+\frac{5}{12} g_1 g_2 C_{BWl,1}^{(d8)}+\frac{5}{12} g_1 g_2 C_{BWl,1}^{(d8)*}-\frac{1}{24} g_1^2 C_{Hl,1}^{(d8)}+\frac{1}{24} g_2^2 C_{Hl,1}^{(d8)}\\&-\frac{1}{24} g_1^2 C_{Hl,2}^{(d8)}+\frac{1}{24} g_2^2 C_{Hl,2}^{(d8)}-\frac{1}{12} g_1^2 C_{Hl,3}^{(d8)}-\frac{1}{12} g_1^2 C_{Hl,4}^{(d8)}+\frac{85}{12} g_1^2 C_{l,1}^{(d8)}+\frac{119}{12} g_2^2 C_{l,1}^{(d8)}\\&-\frac{5}{12} g_1^2 C_{l,2}^{(d8)}+\frac{15}{4} g_2^2 C_{l,2}^{(d8)}-\frac{1}{4} g_1^2 C_{ld,1}^{(d8)}+\frac{5}{12} g_1^2 C_{ld,2}^{(d8)}-\frac{1}{4} g_1^2 C_{le,1}^{(d8)}+\frac{5}{12} g_1^2 C_{le,2}^{(d8)}\\&-\frac{1}{4} g_1^2 C_{lq,1}^{(d8)}+\frac{5}{12} g_1^2 C_{lq,2}^{(d8)}-\frac{1}{8} g_1^2 C_{lq,3}^{(d8)}+\frac{3}{8} g_2^2 C_{lq,3}^{(d8)}+\frac{5}{24} g_1^2 C_{lq,4}^{(d8)}-\frac{5}{8} g_2^2 C_{lq,4}^{(d8)}\\&+\frac{1}{2} g_1^2 C_{lu,1}^{(d8)}-\frac{5}{6} g_1^2 C_{lu,2}^{(d8)}+\frac{5}{12} i g_2^2 C_{Wl,1}^{(d8)}-\frac{5}{4} g_2^2 C_{Wl,2}^{(d8)}-\frac{2}{3} Y_e C_{Hl,1}^{(d8)} Y_e{}^*+\frac{1}{6} Y_e C_{Hl,2}^{(d8)} Y_e{}^*\\&-\frac{2}{3} Y_e C_{Hl,3}^{(d8)} Y_e{}^*+\frac{1}{6} Y_e C_{Hl,4}^{(d8)} Y_e{}^*+2 Y_e C_{l,1}^{(d8)} Y_e{}^*+\frac{1}{6} Y_e C_{le,1}^{(d8)} Y_e{}^*-\frac{5}{6} Y_e C_{le,2}^{(d8)} Y_e{}^*\end{split}\\

\begin{split}\dot{C}_{l,2}^{(d8)}&\supset -\frac{1}{4} g_1^2 C_{Bl,1}^{(d8)}-\frac{3}{4} g_1 g_2 C_{BWl,1}^{(d8)}-\frac{3}{4} g_1 g_2 C_{BWl,1}^{(d8)*}+\frac{1}{24} g_1^2 C_{Hl,1}^{(d8)}+\frac{1}{8} g_2^2 C_{Hl,1}^{(d8)}\\&+\frac{1}{24} g_1^2 C_{Hl,2}^{(d8)}+\frac{1}{8} g_2^2 C_{Hl,2}^{(d8)}+\frac{1}{12} g_1^2 C_{Hl,3}^{(d8)}+\frac{1}{12} g_1^2 C_{Hl,4}^{(d8)}-\frac{3}{4} g_1^2 C_{l,1}^{(d8)}+\frac{27}{4} g_2^2 C_{l,1}^{(d8)}\\&+\frac{17}{12} g_1^2 C_{l,2}^{(d8)}+\frac{17}{4} g_2^2 C_{l,2}^{(d8)}+\frac{1}{4} g_1^2 C_{ld,1}^{(d8)}-\frac{5}{12} g_1^2 C_{ld,2}^{(d8)}+\frac{1}{4} g_1^2 C_{le,1}^{(d8)}-\frac{5}{12} g_1^2 C_{le,2}^{(d8)}\\&+\frac{1}{4} g_1^2 C_{lq,1}^{(d8)}-\frac{5}{12} g_1^2 C_{lq,2}^{(d8)}+\frac{1}{8} g_1^2 C_{lq,3}^{(d8)}+\frac{9}{8} g_2^2 C_{lq,3}^{(d8)}-\frac{5}{24} g_1^2 C_{lq,4}^{(d8)}-\frac{15}{8} g_2^2 C_{lq,4}^{(d8)}\\&-\frac{1}{2} g_1^2 C_{lu,1}^{(d8)}+\frac{5}{6} g_1^2 C_{lu,2}^{(d8)}-\frac{3}{4} i g_2^2 C_{Wl,1}^{(d8)}-\frac{3}{4} g_2^2 C_{Wl,2}^{(d8)}-\frac{1}{2} Y_e C_{Hl,2}^{(d8)} Y_e{}^*+\frac{1}{2} Y_e C_{Hl,4}^{(d8)} Y_e{}^*\\&+2 Y_e C_{l,2}^{(d8)} Y_e{}^*-\frac{1}{2} Y_e C_{le,1}^{(d8)} Y_e{}^*+\frac{1}{2} Y_e C_{le,2}^{(d8)} Y_e{}^*\end{split}
\end{align}
The following are the quadratic contributions.
\begin{align}
\begin{split}\dot{C}_{l,1}^{(d8)}&\supset \frac{1}{3} \left(C_{Hl,1}^{(d6)}\right){}^2+4 C_{Hl,1}^{(d5)*} C_{Hl,1}^{(d7)}+4 C_{Hl,1}^{(d5)} C_{Hl,1}^{(d7)*}+\frac{2}{3} C_{Hl,1}^{(d6)} C_{Hl,2}^{(d6)}+\frac{2}{3} \left(C_{Hl,2}^{(d6)}\right){}^2+\frac{88}{3} \left(C_{l,1}^{(d6)}\right){}^2\\&+2 \left(C_{ld,1}^{(d6)}\right){}^2+\frac{2}{3} \left(C_{le,1}^{(d6)}\right){}^2+4 \left(C_{lq,1}^{(d6)}\right){}^2+4 C_{lq,1}^{(d6)} C_{lq,2}^{(d6)}+2 \left(C_{lq,2}^{(d6)}\right){}^2+2 \left(C_{lu,1}^{(d6)}\right){}^2\end{split}\\

\begin{split}\dot{C}_{l,2}^{(d8)}&\supset \frac{1}{3} \left(C_{Hl,1}^{(d6)}\right){}^2-\frac{2}{3} C_{Hl,1}^{(d6)} C_{Hl,2}^{(d6)}-\frac{2}{3} \left(C_{Hl,2}^{(d6)}\right){}^2-8 \left(C_{l,1}^{(d6)}\right){}^2-2 \left(C_{ld,1}^{(d6)}\right){}^2-\frac{2}{3} \left(C_{le,1}^{(d6)}\right){}^2\\&-4 \left(C_{lq,1}^{(d6)}\right){}^2-4 C_{lq,1}^{(d6)} C_{lq,2}^{(d6)}+2 \left(C_{lq,2}^{(d6)}\right){}^2-2 \left(C_{lu,1}^{(d6)}\right){}^2\end{split}
\end{align}

\subsection{$LL^{\dagger}e_{\mathbb{C}}e_{\mathbb{C}}^{\dagger}$}
We list below the RGEs for the dimension-8 type $LL^{\dagger}e_{\mathbb{C}}e_{\mathbb{C}}^{\dagger}$. The corresponding operators and Wilson coefficients are defined in Table \ref{tab:LLee}.
\begin{table}[htbp]
\begin{align*}
\begin{array}{|c|c|c|}
\hline\hline
\text{abbreviation} & \text{Wilson coefficient} & \text{operator}\\
\hline
C_{le,1}^{(d6)} & C_{e_{\mathbb{C}}Le_{\mathbb{C}}^{\dagger}L^{\dagger} ,1}^{\; prst} & (e_{\mathbb{C}}{}_{p}L_{ri})(e_{\mathbb{C}}^{\dagger}{}_s L^{\dagger}{}_t^i) \\
\hline
C_{le,1}^{(d8)} & C_{e_{\mathbb{C}}Le_{\mathbb{C}}^{\dagger}L^{\dagger} D^2 ,1}^{\; prst} & (e_{\mathbb{C}}{}_{p}L_{ri})(D^{\mu}e_{\mathbb{C}}^{\dagger}{}_s D_{\mu}L^{\dagger}{}_t^i) \\
C_{le,2}^{(d8)} & C_{e_{\mathbb{C}}Le_{\mathbb{C}}^{\dagger}L^{\dagger} D^2 ,2}^{\; prst} & i(e_{\mathbb{C}}{}_{p}\sigma_{\mu\nu}L_{ri})(D^{\mu}e_{\mathbb{C}}^{\dagger}{}_s D^{\nu}L^{\dagger}{}_t^i) \\
\hline
\end{array}
\end{align*}
\caption{List of the $LL^{\dagger }e_{\mathbb{C}}e_{\mathbb{C}}^{\dagger}$-type operators and the corresponding Wilson coefficients in dimension-6 and dimension-8. The leftmost column contains the abbreviations for the Wilson coefficients in the case of $N_f=1$.}
\label{tab:LLee}
\end{table}

\begin{align}
\begin{split}\dot{C}_{le,1}^{(d8)}&\supset -\frac{5}{12} g_1^2 C_{Be,1}^{(d8)}-\frac{5}{3} g_1^2 C_{Bl,1}^{(d8)}-g_1^2 C_{e,1}^{(d8)}-\frac{1}{4} g_1^2 C_{ed,1}^{(d8)}+\frac{5}{12} g_1^2 C_{ed,2}^{(d8)}\\&+\frac{1}{2} g_1^2 C_{eu,1}^{(d8)}-\frac{5}{6} g_1^2 C_{eu,2}^{(d8)}-\frac{1}{12} g_1^2 C_{He,1}^{(d8)}-\frac{1}{12} g_1^2 C_{He,2}^{(d8)}+\frac{1}{12} g_1^2 C_{Hl,1}^{(d8)}+\frac{1}{12} g_1^2 C_{Hl,2}^{(d8)}\\&+\frac{1}{6} g_1^2 C_{Hl,3}^{(d8)}+\frac{1}{6} g_1^2 C_{Hl,4}^{(d8)}-\frac{3}{2} g_1^2 C_{l,1}^{(d8)}+\frac{5}{6} g_1^2 C_{l,2}^{(d8)}+\frac{1}{2} g_1^2 C_{ld,1}^{(d8)}-\frac{5}{6} g_1^2 C_{ld,2}^{(d8)}\\&-\frac{53}{12} g_1^2 C_{le,1}^{(d8)}+\frac{5}{2} g_2^2 C_{le,1}^{(d8)}+\frac{25}{4} g_1^2 C_{le,2}^{(d8)}+\frac{5}{2} g_2^2 C_{le,2}^{(d8)}-6 Y_e Y_u C_{lequ,1}^{(d8)*}-2 Y_e Y_u C_{lequ,2}^{(d8)*}\\&+\frac{1}{2} g_1^2 C_{lq,1}^{(d8)}-\frac{5}{6} g_1^2 C_{lq,2}^{(d8)}+\frac{1}{4} g_1^2 C_{lq,3}^{(d8)}-\frac{5}{12} g_1^2 C_{lq,4}^{(d8)}-g_1^2 C_{lu,1}^{(d8)}+\frac{5}{3} g_1^2 C_{lu,2}^{(d8)}\\&-\frac{1}{4} g_1^2 C_{qe,1}^{(d8)}+\frac{5}{12} g_1^2 C_{qe,2}^{(d8)}-\frac{5}{4} g_2^2 C_{We,1}^{(d8)}+6 Y_e C_{ledq,1}^{(d8)*} Y_d{}^*+\frac{2}{3} Y_e C_{e,1}^{(d8)} Y_e{}^*-\frac{2}{3} Y_e C_{He,1}^{(d8)} Y_e{}^*\\&+\frac{1}{6} Y_e C_{He,2}^{(d8)} Y_e{}^*-\frac{2}{3} Y_e C_{Hl,1}^{(d8)} Y_e{}^*+\frac{1}{6} Y_e C_{Hl,2}^{(d8)} Y_e{}^*-\frac{4}{3} Y_e C_{Hl,3}^{(d8)} Y_e{}^*+\frac{1}{3} Y_e C_{Hl,4}^{(d8)} Y_e{}^*+Y_e C_{l,1}^{(d8)} Y_e{}^*\\&-\frac{5}{3} Y_e C_{l,2}^{(d8)} Y_e{}^*+\frac{22}{3} Y_e C_{le,1}^{(d8)} Y_e{}^*+\frac{5}{3} Y_e C_{le,2}^{(d8)} Y_e{}^*+6 Y_d C_{ledq,1}^{(d8)} Y_e{}^*-6 C_{lequ,1}^{(d8)} Y_e{}^* Y_u{}^*-2 C_{lequ,2}^{(d8)} Y_e{}^* Y_u{}^*\end{split}\\

\begin{split}\dot{C}_{le,2}^{(d8)}&\supset -\frac{1}{4} g_1^2 C_{Be,1}^{(d8)}-g_1^2 C_{Bl,1}^{(d8)}+g_1^2 C_{e,1}^{(d8)}-\frac{3 g_1 Y_e C_{eB,3}^{(d8)*}}{\sqrt{2}}+\frac{1}{4} g_1^2 C_{ed,1}^{(d8)}\\&-\frac{5}{12} g_1^2 C_{ed,2}^{(d8)}-\frac{1}{2} g_1^2 C_{eu,1}^{(d8)}+\frac{5}{6} g_1^2 C_{eu,2}^{(d8)}-\frac{3 g_2 Y_e C_{eW,3}^{(d8)*}}{\sqrt{2}}+\frac{1}{12} g_1^2 C_{He,1}^{(d8)}+\frac{1}{12} g_1^2 C_{He,2}^{(d8)}\\&-\frac{1}{12} g_1^2 C_{Hl,1}^{(d8)}-\frac{1}{12} g_1^2 C_{Hl,2}^{(d8)}-\frac{1}{6} g_1^2 C_{Hl,3}^{(d8)}-\frac{1}{6} g_1^2 C_{Hl,4}^{(d8)}+\frac{3}{2} g_1^2 C_{l,1}^{(d8)}-\frac{5}{6} g_1^2 C_{l,2}^{(d8)}\\&-\frac{1}{2} g_1^2 C_{ld,1}^{(d8)}+\frac{5}{6} g_1^2 C_{ld,2}^{(d8)}+\frac{15}{4} g_1^2 C_{le,1}^{(d8)}+\frac{3}{2} g_2^2 C_{le,1}^{(d8)}+\frac{15}{4} g_1^2 C_{le,2}^{(d8)}+\frac{3}{2} g_2^2 C_{le,2}^{(d8)}\\&-\frac{1}{2} g_1^2 C_{lq,1}^{(d8)}+\frac{5}{6} g_1^2 C_{lq,2}^{(d8)}-\frac{1}{4} g_1^2 C_{lq,3}^{(d8)}+\frac{5}{12} g_1^2 C_{lq,4}^{(d8)}+g_1^2 C_{lu,1}^{(d8)}-\frac{5}{3} g_1^2 C_{lu,2}^{(d8)}\\&+\frac{1}{4} g_1^2 C_{qe,1}^{(d8)}-\frac{5}{12} g_1^2 C_{qe,2}^{(d8)}-\frac{3}{4} g_2^2 C_{We,1}^{(d8)}-2 Y_e C_{e,1}^{(d8)} Y_e{}^*+\frac{3 g_1 C_{eB,3}^{(d8)} Y_e{}^*}{\sqrt{2}}+\frac{3 g_2 C_{eW,3}^{(d8)} Y_e{}^*}{\sqrt{2}}\\&+\frac{1}{2} Y_e C_{He,2}^{(d8)} Y_e{}^*+\frac{1}{2} Y_e C_{Hl,2}^{(d8)} Y_e{}^*+Y_e C_{Hl,4}^{(d8)} Y_e{}^*-3 Y_e C_{l,1}^{(d8)} Y_e{}^*+Y_e C_{l,2}^{(d8)} Y_e{}^*+Y_e C_{le,1}^{(d8)} Y_e{}^*\\&+4 Y_e C_{le,2}^{(d8)} Y_e{}^*\end{split}
\end{align}
The following are the quadratic contributions.
\begin{align}
\begin{split}\dot{C}_{le,1}^{(d8)}&\supset \frac{1}{3} C_{He,1}^{(d6)} C_{Hl,1}^{(d6)}+\frac{2}{3} C_{He,1}^{(d6)} C_{Hl,2}^{(d6)}+2 C_{ed,1}^{(d6)} C_{ld,1}^{(d6)}+\frac{8}{3} C_{e,1}^{(d6)} C_{le,1}^{(d6)}+4 C_{l,1}^{(d6)} C_{le,1}^{(d6)}+\frac{28}{3} \left(C_{le,1}^{(d6)}\right){}^2\\&+12 C_{ledq,1}^{(d6)} C_{ledq,1}^{(d6)*}+12 C_{lequ,1}^{(d6)} C_{lequ,1}^{(d6)*}-6 C_{lequ,1}^{(d6)*} C_{lequ,2}^{(d6)}-6 C_{lequ,1}^{(d6)} C_{lequ,2}^{(d6)*}+3 C_{lequ,2}^{(d6)} C_{lequ,2}^{(d6)*}+2 C_{eu,1}^{(d6)} C_{lu,1}^{(d6)}\\&+4 C_{lq,1}^{(d6)} C_{qe,1}^{(d6)}+2 C_{lq,2}^{(d6)} C_{qe,1}^{(d6)}\end{split}\\

\begin{split}\dot{C}_{le,2}^{(d8)}&\supset \frac{8}{3} C_{eB,1}^{(d6)} C_{eB,1}^{(d6)*}+8 C_{eW,1}^{(d6)} C_{eW,1}^{(d6)*}-\frac{1}{3} C_{He,1}^{(d6)} C_{Hl,1}^{(d6)}-\frac{2}{3} C_{He,1}^{(d6)} C_{Hl,2}^{(d6)}-2 C_{ed,1}^{(d6)} C_{ld,1}^{(d6)}-\frac{8}{3} C_{e,1}^{(d6)} C_{le,1}^{(d6)}\\&-4 C_{l,1}^{(d6)} C_{le,1}^{(d6)}+\frac{4}{3} \left(C_{le,1}^{(d6)}\right){}^2+C_{lequ,2}^{(d6)} C_{lequ,2}^{(d6)*}-2 C_{eu,1}^{(d6)} C_{lu,1}^{(d6)}-4 C_{lq,1}^{(d6)} C_{qe,1}^{(d6)}-2 C_{lq,2}^{(d6)} C_{qe,1}^{(d6)}\end{split}
\end{align}

\subsection{$e_{\mathbb{C}}^2e_{\mathbb{C}}^{\dagger 2}$}
We list below the RGEs for the dimension-8 type $e_{\mathbb{C}}^2e_{\mathbb{C}}^{\dagger 2}$. The corresponding operators and Wilson coefficients are defined in Table \ref{tab:e2e2}.
\begin{table}[htbp]
\begin{align*}
\begin{array}{|c|c|c|}
\hline\hline
\text{abbreviation} & \text{Wilson coefficient} & \text{operator}\\
\hline
C_{e,1}^{(d6)} & C_ {e_{\mathbb{C}}{}_ {[2]}^2 e_{\mathbb{C}}{}^{\dagger}{}_{[2]}^2}^{\; prst} & \left(e_{\mathbb{C}}{}_pe_{\mathbb{C}}{}_r\right)(e_{\mathbb{C}}^{\dagger}{}_s e_{\mathbb{C}}^{\dagger}{}_t) \\
\hline
C_{e,1}^{(d8)} & C_ {e_{\mathbb{C}}{}_ {[2]}^2 e_{\mathbb{C}}{}^{\dagger}{}_{[2]}^2 D^2}^{\; prst} & \left(e_{\mathbb{C}}{}_pe_{\mathbb{C}}{}_r\right)(D^{\mu}e_{\mathbb{C}}^{\dagger}{}_s D_{\mu}e_{\mathbb{C}}^{\dagger}{}_t) \\
 & C_ {e_{\mathbb{C}}{}_ {[1,1]}^2 e_{\mathbb{C}}{}^{\dagger}{}_{[1,1]}^2 D^2}^{\; prst} & \left(e_{\mathbb{C}}{}_p \sigma_{\mu\nu}e_{\mathbb{C}}{}_r\right)(D^{\mu}e_{\mathbb{C}}^{\dagger}{}_s D^{\nu}e_{\mathbb{C}}^{\dagger}{}_t) \\
\hline
\end{array}
\end{align*}
\caption{List of the $e_{\mathbb{C}}^2e_{\mathbb{C}}^{\dagger 2}$-type operators and the corresponding Wilson coefficients in dimension-6 and dimension-8. The leftmost column contains the abbreviations for the Wilson coefficients in the case of $N_f=1$.}
\label{tab:e2e2}
\end{table}

\begin{align}
\begin{split}\dot{C}_{e,1}^{(d8)}&\supset -\frac{5}{3} g_1^2 C_{Be,1}^{(d8)}+\frac{82}{3} g_1^2 C_{e,1}^{(d8)}+\frac{1}{2} g_1^2 C_{ed,1}^{(d8)}-\frac{5}{6} g_1^2 C_{ed,2}^{(d8)}-g_1^2 C_{eu,1}^{(d8)}\\&+\frac{5}{3} g_1^2 C_{eu,2}^{(d8)}+\frac{1}{6} g_1^2 C_{He,1}^{(d8)}+\frac{1}{6} g_1^2 C_{He,2}^{(d8)}-\frac{1}{2} g_1^2 C_{le,1}^{(d8)}+\frac{5}{6} g_1^2 C_{le,2}^{(d8)}+\frac{1}{2} g_1^2 C_{qe,1}^{(d8)}\\&-\frac{5}{6} g_1^2 C_{qe,2}^{(d8)}+4 Y_e C_{e,1}^{(d8)} Y_e{}^*-\frac{4}{3} Y_e C_{He,1}^{(d8)} Y_e{}^*+\frac{1}{3} Y_e C_{He,2}^{(d8)} Y_e{}^*+\frac{1}{3} Y_e C_{le,1}^{(d8)} Y_e{}^*-\frac{5}{3} Y_e C_{le,2}^{(d8)} Y_e{}^*\end{split}
\end{align}
The following are the quadratic contributions.
\begin{align}
\begin{split}\dot{C}_{e,1}^{(d8)}&\supset \frac{80}{3} \left(C_{e,1}^{(d6)}\right){}^2+2 \left(C_{ed,1}^{(d6)}\right){}^2+2 \left(C_{eu,1}^{(d6)}\right){}^2+\frac{2}{3} \left(C_{He,1}^{(d6)}\right){}^2+\frac{4}{3} \left(C_{le,1}^{(d6)}\right){}^2+4 \left(C_{qe,1}^{(d6)}\right){}^2\end{split}
\end{align}

\section{Conclusion}\label{sec:con}

In this work, we have provided the complete one-loop RGEs for four-fermion operators in the dimension-8 SMEFT, which are systematically derived using modern on-shell amplitude techniques. 
The UV divergences in the one-loop integrals are extracted as the two-particle unitarity cuts, which factorized into product of known tree-level amplitudes integrated over the physical phase space.
The calculations build critically upon the previous achievement of on-shell operator basis constructed via the Young Tensor method. We present a systematic algorithm on the unitarity cut method, in which the tree-level amplitudes run over all the complete set of amplitude basis and the cut results are projected onto the operator basis. This algorithm provides an automatic and efficient way to calculate the RGE. Furthermore, the algorithm implements the flavor structures comprehensively, particularly for operators with non-trivial flavor relations, enabling the factorization to be described through simple flavor tensor contractions.
As an crucial validation, our results precisely reproduce all known dimension-6 RGEs. At dimension-8, we computed both the linear and quadratic contributions to the RGEs with general flavor components. 

The phenomenological implications of our results are substantial. Four-fermion operators, particularly at dimension-8, contribute to a wide range of precision observables in electroweak, flavor, and collider physics. Our complete, flavor-general one-loop RGEs enable a systematic study of their renormalization group evolution, including mixing effects. This is essential for accurately interpreting experimental constraints and correlating signals across energy scales. Furthermore, our formulation specifically facilitates comprehensive studies of flavor-violating effects, which are among the most sensitive probes of new physics. To maximize utility, we provide our results in an accompanying Mathematica file, empowering precision phenomenology studies that properly account for the RG evolution of higher-dimensional operators and aiding in the investigation of the UV origins of BSM signals.

This work opens several clear paths for future research. First, the methodology developed here can be directly applied to calculate the remaining sectors of the dimension-8 SMEFT RGEs, including the two-fermion sector and the purely bosonic sector, where careful treatment of length-changing effects is required. 
Second, our framework is well-suited for extension to other effective field theories of current interest, such as the Higgs Effective Field Theory (HEFT). 
Third, we are developing an improved algorithm that fully utilizes angular momentum selection rules, which promises greater algebraic efficiency and is particularly valuable for automated calculations—a step toward systematic exploration of RGEs at even higher loop orders. 
Finally, challenges at higher loops, including the treatment of unitarity cuts with more than two particles, the ultraviolet divergences in higher-loop master integrals, and possible contributions from evanescent operators in dimensional regularization, remain important topics for future investigation.

\section*{Acknowledgments} 

M.-L.X. is supported by the National Natural Science Foundation of China (Grant No.12405123), Fundamental Research Funds for the Central Universities, Sun Yat-sen University (Grant No.25hytd001), Shenzhen Science and Technology Program (JCYJ20240813150911015). J.-H. Y. is supported by the National Science Foundation of China under Grants No. 12347105, No. 12375099 and No. 12447101, and the National Key Research and Development Program of China Grant No. 2020YFC2201501, No. 2021YFA0718304.  Y.-H.Z. is supported by a KIAS Individual Grant (PG096402) through the School of Physics at the Korea Institute for Advanced Study.

\appendix

\bibliographystyle{JHEP}
\bibliography{ref}

\end{document}